\documentclass[12pt,a4paper,twoside]{report}

\usepackage{amssymb,amsmath}
\usepackage{bbm}
\usepackage{mathrsfs}
\usepackage{eufrak}
\usepackage[dvips]{graphicx}
\usepackage{feynmf}
\usepackage{subfigure}
\usepackage{psfrag}
\usepackage{amsfonts}
\usepackage{here}
\usepackage{german}

\newcommand {\odd}{\mathbb{O}}
\newcommand {\alphodd}{\alpha_{\odd}}

\newcommand {\M}{{\cal{M}}}
\newcommand {\Mgo}{\M^{\gamma+\odd}}
\newcommand {\wg}{\omega}
\newcommand {\pb}{\bar{p}}
\newcommand {\al}{\alpha}
\newcommand {\be}{\beta}
\newcommand {\pom}{\mathbb{P}}
\newcommand{\vare}{\varepsilon}
\newcommand {\Lc}{\mathcal{L}}
\newcommand {\Nc}{\mathcal{N}}
\newcommand{\scripts}{\scriptstyle}

\begin{document}
\selectlanguage{\USenglish}
\pagestyle{empty}
\begin{titlepage}
\begin{center}
\huge
Faculty of Physics and Astronomy\\
\vspace{3mm}
\large
University of Heidelberg\\
\vspace{140mm}
\renewcommand{\baselinestretch}{1.3}
{\bf
Diploma thesis\\ in Physics\\ submitted by \\ Tania Robens \\  born in Antwerpen\\ 2002} 
\cleardoublepage
\LARGE
Odderoninduced Pion-Photoproduction at  $e^{+}e^{-}$ Colliders\\
\vspace{140mm}
\large
This diploma thesis has been carried out by Tania Robens at the \\
Institut f\"ur Theoretische Physik\\
under the supervision of\\
Prof. Otto Nachtmann\\
\end{center}
\cleardoublepage
\end{titlepage}
\renewcommand{\baselinestretch}{1}



\pagestyle{empty}

\selectlanguage{\german}
\sc
\begin{center}
Odderoninduzierte Pion-Photoproduktion an $e^{+}e^{-}$ Beschleunigern
\end{center}
\rm
In diffraktiven hadronischen Streuprozessen tr\"agt die aus der Regge Theorie stammende Pomerontrajektorie signifikant zum Verhalten von totalen und differentiellen Wirkungsquerschnitten bei; dies gilt ebenfalls f\"ur die Beschreibung von Proton-Strukturfunktion in tiefinelastischer Streuung. Bisher ist das Odderon als Partner des Pomerons mit ungerader C-Parit\"at nur in Unterschieden zwischen differentielle Wirkungsquerschnitten f\"ur $pp$ und $p\bar{p}$ Streuung  bei kleinem $t$ beobachtet worden. Wir untersuchen die exklusive Photoproduktion pseudoskalarer Mesonen $\gamma\gamma\,\rightarrow\,\pi^{0}\pi^{0}$ an $e^{+}e^{-}$ Beschleunigern; in dieser Reaktion ist Pomeronaustausch aufgrund von Parit\"atserhaltung verboten. Das Odderon wird durch einen effektiven Propagator beschrieben. Der Odderonbeitrag erzeugt signifikante Modifikationen der differentiellen Wirkungsquerschnitte. Wir untersuchen die Effekte der Variation von Trajektorien- und Kopplungsparametern. Wir geben numerische Ergebnisse f\"ur totale und differentielle Wirkungsquerschnitte f\"ur OPAL, BaBar, TESLA und den Photoncollider bei TESLA unter Einbezug von Detektorcuts an.  

\vspace{20mm}

\selectlanguage{USenglish}
\sc
\begin{center}
Odderoninduced pion-photoproduction at $e^{+}e^{-}$ colliders
\end{center}
\rm 
In diffractive hadronic scattering processes, the Pomeron trajectory, originating from Regge theory, significantly contributes to the behavior of the total and differential cross sections; the same holds for the description of the proton structure function in deep inelastic scattering. So far, the Odderon as the odd-parity partner of the Pomeron has only been observed in differences between differential cross sections  for $pp$ and $p\bar{p}$ scattering at low $t$. We investigate exclusive photoproduction of pseudoscalar mesons $\gamma\gamma\,\rightarrow\,\pi^{0}\pi^{0}$ at $e^{+}e^{-}$ colliders; in this reaction, Pomeron exchange is forbidden by C-parity conservation. The Odderon is described by an effective propagator. The Odderon contribution produces significant modifications of the differential cross sections. We investigate the effects of variation of trajectory and coupling parameters. We provide numerical results for total and differential cross sections for OPAL, BaBar, TESLA and the TESLA photon collider including detector cuts.  

\cleardoublepage


\selectlanguage{\USenglish}
\setcounter{page}{1}
\pagestyle{plain}
\tableofcontents

\newpage

\chapter[{\normalsize Introduction and motivation}]{\Large Introduction and motivation}

In modern particle physics, scattering processes are described within the framework of quantum field theory. Depending on the nature of the considered forces, the rules of Quantum Electrodynamics, Quantum Flavordynamics, and Quantumchromodynamics describe interactions due to the electromagnetic, electroweak, and strong interactions. Especially in the perturbative regime, scattering matrices and therefore cross sections can easily be computed.

Before the development of Quantumchromodynamics as a field theory apt to describe strong interactions, Regge theory provided a good framework for the description of high energy hadronic reactions. It was introduced by Regge in the end of the 1950s in connection with non-relativistic potential scattering \cite{Regge:1959mz},\cite{Regge:1960zc}. Today, it is especially valuable in the description of diffractive hadronic interactions where particles interact via an exchange particle corresponding to a color singlet with $P=C=+1$ and therefore carrying the quantum numbers of the vacuum. Regge theory predicts the behavior of the total and differential cross sections according to
\begin{displaymath}
\sigma_{tot}\;\propto\;s^{\al(0)-1}\,;\frac{d\sigma}{dt}\,\propto\,s^{2(\al(t)_1)}\,,
\end{displaymath}
where $\al(0)$ denotes the intercept of the Regge trajectory. For diffractive processes, $\al(0)\,\gtrsim\,1$, leading to an increase of the total cross sections with growing $s$. Furthermore, considering the Regge limit $t\,\ll\,s$, we expect a rapidity gap between the outgoing particles in diffractive scattering.

Experiments in the last 20 years \cite{Hagiwara:2002pw} showed that total hadronic cross sections rise for growing center of mass energies. This behavior can be explained by the introduction of the Pomeron trajectory. Donnachie and Landshoff \cite{Donnachie:1984hf} provided a fit for total and differential elastic $pp$ and $p\bar{p}$ cross sections leading to $\al_{\pom}\,=\,1.08$. However, this region cannot be treated perturbatively; the behavior of the cross sections is therefore ascribed to the soft Pomeron. Additional evidence for the existence of the Pomeron is the behavior of proton structure functions in deep inelastic scattering. This region can be treated perturbatively; the Pomeron intercept is given by $\al_{\pom}\,\approx\,1.4$ \cite{Donnachie:2001xx} and can be derived directly from perturbative QCD. Here, the Pomeron is represented by the exchange of two gluons. However, the unification or transition of the latter so-called hard Pomeron to the soft Pomeron describing the behavior of total and differential cross sections in hadronic scattering is still an open question.

The Odderon as a Regge trajectory with an intercept close to one but with $P=C=-1$ was first introduced by Lukaszuk and Nicolescu in connection with the rise of total $pp$ cross sections \cite{Lukaszuk:1973nt}. In pQCD it can be described by the exchange of three or more gluons and therefore follows as the natural extension of the Pomeron. However, so far effects of Odderon exchange have only been observed in connection with differences between partial cross sections in $pp$ and $p\bar{p}$ scattering for low $t$ \cite{Donnachie:1984hf}, \cite{Dosch:2002ai}. Predictions for cross sections resulting from a nonperturbative approach to QCD \cite{Rueter:1998gj}, \cite{Berger:2000wt} for the diffractive production of pseudoscalar and vector mesons have not been confirmed by experiment \cite{Olsson:2001nm}.

The proof of the existence or non-existence of the Odderon as the $C=P=-1$ partner of the Pomeron as well as its description in QCD would provide valuable insight into the theory of strong interactions.  Therefore, we are investigating exclusive processes with $\gamma\gamma \,\rightarrow\,PS\, \,PS$ where Pomeron exchange is prohibited by parity conservation; here, the influence of Odderon exchange should clearly be visible. We adapt an effective phenomenological description of the non-perturbative Odderon closely following \cite{Kilian:1998ew} and \cite{Donnachie:1984hf}. $e^{+}e^{-}$ colliders such as TESLA, LEP, or BaBar provide an ideal environment for the above reaction. Similar reactions have already been investigated using a different model for the Odderon \cite{Motyka:1998kb}.

In the next chapter, we will give a short review of Regge theory and the perturbative as well as non-perturbative effective description of Pomeron and Odderon. In the third chapter, we will address the $\gamma\,\gamma^{*}\,\pi^{0}$ and $\gamma\,\odd^{*}\,\pi^{0}$ coupling and sketch its derivation from current algebra. Chapter four gives an overview of the kinematics for $2\,\rightarrow\,2$ particle interactions and the general form of the differential cross sections. We placed the calculation of the matrix element for the process $\gamma\gamma\,\rightarrow\,\pi^{0}\pi^{0}$ by photon and Odderon exchange in chapter five. Chapter six gives an overview on the spectra used for the photons produced at a linear $e^{+}e^{-}$ collider as well as a photon collider; chapter seven gives the numerical results for $\frac{d\sigma}{d|k_{t}|^{2}}$ and $\frac{d\sigma}{dk_{l}}$ for several parameter sets and collider environments. We placed the summary and outlook in the last chapter.

\chapter[{\normalsize {Pomeron and Odderon: Motivation and effective description}}] {\Large Pomeron and Odderon: Motivation and effective description}

\section {A short review of Regge theory}\label{sec:Revreg}

\subsection{General quantities in scattering processes}

We will just provide a short list of general quantities describing scattering processes and refer to he literature (e.g. \cite{Nachtmann:1990ta}, \cite{Peskin:1995ev}) for more details.

\begin{itemize}
\item{The $S$ matrix}

The $S$ -matrix provides the connection between in- and outgoing physical states; it is defined by

\begin{eqnarray}
\langle a_{out}|b_{in}\rangle & = & \langle a\,|S\,|b\rangle\;\equiv\;S_{ab},
\end{eqnarray} 
with

\begin{eqnarray*}
|a_{out}\rangle &=& \lim_{t \rightarrow + \infty }U(t,t_{0})|a(t_{0}) \rangle \, ,\\
|b_{in}\rangle &=& \lim_{t \rightarrow - \infty }U(t,t_{0})|b (t_{0})\rangle\;.
\end{eqnarray*}
 $U(t_{1},t_{2})$ is the time-evolution operator

\begin{equation}
U(t_{2},t_{1})|a(t_{1})\rangle\;=\;|a(t_{2})\rangle\;,
\end{equation}
$|a\rangle\,=\,|p_{a_{1}}p_{a_{2}}...p_{a_{n}}\rangle$ and $b\,=\,|k_{b_{1}}k_{b_{2}}...k_{b_{m}}\rangle$ denote any state vector.

Furthermore, $\M$ is defined by

\begin{eqnarray}
S_{a\,b}&=&\mathbbm{1}_{a\,b}\,+i\,(2\pi)^{4}\,\delta^{(4)}(\sum_{i} p_{a_{i}}\,-\sum_{j}k_{b_{j}})\,\M_{a\,b}\;.\label{eq:SMconn}
\end{eqnarray}

\item{Mandelstam variables}\\
2 particle $\rightarrow$ 2 particle scattering processes can be described with the help of the so called Mandelstam variables $s,t,$ and $u$ which are Lorentz invariant and given by

\begin{eqnarray}
s &=&(p_{1}+p_{2})^{2}\;=\;(k_{1}+k_{2})^{2}\;,\nonumber\\
t &=&(p_{1}-k_{1})^{2}\;=\;(p_{2}-k_{2})^{2}\;,\nonumber\\
u &=&(p_{1}-k_{2})^{2}\;=\;(p_{2}-k_{1})^{2}\label{eq:Mandeldef}
\end{eqnarray}
for the reaction described in figure \ref{fig:2to2}. $p_{i},\,k_{j}$ denote four-vectors of the incoming and outgoing particles.

\begin{fmffile}{fd2part}
\begin{figure}
\centering
\fmfframe(0,0)(0,10){
\begin{fmfgraph*}(90,80)
\fmfbottomn{i}{2}
\fmftopn{o}{2}
\fmf{fermion,label=$p_{1}$}{i1,v1}
\fmf{fermion,label=$p_{2}$}{i2,v1}
\fmf{fermion,label=$k_{1}$}{v1,o1}
\fmf{fermion,label=$k_{2}$}{v1,o2}
\fmfblob{0.18w}{v1}
\end{fmfgraph*}}
\caption{2 $\rightarrow$ 2 particle scattering process}
\label{fig:2to2}
\end{figure}
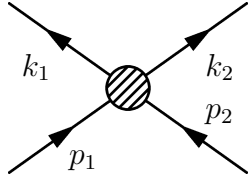
\end{fmffile}
For the Mandelstam-variables, the following relation holds:

\begin{equation}
s+t+u = \sum_{i} m^{2}_{i}\label{eq:Mandelrel}\;\;;
\end{equation}
the sum goes over all particles in the reaction.
Reactions where $p_{1}$ and $p_{2}$ denote the incoming particles are called $s$ channel reactions; here $s\,\geq\,0$ while $t,u\,\leq\,0$. For the description of scattering processes in the Regge-language, we talk about a $t$ channel process if $p_{1}$ and $-k_{1}$ denote the incoming particles, i.e. $t$ takes the role of $s$ and vice versa. Similar considerations hold for reactions in the $u$ channel.
This has to be distinguished from the terminology of an exchange particle being in the $s,\,t\,$ or $u$ channel. For $s\,\geq\,0$, processes in these channels are given by figure \ref{fig:stu}; here, the first diagram corresponds to particle-antiparticle annihilation. 

\begin{fmffile}{fdstu}
\begin{figure}
\centering
\subfigure[$s$ channel]{
\fmfframe(5,5)(25,5){
\begin{fmfgraph*}(70,60)
\fmfbottomn{i}{2}
\fmftopn{o}{2}
\fmf{fermion}{i1,v1}
\fmf{fermion}{v1,i2}
\fmf{fermion}{v2,o1}
\fmf{fermion}{o2,v2}
\fmf{photon}{v1,v2}
\fmflabel{$p_{1}$}{i1}
\fmflabel{$p_{2}$}{i2}
\fmflabel{$k_{1}$}{o1}
\fmflabel{$k_{2}$}{o2}
\end{fmfgraph*}}}
\subfigure[$t$ channel]{
\fmfframe(25,5)(25,5){
\begin{fmfgraph*}(70,60)
\fmfbottomn{i}{2}
\fmftopn{o}{2}
\fmf{fermion}{i1,v1}
\fmf{fermion}{i2,v2}
\fmf{fermion}{v1,o1}
\fmf{fermion}{v2,o2}
\fmf{photon}{v1,v2}
\fmflabel{$p_{1}$}{i1}
\fmflabel{$p_{2}$}{i2}
\fmflabel{$k_{1}$}{o1}
\fmflabel{$k_{2}$}{o2}
\end{fmfgraph*}}}
\subfigure[$u$ channel]{
\fmfframe(25,5)(5,5){
\begin{fmfgraph*}(70,60)
\fmfbottomn{i}{2}
\fmftopn{o}{2}
\fmf{fermion}{i1,v1}
\fmf{fermion}{i2,v2}
\fmf{fermion}{v2,o1}
\fmf{fermion}{v1,o2}
\fmfforce{0.2w,0.5h}{v1}
\fmfforce{0.8w,0.5h}{v2}
\fmf{photon}{v1,v2}
\fmflabel{$p_{1}$}{i1}
\fmflabel{$p_{2}$}{i2}
\fmflabel{$k_{1}$}{o1}
\fmflabel{$k_{2}$}{o2}
\end{fmfgraph*}}}
\caption{exchange particle in the $s,t,u$ channel}
\label{fig:stu}
\end{figure}
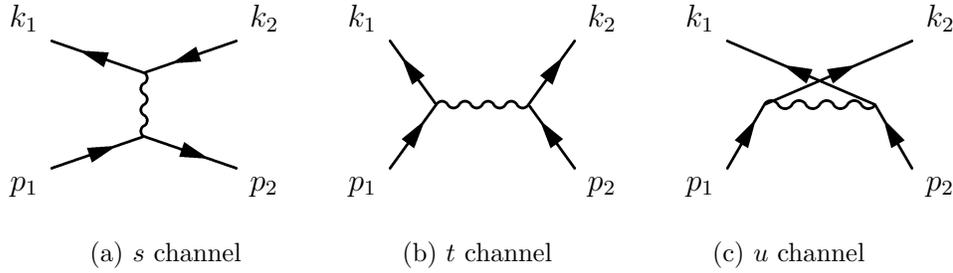
\end{fmffile}

\item{Cross sections for $2\,\rightarrow\,n$ particle reactions}\\
The general cross-section for $2 \to n$ particle processes is given by

\begin{equation}
d\sigma = \frac{1}{2w} \prod_{i=1}^n \frac{d^{3}k_{i}}{(2\pi)^{3} 2 k^{0}_{i}} (2\pi)^{4} \delta^{4}(\sum_{i} k_{i} - p_{1} - p_{2}) \left|\cal M\right|^{2},\label{eq:dsig1}
\end{equation}
(see e.g. \cite{Nachtmann:1990ta}) with
\begin{eqnarray*}
w &= &\sqrt{(s-m_{1}^{2}-m_{2}^{2})^{2}-4m_{1}^{2}m_{2}^{2}}\;,\\
m_{1},m_{2}&:&\;\;\mbox{masses of ingoing particles}\;,\\
\left|\cal M\right|^{2}&:&\;\;\mbox{squared matrix element of the reaction}\;.
\end{eqnarray*}
\end{itemize}

\subsection {General properties of scattering amplitudes} \label{sec:genscat}

In Regge theory, statements about the behavior of scattering amplitudes for $2\,\rightarrow\,2$ particle reactions can be made by assuming  Lorentz-invariance, unitarity, and analyticity of $\M$. Investigating these properties one by one, we obtain:

\begin{itemize}
\item{from Lorentz-invariance:}
\\
The Lorentz-invariant scattering amplitude, here denoted by $\M$,  can be expressed in terms of the Mandelstam-variables:
\begin{eqnarray}
\M&=&\M(s,t,u)\;.\label{eq:Mvar}
\end{eqnarray}
As $s,\,t\,$ and $u\,$ are connected by (\ref{eq:Mandelrel}), $\M$ can be taken as $\M(s,t,m^{2}_{i})$ only. 

\item{from unitarity:}

From the definition of the scattering matrix $S$ and the completeness of the physical states, we obtain the unitarity of $S$ (see e.g. \cite{Nachtmann:1990ta}); combining this with (\ref{eq:SMconn}) leads to

\begin{eqnarray}
2\,Im\,\M_{a\,b}&=&(2\pi)^{4}\,\delta^{(4)}\left(\sum_{i} p_{a_{i}}\,-\sum_{j}k_{b_{j}}\right)\sum_{c}\M_{a\,c}\,\M^{\dag}_{c\,b}\label{eq:cutr}
\end{eqnarray}
known as the Cutkosky rules \cite{Cutkosky:1960sp}.

\begin{figure}
\centering
\psfrag{sum}{$\sum_{c}$}
\psfrag{Mab}{$\M_{ab}$}
\psfrag{Mac}{$\M_{ac}$}
\psfrag{Mcb}{$\M_{cb}^{\dagger}$}
\psfrag{2 Im}{2 Im}
\psfrag{=}{$=$}
\includegraphics[width=0.7\textwidth]{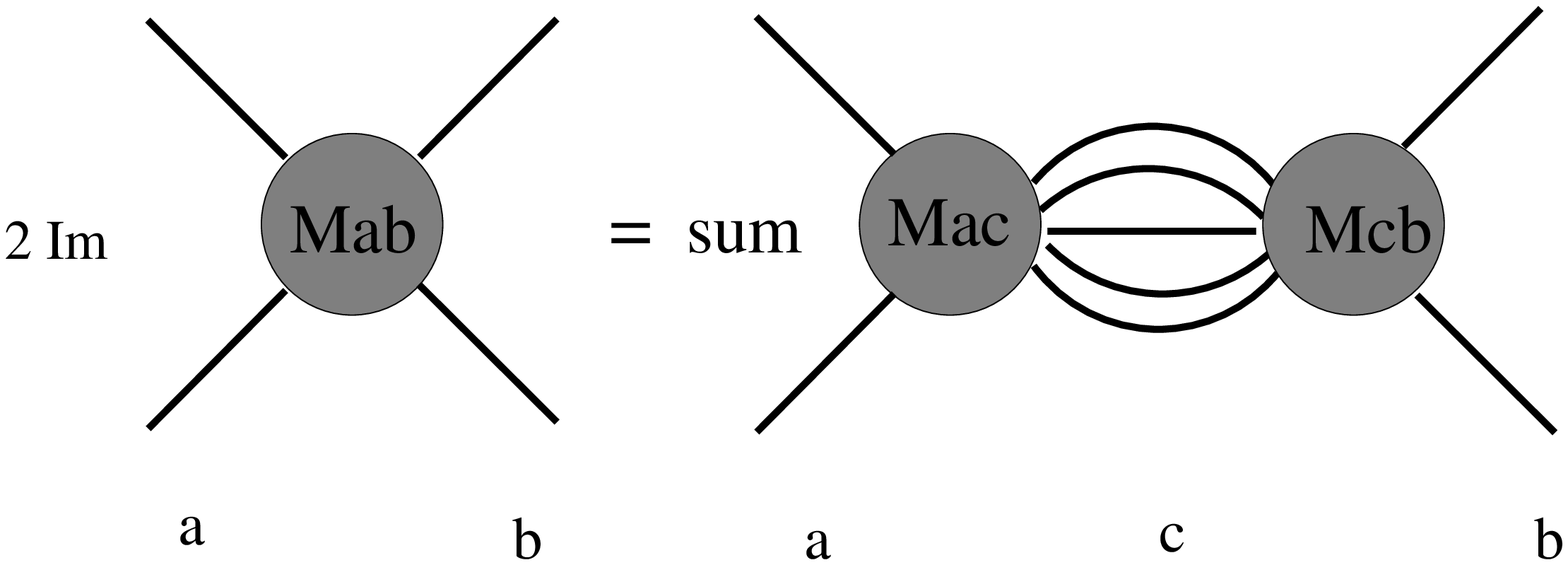}
\caption{Symbolic description of Cutkosky rules}
\label{fig:cutk}
\end{figure}
 For $a=b$ and the case of two incoming particles, this reduces to the optical theorem:

\begin{eqnarray}
Im \M_{aa} & = & w(s,m^{2}_{1},m^{2}_{2})\,\sigma_{tot}(a\rightarrow \mbox{anything})\;. \label{eq:optthem}
\end{eqnarray}
Here, $|a \rangle\,=\,|p(p_{1}),\,p(p_{2})\rangle$, and $p$ denotes the same particle type. $w$ is given according to (\ref{eq:dsig1}).

\item{from analyticity}

From the Cutkosky rules, we can draw conclusions about the value of $Im \M\,$ in dependence of $s\,$ and $t\,$; from (\ref{eq:cutr}), we see that $Im\, \M\, \neq\,0\,$ only if there are states associated with possible exchange particles contributing. In an s-channel process, $Im\,\M\,\neq\,0\,$ for $s=M^{2}\,$, in case of a single-particle and $s\,\geq\,4\,m^{2}$ in the case of multi -particle exchanges  with $M$ and $m$ being the respective masses of the contributing particles. A change from $s$- to $u$-channel- reactions leads to similar cuts on the negative $s$- axis (see figure \ref{fig:sing}).

\begin{figure}
\centering
\psfrag{Re s}{Re s}
\psfrag{Im s}{Im s}
\psfrag{s-plane}{s plane}
\psfrag{C}{C}
\includegraphics[width=0.5\textwidth]{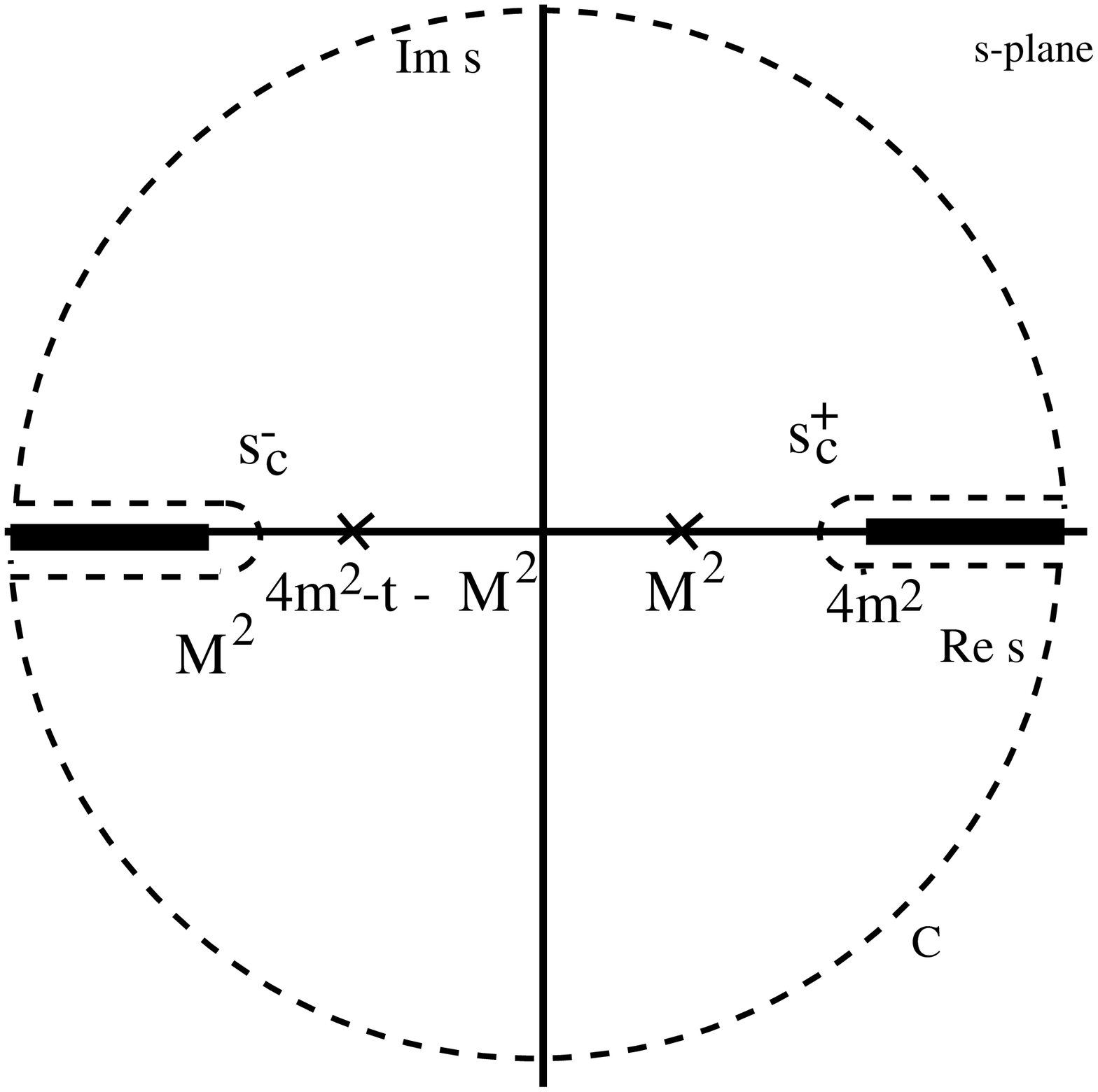}
\caption{Singularity structure in the complex s-plane}
\label{fig:sing}
\end{figure}

Seeing now that $Im\, \M=0$ on some part of the real $s-$axis, we can apply the Schwarz reflection principle\footnote{see e.g. \cite{Bak:1982}}, leading to

\begin{eqnarray}
\M(s^{*},t)&=&\M^{*}(s,t)\;.\label{eq:schwarzref}
\end{eqnarray}

Combining this with $Im\, \M\, \neq\,0\,$ for the values given above, we see that $\M(s,t)$ cannot be analytic in the whole $s-$ plane, as this would require $Im\, \M=0$ for $s$ real. Therefore, we obtain poles and branch-cuts corresponding to $s=M^{2}$ and $s\,\geq\,4m^{2}$; the value of $\M$ in these regions is given by 

\begin{eqnarray}
\M(s,t)&=&\lim_{\varepsilon \rightarrow 0+}\,\M(s+i\varepsilon,t)\label{eq:limM}
\end{eqnarray} 

for an $s$ channel reaction. If we switch to $u$ channel reactions, $i\varepsilon\,\rightarrow\,-i\varepsilon$. Furthermore, we obtain for $Im\,\M(s,t)$  \footnote{Actually, this is only defined for $s\,\geq\,0$, i.e. an $s$ channel reaction. We keep the same notation for $s\,<\,0$ for simplicity.}:

\begin{eqnarray}
Im\,\M(s,t)&=&\frac{1}{2i}\,\lim_{\varepsilon \rightarrow 0}\left[\M(s+i\varepsilon,t)\,-\,\M(s-i\varepsilon,t)\right]\;.\label{eq:ImM}
\end{eqnarray}
A second feature following from analyticity, together with Lorentz-invariance, is crossing symmetry. Taking into account (\ref{eq:Mvar}), we see that we can consider $s < 0$ a physical region in the $s$ plane if we switch from $s$ to $u$ or $t$ channel reaction; in terms of $\M$, this implies

\begin{eqnarray}
\M_{a+b \rightarrow c+d}(s,t,u)&=\M_{a+\bar{c} \rightarrow \bar{b}+d}(t,s,u)=&
\M_{a+\bar{d} \rightarrow \bar{b}+c}(u,t,s)\;.\nonumber\\ \label{eq:Mcross}
\end{eqnarray}
Thirdly, we can use the analyticity of $\M$ to relate its real and imaginary parts using dispersion relations; in short, we use Cauchy's theorem to write

\begin{eqnarray}
\M(s,t)&=&\frac{1}{2\pi i}\,\left(\int_{C}\frac{\M(s',t)}{s'-s}ds'\,+\;\mbox{Poles}\right)\,; \label{eq:Mcauch}
\end{eqnarray}
in the case discussed above, 

\begin{eqnarray*}
\mbox{Poles} &=& \frac{\rho}{s-M^2}\,-\,\frac{\rho '}{s-4m^{2}+M^{2}+t}
\end{eqnarray*}
with $\rho\,$ and $\rho '$ being the respective residua.

In the simplest case, we can rewrite this in the form of a dispersion relation: 

\begin{eqnarray}
\M(s,t)&=& \mbox{Poles} + \frac{1}{\pi} \int^{\infty}_{s^{+}_{c}} \frac{Im \M(s',t)}{s'-s}\,ds'\,+\frac{1}{\pi} \int^{s^{-}_{c}}_{-\infty}\frac{Im \M(s',t)}{s'-s}\,ds'\nonumber\\ \label{eq:Mdec}
\end{eqnarray}

relating the real and imaginary parts of $\M\,$. Here, $s^{\pm}_{c}$ denote the beginning of the branch cuts (see fig.\ref{fig:sing}). $Im \M\,$ is given by (\ref{eq:ImM}).

For more complicated cases, (\ref{eq:Mdec}) is modified by use of subtractions to make the transition from (\ref{eq:Mcauch}) possible.

\end{itemize}

\subsection{Decomposition in partial waves; analytic continuation to the complex angular momentum plane}

In analogy with scattering processes in Quantum Mechanics, the amplitude $\M$ can be decomposed in terms of partial waves in the $s$ channel corresponding to

\begin{eqnarray}
\M(s,t)&=&\sum^{\infty}_{l=0} (2l+1)\,f_{l}(s) P_{l}(z)
\end{eqnarray}
with

\begin{eqnarray*}
z&=&\cos \theta\;,\\
\theta&:&\mbox{scattering angle}\;,\\ 
P_{l}(z)&:& \mbox{Legendre Polynomial}\;,\\
f_{l}(z)&:& \mbox{partial wave amplitude}\;.
\end{eqnarray*}
We are only considering particles with spin 0; generalizations for particles with nonzero spin can be found in the literature.

With Cauchy's theorem we can rewrite this equation by changing into the complex angular momentum plane:

\begin{eqnarray}
\M(s,t)&=&-\frac{1}{2i}\,\int_{C}\frac{(2l+1)f(l,s)P_{l}(-z)}{\sin \pi l} \label{eq:Mint}
\end{eqnarray}
with $f(l,s)\,$ being the continuation of $f_{l}(s)$ into the complex plane. However, uniqueness of $f(l,s)\,$ requires a decomposition into $f^{\pm}_{l}(s)$; these functions are related to $Im \M$ by the inversion of (\ref{eq:Mint}) and double variable dispersion relations similar to (\ref{eq:Mdec}) (for more details, see \cite{Martin:1970}). After a change to integration over $z'$, they are given by

\begin{eqnarray}
f^{\pm}(l,s)&=&\frac{1}{\pi}\,\int^{\infty}_{z_{0}}\left[Im\M_{t}(s,z')\pm Im\M_{u}(s,z')\right]Q'_{l}(z')dz' \label{eq:frosgrib}
\end{eqnarray}
with $Im \M_{t/u}\,$ denoting discontinuities across the $t$-/$u$- channel cuts according to (\ref{eq:ImM}) and $Q'_{l}(z')\,$ reflecting the dependence on the Legendre polynomials $P_{l}(z)$. We see that $f^{+}_{l}(l,s)\,$ is even, $f^{-}_{l}(l,s)\,$ odd under exchange of $u$ and $t$; they are said to be of even and odd signature, respectively. (\ref{eq:frosgrib}) is known as the Froissart-Gribov projection. 

\vspace{3mm}
Taking this into account, we can define matrix-elements $\M^{\pm}(s,t)\,$ by

\begin{eqnarray*}
\M^{\pm}(s,t)&=&\sum^{\infty}_{l=0}(2l+1)f^{\pm}(l,s)P_{l}(z)\;;
\end{eqnarray*}    
$\M(s,t)\,$ is then given by

\begin{eqnarray}
\M(s,t)&=&\frac{1}{2}\left[\M^{+}(s,z)+\M^{+}(s,-z)+\M^{-}(s,z)-\M^{-}(s,-z)\right]\;.\nonumber\\ & &
\end{eqnarray}

\begin{figure}
\centering
\psfrag{l-plane}{l plane}
\psfrag{Im l}{Im l}
\psfrag{Re l}{Re l}
\psfrag{C}{C}
\psfrag{C'}{C'}
\psfrag{ai}{$\al_{i}$}
\includegraphics[width=0.5\textwidth]{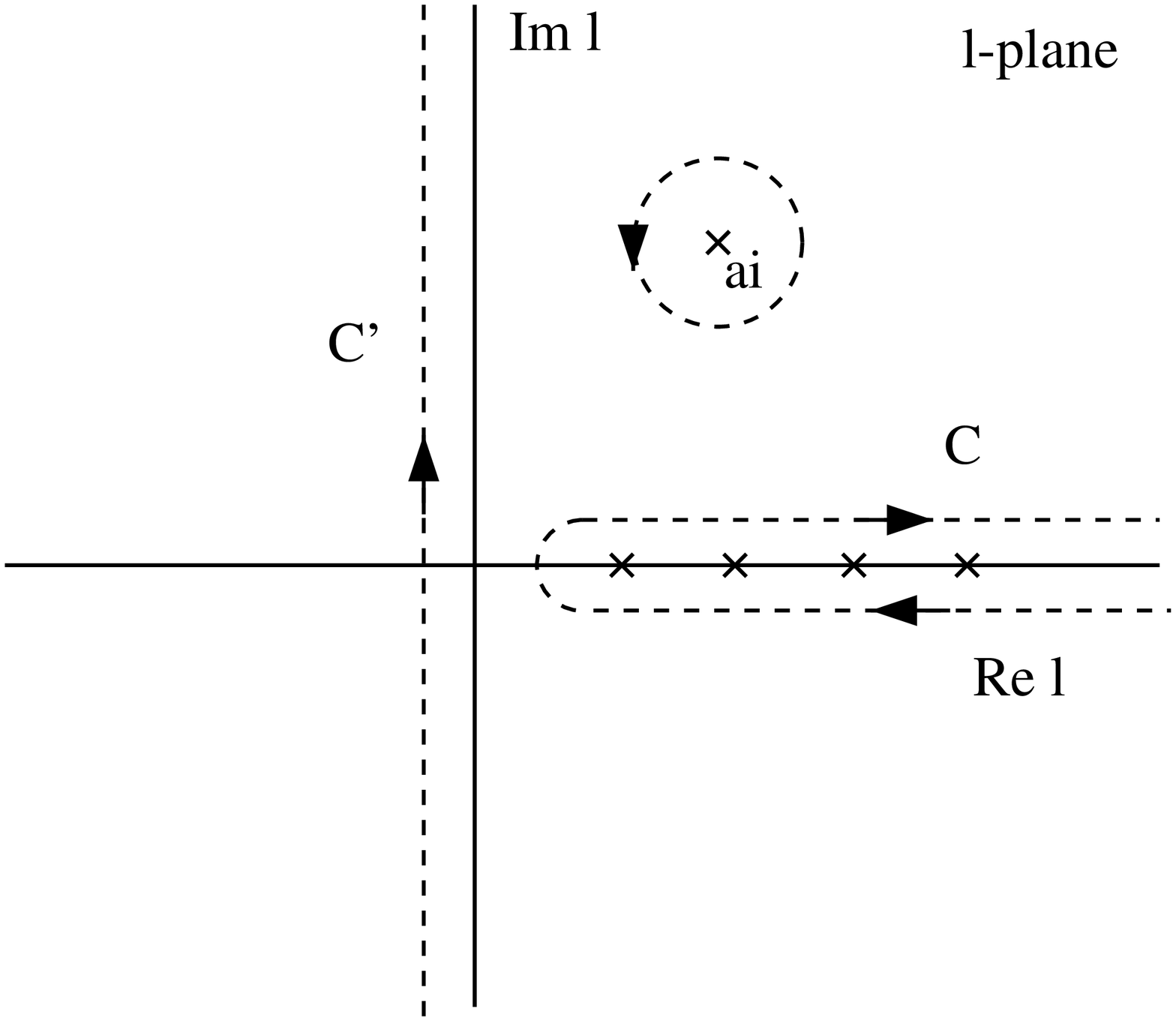}
\caption{contour before and after Mandelstam-Somerfeld-Watson transform}
\label{fig:contour}
\end{figure}
Changing now to contour integration by use of Cauchy's theorem and shifting the contour of integration to a semi-circle with a slightly negative intercept on the real axis by the Mandelstam-Sommerfeld-Watson transform (see fig. \ref{fig:contour}), we finally obtain

\begin{eqnarray}
\lefteqn {\M(s,t) \;=}  & & \nonumber\\
 & &  \frac{-1}{2i}  \int^{-\frac{1}{2}+i \infty}_{-\frac{1}{2}-i \infty}  \frac{2l+1}{2 \sin (\pi l)}\left[f^{+}(l,s)(P_{l}(-z)+P_{l}(z)) +f^{-}(l,s)(P_{l}(-z)-P_{l}(z))\right]dl  \nonumber\\
& &  - \sum_{i}\frac{(2 \alpha^{\pm}_{i}(s)+1)}{2 \sin (\pi \alpha^{\pm}_{i})}\,\beta^{\pm}_{i}(s)(P_{\alpha^{\pm}_{i}}(-z)\pm P_{\alpha^{\pm}_{i}}(z))  \label{eq:Regamp}
\end{eqnarray}
with $\alpha^{\pm}_{i}(s)\,$ and $\beta^{\pm}_{i}(s)\,$ denoting additional poles and the according residues of $f^{\pm}(s,l)$.
For the derivation of (\ref{eq:Regamp}), we furthermore needed the assumption of maximal analyticity of the second kind, i.e. analytic behavior of $\M$ except at a finite number of poles.

\subsection{Regge poles and trajectories; high energy behavior}\label{sec:Regbeh}

Starting with the expression given by (\ref{eq:Regamp}) for $\M\,$, we now consider the asymptotic behavior of $\M(s,t)\,$ for $t\rightarrow\infty$ for a $t$ channel reaction and $s \rightarrow \infty$ for an $s\,$ channel reaction. From crossing symmetry (\ref{eq:Mcross}) of $\M\,$, we know that results for either exchange channel hold true for the other by substituting $ s\, \longleftrightarrow\,t$.

In the $t$ channel it can easily be shown by using the asymptotic behavior of $P_{l}(z)\,$ that the contributions to $\M$ given by the integrals in (\ref{eq:Regamp}) vanish for $z \rightarrow \infty$; we are therefore left with the pole contributions only. Similarly, it can be shown that the sum is dominated by the term with the largest $\alpha_{i}(t)$; this is usually referred to as the leading Regge pole. In total, it can be concluded that for $t \rightarrow \infty$, the behavior of $\M$ is given by

\begin{eqnarray}
\M(s,t)& \sim & \phi(s) t^{\alpha(s)}
\end{eqnarray}
with $\phi(s)\,$ being a function depending on $s$ only and $\alpha(s)$ being the pole position of the leading Regge pole in the complex $l$ plane. Following the argument given above, we obtain for $s$ channel reactions

\begin{eqnarray}
\M(s,t)& \sim & \phi(t) s^{\alpha(t)}\label{eq:Malph}.
\end{eqnarray}
It can also be shown that in $s$- channel reactions, $\alpha(t)$ is purely real if $t$ is below the lowest two-particle threshold; the same statements hold for $t$ channel reactions with $s\,\leftrightarrow\,t$.

Another result from $\M\,$ being written in the form given by (\ref{eq:Regamp}) is the description of pole behavior in dependence of $s/t\,$ described by Chew-Frautschi plots. Looking at $t$ only now (the substitution for $s$ can easily be made following the above argumentation), it can be shown that the contribution from one of the poles behaves like

\begin{eqnarray*}
f_{l}(s)&=&\frac{1}{2}\,\left(1\pm(-1)^{l}\right)\,\left(\frac{2\al(s_{p})+1}{\al(s_{p})+l+1}\right)\,\frac{\be(s_{p})/\al'_{real}(s_{p})}{(s-s_{p})+i\al_{im}(s_{p})/\al'_{real}(s_{p})}
\end{eqnarray*}
near a pole at $s_{p}$. For $\al_{im}$ small, this corresponds to a Breit-Wigner resonance formula. Therefore, we expect physical exchange particles in the $t$ channel exchange whenever $l$ takes an integer value; the particle mass will then be given by $t=m^{2}$. These are so-called $t$-channel resonances.

For extension to $t<0$, i.e. the $s$ channel region, $\alpha(t)$ describes the high energy behavior of the amplitude according to (\ref{eq:Malph}); taking the dominating pole contribution only, $\M$ is given by

\begin{eqnarray}
\M(s,t)_{pole}&=& -\pi\left(2\alpha(t)+1\right)\beta(t)P_{\alpha}(z_{t})\frac{e^{-i \pi\alpha}\pm1}{2 \sin \pi\alpha} + \mathcal{O}(z^{-\alpha-1}_{t}) \nonumber\\
& &\label{eq:domregp}
\end{eqnarray}
$\beta(t)$ is the residue of the pole at $\alpha(t)$.

\vspace{3mm}
Finally, we can investigate the high-energy behavior of cross sections calculated from $\M$; for $s\rightarrow\infty\, , t=0\,$, we obtain

\begin{eqnarray*}
\M(s,t=0)&=&const.\,s^{\alpha(0)}
\end{eqnarray*}
and therefore

\begin{eqnarray}
\sigma_{tot} & \sim & s^{\alpha(0)-1}\;. \label{eq:sigtotr}
\end{eqnarray}
For differential cross sections, we obtain

\begin{eqnarray*}
\frac{d\sigma}{dt}&\sim&\left|\phi(t)\right|^{2}\,s^{2\alpha(t)-2};
\end{eqnarray*}
for small $t$, $\alpha(t)$ can be expanded by

\begin{displaymath}
\alpha(t)\;\approx\;\alpha(0)\,+\alpha' t\;.
\end{displaymath}

\vspace{10mm}

\subsection{Pomeron and Odderon from Regge theory}\label{sec:POReg}

\begin{figure}
\centering
\psfrag{plab=sqrt(2)/2 [GeV]}{$p_{lab}$}
\psfrag{=}{$=$}
\psfrag{sqrt(s)/2}{$\sqrt{s}/2$}
\psfrag{Cross section}{Cross section}
\psfrag{mb}{mb}
\includegraphics[angle=-90.,width=0.5\textwidth]{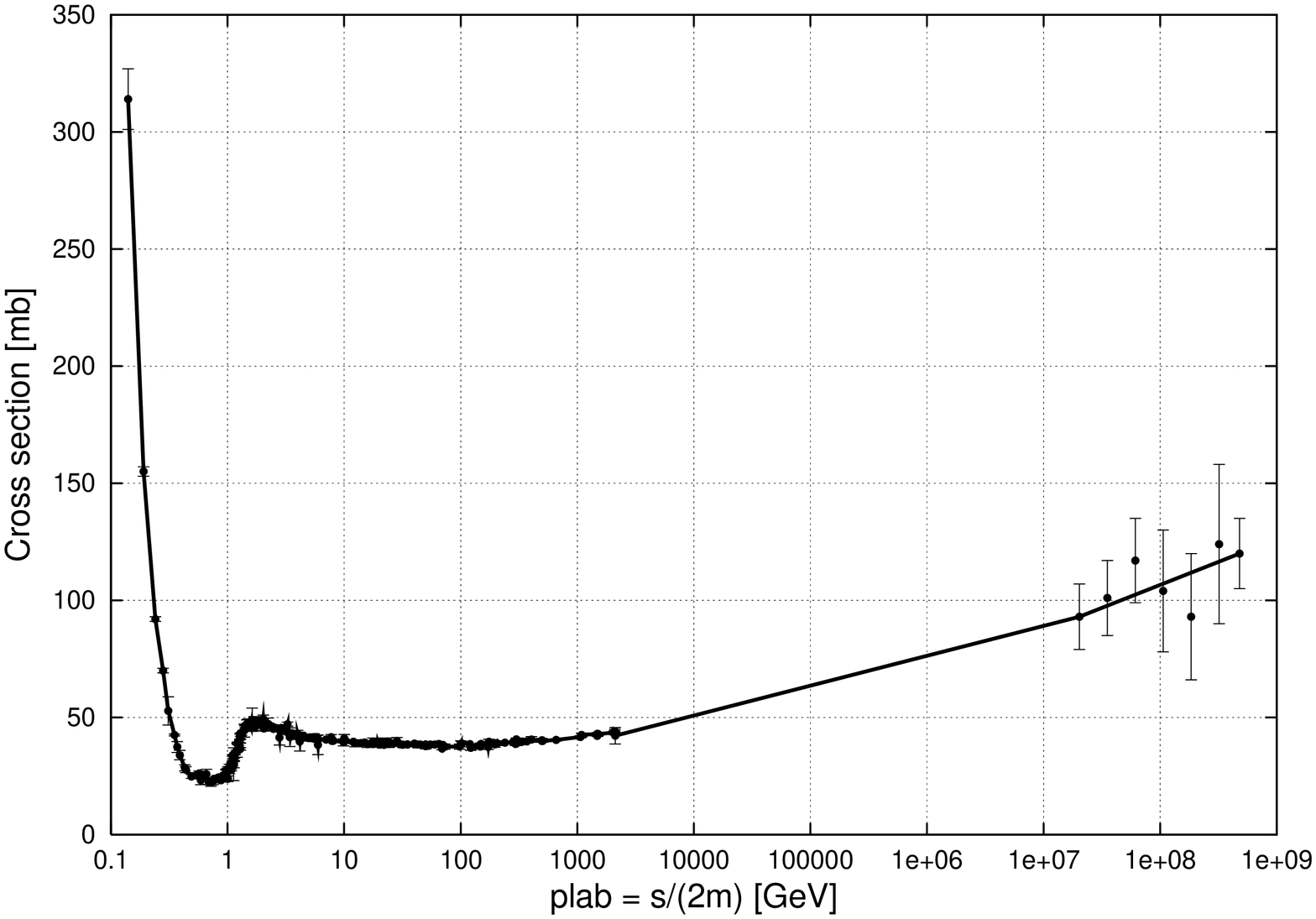}
\caption{total pp cross sections \cite{Hagiwara:2002pw}; not fitted}
\label{fig:pptot}
\end{figure}

Taking into account the estimated behavior of total cross sections described by (\ref{eq:sigtotr}), we can investigate the experimental data for total hadron-hadron cross sections; as can easily be seen from figure \ref{fig:pptot}, cross sections for various final states tend to increase as $s$ is growing.

 Historically, the Pomeron was postulated at a time when experiments did not yet exceed a total squared energy of $\sqrt{s}=10\, \mbox{GeV}\,$; for these values, $\sigma_{tot}$ is approximately constant. Therefore the Pomeron trajectory was originally assumed to have an intercept at $\alpha(0)=1$ according to (\ref{eq:sigtotr}). Recent data, however, even show a rise in total cross sections as $s\rightarrow \infty$; therefore, $\alpha(0)$ has to be bigger than 1. Fitting data from $pp$ and $p\bar{p}$ reactions, $\alpha(0)$ can indeed be shown to have a value around $1.08$ \cite{Barnett:1996hr}.
Foldy and Peierls \cite{Foldy:1963} showed that any exchange leading to a non-falling cross section for rising $s$ has to carry the quantum numbers of the vacuum, i.e. be colorless, have zero isospin, and even parity; moreover, the optical theorem requires even signature. Assuming the dominance of this trajectory in high energy hadronic reactions, the Pomeranchuk theorem \cite{Pomeranchuk:1958} stating that 

\begin{equation}
\sigma^{pp}(s)\;=\;\sigma^{\pb p}(s)\;\;\mbox{for $s\,\rightarrow\,\infty$}
\end{equation}\label{eq:pomthe}
is automatically satisfied. The corresponding trajectory was therefore called Pomeron-trajectory. The intercept at $1.08$ violates the Froissart-Martin bound stating that for $s \rightarrow \infty$ for any hadronic total cross section, 

\begin{eqnarray} 
\sigma_{tot}&<& A \ln^{2}s
\end{eqnarray}
with

\begin{displaymath}
A\;\approx\; 60 mb\;.
\end{displaymath}
However, as this violation only becomes visible at high momenta at the order of the Planck scale, it is not improbable that processes different from the single-Pomeron exchange will enter and preserve this bound. Moreover, it can be argued that the intercept at $\alpha(0)=1.08$ is only an effective intercept actually arising from the exchange of two or more Pomerons (so-called Regge cuts). For more details as well as a  treatment of the Froissart Bound in perturbative QCD we refer to the literature (e.g., \cite{Forshaw:1997dc}).
 
\vspace{5mm}

An amplitude similar to the Pomeron with respect to color, but with odd parity, was first proposed by Joynson et al \cite{Joynson:1975az} in investigation of  $\frac{d\sigma}{dt}|_{t=0}\,$  for $\pi^{-}p \rightarrow \pi^{0}n\,$ as well as differences between  $\pi^{+}p\,$ and $\pi^{-}p\,$ total cross sections in dependence of total lab energy. They work in the framework of so-called helicity-flip and helicity non-flip amplitudes where the dependencies of the quantities investigated are given by

\begin{eqnarray}
\frac{d\sigma}{dt}|_{t=0}&\propto&|\M_{hf}|^{2}\;,\nonumber\\
\sigma_{\pi^{+}p}-\sigma_{\pi^{-}p}&\propto&Im\, \M_{hf}(t=0)\;.
\end{eqnarray}
In their work, they realize that, while each magnitude on its own can be described by the at that time standard Regge-theory, the parameters from both fits are incompatible. They conclude that a new Regge amplitude which is purely real at $t=0$  and has an intercept of $\al(t=0)=1$ is apt to describe both measurements simultaneously; these requirements are fulfilled by a Regge trajectory with odd signature, as we can see from (\ref{eq:domregp}). In addition, it corresponds to a pole for $l=1$ in the complex $l-$plane. Due to the odd signature of this trajectory, it is called the Odderon. Equally, the Odderon has odd parity; for a discussion of the connection between parity and signature, see e.g. \cite{Martin:1970}.\\
 Similar effects were already noticed for comparisons between $N N$ and $N \bar{N}$ total cross sections by Bouquet et al. \cite{Bouquet:1976xz} as well as Lukaszuk and Nicolescu in considerations of $pp$ and $p\bar{p}$ cross sections \cite{Lukaszuk:1973nt}.

\section{Nonperturbative effective propagators and vertices} \label{sec:effpropsec}

There have been various attempts to deduce the behavior of Pomeron and Odderon trajectories from QCD. Here, states with corresponding quantum numbers can easily be represented by two or three reggeized gluons, respectively. However, the perturbative Pomeron and Odderon are not apt to describe current experiments where nonperturbative effects play a major role. Effective Pomeron and Odderon propagators and vertices can be deduced from hadron-hadron scattering experiments by making a Regge-ansatz for the scattering amplitudes. However, the so-derived expressions are only effective descriptions not directly corresponding to real physical particles. 
In the following section, we use standard notations from QFT (see appendix \ref{app:standnot}). 

\subsection{General features of $pp$ and $p\bar{p}$ elastic scattering}

For a motivation as well as deduction of effective Pomeron and Odderon propagators, we consider elastic $pp$ and $p\bar{p}$ processes\footnote{This section is highly indebted to private communication with O. Nachtmann} :
\begin{figure}[h!]
\centering
\subfigure[$pp$-scattering]{
\centering
\psfrag{p1}{$\scripts p(p_{1})$}
\psfrag{p1'}{$\scripts p(p'_{1})$}
\psfrag{p2}{$\scripts p(p_{2})$}
\psfrag{p2'}{$\scripts p(p'_{2})$}
\psfrag{X}{$\scripts X(p_{1}-p'_{1})$}
\includegraphics[width=0.35\textwidth]{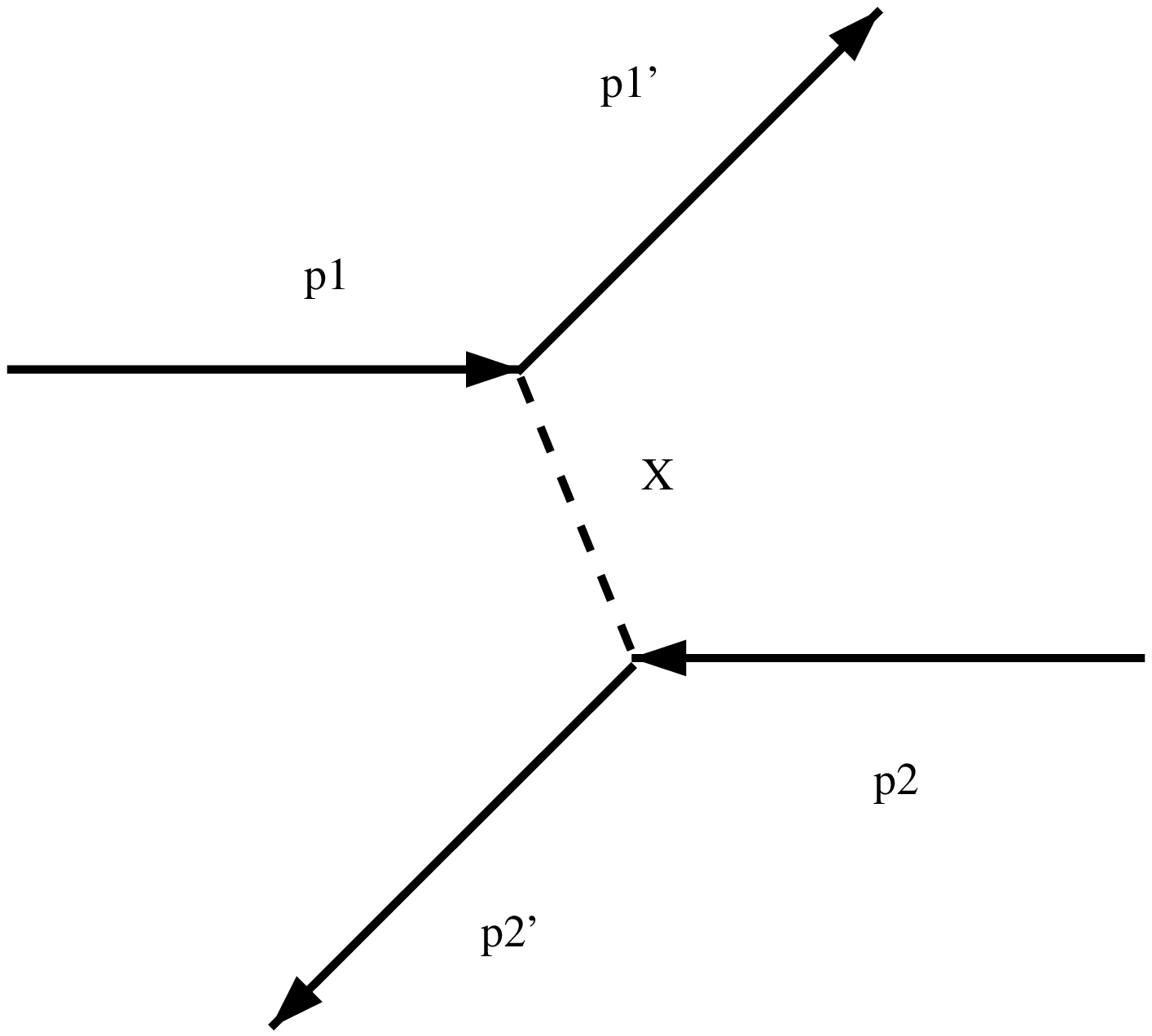}}
\subfigure[$p\pb$-scattering]{
\centering
\psfrag{p1}{$\scripts p(p_{1})$}
\psfrag{p1'}{$\scripts p(p'_{1})$}
\psfrag{p2}{$\scripts \bar{p}(p_{2})$}
\psfrag{p2'}{$\scripts \bar{p}(p'_{2})$}
\psfrag{X}{$\scripts X(p_{1}-p'_{1})$}
\includegraphics[ width=0.35\textwidth]{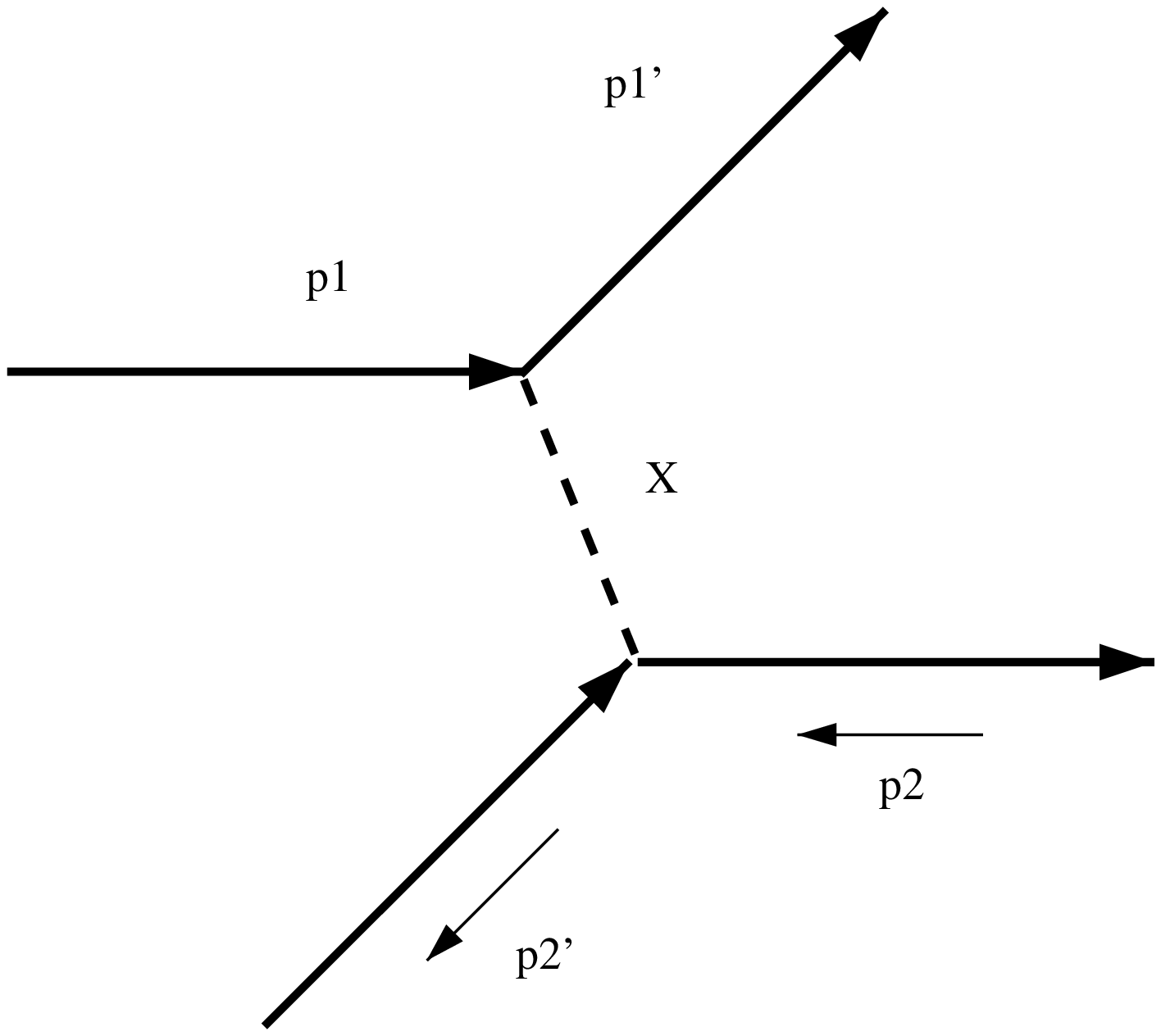}}
\label{fig:ppscatt}
\end{figure}

\noindent
For both reactions, we can decompose $\M_{pp/p\pb}$ in helicity amplitudes using
\indent
\begin{eqnarray}
\lefteqn{{\big\langle} p(p'_{1})\,p(p'_{2})|\M|p(p_{1})\,p(p_{2}){\big \rangle}\;  =} &  & \nonumber\\
& & \bar{u}_{\alpha'}(p'_{1})\,\bar{u}_{\be'}(p'_{2})\,\Gamma^{(pp)}_{\al'\al,\be'\be}(p'_{1},p'_{2},p_{1},p_{2})\,u_{\al}(p_{1})\,u_{\be}(p_{2})\nonumber\\ & & \nonumber\\
\lefteqn{\langle \pb (p'_{1})\,p(p'_{2})|\M|\pb (p_{1})\,p(p_{2})\rangle 
\;  =} &  & \nonumber\\
& & \bar{v}_{\alpha}(p_{1})\,\bar{u}_{\be'}(p'_{2})\,\Gamma^{(\pb p)}_{\al\al',\be'\be}(p'_{1},p'_{2},p_{1},p_{2})\,v_{\al'}(p'_{1})\,u_{\be}(p_{2})\nonumber\\
& &
\end{eqnarray}
with

\begin{eqnarray}
\lefteqn{\Gamma^{(pp)}_{\al'\al,\be'\be}(p'_{1},p'_{2},p_{1},p_{2}) \; =  \;\left(\frac{1}{2m}\right)^{4}\, u_{\alpha'}(p'_{1},s'_{1})\,u_{\be'}(p'_{2},s'_{2})\,\times}& & \nonumber \\
& & {\big\langle} p(p'_{1},s'_{1})\,p(p'_{2},s'_{2})|\M|p(p_{1},s_{1})\,p(p_{2},s_{2}){\big\rangle}\,\bar{u}_{\al}(p_{1},s_{1})\,\bar{u}_{\be}(p_{2},s_{2})\nonumber \\
& & \nonumber \\
\lefteqn{\Gamma^{(\pb p)}_{\al\al',\be'\be}(p'_{1},p'_{2},p_{1},p_{2})\;  =  \; \left(\frac{1}{2m}\right)^{4}\, v_{\alpha}(p_{1},s_{1})\,u_{\be'}(p'_{2},s'_{2})\,\times} & & \nonumber \\
& &  {\big\langle} \pb(p'_{1},s'_{1})\,p(p'_{2},s'_{2})|\M|\pb (p_{1},s_{1})\,p(p_{2},s_{2}){\big\rangle}\,\bar{v}_{\al'}(p'_{1},s'_{1})\,\bar{u}_{\be}(p_{2},s_{2}) \nonumber \\ & & \label{eq:Gammas}
\end{eqnarray}
where we sum over the spins $s_{i},\,s'_{i}$. From the LSZ reduction formalism (see e.g.\cite{Itzykson:1980rh}), we know that incoming particles and outgoing antiparticles are equivalent by the substitution

\begin{eqnarray}
u(p,s)e^{-ipx} &\longleftrightarrow& -v(p,s)e^{ipx}
\end{eqnarray}
in the corresponding reduction formula for fermions; taking this as well as (\ref{eq:Mcross}) into account, we can easily see that

\begin{eqnarray}
\Gamma^{(\pb p)}_{\al\al',\be'\be}(p'_{1},p'_{2},p_{1},p_{2})&=&-\;\Gamma^{(p p)}_{\al\al',\be'\be}(-p_{1},p'_{2},-p'_{1},p_{2})
\end{eqnarray}
In general, for $2 \rightarrow 2$ particle reactions with fermions, we can decompose $\Gamma$ into 16 helicity amplitudes; however, as we are interested in reactions with helicity conservation for $s\rightarrow \infty$ and $|t|$ small, we will solely investigate the amplitude corresponding to $\gamma^{\mu}\otimes \gamma_{\mu}$. Helicity conservation in connection with diffractive scattering has been proposed e.g by Gilman \cite{Gilman:1970vi}  and treated theoretically by Landshoff \cite{Landshoff:1971pw} and Jones \cite{Jones:1970}; for a closer treatment, see \cite{Landshoff:1971pw} and references therein.

In \cite{Landshoff:1971pw}, the proton-Pomeron coupling is decomposed according to

\begin{eqnarray}
\bar{u}(k')\gamma_{\mu}\,u(k)\,[F^{p}_{1}(t)+F^{n}_{1}(t)]+i \bar{u}(k')\sigma_{\mu\nu}u(k)\,q^{\nu}[F^{p}_{2}(t)+F^{n}_{2}(t)]\label{eq:protc}
\end{eqnarray}
where $F^{i}_{1}\,$ and $F^{i}_{2}$ denote couplings to charge and magnetic moment and $q\,=\,k-k'$. $i=p,n\,$ corresponds to contributions from p and n quarks (corresponding to u and d quarks in contemporary notation); the first term in (\ref{eq:protc}) corresponds to helicity conservation, the second to a helicity flip. See \cite{Landshoff:1971pw} and references therein for more details.

It is then shown that the coupling is dominated by the helicity-conserving part; therefore, we only have to look at the  $\gamma^{\mu}\otimes \gamma_{\mu}$ like contribution to $\M$.
Taking this into account, we can write $\Gamma^{(pp)}$ and $\Gamma^{(\pb p)}$ as 

\begin{eqnarray}
 & & \Gamma^{(pp)}_{\al'\al,\be'\be}(p'_{1},p'_{2},p_{1},p_{2}) = \nonumber\\
& & T_{pp}(\nu,t) \left(\frac{1}{2m}\right)^{4}\left[(p'_{1}\mkern-15mu / +m)\gamma^{\mu}(p_{1}\mkern-15mu / +m)\right]_{\al'\al}\,\left(\frac{1}{2m}\right)^{4}\left[(p_{2}'\mkern-15mu / +m)\gamma_{\mu}(p_{2}\mkern-15mu / +m)\right]_{\be'\be}\;,\nonumber\\
& & \nonumber \\
 & & \Gamma^{(\pb p)}_{\al\al',\be'\be}(p'_{1},p'_{2},p_{1},p_{2}) = \nonumber\\
& & T_{\pb p}(\nu,t) (\frac{1}{2m})^{4}\left[(-p'_{1}\mkern-15mu / +m)\gamma^{\mu}(-p_{1}\mkern-15mu / +m)\right]_{\al\al'}\,(\frac{1}{2m})^{4}\left[(p_{2}'\mkern-15mu / +m)\gamma_{\mu}(p_{2}\mkern-15mu / +m)\right]_{\be'\be}\nonumber\\
& & \label{eq:Gammag}
\end{eqnarray}
where $\nu$ is given by

\begin{displaymath}
\nu\;=\;\frac{s-u}{4}\;.
\end{displaymath}
Considerations closely following the ones given in (\ref{sec:genscat}) show that $T_{pp}$ and $T_{\pb p}$ are limits of an analytic function $ A(\nu,t)$ with cuts and singularities in the complex $\nu$ plane given by 

\begin{eqnarray*}
\nu_{1}&=&m^{2}-\frac{1}{2}m^{2}_{\pi}-\frac{1}{4}t\;\; p\pb\mbox{ -scattering}\;, \\
\nu_{2}&=&m^{2}+\frac{1}{4}t\;\;  pp \mbox{-scattering}\;.
\end{eqnarray*}
Furthermore, we can conclude that

\begin{eqnarray}
A^{*}(\nu^{*},t)&=&A(\nu,t)
\end{eqnarray}
(see (\ref{eq:schwarzref})) and

\begin{eqnarray}
T_{pp}(\nu,t)&=&\lim_{\varepsilon\rightarrow 0}\,A(\nu+i\varepsilon,t)\;,\nonumber\\
T_{\pb p}(\nu,t)&=&-\lim_{\varepsilon\rightarrow 0}\,A(-\nu-i\varepsilon,t)\end{eqnarray}
(see (\ref{eq:limM})).
We now define 

\begin{eqnarray}
A_{\pm}(\nu,t)&=&\frac{1}{2}\left[A(\nu,t)\mp A(-\nu,t)\right]\label{eq:amplpo}
\end{eqnarray}
and obtain

\begin{eqnarray}
T_{pp}&=&\lim_{\varepsilon \rightarrow 0}\left[A_{+}(\nu+i\varepsilon,t)+A_{-}(\nu+i\varepsilon,t)\right]\;, \nonumber \\
T_{\pb p}&=& \lim_{\varepsilon \rightarrow 0} \left[A_{+}(\nu+i\varepsilon,t)-A_{-}(\nu+i\varepsilon,t)\right]\;. \nonumber\\ \label{eq:expT}
\end{eqnarray}
Having in mind that the Pomeron is even under charge conjugation and therefore  couples to $p$ and $\pb$ in the same manner while the Odderon is odd leading to a sign change between $p$ and $\pb$-Odderon couplings, we can interpret $A_{+}$ as the Pomeron, $A_{-}$ as the Odderon amplitude. Combining (\ref{eq:Gammas}), (\ref{eq:Gammag}) and (\ref{eq:expT}), we get 
\begin{eqnarray}
& &  {\big\langle}p(p'_{1} ,s'_{1})\, p(p'_{2},s'_{2})\left|\M\right|p(p_{1},s_{1})\, p(p_{2},s_{2}) {\big\rangle} \nonumber\\
& &= \; T_{pp}(\nu,t)\, \bar{u}(p'_{1},s'_{1})\,\gamma^{\mu}\, u(p_{1},s_{1})\,\bar{u}(p'_{2},s'_{2})\,\gamma_{\mu}\, u(p_{2},s_{2}) \nonumber\\
& &  \longrightarrow  T_{pp}(\frac{s}{2},t)\,2s\,\delta_{s'_{1}s_{1}}\delta_{s'_{2}s_{2}}\label{eq:Tpp}
\end{eqnarray}
in the limit $s\rightarrow \infty$ and $|t|$ small.
Likewise,

\begin{eqnarray}
 {\big\langle}\pb(p'_{1} ,s'_{1})\, p(p'_{2},s'_{2})\left|\M\right|\pb(p_{1},s_{1})\, p(p_{2},s_{2}) {\big\rangle} &  \longrightarrow & T_{\pb p}(\frac{s}{2},t)\,2s\,\delta_{s'_{1}s_{1}}\delta_{s'_{2}s_{2}}\nonumber\\ \label{eq:Tpbp}
& &
\end{eqnarray}
in the same limits.

\subsection{Effective Pomeron and Odderon propagators}\label{sec:effprop}

Closely following \cite{Donnachie:1984hf}, we will now make a Regge pole ansatz for the amplitudes for Pomeron and Odderon exchange given by (\ref{eq:amplpo}); for the Pomeron, we use

\begin{eqnarray}
A_{+}(\nu+i\varepsilon,t)&=&  i\,(3 \be_{\pom} F_{1}(t))^{2} \,\exp \left\{(\al_{\pom}(t)-1) \left[\ln(\frac{\nu}{\nu_{0}})-i\varphi_{\pom}\right]\right\}\nonumber\\
\end{eqnarray}
with

\begin{eqnarray}
\be_{\pom}&:&\mbox{Pomeron-quark-coupling}\;, \nonumber \\
\al_{\pom}(t)&:&\mbox{Pomeron trajectory}\;, \nonumber \\
F_{1}(t)&:&\mbox{form-factor of the nucleon}\;.\label{eq:parpom}
\end{eqnarray}
The phase factor $\varphi_{\pom}$ can be determined by looking at the analyticity properties of $A_{+}(\nu, t)$ as described in section (\ref{sec:genscat}); taking these into account, we can determine $\varphi_{\pom}$ to take the value $\frac{\pi}{2}$ and get as a final expression

\begin{eqnarray}
A_{+}(\nu+i\varepsilon,t)&=&  i\,(3 \be_{\pom} F_{1}(t))^{2} \,\exp \left\{(\al_{\pom}(t)-1) \left[\ln(\frac{\nu}{\nu_{0}})- \frac{i}{2}\pi\right]\right\}\nonumber\\ \label{eq:amppom}
\end{eqnarray}
in the limit $\nu\rightarrow\infty$.
Similar considerations for the Odderon amplitude lead to

\begin{eqnarray}
A_{-}(\nu+i\varepsilon,t)&=&  (3 \be_{\odd} F_{1}(t)) ^{2} \, \eta_{\odd} \exp \left\{(\al_{\odd}(t)-1) \left[\ln(\frac{\nu}{\nu_{0}})- \frac{i}{2}\pi\right]\right\}\nonumber\\ \label{eq:ampodd}
\end{eqnarray}
with  $\be_{\odd}\,$ and $\al_{\odd}$ being the analogous quantities of (\ref{eq:parpom}) in the case of Odderon exchange.
From the optical theorem (\ref{eq:optthem}) as well as the observations leading to postulations of the Pomeron and Odderon in the first place (see section \ref{sec:POReg}), we can conclude that 

\begin{eqnarray}
A_{+}(\nu+i\varepsilon,t=0)&\propto&+c_{im}\;\;\mbox{purely imaginary and positive}\;,\nonumber\\
A_{-}(\nu+i\varepsilon,t=0)&\propto&\pm\,c_{real}\;\;\mbox{purely real}\;.
\end{eqnarray}
We see that the expressions given above fulfill these requirements; $\eta_{\odd}\,=\pm1\,$ reflects the sign ambiguity in $A_{-}$. 
Combining now (\ref{eq:expT}), (\ref{eq:Tpp}), (\ref{eq:Tpbp}), (\ref{eq:amppom}), and (\ref{eq:ampodd}), we arrive at

\begin{eqnarray}
& &  {\Big\langle}p(p'_{1} ,s'_{1})\, p(p'_{2},s'_{2})\left|\M\right|p(p_{1},s_{1})\, p(p_{2},s_{2}) {\Big\rangle} \nonumber\\
& & \longrightarrow 2s\:(3 \be_{\pom} F_{1}(t))^{2}\,\left\{i\,(-i\frac{s}{2\nu_{0}})^{\al_{\pom}(t)-1}+\eta_{\odd}(\frac{\be_{\odd}}{\be_{\pom}})^{2}\,(-i\frac{s}{2\nu_{0}})^{\al_{\odd}(t)-1}\right\}\,\delta_{s'_{1}s_{1}}\,\delta_{s'_{2}s_{2}}\;,\nonumber\\
& & \nonumber \\
& &  {\Big\langle}\pb(p'_{1} ,s'_{1})\, p(p'_{2},s'_{2})\left|\M\right|\pb(p_{1},s_{1})\, p(p_{2},s_{2}) {\Big\rangle} \nonumber\\
& & \longrightarrow 2s\:(3 \be_{\pom} F_{1}(t))^{2}\,\left\{i\,(-i\frac{s}{2\nu_{0}})^{\al_{\pom}(t)-1}-\eta_{\odd}(\frac{\be_{\odd}}{\be_{\pom}})^{2}\,(-i\frac{s}{2\nu_{0}})^{\al_{\odd}(t)-1}\right\}\,\delta_{s'_{1}s_{1}}\,\delta_{s'_{2}s_{2}}\;.\nonumber\\
& &
\end{eqnarray}
These expressions are related to the total cross sections by the optical theorem (\ref{eq:optthem}).

Using now the fits from \cite{Barnett:1996hr} and following \cite{Donnachie:1984hf} for the parameters given in (\ref{eq:parpom}) for Pomeron and Odderon respectively, we obtain 

\begin{eqnarray}
\al_{\pom}(t)&=&1+\varepsilon+\al'_{\pom}t\;,\nonumber\\
\frac{1}{2\nu_{0}}\;=\;\frac{1}{s_{0}}&=&\frac{1}{m^{2}_{p}}\label{eq:pompar}
\end{eqnarray}
with 

\begin{eqnarray*}
\varepsilon\;=\;0.08 &,\al'_{\pom}\;=\;0.25 \mbox{GeV}^{-2},\;m_{p}\,=\,0.95\;\mbox{GeV}&\mbox{and}\;\; \be_{\pom}\;=\;1.87 \mbox{GeV}^{-1} 
\end{eqnarray*}  
and for the Odderon
\begin{eqnarray}
\al_{\odd}(t)&=&1+\varepsilon'+\al'_{\odd}t\nonumber\\
& & \label{eq:oddpar}
\end{eqnarray}
with $\varepsilon'\;=\;0$.
For the proton form factor, we use \cite{Donnachie:1984hf}

\begin{eqnarray}
F^{p}(t)\;= &  F^{p}_{1}(t) & =  \;\frac{4m^{2}_{p}-t\,\frac{\mu_{p}}{\mu_{N}}}{4m^{2}_{p}-t}\;G_{D}(t)\label{eq:formnuc}
\end{eqnarray}
with

\begin{eqnarray}
\mu_{p}\;\simeq\; 2.8 \mu_{N}&;&  
\mu_{N}\;=\;\frac{e}{2\,m_{p}}\; \mbox{nuclear magneton}\nonumber \\
\end{eqnarray}
and $G_{D}(t)$ the electric form factor fitted in a dipole form 

\begin{equation}
G_{D}\;=\;\frac{1}{(1-t/m^{2}_{D})^{2}}
\end{equation}
with $m^{2}_{D}\,=\,0.71\;\mbox{GeV}^{2}$ \cite{Bosted:1992rq}, \cite{Arnold:1988us}.

\vspace{3mm} 

As a final step, we will consider the total $pp$ and $\pb p$ cross sections as well as the $\rho-\,$parameter for the case $t=0\,$. The latter is defined by

\begin{equation}
\rho\;=\;\frac{Re\, \M}{Im\, \M}\;.
\end{equation}
For the cross sections in the case of $t=0$, we arrive at the following expressions

\begin{eqnarray}
\sigma^{\kappa}_{tot}&=&2(3\be_{\pom})^{2}\left\{\left(s/s_{0}\right)^{\vare}\cos(\frac{\vare}{2}\pi)+f_{\kappa}\eta_{\odd}(\frac{\be_{\odd}}{\be_{\pom}})^{2}\,(s/s_{0})^{\vare'}\sin(\frac{\vare'}{2}\pi) \right\}\nonumber\\
& & \label{eq:siglim}
\end{eqnarray}
with

\begin{eqnarray*}
f_{\kappa}&=&\left\{ \begin{array}{ll} 
-1 & \mbox{for}\; \kappa = pp\\
+1 & \mbox{for}\; \kappa = \pb p
\end{array} \right.
\end{eqnarray*}
For $\rho$, we obtain in the same limit

\begin{eqnarray}
\rho^{(pp)}(s)-\rho^{(\pb p)}(s) & \longrightarrow & -\;\frac{2 \eta_{\odd}\,(\frac{\be_{\odd}}{\be_{\pom}})^{2} \cos(\frac{\vare'}{2}\pi)}{(s/s_{0})^{\vare-\vare'}\cos(\frac{\vare}{2}\pi)}\;. \label{eq:rholim}
\end{eqnarray}
With $\vare'=0$ according to (\ref{eq:oddpar}), we see that $\sigma^{(pp)}\rightarrow \sigma^{(\pb p)}$ in the limit given above.


Taking the results given by (\ref{eq:siglim}) and (\ref{eq:rholim}), we can now extract effective Feynman propagators and couplings for the Pomeron and Odderon respectively. In case of the Pomeron, we obtain

\begin{eqnarray}
\pom-\mbox{propagator}&:& -i\;\left[i\left(-is/s_{0}\right)^{(\al_{\pom}(t)-1)}\right]\;,\nonumber\\
\pom\; pp-\mbox{vertex}&:&-i\;3\,\be_{\pom}\;F_{1}[(p'-p)^{2}]\;\gamma^{\mu}\;,\nonumber\\
\pom\; \pb p-\mbox{vertex}&:&-i3\,\be_{\pom}\;F_{1}[(p'-p)^{2}]\;\gamma^{\mu}\;,\nonumber\\
& & \label{eq:Feynpom}
\end{eqnarray}
and for the Odderon

\begin{eqnarray}
\odd-\mbox{propagator}&:& -i\;\eta_{\odd}\left[\left(-is/s_{0}\right)^{(\al_{\odd}(t)-1)}\right]\;,\nonumber\\
\odd\; pp-\mbox{vertex}&:&-i\;3\,\be_{\odd}\;F_{1}[(p'-p)^{2}]\;\gamma^{\mu}\;,\nonumber\\
\odd\; \pb p-\mbox{vertex}&:&+i3\,\be_{\odd}\;F_{1}[(p'-p)^{2}]\;\gamma^{\mu}\;.\nonumber\\
& & \label{eq:Feynodd}
\end{eqnarray}
However, when using the expressions given by (\ref{eq:Feynpom}) and (\ref{eq:Feynodd}) for calculations within the framework of field theory, it has to be taken into account that they only correspond to effective expressions.

\subsection{Vertex for $\gamma \,\odd^{*} \,\pi^{0}\,$ coupling}\label{sec:effver}

For completeness, we already list the result for the $\gamma \, \odd^{*} \,\pi^{0}$ coupling; see chapter \ref{sec:anocoup} for a closer discussion.
From (\ref{eq:gammacoup}), the $\gamma\gamma^{*}\,\pi^{0}\,$ vertex is given by

\begin{eqnarray}
\Gamma_{\mu\nu}&=&\frac{\al}{\pi\,f_{\pi^{0}}}\,\vare_{\mu\nu\rho\sigma}\, k^{\rho}_{1} k^{\sigma}_{2}\,T(k^{2}_{1},k^{2}_{2})\label{eq:gammunu}
\end{eqnarray}
with 

\begin{eqnarray*}
T(k^{2}_{1},0)\;=\;\frac{1}{1-k^{2}_{1}/8\pi^{2}f_{\pi^{0}}}&,&f_{\pi}\,=\,93\;\mbox{GeV}\;.
\end{eqnarray*}
(see (\ref{eq:fincouppgg})).
Following the argumentation given in section \ref{sec:realver}, the $\pi^{0}\,\gamma\,\odd^{*}\,$-vertex can be obtained by simply substituting $e\,\rightarrow\,\be_{\odd}\,$ and also assuming  color-blindness of the Odderon; taking this into account, we obtain

\begin{equation}
\frac{T^{\gamma\odd}}{T^{\gamma\gamma}}\;=\;\frac{\be_{\odd}}{e}\,r_{PS}\label{eq:ratT}
\end{equation}
with $ r_{\pi^{0}}\,=3\,$. Here, we closely follow \cite{Kilian:1998ew}.

Combining (\ref{eq:gammunu}) and (\ref{eq:ratT}), we obtain for the $\gamma\,\odd^{*}\,\pi^{0}$ coupling

\begin{equation}
\Gamma^{\gamma\,\odd^{*}\,\pi^{0}}_{\mu\nu}\;=\;\frac{\al}{\pi\,f_{\pi^{0}}}\,\vare_{\mu\nu\rho\sigma}\, k^{\rho}_{1} k^{\sigma}_{2}\,\frac{1}{1-k^{2}_{1}/8\pi^{2}f_{\pi^{0}}}\,\frac{\be_{\odd}}{e}\,r_{\pi^{0}}\;.
\end{equation}

\section{Pomeron and Odderon from pQCD}

In the perturbative regime, the Pomeron and Odderon can be described as states composed of two or three reggeized gluons, respectively, leading to a Regge-behavior in the form of

\begin{equation}
\M(s,t)\;\sim\,\phi(t)\,s^{\al(t)}\;. \label{eq:Malph2}
\end{equation}
However, due to dominance of nonperturbative processes, the perturbative expressions are not apt to describe the behavior of total hadronic cross sections adequately. There is extensive literature on the field of applicability of perturbative and non-perturbative approaches; we will only give a short sketch of the derivation of the perturbative Pomeron and Odderon and refer to the literature for more discussion (see e.g. \cite{Forshaw:1997dc} and references therein).

\subsection{Reggeization of particles}

A particle is said to reggeize if an exchange in an $s$ or $t$ channel reaction with the particles quantum numbers leads to a Regge-behavior as described in section \ref{sec:Regbeh}. Furthermore, we require $\al(m^2)=J\,$, with $m$ and $J$ being the particles mass and spin \footnote{For a more detailed discussion of the relations between the angular momentum introduced in section \ref{sec:Revreg} and the spin of the particles, see \cite{Collins:1977}, \cite{Martin:1970}.}, respectively; therefore, the particle has to lie on the Regge-trajectory described by $\al(t)$.

Reggeization of particles was first proposed by Gell-Mann \cite{Gell-Mann:1962}. The possibility of reggeization of fermions was first shown by McCoy and Wu \cite{McCoy:1976ff} for QED and later extended to non-abelian theories by Mason \cite{Mason:1976sq} and Sen \cite{Sen:1983xv}. Similarly, the reggeization of the gluon was first shown by Tyburski \cite{Tyburski:1976mr}, Frankfurt and Sherman \cite{Frankfurt:1975}, and Lipatov \cite{Lipatov:1976zz}. 

In order to arrive at a behavior described by (\ref{eq:Malph2}), we expand $\M$ according to

\begin{eqnarray}
\M\; \propto \;s^{\al(t)}\;=&\exp(\al(t)\,\ln(s))\;=&\sum^{\infty}_{n=0}\,\frac{1}{n!}\,(\al(t)\,\ln(s))^{n}
\end{eqnarray}
and therefore
\begin{equation}
\M\;=\;\sum^{\infty}_{n}\,\M_{n}
\end{equation}
with $\M_{n}\,=\,\mathcal{O}((\al_{s}\ln(s))^{n})$.
Starting from perturbation theory, we will can relate $g$ (the coupling constant of the respective theory) to $\al(t)$; therefore, the expansion will be done in $(g^{2}\,\ln s)^{n}$.

We define the functions $\mathcal{F}_{n}$ via Mellin transformations of $\M_{n}$ such that

\begin{eqnarray}
\int\,Im\M_{n}\,\left(\frac{s}{k^{2}}\right)^{-\wg-1}\,d(\frac{s}{k^2})&\propto&\int \prod_{i}\,\frac{d^{2}k_{i}}{(k^{2}_{i}-m^{2})\,((k_{i}-q)^{2}-m^{2})}\mathcal{F}_{n}(\wg,k_{i},q)\;,\nonumber\\
& &  \label{eq:melrel}
\end{eqnarray}
where $k_{i}$ denote momenta and $m$ masses of internal exchange particles; the exact form of (\ref{eq:melrel}) as well as the meaning of $k_{i}$ depend on the case investigated.
If summing up $\mathcal{F}_{n}$ leads to a final expression according to

\begin{equation}
\mathcal{F}(\wg)\;=\;\sum_{n=0}^{\infty}\,\mathcal{F}_{n}(\wg)\;\propto\;\frac{1}{(\wg-\wg_{pole})^{\kappa}},\label{eq:Fmel}
\end{equation}
with $\kappa$ denoting the type of pole, we obtain

\begin{eqnarray*}
\M & \propto &  \left(\frac{s}{t}\right)^{\wg_{pole}}
\end{eqnarray*}
as required by (\ref{eq:Malph2}).

\begin{figure}
\centering
\includegraphics[width=0.5\textwidth]{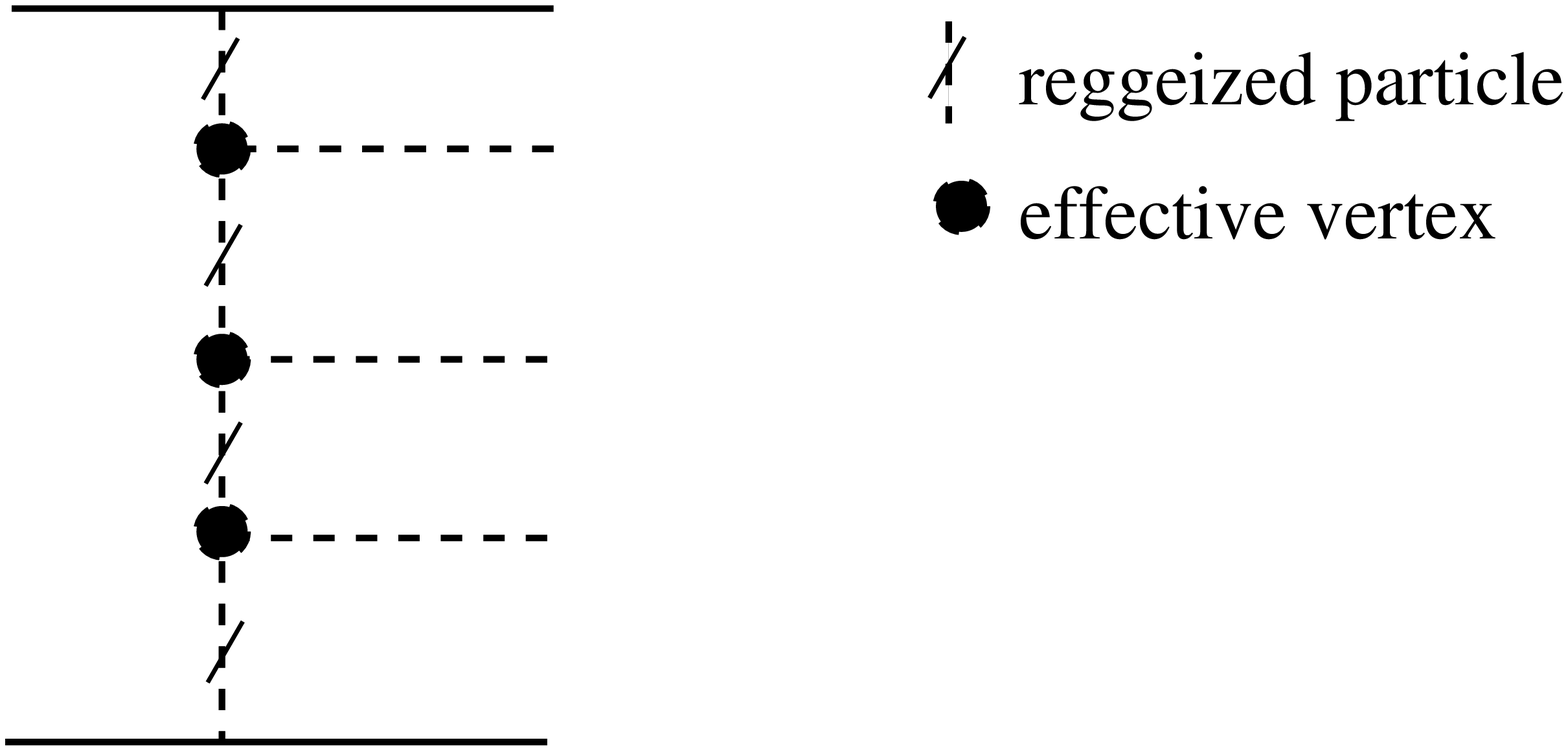}
\caption{3-runged ladder diagram for reggeized particle exchange}
\label{fig:ladder}
\end{figure}

\begin{figure}[b]
\centering
\psfrag{F}{$\mathcal{F}$}
\includegraphics[width=0.5\textwidth]{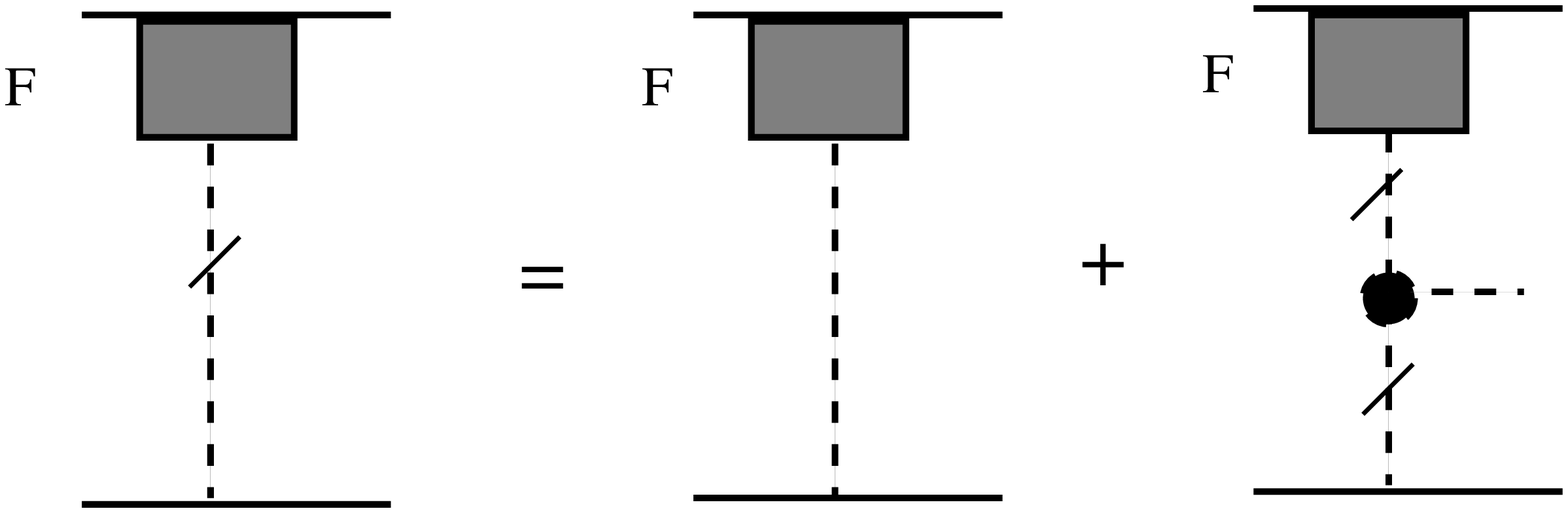}
\caption{symbolic form of integral equation corresponding to (\ref{eq:intequ})}
\label{fig:integr}
\end{figure}

One way of calculating this is to express the contributions to $\M_{n}$ by the use of $n^{th}$-rung ladder diagrams (see figure \ref{fig:ladder}) to which we can relate the imaginary part of $\M$ with the help of the Cutkosky rules (\ref{eq:cutr}). 
Non-laddertype contributions are included by defining effective vertices; this is e.g. the case in the reggeization of the gluon in QCD. Finally, calculating the  functions $f_{n}$ defined by the inverse Mellin transform of $\mathcal{F}_{n}$ such that

\begin{eqnarray}
Im\M_{n}&\propto&\int d^{2}k_{n}\,f_{n}(s,k_{n},k_{i},...)\label{eq:Mprop}
\end{eqnarray}
leads to an recursive integral equation  for $\mathcal{F}$. Symbolically, this is given by

\begin{eqnarray}
\mathcal{F}&=&\mbox{lowest order term}\,+\,(\mathcal{F}+\mbox{extra rung})\label{eq:intequ}
\end{eqnarray}
(see figure \ref{fig:integr}).

Analytically, (\ref{eq:intequ}) corresponds to 

\begin{equation}
\wg\,\mathcal{F}(\wg)\;=\;\mathcal{F}^{0}(\wg)+\mathcal{K}_{0}\bullet\mathcal{F}(\wg) \label{eq:intequ2}
\end{equation}
with $\mathcal{F}^{0}(\wg)$ denoting the propagation of a reggeized particle only and $\mathcal{K}_{0}$ the corresponding kernel. In derivation of the perturbative Pomeron, this is known as the BFKL equation; a similar relation called BKP equation holds for the derivation of the perturbative Odderon.

The formulas derived so far hold for the case of scattering off on-shell particles; in this case, $\mathcal{F}$ includes the couplings to the external particles. In the case of scattering including hadrons, (\ref{eq:melrel}) has to be modified to take the quark distribution inside the hadrons into account; this is done via the definition of impact factors. In this case, we obtain

\begin{eqnarray}
\lefteqn{\int\,Im\M_{n}\,(\frac{s}{k^{2}})^{-\wg-1}\,d(\frac{s}{k^2})\;\propto}& & \nonumber \\
 & & \int \prod_{i}\,\frac{d^{2}k_{i}}{(k^{2}_{i}-m^{2})\,((k_{i}-q)^{2}-m^{2})}\,\Phi^{A}(k_{i},q)\,\mathcal{F}_{n}(\wg,k_{i},q)\,\Phi^{B}(k_{i},q) \nonumber\\
& &  \label{eq:melrel2}
\end{eqnarray}
with $\Phi^{A}(k_{i},q)$ and $\Phi^{B}(k_{i},q)$ denoting the impact factors. For more details, see \cite{Forshaw:1997dc} and references therein. 

\subsection{The reggeized gluon}\label{sec:regglu} 

In \cite{Forshaw:1997dc}, the gluon in reggeized using the methods described in the previous section; effective vertices include the non-laddertype contributions to $\M_{n}$, while effective propagators corresponding to the reggeized gluon include higher order contributions; in fact, a reggeized propagator consists of an n-runged ladder which vertical parts consist of n-runged ladders etc. For$\mathcal{F} (\wg)$, we obtain

\begin{eqnarray}
\mathcal{F}(\wg)&=&\frac{\pi}{2}\,\frac{\al_{s}N q^{2}_{\bot}}{4\pi^{2}}\,\frac{1}{(\wg-\epsilon_{G}(-q^{2}_{\bot}))}\end{eqnarray}
with

\begin{eqnarray*}
N\,=\,3\,:\;\mbox{number of colors}&,&\al_{s}\,=\,\frac{g^{2}}{4\pi}\;,\\
\epsilon_{G}\;\propto\;\int\frac{d^{2}k_{\perp}}{k^{2}_{\perp}(k^{2}_{\perp}-q^{2}_{\perp})}\;. & &
\end{eqnarray*}
This leads to 

\begin{eqnarray}
\M&\propto&\frac{k^{2}_{\bot}}{t}\,(\frac{s}{k^{2}_{\bot}})^{\al_{G}(t)}\,\frac{1-e^{i\pi\al_{G}(t)}}{2}
\end{eqnarray}
corresponding to a Regge trajectory of odd signature. Here, $\al_{G}(t)\,=\,1\,+\,\epsilon_{G}(t)$ and $t\,=\,-q^{2}$.

\subsection{pQCD Pomeron}

With the help of the reggeized gluon from section \ref{sec:regglu}, and again closely following \cite{Forshaw:1997dc}, it is now easy to construct the  Pomeron from pQCD: the lowest order contribution simply corresponds to the exchange of two reggeized gluons, higher orders can successively be built from n-rung ladder diagrams. Here, the use of the modified effective propagators and vertices ensure again the inclusion of non-ladder-type diagrams and higher order contributions to the propagator. The integral equation (known as the BFKL equation, \cite{Kuraev:1976ge}, \cite{Balitsky:1978ic}) is now related to the propagation of two gluons representing the Pomeron. Summarizing, we obtain the following behavior

\begin{eqnarray}
\mathcal{F}(\wg,k_{1\,\bot},k_{2\,\bot},0)&\propto&\frac{1}{\sqrt{\wg-\wg_{0}}}\label{eq:fpom}\label{eq:Melpom}
\end{eqnarray}
with
\begin{displaymath}
\wg_{0}\;=\;4\,\frac{N\al_{s}}{\pi}\,\ln2\;.
\end{displaymath}
The calculation resulting in (\ref{eq:fpom}) is done for $q=0\,$, i.e. zero momentum transfer. However, it can be shown that calculations done for $q\neq 0$ lead to the same high-energy behavior. In total, we obtain an $\M$ dependence in the form of

\begin{eqnarray}
\M&\propto& s^{\al_{\pom}(t)}\label{eq:MpropP}
\end{eqnarray}
with $\al_{\pom}(t)=1+\wg_{0}$.

\vspace{3mm}
We can already distinguish the behavior of this hard Pomeron from the soft Pomeron considered in section \ref{sec:effpropsec}:

\begin{itemize}
\item $\al_{\pom}(t)\,$ as given by (\ref{eq:MpropP}) does not depend on $t$ as expected by Regge theory; furthermore, the intercept at $t=0$ is much too large to describe a Regge-type behavior of cross sections. The behavior can slightly be improved by taking into account the running of $\al_{s}(t)$, the strong coupling constant; however, these modifications do not suffice to deduce the soft Pomeron behavior.
\item The singularity of the Mellin transform given by (\ref{eq:Melpom}) is a cut rather than a pole; this can be fixed by taking running coupling into account.
\item As a consequence of the large intercept, unitarity corresponding to the Froissart-Martin bound seems to be violated for small values of $s$.
\end{itemize}

Therefore, we conclude that the hard Pomeron from pQCD cannot describe the soft-Pomeron effects of hadron-hadron scattering.

\subsection{pQCD Odderon}

Analogously to the Pomeron, the Odderon can be described in QCD as the exchange of three reggeized gluons. Bartels \cite{Bartels:1980pe} and Kwiecinski et al \cite{Kwiecinski:1980wb} suggested an integral equation corresponding to (\ref{eq:intequ}) describing the exchange of $n$ (Bartels) and 3 gluons (K and L) respectively as early as 1980. In both cases, the authors build the 3-gluon exchange out of reggeized gluons. Effective 3- and 4-reggeon vertices are defined. The integral-equation similar to (\ref{eq:intequ2}) known as the BKP equation then relates the equivalent of $\mathcal{F}$ as given in (\ref{eq:Fmel}) to the lowest order state + extra contributions; however, the ``extra rung'' piece has to be summed over possible interactions between two of the three gluons, respectively. The two-gluon interactions are again described by the BFKL equation derived for the Pomeron case.

The intercept and therefore the pole of the reggeized Odderon propagator has been studied extensively by several authors (see e.g. \cite{Braun:1998mg}, \cite{Janik:1999ae}, \cite{Bartels:1999yt}); recent results give $\al(0)\approx\,0.96\,$ \cite{Janik:1999ae}  and $\al(0)\approx 1$ \cite{Bartels:1999yt}; however, the two values have been calculated using different techniques and confirmation or exclusion from experiment remains open.


\chapter[{\normalsize The $\pi^{0}-\gamma\gamma\,$ and $\pi^{0}-\gamma^{*}\gamma\,$ coupling}]{\LARGE The $\pi^{0}-\gamma\gamma\,$ and $\pi^{0}-\gamma^{*}\gamma\,$ coupling}\label{sec:anocoup}

\section{Noether currents in field theory}\label{sec:noethcurr}

As the $\pi^{0}-\gamma\gamma\,$ coupling is mainly described by a model based on partially conserved axial current, we give a short introduction to current description of electroweak interactions; for a detailed discussion, see e.g. \cite{deAlfaro:1973}.\\
We consider a field theory where the Lagrangian is given by 

\begin{eqnarray}
\mathcal{L}&=&\mathcal{L}(\phi_{i}(x),\partial_{\mu}\phi_{i}(x))
\end{eqnarray}
with $\phi_{i}(x)\,$ being the quantized fields and $x$ spacetime 4-vectors.
We now consider an infinitesimal transformation of the fields: 

\begin{eqnarray}
\phi_{i}(x)&\rightarrow&\phi'_{i}(x)\,=\,\phi_{i}(x)+\delta\phi_{i}(x),\nonumber\\
\delta\phi_{i}(x)&=&i\vare^{a}t^{a}_{ij}\phi_{j}(x)\label{eq:lietra}
\end{eqnarray}
where $t^{a}_{ij}$ are the generators of the corresponding group of transformations obeying

\begin{equation}
[t^{a},t^{b}]\;=\;iC^{abc}t^{c}\;.\label{eq:Lieal}
\end{equation}
$C^{abc}\,$ are the structure constants of the group. A group of transformations where the group elements $U(\vare^{a})$ can be described by infinitesimal variations around unity as in (\ref{eq:lietra}) is called Lie-group; we can express the corresponding transformations according to 

\begin{eqnarray*}
\phi'_{i}(x)\,=&U(\vare^{a})\,\phi_{i}(x)&=\,e^{i\vare^{a}t^{a}_{ij}}\,\phi_{j}(x),
\end{eqnarray*}
i.e.

\begin{displaymath}
U(\vare)\;=\;e^{i\vare^{a}\,t^{a}_{ij}}\;.
\end{displaymath}
The algebra (\ref{eq:Lieal}) is called Lie-algebra. 

We can now calculate $\delta\Lc(\phi_{i}(x),\partial_{\mu}\phi_{i}(x))\,$ resulting from an infinitesimal transformation of $\phi_{i}\,$:

\begin{eqnarray}
\delta\Lc (\phi_{i}(x),\partial_{\mu}\phi_{i}(x))\;=  &\vare^{a} \partial_{\mu}\left[\frac{\delta\Lc}{\delta(\partial_{\mu}\phi_{i})}\,it^{a}_{ij}\phi_{j}\right]& =\vare^{a}\partial_{\mu}J^{\mu\,,a}\;,
\end{eqnarray}
where we defined the current $J^{\mu\,,a}$ according to 

\begin{eqnarray}
J^{\mu\,,a}&\equiv& i\frac{\delta\Lc}{\delta(\partial_{\mu}\phi_{i})}\,t^{a}_{ij}\phi_{j}\;.
\end{eqnarray}
If now $\delta\Lc\,=\,0\,$ under the transformation, $\partial_{\mu}J^{\mu\,,a}\,=\,0\,$, i.e. $J^{\mu\,a}$ is conserved. We obtain corresponding constant charges $Q^{a}$ with

\begin{eqnarray}
Q^{a}(x^{0})&=&\int d^{3}x\,J^{a}_{0}(x)\label{eq:QNoet}
\end{eqnarray}
which can be shown to obey the Lie-algebra (\ref{eq:Lieal}) and can therefore be identified with a generator of the Lie-group. Even in the case of unconserved currents, the charges $Q^{a}$ still obey (\ref{eq:Lieal}) if taken at the same time $x^{0}$.

\vspace{3mm}

In the $V-A$ theory of weak interactions\footnote{The $V-A$ theory is nowadays embedded in the standard model and represented by te exchange of $W$ and $Z$ gauge bosons.}, the Lagrangians are expressed in terms of conserved or unconserved Noether currents; for a more detailed discussion, see e.g. \cite{deAlfaro:1973}, \cite{Halzen:1984mc}. For different interactions, we obtain:

\begin{itemize}
\item{Electromagnetic coupling to leptons}\\

The Lagrangian of a free leptonic theory is given by

\begin{equation}
\Lc_{0}\;=\bar{\psi}(i\gamma_{\mu}\partial^{\mu}-m)\psi
\end{equation}
where $\psi\,$ denotes the fermion-field and $m\,$ the associated mass (see appendix \ref{app:standnot} for notational conventions). From gauge invariance of the electromagnetic field, we can perform the minimal substitution $\partial_{\mu}\,\rightarrow\,\partial_{\mu}+ ieA_{\mu}(x)$, thereby introducing the interaction Lagrangian

\begin{equation}
\Lc_{em}\;=-\,e\,\bar{\psi}(x)\,\gamma_{\mu}\psi(x)\,A^{\mu}(x)\;.\label{eq:Lem}
\end{equation}
$\Lc_{tot}=\Lc_{0}+\Lc_{em}$ is invariant under 

\begin{eqnarray}
\psi(x) & \rightarrow & e^{-ie\vare(x)}\,\psi(x)\nonumber\\
A_{\mu}(x) & \rightarrow & A_{\mu}(x)+\partial_{\mu}\vare(x)\label{eq:transem}
\end{eqnarray}
leading to the conserved current

\begin{equation}
 j^{(em)}_{\mu}(x)\;=\;e\,\bar{\psi}(x)\,\gamma_{\mu}\,\psi(x)\label{eq:curem}
\end{equation} 
and the corresponding charge operator $Q$ given by (\ref{eq:QNoet}). We can now rewrite (\ref{eq:Lem}):

\begin{equation}
\Lc_{em}\;=-\,j^{(em)}_{\mu}(x)\, A^{\mu}(x)\;.\label{eq:Lem2}
\end{equation}

\item{purely leptonic interactions}\\

Motivated by (\ref{eq:Lem2}), the lepton-lepton interaction Lagrangian is given by

\begin{eqnarray}
\Lc_{w}&=&-\frac{G}{\sqrt{2}}\,l^{\mu\,\dagger}(x)l_{\mu}(x),\nonumber\\
l^{\mu}(x)&=&\bar{\psi}_{e}(x)\gamma^{\mu}(1-\gamma^{5})\psi_{\nu_{e}}(x)\,+\,
\bar{\psi}_{\mu}(x)\gamma^{\mu}(1-\gamma^{5})\psi_{\nu_{\mu}}(x)\;.\nonumber\\ \label{eq:llint}
\end{eqnarray}
$G\,=1.03\,\times\,10^{-5}m^{-2}_{p}\,$ is Fermis constant. From experimental observations it has been seen that only the left-handed parts of the leptonic currents contribute. We split $l^{\mu}(x)\,$ into vector and axial vectors parts corresponding to their properties under Lorentz-transformations:

\begin{eqnarray}
l^{\mu}(x)&=&l^{\mu}_{V}(x)+l^{\mu}_{A}(x)\nonumber\\
&= & \sum_{i}\bar{\psi}_{i}(x)\gamma^{\mu}\psi_{\nu_{i}}(x)\,-\sum_{i}\bar{\psi}_{i}(x)\gamma^{\mu}\gamma^{5}\psi_{\nu_{i}}(x)\label{eq:splitl}
\end{eqnarray}
with $i=\{e,\mu\}\,$.

We now consider two different kinds of transformations and the associated currents and charges:

\begin{enumerate}

\item

\begin{eqnarray}
j^{\mu}_{i}&=&\bar{\psi}_{i}\gamma^{\mu}\psi_{i}+\bar{\psi}_{\nu_{i}}\gamma^{\mu}\psi_{\nu_{i}}\label{eq:curlep}
\end{eqnarray}

from the transformation
\begin{eqnarray}
\binom{\psi_{i}}{\psi_{\nu_{i}}}&\rightarrow&e^{-i\vare}\,\binom{\psi_{i}}{\psi_{\nu_{i}}}\;;
\end{eqnarray}


\begin{displaymath}
Q^{i}\;=\;(N_{i}+N_{\nu_{i}})-(N_{\bar{i}}+N_{\bar{\nu}_{i}})
\end{displaymath}
corresponds to the conserved lepton number.

\item 
We can introduce an isospin-like formalism, denoting

\begin{eqnarray}
\Psi^{i}&=&\binom{\psi_{i}}{\psi_{\nu_{i}}}\;.\label{eq:isolep}
\end{eqnarray}
In this formalism, the currents $\bar{\Psi}^{i}\,\gamma_{\mu}\,\Psi^{i}$ corresponding to the transformation (\ref{eq:transem}) are conserved, while the currents corresponding to the vector and axial currents in the isospin formalism,

\begin{eqnarray}
l^{\al}_{V,\mu}&=&\sum_{i}\bar{\Psi}^{i}\gamma_{\mu}\frac{1}{2}\tau^{\al}\,\Psi^{i}\nonumber\\
l^{\al}_{A,\mu}&=&\sum_{i}\bar{\Psi}^{i}\gamma_{\mu}\gamma_{5}\frac{1}{2}\tau^{\al}\,\Psi^{i}\nonumber\
\end{eqnarray}
are only conserved in the limit $m_{e}\,=m_{\mu}\,\rightarrow 0$. Actually, it can be shown that the violation of current conservation is proportional to $m_{i}$.
\end{enumerate}

\item{quarks and associated currents}

 We can define quark currents similar to (\ref{eq:curlep}), i.e. consider conserved currents of the form 

\begin{equation}
j^{i}_{\mu}\;=\;\bar{\psi}^{i}\gamma_{\mu}\psi^{i}
\end{equation}

the corresponding conserved charges are the quark numbers; $i$ denotes different types of quarks; we limit our considerations to $u$, $d$ and $s$ quarks.

In addition, we define vector and axial vector currents corresponding to the transformations under the SU(3)-flavor group; they are given by

\begin{eqnarray}
V^{a}_{\mu}&=&\bar{\psi}_{q}\gamma_{\mu}\frac{1}{2}\lambda^{a}\psi_{q},\nonumber\\
A^{a}_{\mu}&=&\bar{\psi}_{q}\gamma_{\mu}\gamma_{5}\frac{1}{2}\lambda^{a}\psi_{q}.\label{eq:curquar}
\end{eqnarray}
where

\begin{displaymath}
 \psi_{q}\,=\,\left(\begin{array}{c}u\\d\\s\end{array}\right)\,;
\end{displaymath}
$\lambda^{a}$ are the Gell-Mann matrices.

The vector current is conserved if all quark-masses are equal; for conservation of the axial vector current, we additionally have to require $m_{q}=0$.

A closed Lie-algebra corresponding to (\ref{eq:Lieal}) for the axial currents can only be constructed by combining the charges associated with $V^{a}_{\mu}\,$ and $A^{a}_{\mu}$

\item{electromagnetic coupling to hadrons}

The substitutions of the form $\partial_{\mu}\,\rightarrow\,\partial_{\mu}+ie_{q}\,A_{\mu}\,$ leading to (\ref{eq:Lem}) in the leptonic case correspond to 

\begin{equation}
\partial_{\mu}\,\rightarrow\,\partial_{\mu}+ie\frac{1}{2}\left(\lambda^{3}+\frac{1}{\sqrt{3}}\lambda^{8}\right)A_{\mu}
\end{equation}

for the single quarks; the current describing the electromagnetic coupling of hadrons is then given by

\begin{eqnarray}
J^{em}_{\mu}&=&V^{3}_{\mu}+\frac{1}{2}\,V^{Y}_{\mu}\nonumber\\
V^{Y}_{\mu}&=&\frac{1}{3}V^{8}_{\mu}
\end{eqnarray}

and $V^{i}_{\mu}$ given by (\ref{eq:curquar}).

Therefore, we obtain

\begin{equation}
\Lc^{(h)}_{em}\;=\;-e\,J^{em}_{\mu}A^{\mu}\;.
\end{equation}

\item{hadron-lepton interactions}

Similarly to the current describing electromagnetic interactions, currents corresponding to hadron-lepton interactions can be constructed using the Gell-Mann matrices of the SU(3)$_{flavour}$ group.

Basically, the hadronic currents are divided into parts conserving and parts violating strangeness $S$; both currents are split into vector and axial vector parts. The distinction between $\Delta S=0$ and $\Delta S\neq0$ is done by the Cabibbo-angle $\theta_{C}$; in general, the derivation follows the argument given above, so we just quote the results:

\begin{eqnarray}
J^{(h)}_{\mu}&=&J^{\Delta S=0}_{\mu}\,\cos\theta_{C}+J^{\Delta S\neq0}_{\mu}\,\sin\theta_{C}\;,\nonumber\\
J^{\Delta S=0}_{\mu}&=&V^{\Delta S=0}_{\mu}-A^{\Delta S=0}_{\mu}\;,\nonumber\\
J^{\Delta S\neq0}_{\mu}&=&V^{\Delta S\neq0}_{\mu}-A^{\Delta S\neq0}_{\mu}\;.\nonumber\\
\end{eqnarray}

In terms of the vector and axial vectors (\ref{eq:curquar})  associated with the Gell-Mann matrices, we obtain

\begin{eqnarray}
V^{\Delta S=0}_{\mu}&=&V^{1}_{\mu}+i\,V^{2}_{\mu}\;,\nonumber\\
V^{\Delta S\neq0}_{\mu}&=&V^{4}_{\mu}+i\,V^{5}_{\mu}\;,\nonumber\\
\end{eqnarray}

we get the corresponding expressions for the axial vector currents $A_{\mu}\,$ by substituting $\gamma_{\mu}\,\rightarrow\,\gamma_{\mu}\gamma_{5}$.

Summing everything up, $J^{(hl)}_{\mu}$ can be written in the form

\begin{equation}
J^{(h)}_{\mu}\;=\;\bar{\psi}_{u}\gamma_{\mu}(1-\gamma_{5})\psi_{d}\cos\theta_{C}+\bar{\psi}_{u}\gamma_{\mu}(1-\gamma_{5})\psi_{s}\sin\theta_{C}\;. \label{eq:hcur}
\end{equation}
The hadron-lepton interaction Lagrangian is then given by

\begin{equation}
\Lc^{(hl)}\;=\;-\frac{G}{\sqrt{2}}\,J^{(h)\,\mu}\,l^{\dagger}_{\mu}\,+h.c.\;.
\end{equation}
The Cabbibo angle $\theta_{C}$ can be determined from the Cabbibo-Kobayashi-Maskara matrix; \cite{Hagiwara:2002pw} gives limits according to $0.219\,\leq\,\cos\theta_{C}\,\leq\,0.226$.
\item{hadron-hadron interactions}\\
Following (\ref{eq:llint}), we can express purely hadronic interactions by 

\begin{equation}
\Lc_{hw}\;=\;-\frac{G}{\sqrt{2}}\,J^{(h)\,\mu\,\dagger}(x)\,J^{(h)}_{\mu}(x)
\end{equation}
with $J^{(h)\,\mu}$ given by (\ref{eq:hcur}).
\end{itemize}

\section{Partially conserved axial current}
The assumption of the partially conserved axial current was derived in an $SU_{L}(2)\;\times\; SU_{R}(2)\; \sigma$ model suggested by Gell-Mann and Levy \cite{Gell-Mann:1960np}. They consider a system consisting of an isospin doublet of massless fermions, a triplet of pseudoscalar pions, and a scalar field. The symmetry of the system is spontaneously broken and the pions are derived as massless Golstone bosons. The pion mass is then generated by putting $m_{f}\,\neq\,0$. In this model,

\begin{eqnarray}
\langle 0|A^{k}_{\mu}(x)|\pi^{j}\rangle &=&i\,\delta_{jk}f_{\pi}p_{\mu}e^{-ixp}\;,
\end{eqnarray}
where $A^{k}_{\mu}(x)$ is the axial vector current associated with an axial $SU(2)$ transformation; $p_{\mu}$ is the momentum-vector of the pion and $f_{\pi}\,$  the pion decay constant (see e.g. \cite{Itzykson:1980rh} for details). 
Taking the derivative, we now obtain
\begin{eqnarray}
\langle 0|\partial^{\mu} A^{k}_{\mu}(x)|\pi^{j}\rangle&=&\delta_{jk}f_{\pi}m^{2}_{\pi}\,e^{-ixp}\;.\label{eq:derpcac}
\end{eqnarray}
If we define

\begin{equation}
\phi^{k}_{\pi}(x)\;=\;\frac{1}{m^{2}_{\pi}\,f_{\pi}}\,\partial^{\mu}A^{k}_{\mu}(x)\label{eq:pcac}
\end{equation}
and combine this with (\ref{eq:derpcac}), we see that we can associate $\phi^{k}_{\pi}(x)$ with the pion field operator, as

\begin{equation}
\langle 0|\phi^{k}_{\pi}(x)|\pi^{j}\rangle\;=\;\delta_{jk}\,e^{-ixp}\;.
\end{equation}
The generalization of (\ref{eq:pcac}), 
\begin{equation}
\partial_{\mu}A^{\mu}\;\propto\;m^{2}\;,
\end{equation} 
 is called the hypothesis of the partially conserved axial current. 
It was first suggested by Nambu \cite{Nambu:1960xd}, Chou \cite{Chou:1961}, and Gell-Mann and Levy \cite{Gell-Mann:1960np}.
The decay constant $f_{\pi}$ was approximately determined as $f_{\pi}\,=\,93\; \mbox{MeV}$ in the process $\pi\rightarrow\mu\,\nu_{\mu}$.

\section{The axial anomaly}

\begin{fmffile}{fdan}
\begin{figure}
\centering
\subfigure[]{
\fmfframe(10,15)(15,10){
\begin{fmfgraph*}(120,50)
\fmfleftn{i}{1}
\fmfrightn{o}{2}
\fmf{scalar,label=$k_{1}+k_{2}$ }{i1,v1}
\fmf{fermion,label=$p+k_{1}$,l.side=left }{v1,v2}
\fmf{fermion,label=$p$}{v2,v3}
\fmf{fermion,label=$p-k_{2}$,l.side=left}{v3,v1}
\fmf{photon,label=$k_{1}\al$}{v2,o2}
\fmf{photon,label=$k_{2}\be$}{v3,o1}
\fmfforce{(0.3w,0.5h)}{v1}
\fmfforce{(0.7w,1.h)}{v2}
\fmfforce{(0.7w,0.h)}{v3}
\end{fmfgraph*}}}
\subfigure[]{
\fmfframe(10,15)(15,10){
\begin{fmfgraph*}(120,50)
\fmfleftn{i}{1}
\fmfrightn{o}{2}
\fmf{scalar,label=$k_{1}+k_{2}$ }{i1,v1}
\fmf{fermion,label=$p+k_{2}$,l.side=left }{v1,v2}
\fmf{fermion,label=$p$}{v2,v3}
\fmf{fermion,label=$p-k_{1}$,l.side=left}{v3,v1}
\fmf{photon,label=$k_{2}\be$}{v2,o2}
\fmf{photon,label=$k_{1}\al$}{v3,o1}
\fmfforce{(0.3w,0.5h)}{v1}
\fmfforce{(0.7w,1.h)}{v2}
\fmfforce{(0.7w,0.h)}{v3}
\end{fmfgraph*}}}
\caption{Diagrams contributing to triangle anomaly}
\label{fig:trian}
\end{figure}
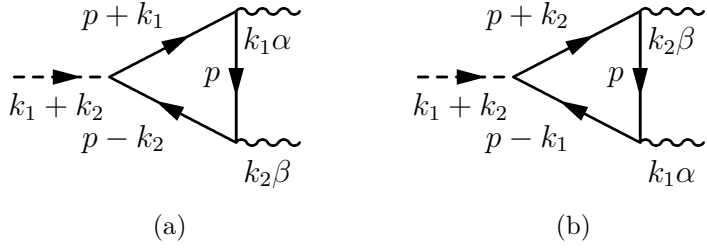
\end{fmffile}

We consider a process described by a triangle diagram figure \ref{fig:trian}, where two vector currents and an axial vector current couple in lowest order in perturbation theory. We will consider the matrix element

\begin{eqnarray}
\int d^{4}x\,e^{-iqx}\,\langle k_{1}k_{2}|\,A_{\mu}(x)|0\rangle&=&(2\pi)^{4}\,\delta^{(4)}(q-k_{1}-k_{2})\,(-ie)^{2}\times\nonumber\\
& & \epsilon^{*\,\al}(k_{1})\epsilon^{*\,\be}(k_{2})t_{\al\be\mu}(k_{1}k_{2})\;.\nonumber\\ \label{eq:axrel}
\end{eqnarray} 
The Feynman-rules for $t_{\al\be\mu}$ are given by figure \ref{fig:trian}:

\begin{eqnarray}
t_{\al\be\mu}(k_{1},k_{2})&=&i\,\int\frac{d^{4}p}{(2\pi)^{4}}\,Tr\left[\frac{1}{p\mkern-8mu/-k_{2}\mkern-16mu/-m}\,\gamma_{\be}\,\frac{1}{p\mkern-8mu/-m}\,\gamma_{\al}\,\frac{1}{p\mkern-8mu/+k_{1}\mkern-16mu/-m}\,\gamma_{\mu}\,\gamma_{5} \right]\nonumber\\
 & + & \left\{\begin{array}{ccc}\al&\longleftrightarrow&\be\\
                                 k_{1}&\longleftrightarrow&k_{2}\\
\end{array}\right\}\;.
\end{eqnarray}
Additionally, from the equation of motion for Heisenberg fields, we obtain

\begin{equation}
\partial_{\mu}A^{\mu}(x)\;=\;2\,i\,m_{A}\,P(x) \label{eq:axnaiv}
\end{equation}
with

\begin{equation}
P(x)\;=\;\bar{\psi}(x)\,\gamma_{5}\,\psi(x)
\end{equation}
and $m_{A}$ denoting the mass of the particle associated with the axial vector current. (\ref{eq:axnaiv}) corresponds to conservation of the axial current for $m_{A}\,\rightarrow\,0$.

Naively, we expect the following Ward-identities to hold (for a derivation, see e.g. \cite{Bertlmann:1996xk}) :

\begin{eqnarray}
k^{\al}_{1}\,t_{\al\be\mu}(k_{1},k_{2})\,=&k^{\be}_{2}\,t_{\al\be\mu}(k_{1},k_{2})\;=0&\;\;\mbox{(Vector Ward Identity)},\nonumber\\
(k_{1}+k_{2})^{\mu}t_{\al\be\mu}=&2\,m_{A}\,\nu_{\al\be}(k_{1},k_{2})&\;\;\mbox{(Axial Ward Identity)}\nonumber\\
& &\label{eq:wardids}
\end{eqnarray}
with

\begin{eqnarray*}
\nu_{\al\be}(k_{1},k_{2})&=&i\,\int\frac{d^{4}p}{2\pi^{4}}\,Tr\left[\frac{1}{p\mkern-8mu/-k_{2}\mkern-16mu/-m}\,\gamma_{\be}\,\frac{1}{p\mkern-8mu/-m}\,\gamma_{\al}\,\frac{1}{p\mkern-8mu/+k_{1}\mkern-16mu/-m}\,\gamma_{\mu}\, \right]\nonumber\\
 & + & \left\{\begin{array}{ccc}\al&\longleftrightarrow&\be\\
                                 k_{1}&\longleftrightarrow&k_{2}\\
\end{array}\right\}\;.
\end{eqnarray*}

However, when calculating the Feynman diagrams, we see that the Ward identities (\ref{eq:wardids}) cannot be fulfilled simultaneously\footnote{Actually, the fulfillment of (\ref{eq:wardids}) depends on the choice of regularization used for the calculation of the Feynman diagrams given in figure \ref{fig:trian}. While a naive linear shift of the integration parameter as well as a Pauli-Villars regularization violates the Vector Ward Identities, a dimensional regularization only implies a modification of the Axial Ward Identity. However, the axial anomaly given by (\ref{eq:axc2}) is independent of the regularization method if we require the Ward identities for the photon vertices to hold. For further discussion, see \cite{Bertlmann:1996xk}. Here, we pursue the method of simple integral shifting.}. If we require the vector Ward identities to hold, we have to modify $t_{\al\be\mu}(k_{1},k_{2})$ according to 
\begin{eqnarray}
t_{\al\be\mu}(k_{1},k_{2})\;\longrightarrow\;T_{\al\be\mu}(k_{1},k_{2})&=&{t}_{\al\be\mu}(k_{1},k_{2})-\frac{i}{8\pi^{2}}\,(k_{1}-k_{2})^{\nu}\,\vare_{\al\be\mu\nu}\;.\nonumber\\
 & &
\end{eqnarray}
We then obtain the anomalous axial Ward identity
\begin{equation}
q^{\mu}T_{\al\be\mu}(k_{1},k_{2})\;=\;2\,m_{A}\,\nu_{\al\be}-\frac{i}{2\pi^{2}}\,\vare_{\al\be\lambda\nu}\,k^{\lambda}_{1}\,k^{\nu}_{2}\;.\label{eq:axc2}
\end{equation} 
This also implies a modification of (\ref{eq:axnaiv}):

\begin{eqnarray}
\partial_{\mu}A^{\mu}&=&2\,i\,m_{A}\,P\;+\frac{e^{2}}{(4\pi)^{2}}\,\vare^{\al\be\rho\sigma}\,F_{\al\be}(x)F_{\rho\sigma}(x) \label{eq:modan}
\end{eqnarray}
with $F_{\mu\nu}(x)\,$ being the usual electromagnetic field strength tensor. The considerations above hold in the simple Abelian case where

\begin{equation}
A_{\mu}(x)\;=\;\bar{\psi}(x)\gamma_{5}\gamma_{\mu}\psi(x)\;;
\end{equation}
for non-Abelian cases, (\ref{eq:axc2}) has to be modified according to

\begin{eqnarray}
q^{\mu}\,T^{abc}_{\al\be\mu}(k_{1},k_{2})&=&2\,m_{A}\,\nu_{\al\be}-\frac{i}{2\pi^{2}}\,\vare^{abc}_{\al\be\lambda\nu}\,k^{\lambda}_{1}\,k^{\nu}_{2}\,\,D^{abc} \label{eq:axc3}
\end{eqnarray} 
with
\begin{equation}
D^{abc}\;=\;\frac{1}{2}\,Tr \left(\left\{K^{a},K^{b}\right\}K^{c}\right)\;.\label{eq:Dmod}
\end{equation}
The matrices $K^{a}$ reflect the internal symmetry and are related to the currents by
\begin{eqnarray}
V^{a}_{\mu}(x)&=&\bar{\psi}(x)\,K^{a}\,\gamma_{\mu}\,\psi(x),\nonumber\\
A^{a}_{\mu}(x)&=&\bar{\psi}(x)\,K^{a}\,\gamma_{\mu}\,\gamma_{5}\,\psi(x)\;.
\end{eqnarray}
In the chiral limit $m_{A}\,\rightarrow\,0$, we see from (\ref{eq:axc2}) to (\ref{eq:axc3}) that $q_{\mu}T^{\al\be\mu}$ and $d^{\mu}A_{\mu}$ are dominated by the anomaly.

\section{The processes $\pi^{0} \rightarrow \gamma\gamma$ and $\pi^{0} \rightarrow \gamma\gamma^{*}$ }\label{sec:realver}

For the process $\pi^{0} \rightarrow \gamma\gamma$, we consider the matrix element

\begin{eqnarray}
\langle \gamma(k_{1})\,\gamma(k_{2})|\pi^{0}(q)\rangle &=&i\int d^{4}x\,e^{-iqx}(m^{2}-q^{2})\,\langle \gamma(k_{1})\,\gamma(k_{2})|\phi_{\pi}(x)|0\rangle \nonumber\\
 & = & (2\pi)^{4}\,\delta^{(4)}(q-k_{1}-k_{2})\,\vare^{*\,\al}(k_{1})\vare^{*\,\be}(k_{2})\Gamma_{\al\be}(k_{1},k_{2})\;.\nonumber\\
\end{eqnarray}
Comparing this with (\ref{eq:axrel}), and taking into account PCAC as given by (\ref{eq:pcac}), we naively expect

\begin{eqnarray}
\int d^{4}x\,\partial^{\mu}e^{-iqx}\,\langle\gamma(k_{1})\,\gamma(k_{2})|A_{\mu}(x)|0\rangle & =
  & m^{2}_{\pi}\,f_{\pi}\int d^{4}x\,e^{-iqx}\langle\gamma(k_{1})\,\gamma(k_{2})|\phi_{\pi}(x)|0\rangle\nonumber\\
& & 
\end{eqnarray}
and therefore

\begin{eqnarray}
\lefteqn{(2\pi)^{4}\,\delta^{(4)}(q-k_{1}-k_{2})\,\vare^{*\,\al}(k_{1})\,\vare^{*\,\be}(k_{2})\,(-ie)^{2}\,q^{\mu}\,t_{\al\be\mu}(k_{1},k_{2})\;=\;} & &\nonumber\\
 & = & \frac{-i\,m^{2}\,f_{\pi}}{m^{2}-q^{2}}\,(2\pi)^{4}\,\delta^{(4)}(q-k_{1}-k_{2})\,\vare^{*\,\al}(k_{1})\vare^{*\,\be}(k_{2})\,\Gamma_{\al\be}(k_{1},k_{2})\nonumber \\
& &
\end{eqnarray} 
leading to

\begin{equation}
q^{\mu}t_{\al\be\mu}\;=\;\frac{-i\,m^{2}\,f_{\pi}}{(-ie)^{2}(m^{2}-q^{2})}\Gamma_{\al\be}(k_{1},k_{2})\;.\label{eq:naivrela}
\end{equation}
However, this would lead to $\langle \gamma(k_{1})\,\gamma(k_{2})|\pi^{0}(q)\rangle\,=\,0$ for $q^{\mu}\rightarrow\,0$. Adler \cite{Adler:1969gk} suggested that (\ref{eq:naivrela}) has to be modified according to the axial anomaly (\ref{eq:axc3}); then, we obtain

\begin{equation}
q^{\mu}T_{\al\be\mu}\;=\;\frac{-i\,m^{2}\,f_{\pi}}{(-ie)^{2}(m^{2}-q^{2})}\Gamma_{\al\be}(k_{1},k_{2})-\frac{i\,c}{2\pi^{2}}\epsilon_{\al\be\mu\nu}k^{\al}_{1}k^{\be}_{2}
\end{equation}
and therefore in the soft pion limit $q^{\mu}\,\rightarrow\,0$:

\begin{equation}
\Gamma_{\al\be}(k_{1},k_{2})\;=\;\frac{e^{2}\,c}{2\pi^{2}}\,\epsilon_{\al\be\mu\nu}k^{\al}_{1}k^{\be}_{2}\;.\label{eq:gammacoup}
\end{equation}
The constant $c$ reflects the quark-content of the $\pi^{0}$. In a simple quark model where

\begin{eqnarray}
j_{\mu}(x)&=&\bar{\psi}(x)\,\gamma_{\mu}\,Q\,\psi(x)\;,\nonumber\\A_{\mu}(x)&=&\bar{\psi}(x)\,\gamma_{\mu}\,\gamma_{5}\,\frac{\lambda^{3}}{2}\,\,\psi(x)\;,
\end{eqnarray}
with
\begin{eqnarray}
Q\;=\;\frac{1}{3}\left(\begin{array}{ccc} 2 & 0 & 0 \\ 0 & -1 & 0 \\ 0 & 0 & -1 \end{array}\right) &,& \lambda^{3}\;=\;\left(\begin{array}{ccc} 1 & 0 & 0 \\ 0 & -1 & 0 \\ 0 & 0 & 0 \end{array}\right),
\end{eqnarray}
$\psi\;=\;\left(\begin{array}{c}u\\d\\s \end{array} \right)$, and the inclusion of color, we can determine $c$ using (\ref{eq:Dmod}) as $c\,=\,\frac{1}{2}$ and therefore 

\begin{eqnarray}
\langle \gamma(k_{1})\gamma(k_{2})|\,\pi^{0}\rangle & = & \frac{e^{2}}{4\pi^{2}\,f_{\pi}}\,\vare_{\al\be\mu\nu}\,k^{\mu}_{1}k^{\nu}_{2}\,\epsilon^{*\,\al}(k_{1})\,\epsilon^{*\,\be}(k_{2})\, (2\pi)^{4}\,\delta^{(4)}(q-k_{1}-k_{2})\;.\nonumber\\ \label{eq:pigg}
\end{eqnarray}

For the process $\pi^{0}\,\rightarrow\,\gamma\,\gamma^{*}\,$, we consider the meson transition amplitude $F_{\gamma^{*}\gamma^{*}\pi}(-k^{2}_{1},-k^{2}_{2})$ defined by

\begin{eqnarray}
\lefteqn{\langle \gamma^{*}(k_{1})\gamma^{*}(k_{2})|\,\pi^{0}\rangle} & & \nonumber \\
 & = & e^{2}\,\vare_{\al\be\mu\nu}\,k^{\mu}_{1}k^{\nu}_{2}\,\epsilon^{*\,\al}(k_{1})\,\epsilon^{*\,\be}(k_{2})\,F_{\gamma^{*}\gamma^{*}\pi}(-k^{2}_{1},-k^{2}_{2})\,(2\pi)^{4}\,\delta^{(4)}(q-k_{1}-k_{2})\nonumber\\
& &
\end{eqnarray}
for off-shell photons, i.e. $k^{2}_{i}\,\neq\,0$.
For two on-shell photons, we immediately obtain from (\ref{eq:pigg})

\begin{equation}
F_{\gamma\pi}\,\equiv\,F_{\gamma^{*}\gamma^{*}\pi}(0,0)\,=\,\frac{1}{4\pi^{2}\,f_{\pi}}\;.
\end{equation}
On the other hand, pQCD gives for the case of large $Q^{2}_{i}\,=\,-k^{2}_{i}$ 
 
\begin{equation}
F_{\gamma^{*}\gamma^{*}\pi}(Q^{2}_{1},Q^{2}_{2})\,=\,\frac{\sqrt{2}}{3}\,\int^{1}_{0}\frac{\phi_{\pi}(x)\,dx}{xQ^{2}_{1}+(1-x)Q^{2}_{2}}
\end{equation}
with

\begin{displaymath}
\phi_{\pi}(x)\,=\,6\,f_{\pi}\,x(1-x)
\end{displaymath}
being the pion distribution amplitude \cite{Lepage:1979zb}, \cite{Lepage:1980fj}, \cite{Brodsky:1981rp}. 

For the case $k^{2}_{2}\,=\,0$, i.e. one on-shell photon, \cite{Brodsky:1981rp} gives an interpolating dipole-form for $F_{\gamma^{*}\gamma\pi}(Q^{2}_{1},0)$:

\begin{equation}
F_{\gamma^{*}\gamma\pi}(Q^{2}_{1},0)\;=\;\frac{1}{4\pi^{2}\,f_{\pi}}\,\frac{1}{1+Q^{2}_{1}/8\pi^{2}\,f^{2}_{\pi}}\;.\label{eq:fincouppgg}
\end{equation}
This formula has been confirmed by calculations using the constituent quark model \cite{Anisovich:1997hh} as well as QCD sum rules \cite{Radyushkin:1996pm}.

Values for other pseudoscalar mesons can be simply derived by substituting the respective decay constant $f_{VM}$ and quark content (see e.g. \cite{Kilian:1998ew}). Similarly, coupling to other vector mesons than the photons can be considered by substituting $e\,\rightarrow\,c_{VM}\,$ with $c_{VM}$ reflecting the coupling of the vector meson.


\chapter[{\normalsize Kinematics for 2 particle $\to$ 2 particle scattering processes}]{\LARGE Kinematics for 2 particle $\to$ 2 particle scattering processes}\label{chap:kin}

\section {General kinematics}

The general cross-section for $2 \to n$ particle processes is given by (\ref{eq:dsig1}):

\begin{equation}
d\sigma = \frac{1}{2w} \prod_{i=1}^n \frac{d^{3}k_{i}}{(2\pi)^{3} 2k^{0}_{i}} (2\pi)^{4} \delta^{4}(\sum_{i} k_{i} - p_{1} - p_{1}) \left|\cal M\right|^{2},\label{eq:dsig1b}
\end{equation}
with
\begin{eqnarray}
p_{1}, p_{2}& :&\;\;\mbox{4-momenta of ingoing particles,}\nonumber \\
 k_{i} &:&\;\;\mbox{4-momenta of outgoing particles,}\nonumber \\
w(s, m^{2}_{1},m^{2}_{2}) &= &\sqrt{(s-m_{1}^{2}-m_{2}^{2})^{2}-4\,m_{1}^{2}\,m_{2}^{2}},\nonumber \\
m_{1},m_{2}&:&\;\;\mbox{masses of ingoing particles},\nonumber \\
\left|\cal M\right|^{2}&:&\;\;\mbox{squared matrix element of the reaction}.\label{eq:kinpar}
\end{eqnarray}
For $n=2$, this reduces to

\begin{eqnarray}
d\sigma & = & \frac{1}{2w} \frac{d^{3}k_{1}}{(2\pi)^{3} 2k^{0}_{1}}\quad \frac{d^{3}k_{2}}{(2\pi)^{3}\,2\,k^{0}_{2}}\quad (2\pi)^{4} \delta^{4}( k_{1}+k_{2} - p_{1} - p_{2}) \left|\cal M\right|^{2} \nonumber  \\
& = &\frac{1}{2w}  \frac{d^{3}k_{1}\quad d^{3}k_{2} }{16\,\pi^{2}\, k^{0}_{1}\, k^{0}_{2}} \delta^{4}( k_{1}+k_{2} - p_{1} - p_{1}) \left|\cal M\right|^{2}.\label{eq:dsig2} 
\end{eqnarray}
Using

\begin{equation}
\frac {d^{3}k_{i}}{2k^{0}_{i}}=d^{4}k_{i}\, \delta(k^{2}_{i}-m'^{2}_{i})\;\Theta(k^{0}_{i})\label{eq:dp1}
\end{equation}
we obtain
\begin{equation}
d\sigma = \frac{1}{2w} \frac{d^{3}k_{1}}{ 2k^{0}_{1}}\;\frac{dk_{2}}{(2\pi)^{2}}\;\delta(k^{2}_{2}-m'^{2}_{2})\;\Theta(k^{0}_{2}) \delta^{4}( k_{1}+k_{2} - p_{1} - p_{2}) \left|\cal M\right|^{2}\label{eq:dsig3} 
\end{equation}
and integration over $d^{4}k_{2}$ gives

\begin{equation}
d\sigma = \frac{1}{2w} \frac{d^{3}k_{1}}{ 2k^{0}_{1}}\, \frac{1}{(2\pi)^{2}}\,\delta(k^{2}_{2}-m'^{2}_{2})\,\Theta(k^{0}_{2}) \left|\cal M\right|^{2} \label{eq:sig22}
\end{equation}
where $k^{\mu}_{2}=p^{\mu}_{1}+p^{\mu}_{2}-k^{\mu}_{1}$.
\vspace{5mm}

The total cross section $\sigma_{tot}$ derived from the integral over (\ref{eq:sig22}) corresponds to the inclusive cross section if the two outgoing particles are different. In case of identical particles, however, we have to take their indistinguishability into account; here, $\sigma_{tot}$ from (\ref{eq:sig22}) corresponds to the exclusive cross section. The inclusive total cross section is then given by

\begin{eqnarray}
\sigma_{inc}&=&\frac{1}{n_{i}\,!}\,\sigma_{exc}.\label{eq:siginc}
\end{eqnarray}
For $\gamma \gamma\,\rightarrow\,\pi_{0}\pi_{0}\,$,  $n_{i}\,=\,2$. 

\begin{figure}
\centering
\psfrag{p1}{$ p_{1}$}
\psfrag{p2}{$ p_{2}$}
\psfrag{k1}{$k_{1}$}
\psfrag{k2}{$k_{2}$}
\psfrag{ez}{${\scripts \hat{e}_{z}}$}
\includegraphics[ width=0.4\textwidth]{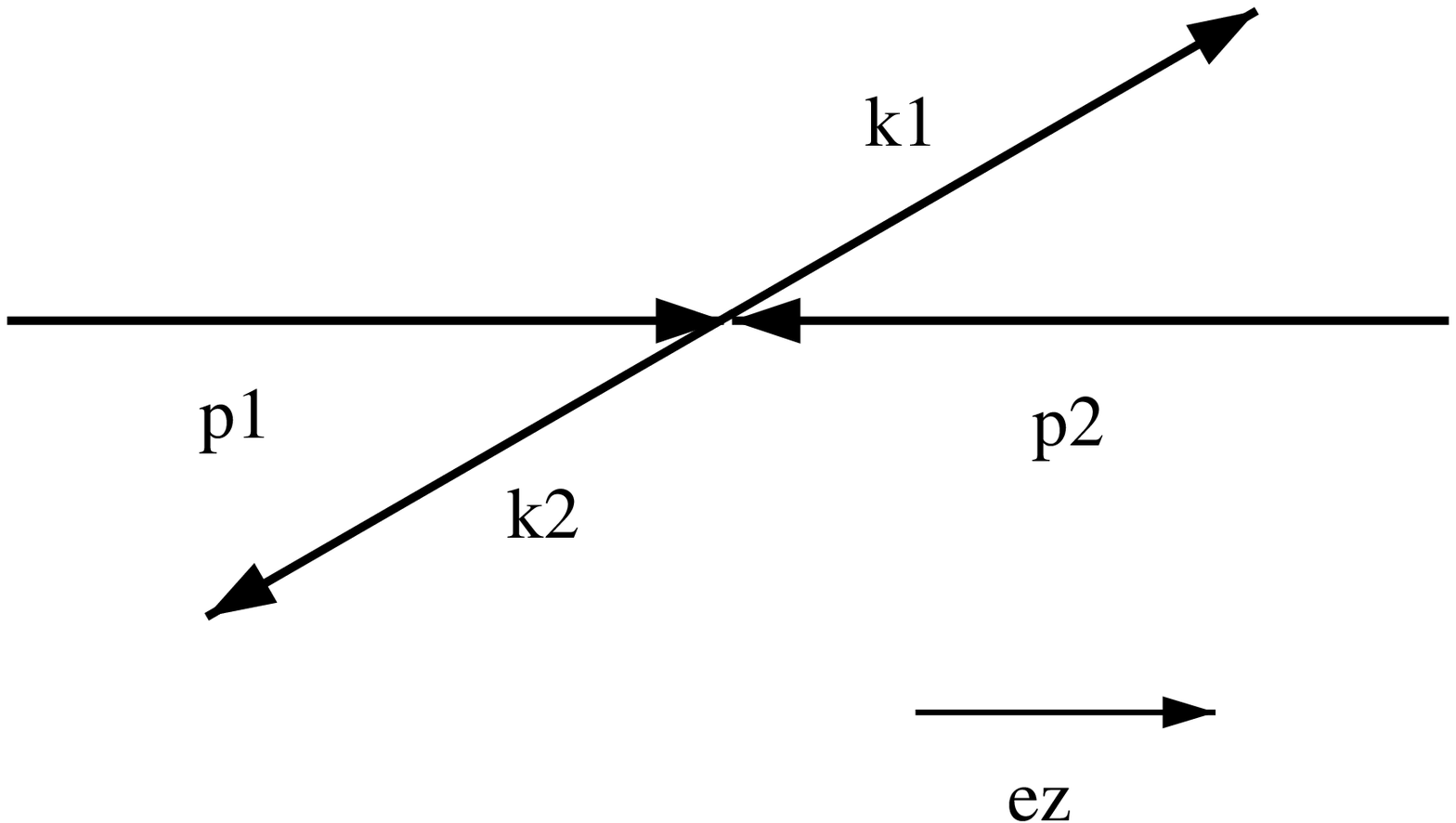}
\caption{$2\,\rightarrow\,2\,$ particle reaction in cm frame } 
\label{fig:2partkin}                                      
\end{figure}

\section{Kinematics for on-shell photons as incoming particles}

For on-shell photons, $q^{2}_{i}=m^{2}_{i}=0$; from (\ref{eq:kinpar}), we see that this implies $w=s$.\\

In the system described by figure \ref{fig:2partkin} with photons as incoming particles, we obtain

\begin{eqnarray}
q^{\mu}_{1}\equiv\,p^{\mu}_{1}=\left( \begin{array}{c}\omega_{1}\\0\\0\\ \omega_{1}\end{array} \right) &,&
q^{\mu}_{2}\equiv\,p^{\mu}_{2}=\left( \begin{array}{c}\omega_{2}\\0\\0\\-\omega_{2}\end{array} \right)\;, \nonumber\\
k^{\mu}_{1}=\left( \begin{array}{c}k^{0}_{1}\\ \vec{k}_t \\k_{1,l} \end{array} \right)&,&
k^{\mu}_{2}=q^{\mu}_{1}+q^{\mu}_{2}-k^{\mu}_{1}
\end{eqnarray}\label{eq:4vec}
for the photons and the outgoing particles respectively.
Furthermore,

\begin{eqnarray}
w = s & = & (q_{1}+q_{2})^{2}=4\,\wg_{1}\,\wg_{2},\nonumber\\
k^{0}_{2} &= & \wg_{1}+\wg_{2}-k^{0}_{1},\nonumber\\
k^{2}_{2} & = & (q_{1}+q_{2}-k_{1})^{2}\nonumber\\
& = & 2\, \big( 2\,\wg_{1}\,\wg_{2}-(\wg_{1}+\wg_{2})\,k^{0}_{1}+(\wg_{1}-\wg_{2})\,k_{1,l}\big)+m^{2}_{1} \label{eq:zvar}
\end{eqnarray}
where
\begin{displaymath}
k^{0}_{1}=\sqrt{m^{2}_{1}+\vec{k_{1}}^{2}}
\end{displaymath}
with $\vec{k_{1}}\;=\;\binom{\vec{k_{t}}}{k_{l}}$.

\subsection{Constraints due to the $\delta$ and $\Theta$-functions}\label{sec:constr}

From (\ref{eq:sig22}) we see that the kinematics of the explored system are restricted; in order to evaluate the effects, we consider two different cases:

\begin{itemize}
\item  Calculation of {\large$\frac {d\sigma}{dk_{t}}$}:

In this case, $k_{t}$ is kept as a free parameter. Therefore, we have to solve the $\delta$ and $\Theta$ function appearing in (\ref{eq:sig22}) in terms of $k_{t}$ using (\ref{eq:zvar}):\\
 
\begin{equation}
\delta(k^{2}_{2}-m^{2}_{2})  =  \left(\frac{1}{\left|f'(k_{l})\right|_{k_{l}=k_{l+}}}\,\delta\left(k_{l}-k_{l+}\right) + \frac{1}{\left|f'(k_{l})\right|_{k_{l}=k_{l-}}}\,\delta\left(k_{l}-k_{l-}\right) \right)
\end{equation}
with
\begin{eqnarray}
k_{l+} & = & \frac{1}{2}\,(\wg_{1}-\wg_{2})+\frac{(\wg_{1}+\wg_{2})\,\sqrt{\wg_{1}\,\wg_{2}-m^{2}-k^{2}_{t}}}{2\,\sqrt{\wg_{1}\,\wg_{2}}},\nonumber\\
k_{l-} & = & \frac{1}{2}\,(\wg_{1}-\wg_{2})-\frac{(\wg_{1}+\wg_{2})\,\sqrt{\wg_{1}\,\wg_{2}-m^{2}-k^{2}_{t}}}{2\,\sqrt{\wg_{1}\,\wg_{2}}},\nonumber\\
f'(k_{l}) & = &2\,\left( \wg_{1}-\wg_{2} - \frac{kl\,(\wg_{1}+\wg_{2})}{\sqrt{k^{2}_{l}+k^{2}_{t}+m^{2}}}\right)=\nonumber\\
 & = &2\,\left(\wg_{1}-\wg_{2} - \frac{kl\,(\wg_{1}+\wg_{2})}{k^{0}_{1}}\right).\label{eq:klval}
\end{eqnarray}
We used  $m'_{1} = m'_{2}\equiv\,m$ for $\gamma\,\gamma\to\,\pi_{0}\,\pi_{0}$. 

The equation above gives a restriction on $k_{t}$:

\begin{equation}
k^{2}_{t}\,\le\,\wg_{1}\,\wg_{2}-m^{2}.
\end{equation}

$\Theta(k^{0}_{2})$ leads to

\begin{equation}
k^{2}_{l}\,\le\,(\wg_{1}+\wg_{2})^{2}-m^{2}-k^{2}_{t}. \label{eq:klres}
\end{equation}

By going into the center of mass frame of the two photons, we can easily see that (\ref{eq:klres}), taken $k_{l}$ to be given according to (\ref{eq:klval}), is always fulfilled: $k_{l+/-}$ reduce to

\begin{eqnarray}
k_{l+} & = & \sqrt{\omega^{2}-m^{2}-k^{2}_{t}},\nonumber\\
k_{l-} & = & -\sqrt{\omega^{2}-m^{2}-k^{2}_{t}}\label{eq:klvalcmp}
\end{eqnarray}

with

\begin{displaymath}
\omega_{1}\,=\,\omega_{2}\,\equiv\,\omega.\\
\end{displaymath}

The restriction arising from the $\Theta$-function in this system implies

\begin{displaymath}
k_{l}^{2}\,\leq\,4\,\omega^{2}-m^{2}-k^{2}_{t}  ;\\
\end{displaymath}
taking the values for $k_{l\,\pm}$ according to (\ref{eq:klvalcmp}), we see that this condition is always fulfilled:
\begin{displaymath}
k_{l}^{2}\,=\,\omega^{2}-m^{2}-k^{2}_{t}\,\leq\,4\,\omega^{2}-m^{2}-k^{2}_{t}.
\end{displaymath}

Solving the $\delta$ - function, we obtain

\begin{displaymath}
k^{2}_{l} \,=\,\omega^{2}-m^{2}-k^{2}_{t}\,\leq\,\omega^{2}-m^{2}.  
\end{displaymath}

\item  Calculation of {\large$\frac {d\sigma}{dk_{l}}$}:

Analogous treatment for $\frac {d\sigma}{dk_{l}}$ gives
 
\begin{equation}
\delta(k^{2}_{2}-m^{2}_{2})  =  \frac{1}{\left|f'(k_{t})\right|_{k_{t}=k_{t_{0}}}}\,\delta\left(k_{t}-k_{t_{0}}\right)\\
\end{equation}
with
\begin{eqnarray}
k_{t_{0}} & = & \frac{\sqrt{4\,(\wg_{1}-k_{l})\,\wg_{1}\,\wg_{2}\,(k_{l}+\wg_{2})-\, m^{2}\,(\wg_{1}+\wg_{2})^{2}}}{\wg_{1}+\wg_{2}},\nonumber\\
f'(k_{t}) & = &  \frac{2\,k_{t}\,(\wg_{1}+\wg_{2})}{k^{0}_{1}}.\label{eq:ktval}
\end{eqnarray}

In this case, we get a restriction on $k_{l}$:

\begin{displaymath}
k_{l}\,\in\,\left[k_{l_{1}},k_{l_{2}}\right]
\end{displaymath}

with

\begin{displaymath}
k_{l_{1/2}}=\frac{1}{2}\,(\wg_{1}-\wg_{2})\,\pm\,\frac{(\wg_{1}+\wg_{2})\,\sqrt{\wg_{1}\,\wg_{2}-m^{2}}}{2\,\sqrt{\wg_{1}\,\wg_{2}}}.
\end{displaymath}

$\Theta(k^{0}_{2})$ leads to

\begin{displaymath}
k^{2}_{t}\,\le\,(\wg_{1}+\wg_{2})^{2}-m^{2}-k^{2}_{l}.
\end{displaymath} 

Considerations similar to those following (\ref{eq:klval}) again show that this condition is always fulfilled by the values of $k_{t}$ given by (\ref{eq:ktval}).

\end{itemize}

\subsection{Mandelstam variables}\label{sec:MV}

With two onshell photons as incoming particles, we obtain the following relations between the 4-vectors of the particles and the Mandelstam variables:

\begin{eqnarray}
s & = & (q_{1}+q_{2})^{2} = (k_{1}+k_{2})^{2} \nonumber\\
 & = & 2q_{1}q_{2} = 2(m^{2}+k_{1}k_{2}),\nonumber\\
t & = & (q_{1}-k_{1})^{2} = (q_{2}-k_{2})^{2} \nonumber\\
 & = & -2q_{1}k_{1}+m^{2} = -2q_{2}k_{2}+m^{2},\nonumber\\
u & = & (q_{1}-k_{2})^{2} = (q_{2}-k_{1})^{2} \nonumber\\
 & =  & -2q_{1}k_{2}+m^{2} = -2q_{2}k_{1}+m^{2}.\label{eq:mandels}  
\end{eqnarray}
This implies 

\begin{eqnarray}
q_{1}q_{2} & = & \frac{s}{2},\nonumber\\
q_{1}q_{3} & = & \frac{1}{2}\,(t-m^{2}),\nonumber\\
q_{2}q_{3} & = & \frac{1}{2}\,(s+u-m^{2}),\nonumber\\
q^{2}_{3} & = & t \label{eq:conmandelst}
\end{eqnarray}
with $q^{\mu}_{3}= q^{\mu}_{1}- k^{\mu}_{1} $ for the $t$- channel reaction.

\vspace{5mm}
For a process in the $u$- channel, we obtain

\begin{eqnarray}
q_{1}q_{2} & = & \frac{s}{2},\nonumber\\
q_{1}q'_{3} & = & \frac{1}{2}\,(u-m^{2}),\nonumber\\
q_{2}q'_{3} & = & \frac{1}{2}\,(s+t-m^{2}),\nonumber\\
q^{2}_{3} & = & u \label{eq:conmandelsu}
\end{eqnarray}
with $q'^{\mu}_{3} = q^{\mu}_{1}- k^{\mu}_{2}$. Here, the $s$ and $u$ channel denote exchanges for an $s$ channel reaction as given in figure \ref{fig:stu}.

\vspace{5mm}
We see that switching between $t$ and $u$ channel implies the substitution

\begin{eqnarray}
t & \longleftrightarrow\ &u \label{eq:subsm2}
\end{eqnarray}
in (\ref{eq:conmandelst}) and (\ref{eq:conmandelsu}), respectively.

\section{Expressions for $\frac{d\sigma}{dk_{t}},\,\frac{d\sigma}{dk_{l}}$ in terms of $\wg_{1}\,$ and $\wg_{2}$ }

In order to arrive at final expressions for the differential cross sections, we still need to solve $d^{3}k_{1}$; this can easily be done using

\begin{equation}
d^{3}k_{1}=\left|k_{t}\right|\,dk_{t}\,dk_{l}\,d\varphi\,\to\,\left|k_{t}\right|\,dk_{t}\,dk_{l}\,2\,\pi\\ \label{eq:dsigint}
\end{equation}
after integration over $\varphi$.

Taking this into account and combining (\ref{eq:sig22}),(\ref{eq:siginc}),(\ref{eq:klval}), and (\ref{eq:ktval}), we obtain:

\begin{eqnarray}
\frac{d\sigma}{dk_{t}}(k_{t}) & = &\frac{1}{n_{i}\, !}\, \frac{1}{32\,\pi\,\wg_{1}\,\wg_{2}}\,\left|k_{t}\right|\,\times\nonumber \\
 & &\left(\frac{1}{k^{0}_{1}{\left(k_{l+}\right)}}\,\frac{1}{\left|f'(k_{l+})\right|}\,\left|\cal M\right|^{2}{(k_{l+})}\,+\,\frac{1}{k^{0}_{1}(k_{l-})}\,\frac{1}{\left|f'{ (k_{l-})}\right|}\,\left|\cal M\right|^{2}{(k_{l-})}\right),\nonumber\\
\frac{d\sigma}{dk_{l}}(k_{l}) & = &\frac{1}{n_{i}\, !} \frac{1}{32\,\pi\,\wg_{1}\,\wg_{2}}\,\left|k_{t_{0}}\right|\,\frac{1}{k^{0}_{1}{ (k_{t_{0}})}}\,\frac{1}{\left|f'{ (k_{t_{0}})}\right|}\,\left|\cal M\right|^{2}{ (k_{t_{0}})}\label{eq:exprds}
\end{eqnarray}
where $k_{l\pm}\,=\,k_{l\pm}(k_{t})$ for the calculation of $ \frac{d\sigma}{dk_{t}}$ and $k_{t_{0}}\,=\,k_{t_{0}}(k_{l})$ for calculation of $\frac{d\sigma}{dk_{l}}$; see section \ref{sec:constr}.
$k_{l}$ and $k_{t}$ have to be kept within the kinematical limits described in the previous section.

$\frac{d\sigma}{dk^{2}_{t}}$ can be easily derived from $\frac{d\sigma}{dk_{t}}$ by

\begin{equation}
\frac{d\sigma}{dk^{2}_{t}}= \frac{1}{2\,\left|k_{t}\right|}\,\frac{d\sigma}{dk_{t}},\label{eq:skt2}
\end{equation}
it is therefore given by

\begin{eqnarray}
\lefteqn{\frac{d\sigma}{d\left|k_{t}\right|^{2}}\; = \;\frac{1}{n_{i}\, !} \frac{1}{64\,\pi\,\wg_{1}\,\wg_{2}}\,\times} & & \nonumber \\
 & &\left(\frac{1}{k^{0}_{1}{\left(k_{l+}\right)}}\,\frac{1}{\left|f'(k_{l+})\right|}\,\left|\cal M\right|^{2}{ (k_{l+})}\,+\,\frac{1}{k^{0}_{1}(k_{l-})}\,\frac{1}{\left|f'{ (k_{l-})}\right|}\,\left|\cal M\right|^{2}{ (k_{l-})}\right).\nonumber\\
& & \label{eq:exprds2}
\end{eqnarray}

\section {Kinematic variables and cross sections in terms of $x_{1},\,x_{2}\,$, and $S$}

\begin{figure}
\centering
\psfrag{a}{$e^{-}{\scripts (E)}$}
\psfrag{b}{$e^{-}{\scripts (E-\wg_{1})}$}
\psfrag{d}{$e^{+}{\scripts (E)}$}
\psfrag{e}{$e^{+}{\scripts (E-\wg_{2})}$}
\psfrag{g1}{$\gamma\,{\scripts (\wg_{1})}$}
\psfrag{g2}{$\gamma\,{\scripts (\wg_{2})}$}
\psfrag{c}{$\pi^{0}$}
\psfrag{ez}{${\scripts \hat{e}_{z}}$}
\includegraphics[width=0.9\textwidth]{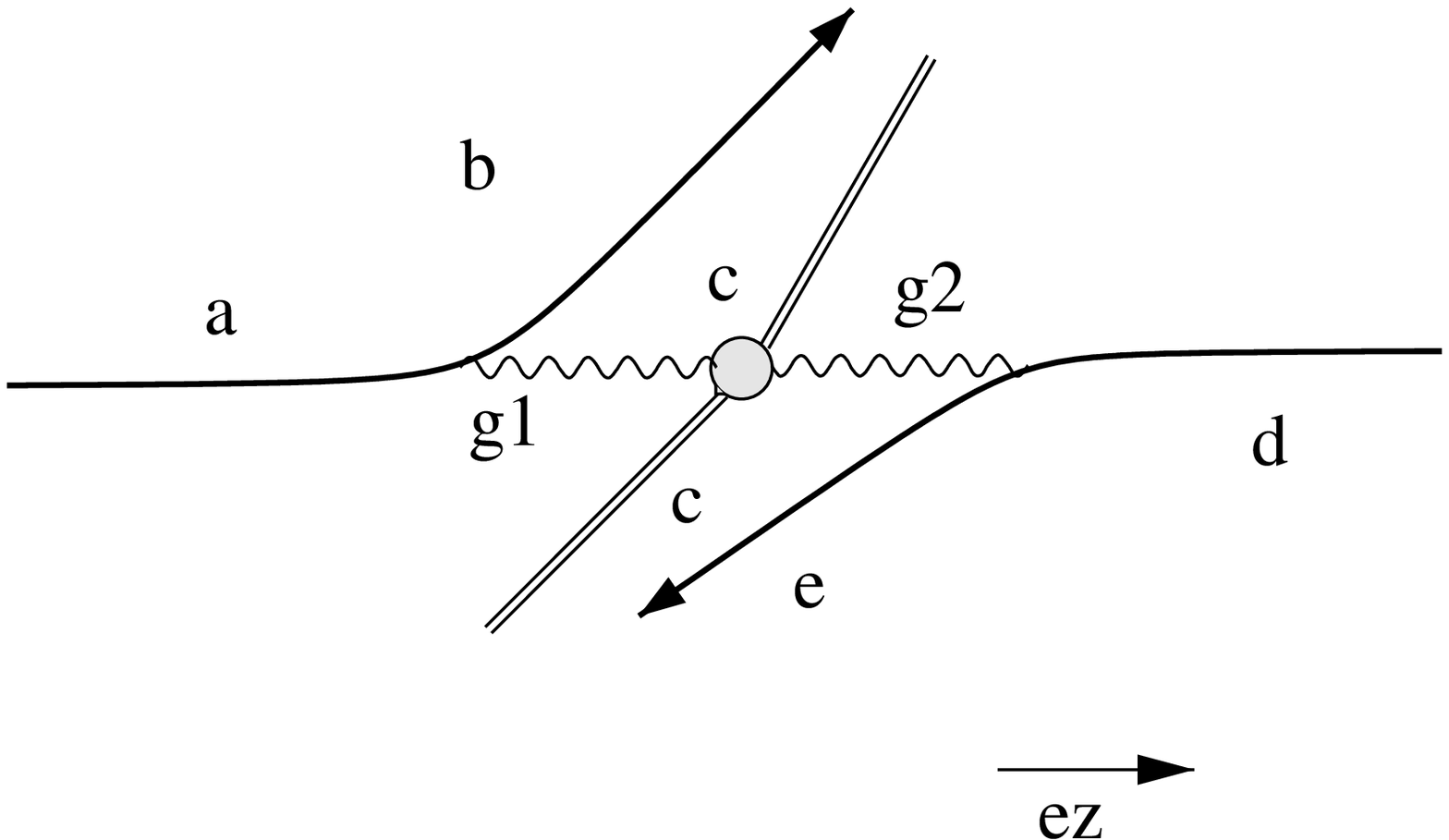}
\caption{$\gamma\gamma\rightarrow\pi^{0}\pi^{0}$ in lab frame} 
\label{fig:totalreact}                                      
\end{figure}

In the frame of a linear collider experiment, the measurable cross sections for $\gamma\gamma$ reactions depend on photon fluxes according to the experiment; therefore, in these reactions all quantities given above depend on the kinematic variables $x_{1},\,x_{2}\,$ and $S$. They are defined by

\begin{eqnarray}
x_{i} & = & \frac{\wg_{i}}{E_{i}},\nonumber\\
S & = & 4\,E^{2}\label{eq:xval}
\end{eqnarray}
with 

\begin{displaymath}
E_{1}\;=\;E_{2}\;\equiv\;E\;.
\end{displaymath}
$E$ is the center of mass energy of each particle participating in the outer process and $S$ the corresponding Mandelstam variable (see figure \ref{fig:totalreact}). 

Using these relations, we obtain:

\begin{itemize}

\item for the kinematic variables in the calculation of {\large $\frac{d\sigma}{dk_{t}}$}:

\begin{eqnarray}
k_{l+} & = & \frac{\sqrt{S}}{4}\,(x_{1}-x_{2})+\frac{(x_{1}+x_{2})\,\sqrt{S\,x_{1}\,x_{2}/4-m^{2}_{1}-k^{2}_{t}}}{2\,\sqrt{x_{1}\,x_{2}}},\nonumber\\
k_{l-} & = & \frac{\sqrt{S}}{4}\,(x_{1}-x_{2})-\frac{(x_{1}+x_{2})\,\sqrt{S\,x_{1}\,x_{2}/4-m^{2}_{1}-k^{2}_{t}}}{2\,\sqrt{x_{1}\,x_{2}}},\nonumber\\
f'(k_{l}) & = &\sqrt{S}\left( x_{1}-x_{2} - \frac{kl\,(x_{1}+x_{2})}{k^{0}_{1}}\right)\label{eq:klvalx}
\end{eqnarray}
and the corresponding restriction

\begin{equation}
k^{2}_{t} \; \le\ \; \frac{S}{4}\,x_{1}\,x_{2}-m^{2}. \label{eq:limkt}
\end{equation} 

\item for the kinematic variables in calculation of {\large $\frac{d\sigma}{dk_{l}}$}:

\begin{eqnarray}
k_{t_{0}} & = & \frac{\sqrt{4\,(\sqrt{S}x_{1}/2-k_{l})\,x_{1}\,x_{2}\,(k_{l}+\sqrt{S}x_{2}/2)-\, m^{2}\,(x_{1}+x_{2})^{2}}}{x_{1}+x_{2}},\nonumber\\
f'(k_{t}) & = & \frac{\sqrt{S}\,k_{t}\,(x_{1}+x_{2})}{k^{0}_{1}}.\label{eq:ktvalx}
\end{eqnarray}

\begin{eqnarray}
k_{l_{1/2}} & = &\frac{\sqrt{S}}{4}\,(x_{1}-x_{2})\,\pm\,\frac{(x_{1}+x_{2})\,\sqrt{S\,x_{1}\,x_{2}/4-m^{2}}}{2\,\sqrt{x_{1}\,x_{2}}}.\label{restktx}
\end{eqnarray} 

\item for the differential cross sections:

\begin{eqnarray}
\frac{d\sigma}{dk_{t}}(k_{t}) & = & \frac{1}{8\,S\,\pi\,x_{1}\,x_{2}}\,\left|k_{t}\right|\,\times \nonumber\\
 & & \left(\frac{1}{k^{0}_{1}{\left(k_{l+}\right)}}\,\frac{1}{\left|f'(k_{l+})\right|}\,\left|\cal M\right|^{2}_{t}{ (k_{l+})}\,+\,\frac{1}{k^{0}_{1}(k_{l-})}\,\frac{1}{\left|f'{ (k_{l-})}\right|}\,\left|\cal M\right|^{2}_{t}{ (k_{l-})}\right), \nonumber\\ \nonumber \\
\frac{d\sigma}{dk_{l}}(k_{l}) & = & \frac{1}{8\,S\,\pi\,x_{1}\,x_{2}}\,\left|k_{t_{0}}\right|\, \frac{1}{k^{0}_{1}{ (k_{t_{0}})}}\,\frac{1}{\left|f'{ (k_{t_{0}})}\right|}\,\left|\cal M\right|^{2}_{t}{ (k_{t-{0}})}, \nonumber \\ \nonumber \\
\frac{d\sigma}{d\left|k_{t}\right|^{2}}(k_{t}) & = & \frac{1}{16\,S\,\pi\,x_{1}\,x_{2}}\,\times \nonumber \\
 & & \left(\frac{1}{k^{0}_{1}{\left(k_{l+}\right)}}\,\frac{1}{\left|f'(k_{l+})\right|}\,\left|\cal M\right|^{2}_{t}{ (k_{l+})}\,+\,\frac{1}{k^{0}_{1}(k_{l-})}\,\frac{1}{\left|f'{ (k_{l-})}\right|}\,\left|\cal M\right|^{2}_{t}{ (k_{l-})}\right).\nonumber\\ \label{eq:exprdsx}
\end{eqnarray}

\end{itemize}

The Mandelstam-variables $s,t,\,$and$\,u$ in this system are given by

\begin{eqnarray}
s &  = &  2\,q_{1}q_{2}  =  4\,\wg_{1}\wg_{2} = S\,x_{1}x_{2},\nonumber\\
t & = &  -2\,q_{1}k_{1}+m^{2}  =  -2\,\wg_{1}(k^{0}_{1}-k_{l})+m^{2} =-\sqrt{S}\,x_{1}(k^{0}_{1}-k_{l})+m^{2},\nonumber\\
u &  = & 2\,m^{2}-s-t  =  m^{2}-S\,x_{1}x_{2}+\sqrt{S}\,x_{1}(k^{0}_{1}-k_{l}).\label{eq:mandelsx}  
\end{eqnarray}

\subsection{Modification for boosted lab-systems}\label{sec:boostlab}

If the lab-system is boosted with respect to the center of mass system of the external incoming particles, we obtain

\begin{displaymath}
E_{1}\;\neq\;E_{2}.
\end{displaymath}
In this case, we have to modify (\ref{eq:xval}) according to

\begin{eqnarray}
x_{i} & = & \frac{\wg_{i}}{E_{i}},\label{eq:xvalboost}
\end{eqnarray}
$S$ is boost-invariant. For a Lorentz-boost of the form 

\begin{eqnarray}
\binom{x^{0}}{x^{3}}_{lab}&=&\left(\begin{array}{cc} \cosh\eta & \sinh \eta \\
\sinh \eta & \cosh\eta \\ \end{array} \right)\,\binom{x^{0}}{x^{3}}_{cm},\nonumber\\
\binom{x^{1}}{x^{2}}_{lab}&=&\binom{x^{1}}{x^{2}}_{cm}\label{eq:Lboost}
\end{eqnarray}
for any 4-vector $x^{\mu}\,$ with a given rapidity $\eta$ , we can relate $E_{i}\,$ and $E$:

\begin{eqnarray}
E_{1}&=&E\,\cosh\eta+\sqrt{E^{2}-m^{2}}\,\sinh\eta ,\nonumber\\
E_{2}&=&E\,\cosh\eta-\sqrt{E^{2}-m^{2}}\,\sinh\eta .\label{eq:Econv}
\end{eqnarray}
In writing (\ref{eq:Econv}), we defined $k_{1}$ and $k_{2}$ such that

\begin{displaymath}
 \vec{k^{3}_{1}}\,=\,|k^{3}_{1}|\,\hat{e}_{z}\;,\; \vec{k^{3}_{2}}\,=\,-|k^{3}_{2}|\,\hat{e}_{z}\;.
\end{displaymath}
For  $m\,\ll\,E\,$, we can rewrite (\ref{eq:xvalboost}) according to

\begin{eqnarray}
\wg_{1}&=&E\,x_{1}\,e^{\eta},\nonumber\\
\wg_{2}&=&E\,x_{2}\,e^{-\eta}.
\end{eqnarray}
We see that calculations for a lab system boosted according to (\ref{eq:Lboost}) in the limit of $m\,\ll\,E\,$ just require the substitutions

\begin{equation}
x_{1}\,\rightarrow\,x_{1} \,e^{\eta}\;;\;x_{2}\,\rightarrow\,x_{2}\,e^{-\eta}.
\end{equation}
$\frac{d\sigma}{d|k_{t}|^{2}}\,$ is boost-invariant; we only have to modify the calculation for  $\frac{d\sigma}{dk_{l}}$.


\newpage

\chapter [\normalsize { The process $\gamma\,\gamma\,\to\,\pi^{0}\,\pi^{0}$ by photon and Odderon exchange}]{\LARGE The process $\gamma\,\gamma\,\to\,\pi^{0}\,\pi^{0}$ by photon and Odderon exchange}\label{chap:Mel}

\section{Matrix element according to photon exchange} \label{sec:photcon}

\begin{figure}[h!]
\centering
\subfigure[$t$-channel exchange,$\,i\M_{1}$]{\label{fig:Feyngt}
\fmfframe(15,20)(25,10){
\begin{fmfgraph*}(110,90)
\fmfbottomn{i}{2}
\fmftopn{o}{2}
\fmf{photon}{i1,v1}
\fmf{photon}{i2,v2}
\fmf{photon,label=$q_{3}$}{v1,v2}
\fmf{plain}{o1,v1}
\fmf{plain}{o2,v2}
\fmflabel{$q_{1}$}{i1}
\fmflabel{$q_{2}$}{i2}
\fmflabel{$k_{1}$}{o1}
\fmflabel{$k_{2}$}{o2}
\end{fmfgraph*}}}
\subfigure[$u$-channel exchange,$\,i\M_{2}$ ]{\label{fig:Feyngu}
\fmfframe(25,20)(15,10){
\begin{fmfgraph*}(110,90)
\fmfbottomn{i}{2}
\fmftopn{o}{2}
\fmf{photon}{i1,v1}
\fmf{photon}{i2,v2}
\fmf{photon,label=$q'_{3}$}{v1,v2}
\fmf{plain}{o1,v2}
\fmf{plain}{v1,o2}
\fmfforce{(.2w,0.5h)}{v1}
\fmfforce{(.8w,0.5h)}{v2}
\fmflabel{$q_{1}$}{i1}
\fmflabel{$q_{2}$}{i2}
\fmflabel{$k_{1}$}{o1}
\fmflabel{$k_{2}$}{o2}
\end{fmfgraph*}}}
\caption{contributions due to photon exchange}
\label{fig:Gamcon}
\end{figure}
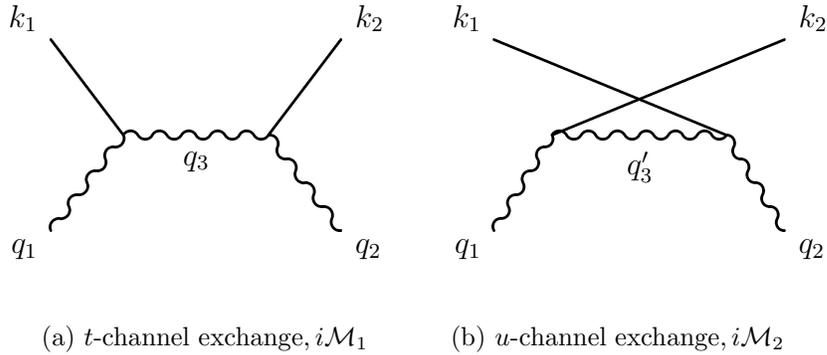

In investigating the process  $\gamma\,\gamma\,\to\,\pi^{0}\,\pi^{0}$, exchange in the $t$- as well as in the $u$ channel has to be taken into account; the corresponding Feynman-diagrams are given in figure \ref{fig:Gamcon}

\begin{fmffile}{fd}
\end{fmffile}
%
Here, $q_{1}$ and $q_{2}$ denote the momenta of the incoming photons, $k_{1}$ and $k_{2}$ the momenta of the outgoing pions. Furthermore, $q_{3}\,=\,q_{1}-k_{1}$ and $q'_{3}\,=\,q_{1}-k_{2}$ as used in chapter \ref{chap:kin}. $t$ and $u$ are given by (\ref{eq:mandels}).

We obtain 4 contributions to $\left|\cal M\right|^{2}$: 

\begin{equation}
\left|\cal M\right|^{2}\,=\,\left|{\cal M}_{1}\right|^{2}\,+\,\left|{\cal M}_{2}\right|^{2}\,+\,{\cal M}^{\ast}_1\,{\cal M}_{2}\,+\,{\cal M}_{1}\,{\cal M}^{\ast}_{2}\;.\label{eq:summs}
\end{equation}

\subsection{Feynman propagators and couplings}

For the calculation of $\left|\cal M\right|^{2}$, the following propagators and coupling constants in momentum-space were used:

\begin{eqnarray}
\mbox{for ingoing photons} & :  & \epsilon^{\mu}(q_{i})\;,\nonumber\\
\mbox{for outgoing mesons} & : & 1\;,\nonumber\\
\mbox{for the photon propagator} & : & \frac{-i g_{\mu\nu}}{q^{2}_{i}-i\,\varepsilon}\;,\nonumber\\
\mbox{for the $\gamma\,\gamma^{*}\,\pi^{0}$ coupling} & : &  g \epsilon_{\mu\nu\rho\sigma} q^{\rho}_{1}q^{\sigma}_{2}\,T(q^{2}_{1},q^{2}_{2})\label{feynpar}
\end{eqnarray}
with

\begin{eqnarray}
T(0,q^{2})&  = & \frac{1}{1-q^{2}/8\pi^{2}\,f^{2}_{\pi^{0}}}\;,\nonumber\\
g \; = \;\frac{\al}{\pi\,f_{\pi^{0}}}&;& f_{\pi^{0}} \,= \, 0.093\; \mbox{GeV}\ \label{eq:Tval}
\end{eqnarray}
(see (\ref{eq:fincouppgg})). We used $g\,=\,0.025\,\mbox{GeV}^{-1}$ according to \cite{Barnett:1996hr}.

\subsection{Computation of ${\cal M}_{1},|{\cal M}_{1}|^{2}$}

Evaluating the Feynman-diagram corresponding to $t$-channel-exchange as given by figure \ref{fig:Feyngt}, we obtain

\begin{eqnarray}
i{\cal M}_{1} & = &  g\,\varepsilon_{\mu\rho\alpha\beta}\,q^{\alpha}_{1}\,q^{\beta}_{3}\,T(0,q^{2}_{3})\,\epsilon^{\mu}(q_{1})\,\frac{-i\,g_{\rho\sigma}}{q^{2}_{3}-i\varepsilon}\,g\,\varepsilon_{\sigma\nu\delta\epsilon}\,q^{\delta}_{1}\,q^{\epsilon}_{3}\,T(q^{2}_{3},0)\,\epsilon^{\nu}(q_{2}) \nonumber\\
 & = &  \frac{-i\,g^{2}}{q^{2}_{3}-i\varepsilon}\,T^{2} (q^{2}_{3},0) \nonumber \\  &  & \Big\{ (\epsilon_{1}q_{3})\,\big( (\epsilon_{2}q_{3})\,(q_{1}q_{2})-(\epsilon_{2}q_{1})\,(q_{2}q_{3}) \big) + (\epsilon_{1}q_{2}) \big( (\epsilon_{2}q_{1})\,q^{2}_{3} - (\epsilon_{2}q_{3})\,(q_{1}q_{3}) \big) \nonumber\\ & & + (\epsilon_{1}\epsilon_{2}) \big( (q_{1}q_{3})\,(q_{2}q_{3}) - (q_{1}q_{2})\,q^{2}_{3} \big) \Big\}\,. \label{eq:M1}
\end{eqnarray}
Making use of the relations given by (\ref{eq:Mandelrel}) and (\ref{eq:mandels}), we can simplify (\ref{eq:M1}):

\begin{eqnarray}
i{\cal M}_{1} & = & \frac{-i\,g^{2}}{t-i\varepsilon}\,T^{2} (t,0)\times \nonumber \\
 &  &\bigg\{ \frac{1}{4}\, \Big[ m^{4}+t(u-s)-m^{2}(s+t+u)\Big] (\epsilon_{1}\epsilon_{2})\nonumber\\ & & +2\Big[(\epsilon_{1}q_{3})\big((m^{2}-s-u)(\epsilon_{2}q_{1})+s(\epsilon_{2}q_{3})\big)\nonumber\\
& &+ (\epsilon_{1}q_{2})\big(2\,t\,(\epsilon_{2}q_{1})+(m^{2}-t)\,(\epsilon_{2},q_{3})\big)\Big]\bigg\}\;.\nonumber\\ \label{eq:M1simpl}
\end{eqnarray}
We now obtain ${\cal M}^{\ast}_{1}$ by putting $\epsilon_{i} \to\,\epsilon^{\ast}_{i}$; furthermore, we use $\sum \epsilon^{\mu}_{i}\,\epsilon^{\ast\nu}_{i} = - g^{\mu\nu}$ and get for $\left|{\cal M}_{1}\right|^{2}_{t}$:

\begin{equation}
\left|{\cal M}_{1}\right|^{2}_{t}  = \frac{1}{4}\,\frac{g^{4}}{t^{2}+\varepsilon^{2}}\,T^{4} (t,0)\frac{1}{8}\,\left(m^{8}-2m^{4}tu+t^{2}\left(s^{2}+u^{2}\right)\right)\;.\label{eq:M1sq}
\end{equation}
Here, $\left|{\cal M}\right|^{2}_{t}$ signifies the total matrix element after averaging over incoming photon polarizations.

\subsection{Computation of ${\cal M}_{2},|{\cal M}_{2}|^{2}$}

It is obvious that we get the same result for ${\cal M}_{2}$ as for ${\cal M}_{1}$ with the substitution

\begin{eqnarray*}
t & \longleftrightarrow\ &u\;. 
\end{eqnarray*}
Therefore, we obtain

\begin{eqnarray}
i{\cal M}_{2} & = & \frac{-i\,g^{2}}{u-i\varepsilon}\,T^{2} (u,0)\times \nonumber \\
 &  &\bigg\{ \frac{1}{4}\, \Big[ m^{4}+u(t-s)-m^{2}(s+t+u)\Big] (\epsilon_{1}\epsilon_{2})\nonumber\\
 & & +2\Big[(\epsilon_{1}q'_{3})\big((m^{2}-s-t)(\epsilon_{2}q_{1})+s(\epsilon_{2}q'_{3})\big) \nonumber\\
 & &+ (\epsilon_{1}q_{2})\big(\,2\,u\,(\epsilon_{2}q_{1})+(m^{2}-u)\,(\epsilon_{2},q'_{3})\big)\Big]\bigg\}\nonumber\\ \label{eq:M2}
\end{eqnarray}
and

\begin{equation}
\left|{\cal M}_{2}\right|^{2}_{t}  =\frac{1}{4}\, \frac{g^{4}}{u^{2}+\varepsilon^{2}}\,T^{4} (u,0) \frac{1}{8}\,\left(m^{8}-2m^{4}tu+u^{2}\left(s^{2}+t^{2}\right)\right)\,.  \label{eq:M2sq}
\end{equation}

\subsection {Computation of ${\cal M}^{\ast}_{1}\,{\cal M}_{2}$ and ${\cal M}_{1}\,{\cal M}^{\ast}_{2}$}

The calculation of ${\cal M}^{\ast}_{1}\,{\cal M}_{2}$ and ${\cal M}_{1}\,{\cal M}^{\ast}_{2}$ can be easily done using (\ref{eq:M1}) and (\ref{eq:M2}) as well as $\sum \epsilon^{\mu}_{i}\,\epsilon^{\ast\nu}_{i} = - g^{\mu\nu}$. The two mixed terms are given by

\begin{eqnarray}
\left({\cal M}^{\ast}_{1}\,{\cal M}_{2}\right)_{t} & = & \left({\cal M}_{1}\,{\cal M}^{\ast}_{2}\right)_{t} = \frac{1}{4}\,\frac{g^{4}}{16\,t\,u}\,T^{2}(t,0)\,T^{2}\,(u,0) \times \nonumber \\
  &  & \Big\{2\,m^{8}+s^{4}+t^{4}+u^{4}+4m^{6}\left(3s-t-u\right)-2s^{2}\left(t^2-tu+u^2\right) \nonumber \\ 
 &  & -2m^{4}\left(s^{2}+6s\left(t+u\right)-3\left(t^{2}+u^{2}\right)\right)\nonumber \\ & & -4m^{2}\left(s^{3}+t^{3}+u^{3}-s^{2}\left(t+u\right) -s\left(t^{2}+tu+u^{2}\right)\right)\Big\}\;.\label{eq:Mmsq}
\end{eqnarray}

\subsection {Total $\left|{\cal M}\right|^{2}$}

For $\left|{\cal M}\right|_{t}^{2}$, we obtain from (\ref{eq:summs}) 

\begin{equation}
\left|\cal M\right|^{2}\,=\,\left|{\cal M}_{1}\right|^{2}\,+\,\left|{\cal M}_{2}\right|^{2}\,+\,{\cal M}^{\ast}_1\,{\cal M}_{2}\,+\,{\cal M}_{1}\,{\cal M}^{\ast}_{2}\,.\label{eq:summs2}
\end{equation}
This also holds in cases of averaging and summing over photon polarizations.
Therefore, combining the results from (\ref{eq:M1sq}), (\ref{eq:M2sq}), and (\ref{eq:Mmsq}), we obtain:

\begin{eqnarray}
\left|\cal M\right|^{2}_{t} & = & \left|{\cal M}_{1}\right|^{2}_{t}\,+\,\left|{\cal M}_{2}\right|^{2}_{t}\,+\,\left({\cal M}^{\ast}_1\,{\cal M}_{2}\right)_{t}\,+\,\left({\cal M}_{1}\,{\cal M}^{\ast}_{2}\right)_{t} = \nonumber\\
 & &\frac{1}{4}\, \frac{1}{8}\,g^{4}\Big\{s^{2}\left(T^{4}_{t}+T^{4}_{u}\right)+\left(m^{4}-tu\right)^{2}\,\left(\frac{T^{4}_{u}}{u^{2}}+\frac{T^{4}_{t}}{t^{2}}\right) + \nonumber\\
 & & \frac{T^{2}_{t}\,T^{2}_{u}}{tu}\big\{2\,m^{8}+s^{4}+t^{4}+u^{4}+4m^{6}\left(3s-t-u\right)-2s^{2}\left(t^2-tu+u^2\right) \nonumber \\ 
 &  & -2m^{4}\left(s^{2}+6s\left(t+u\right)-3\left(t^{2}+u^{2}\right)\right)\nonumber \\ & & -4m^{2}\left(s^{3}+t^{3}+u^{3}-s^{2}\left(t+u\right) -s\left(t^{2}+tu+u^{2}\right)\right)\big\}\Big\}\label{eq:totm}
\end{eqnarray}
with $T_{t}\,\equiv\,T(t,0)$ and $T_{u}\,\equiv\,T(u,0)$.

\section{Inclusion of the Odderon contribution}\label{sec:incluodd}

\subsection{Feynman propagator and $\gamma\,\odd^{*}\pi^{0}\,$-coupling;\\$i{\cal{M}}_{1/2}$ due to pure Odderon-exchange }

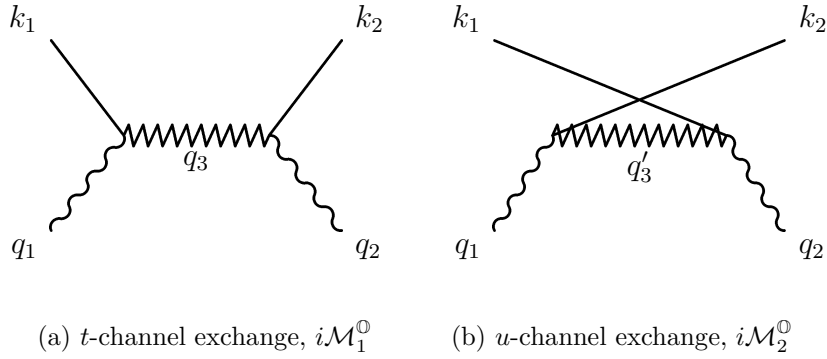
\begin{figure}[h!]
\centering
\subfigure[$t$-channel exchange, $i\M^{\odd}_{1}$]{\label{fig:Feynot}
\fmfframe(15,20)(25,10){
\begin{fmfgraph*}(110,90)
\fmfbottomn{i}{2}
\fmftopn{o}{2}
\fmf{photon}{i1,v1}
\fmf{photon}{i2,v2}
\fmf{zigzag,label=$q_{3}$,zigzag_width=2.}{v1,v2}
\fmf{plain}{o1,v1}
\fmf{plain}{o2,v2}
\fmflabel{$q_{1}$}{i1}
\fmflabel{$q_{2}$}{i2}
\fmflabel{$k_{1}$}{o1}
\fmflabel{$k_{2}$}{o2}
\end{fmfgraph*}}}
\subfigure[$u$-channel exchange, $i\M^{\odd}_{2}$]{\label{fig:Feynou}
\fmfframe(25,20)(15,10){
\begin{fmfgraph*}(110,90)
\fmfbottomn{i}{2}
\fmftopn{o}{2}
\fmf{photon}{i1,v1}
\fmf{photon}{i2,v2}
\fmf{zigzag,label=$q'_{3}$,zigzag_width=0.1}{v1,v2}
\fmf{plain}{o1,v2}
\fmf{plain}{v1,o2}
\fmfforce{(.2w,0.5h)}{v1}
\fmfforce{(.8w,0.5h)}{v2}
\fmflabel{$q_{1}$}{i1}
\fmflabel{$q_{2}$}{i2}
\fmflabel{$k_{1}$}{o1}
\fmflabel{$k_{2}$}{o2}
\end{fmfgraph*}}}
\caption{contributions due to Odderon exchange}
\label{fig:Oddcon}
\end{figure}

Inclusion of the Odderon contribution implies the evaluation of two additional Feynman-diagrams describing the Odderon contribution; they are given by figure \ref{fig:Oddcon}.

\begin{fmffile}{fd2}
\end{fmffile}
Therefore, we obtain 12 additional contributions to $|\M|_{t}^{2}$. We use the following Feynman propagator and $\gamma\,\odd^{*}\pi^{0}\,$-coupling (see sections \ref{sec:effprop} and \ref{sec:effver}):

\begin{eqnarray}
\mbox{Feynman-propagator}&:& \left(-i\right)\eta_{\mathbb{O}}\left(-i s/s_{0}\right)^{\alpha_{\mathbb{O}}(t)-1}g^{\mu\nu}\nonumber\\
\gamma\,\odd^{*}\pi^{0}\,-\mbox{coupling}&:& T^{\gamma\,\odd^{*}\pi^{0}}\,=\,T^{\gamma\gamma^{*}\pi^{0}}\,\frac{\beta_{\mathbb{O}}}{e}\,r_{\pi^{0}}\label{eq:oddpropcoup}
\end{eqnarray}
with 

\begin{eqnarray*}
\alphodd(t)\;=\;\alphodd(0)+\alphodd'\,t&,&\alphodd(0)\,=\,1+\vare'\,.
\end {eqnarray*}
The values of the parameters are taken according to \cite{Kilian:1998ew}:

\begin{eqnarray}
\beta_{\mathbb{P}}\,=\,1.8\,\mbox{GeV}^{-1},& r_{\pi^{0}}\,=\,3,& s_{0}\,=\,1\;\mbox{GeV}^{-2}\nonumber\\
\beta^{2}_{\odd}\,=\,0.05\,\beta^{2}_{\mathbb{P}}\,,&\vare'\,=\,0 & \label{eq:oddval}
\end{eqnarray}
and $\al\,=\,\frac{e^{2}}{4\pi}$. $T^{\gamma\gamma^{*}\pi^{0}}\,\equiv\,T$ as given by (\ref{eq:Tval}).

\vspace{6mm}

Keeping in mind that both propagator and coupling always appear in the combination coupling-propagator-coupling in the process investigated, we can simplify the calculation of the 12 additional contributions to $|\M|_{t}^{2}$.
We see that the substitution

\begin{eqnarray}
\frac{1}{q^{2}_{i}}\,T^{2} &  \longrightarrow & \eta_{\odd}\left(-i s/s_{0}\right)^{\alpha_{\odd}(t)-1}\, (\frac{\beta_{\odd}}{e}\,r_{\pi^{0}})^{2}\,T^{2}\nonumber\\
 & & = \frac{1}{q^{2}_{i}}\,\kappa\,q^{2}_{i}\,\left(-i s/s_{0}\right)^{\alpha_{\odd}(t)-1}\,T^{2}\label{eq:oddsub}
\end{eqnarray}
with 

\begin{eqnarray}
q^{2}_{i} & = &\left\{ \begin{array}{ll} q^{2}_{3}=t\;\; \mbox{for $t$-channel exchange}\\ 
q'^{2}_{3}=u\;\;\mbox{for $u$-channel exchange}\\
\end{array} \right.\nonumber\\
\kappa & = & \eta_{\odd}\,\left(\frac{\beta_{\odd}}{e}\,r_{\pi^{0}}\right)^{2}
\, =\,  \eta_{\odd}\,\frac{0.05\beta^{2}_{\mathbb{P}}}{4\pi\alpha}\,r^{2}_{\pi^{0}}\label{eq:kappa}
\end{eqnarray} 
inserted in the calculation for $i{\cal{M}}_{1}$ (\ref{eq:M1}) and $i{\cal{M}}_{2}$ (\ref{eq:M2}) correspond to the contributions for the process $\gamma\gamma\,\rightarrow\,\pi^{0}\pi^{0}$ due to pure Odderon-exchange; therefore, we use a substitution in the form of

\begin{equation}
\Mgo\;=\;\M^{\gamma}\,+\be\M^{\gamma}\label{eq:Mmodbyodd}
\end{equation}
for the calculation of matrix elements including contributions from ``photon + Odderon'' exchange.

\subsection{Computation of $i{\cal{M}}^{\gamma+\odd}_{1}$ and $\left|\M^{\gamma+\odd}_{1}\right|^{2}$}

Following the argumentation given above, $i{\cal{M}}^{\gamma+\odd}_{1}$ is given by:

\begin{equation}
i{\cal{M}}^{\gamma+\odd}_{1} \; = \; i\left(\M^{\gamma}_{1}\,+\M^{\odd}_{1}\right)
\; = \; i\left(1+\beta'\right)\,\M^{\gamma}_{1}\label{eqn:m1godd}
\end{equation}
with

\begin{eqnarray*}
\beta'&=&\kappa\,t\,\left(-i s/s_{0}\right)^{\alpha_{\odd}(t)-1}\;.
\end{eqnarray*} 
We therefore obtain for $\left|\M^{\gamma+\odd}_{1}\right|^{2}$:

\begin{eqnarray}
\left|\M^{\gamma+\odd}_{1}\right|^{2} & = &  (1+\beta')\,(1+\beta')^{\ast}\,\left|\M^{\gamma}_{1}\right|^{2}\nonumber\\
 &= & \left(1+2\,{\mathit{Re}}(\beta')+\left|\beta'\right|^{2}\right)\,\left|\M^{\gamma}_{1}\right|^{2}\;.\label{eq:m1goddtot}
\end{eqnarray}
Looking at the single terms of (\ref{eq:m1goddtot}), we obtain

\begin{eqnarray*}
2\,{\mathit{Re}}(\beta') & = & 2\,\kappa\, t \left(s/s_{0}\right)^{\alphodd(t)-1}\,\mbox{cos}\left(\frac{\pi}{2}\,\left(\alphodd(t)-1\right)\right)\;,\\
\left|\beta'\right|^{2} & = & \kappa^{2}\, t^{2} \left(s/s_{0}\right)^{2(\alphodd(t)-1)}\;,
\end{eqnarray*} 
and therefore

\begin{eqnarray}
\left|\M^{\gamma+\odd}_{1}\right|^{2} & = & \bigg\{ 1+\kappa\,t\left(s/s_{0}\right)^{\alphodd(t)-1}\,\times \nonumber\\ 
 & &\Big[2\,\mbox{cos}\left(\frac{\pi}{2}\,\left(\alphodd(t)-1\right)\right)+\kappa\,t\left(s/s_{0}\right)^{\alphodd(t)-1}\Big]\bigg\}\,\left|\M^{\gamma}_{1}\right|^{2}\,.\nonumber\\
 & & \label{eq:m1goddtot2}
\end{eqnarray}
 
\subsection{Computation of $i{\cal{M}}^{\gamma+\odd}_{2}$ and $\left|\M^{\gamma+\odd}_{2}\right|^{2}$}

Similar considerations for the calculation of $i{\cal{M}}^{\gamma+\odd}_{2}$ lead to

\begin{eqnarray}
i{\cal{M}}^{\gamma+\odd}_{2} & = &  i\left(1+\beta''\right)\,\M^{\gamma}_{2}\label{eqn:m2godd}
\end{eqnarray}
and

\begin{eqnarray}
\left|\M^{\gamma+\odd}_{2}\right|^{2} & = &  (1+\beta'')\,(1+\beta'')^{*}\,\left|\M^{\gamma}_{1}\right|^{2}\label{eq:m2goddtot}
\end{eqnarray}
with

\begin{eqnarray*}
\beta''&=&\kappa\,u\,\left(-i s/s_{0}\right)^{\alpha_{\odd}(t)-1}\,.
\end{eqnarray*} 
As a final result, we obtain

\begin{eqnarray}
\left|\M^{\gamma+\odd}_{2}\right|^{2} & = & \bigg\{ 1+\kappa\,u\left(s/s_{0}\right)^{\alphodd(t)-1}\,\times \nonumber\\
& &\Big[2\,\mbox{cos}\left(\frac{\pi}{2}\,\left(\alphodd(t)-1\right)\right)+\kappa\,u\left(s/s_{0}\right)^{\alphodd(t)-1}\Big]\bigg\} \,\left|\M^{\gamma}_{2}\right|^{2} \, .\nonumber\\
 & & \label{eq:m2goddtot2}
\end{eqnarray}
 
\subsection{Computation of $(\Mgo_{1})^{*}\Mgo_{2}\;$ and $\Mgo_{1}(\Mgo_{2})^{*}\;$} 
As we see from (\ref{eq:Mmsq}), $\M^{\gamma\, \ast}_{1}\M^{\gamma}_{2} = \M^{\gamma}_{1}\M^{\gamma\, \ast}_{2} $. Using this, we obtain for the mixed matrix elements including photon and Odderon exchange

\begin{eqnarray}
\lefteqn{\left(\Mgo_{1}\right)^{\ast}\Mgo_{2}\, +\,\Mgo_{1}\left(\Mgo_{2}\right)^{*}\;=\;} & &  \nonumber\\
& = & \left((1+\be'^{*})\,(1+\be'')+(1+\be')\,(1+\be''^{*})\right)\,\M_{1}^{\gamma\,*}\,\M^{\gamma}_{2} \nonumber \\
 & = & 2\,(1+{\mathit{Re}}(\beta')+{\mathit{Re}}(\beta'')+{\mathit{Re}}(\beta'^{*}\beta''))\,\M^{\gamma\,*}_{1}\M^{\gamma}_{2} \label{eq:mmgoddtot}
\end{eqnarray}
with

\begin{eqnarray*}
{\mathit{Re}}(\beta') & = & \kappa\, t \left(s/s_{0}\right)^{\alphodd(t)-1}\,\mbox{cos}\left(\frac{\pi}{2}\,\left(\alphodd(t)-1\right)\right)\,,\\
{\mathit{Re}}(\beta'') & = & \kappa\, u \left(s/s_{0}\right)^{\alphodd(t)-1}\,\mbox{cos}\left(\frac{\pi}{2}\,\left(\alphodd(t)-1\right)\right)\,,\\
{\mathit{Re}}(\beta'^{*}\beta'') & = & \kappa^{2}\,tu\,\left(s/s_{0}\right)^{2(\alphodd(t)-1)}\\
\end{eqnarray*}
and therefore
\begin{eqnarray}
 \lefteqn{ \left(\Mgo_{1}\right)^{\ast}\Mgo_{2}\, +\,\Mgo_{1}\left(\Mgo_{2}\right)^{*}} & & \nonumber\\
& = & 2\,\bigg\{1+\kappa \left(s/s_{0}\right)^{\alphodd(t)-1}\Big[\left(t+u\right)\,\mbox{cos}\left(\frac{\pi}{2}(\alphodd(t)-1)\right)\,\nonumber\\ 
& &+\,tu\, \kappa\,\left(s/s_{0}\right)^{\alphodd(t)-1}\Big]\bigg\}\,\M^{\gamma\,*}_{1}\M^{\gamma}_{2}\,. \label{eq:mmgoddtot2}
\end{eqnarray}

\subsection{Total $\left|\Mgo\right|^{2}$}

Combining now the results of (\ref{eq:m1goddtot2}), (\ref{eq:m2goddtot2}), and (\ref{eq:mmgoddtot2}), and inserting this in the result for $\left|\M^{\gamma}\right|_{t}^{2}$ given by (\ref{eq:totm}), we obtain

\begin{eqnarray}
\left|\Mgo\right|_{t}^{2} & = & \frac{1}{4}\, \frac{1}{8}\,g^{4}\Big\{s^{2} \left(T^{4}_{t}\,C^{\gamma+\odd}_{1}+ T^{4}_{u}\,C^{\gamma+\odd}_{2}\right)+\left(m^{4}-tu\right)^{2}\,\nonumber\\
 & & \left(\frac{T^{4}_{u}\,C^{\gamma+\odd}_{1}}{u^{2}} +\frac{T^{4}_{t}\,C^{\gamma+\odd}_{2}}{t^{2}}\right) + \frac{T^{2}_{t}\,T^{2}_{u}}{tu}\,C^{\gamma+\odd}_{3} \big\{2\,m^{8}+s^{4}+t^{4}+u^{4} \nonumber\\
 & &+4m^{6}\left(3s-t-u\right)-2s^{2}\left(t^2-tu+u^2\right)-2m^{4}(s^{2}+6s\left(t+u\right)\nonumber\\
 & &-3\left(t^{2}+u^{2}\right)) -4m^{2}\left(s^{3}+t^{3}+u^{3}-s^{2}\left(t+u\right) -s\left(t^{2}+tu+u^{2}\right)\right)\big\}\Big\}\nonumber\\
 & &\label{eq:totmgodd}
\end{eqnarray}
with

\begin{eqnarray*}
C^{\gamma+\odd}_{1} & = & 1+\kappa\,t\left(s/s_{0}\right)^{\alphodd(t)-1}\,\left[2\,\mbox{cos}\left(\frac{\pi}{2}\,\left(\alphodd(t)-1\right)\right)+\kappa\,t\left(s/s_{0}\right)^{\alphodd(t)-1}\right]\,,\\
C^{\gamma+\odd}_{2} & = & 1+\kappa\,u\left(s/s_{0}\right)^{\alphodd(t)-1}\,\left[2\,\mbox{cos}\left(\frac{\pi}{2}\,\left(\alphodd(t)-1\right)\right)+\kappa\,u\left(s/s_{0}\right)^{\alphodd(t)-1}\right]\,,\\
C^{\gamma+\odd}_{3} & = & 1+\kappa \left(s/s_{0}\right)^{\alphodd(t)-1}\left[\left(t+u\right)\,\mbox{cos}\left(\frac{\pi}{2}(\alphodd(t)-1)\right)\,+\,tu\, \kappa\,\left(s/s_{0}\right)^{\alphodd(t)-1}\right]\,.
\end{eqnarray*}
$\kappa$ and the parameters for the Odderon-propagator and coupling are given by (\ref{eq:oddval}) and (\ref{eq:kappa}).

\newpage


\chapter[\normalsize {Investigation at a linear $e^{+}\,e^{-}$ collider}]{\LARGE Investigation at a linear $e^{+}\,e^{-}$ collider}\label{sec:lincol}

\section{General options}

We derived the formulas in the last chapter for fixed photon energies $\wg_{1}$ and $\wg_{2}$; actually, we have to take the photon production mechanism into account. This is done by using equivalent spectra for the photon fluxes depending on the experiment. 

\begin{itemize}
\item The emission of nearly on-shell virtual photons from the collided particles can be described by the Double Equivalent Photon Approximation (DEPA) \cite{Budnev:1975}. We will refer to this as the direct $e^{+}e^{-}$ option.
\item At a photon collider with on-shell photons \cite{Ginzburg:1983vm}, \cite{Ginzburg:1984yr}, the spectra are mainly given by Compton scattering. A first realization could take place as an extension of the TESLA experiment \cite{Badelek:2001xb}.
\end{itemize} 
In both cases, a relation between the $\gamma\,\gamma$ and the $e^{+}\,e^{-}$ luminosity holds:

\begin{equation}
{\cal L}_{\gamma\gamma} \propto {\cal L}_{e_{+}e_{-}}\,, \label{eq:Lrel1}
\end{equation}
or, more specifically

\begin{equation}
d{\cal L}_{\gamma\gamma}\,=\, {\cal L}_{e_{+}e_{-}}\,N(x_{1})\,N(x_{2})\,dx_{1}\,dx_{2} \label{eq:Lrel2}
\end{equation}
where $x_{i}$ is given by (\ref{eq:xval}):

\begin{displaymath}
x_{i} = \frac{\wg_{i}}{E}\,.
\end{displaymath}

\section{Equivalent photon flux given by the DEPA; direct $e^{+}\,e^{-}$- option} 

\subsection{General relations for polarization vectors of virtual photons}

In general, the polarization vectors of free photons are taken to be the independent solutions of the Maxwell equation for a free electromagnetic field:

\begin{eqnarray*}
\Box\,A^{\mu}&=&0\,,
\end{eqnarray*}
in the sense that $\epsilon^{\mu}\,e^{-i qx}\,$ with $q^{2}\,=\,0$ solves the equation given above; here $\Box\,=\frac{\partial^{2}}{\partial t^{2}}-\triangle\,$ is the d'Alembert operator. This seems to allow for four independent solutions of the equation; taking gauge invariance into account, however, reduces the number of independent solutions to three (in case of a virtual photon) and two (in case of on-shell photons) respectively.\\
As a basis, we can choose the polarization vectors $\epsilon^{\mu}_{i}$ with

\begin{eqnarray}
\epsilon^{\mu}_{\pm 1} & : & \mbox{transverse polarizations} \nonumber\\
\epsilon^{\mu}_{0} & : & \mbox{scalar polarization} \nonumber\\
\epsilon^{\mu}_{s} & : & \mbox{longitudinal polarization} \label{eq:gamrel}
\end{eqnarray}
Due to gauge invariance of the electric field, we can choose Lorentz gauge $\partial_{\mu}A^{\mu}\,=\,0$ and perform a Fourier transformation leading to

\begin{equation}
\epsilon_{\mu}q^{\mu} \; = \; 0 \label{eq:twodeg}
\end{equation}
which reduces the degrees of freedom to three.

\vspace{3mm}
For on-shell photons, equation (\ref{eq:twodeg}) holds true for all choices of gauge as we can easily see by performing a gauge transformation given by $ \epsilon^{\mu}\, \rightarrow \, \epsilon^{\mu}+cq^{\mu}$. We then obtain

\begin{displaymath}
q'_{\mu} \epsilon^{\mu}\;= \; q_{\mu} \epsilon^{\mu}+cq^{2}\,=\,q_{\mu} \epsilon^{\mu}\;=\;0\,.
\end{displaymath}
By choosing  $c\,=\,- \epsilon^{0}/q^{0}$, we can always reduce the condition derived above to $\vec{\epsilon}\,\vec{q}\,=\,0$. This leads to two degrees of freedom for on-shell photons.
For virtual photons, however, the above argumentation does not hold as $q^{2}\,\ne\,0$; we are left with three degrees of freedom.

Taking $\epsilon^{\mu}_{+1}\,$, $\epsilon^{\mu}_{-1}\,$, and $\epsilon^{\mu}_{0}\,$ as independent polarization vectors, the following relations hold:

\begin{eqnarray}
\sum_{a=\left\{ 0,\pm 1\right\}}(-1)^{a}\,\epsilon^{*\,\mu}_{a}\epsilon^{\nu}_{a}&=&g^{\mu\nu}-q^{\mu}q^{\nu}/q^{2}\label{eq:relsumhel}\\
\epsilon^{*\,\mu}_{a}\epsilon^{\nu}_{b}&=&(-1)^{a}\,\delta_{ab}\label{eq:normhel}
\end{eqnarray}
For a derivation of these expressions, see e.g.\cite{Aitchison:1972}; the notation above is closely following \cite{Budnev:1975}.\\
Common choices for $\epsilon^{\mu}_{\pm1}\,$ and $\epsilon^{\mu}_{0}\,$ are given by

\begin{eqnarray}
\epsilon^{\mu}_{\pm 1} = \mp \frac{1}{\sqrt{2}}  \left( \begin{array}{c} 0 \\ 1\\ {\pm i} \\ 0 \end{array} \right) &,& 
\epsilon^{\mu}_{0}=  \frac{1}{Q} \left( \begin{array}{c} {\left|\vec{q}\right|} \\ 0 \\ 0 \\ q^{0}  \end{array} \right)\label{eq:helba}
\end{eqnarray}
with $\,Q^{2} = -q^{2}$.

\subsection{The Double Equivalent Photon Approximation: general idea} 

Although in the process investigated we are interested in photons coming from the electron and positron beam respectively, it is easier to understand the principle of the DEPA by looking at the Equivalent Photon Approximation (EPA); deriving formulas valid for the DEPA goes along the same lines. This section will not provide a complete derivation of the spectra used but only a short overview of the general idea; for more details, see \cite{Budnev:1975}.

\begin{figure}[h!]
\centering
\fmfframe(15,20)(20,20){
\begin{fmfgraph*}(110,90)
\fmfstraight
\fmfleftn{i}{2}
\fmfrightn{o}{4}
\fmf{fermion,label=$p $ }{i2,v1}
\fmf{fermion,label=$p'$}{v1,o4}
\fmf{photon,label=$q$ }{v1,v2}
\fmfv{decor.shape=circle,decor.filled=shaded,decor.size=.15w,label=$\widetilde{\M}^{\mu}$,label.dist=100.}{v2}
\fmf{fermion,label=$P$}{i1,v2}
\fmf{fermion}{v2,o1}
\fmffreeze
\fmf{fermion}{v2,o2}
\fmf{fermion}{v2,o3}
\fmfforce{1.w,0.4h}{o3}
\fmfforce{1.w,0.2h}{o2}
\fmflabel{$k$}{o2}
\end{fmfgraph*}}
\caption{EPA basic idea}
\label{fig:basepa}
\end{figure}
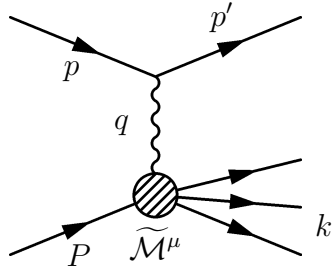

The matrix element of any process calculated with the help of EPA will have the following basic structure (see figure \ref{fig:basepa}):

\begin{eqnarray}
\M & = & {\bar{u}}^{s'}(p')\,(-i e \gamma^{\mu})\,u^{s}(p)\,\left(\frac{-i g_{\mu\nu}}{q^{2}+i\varepsilon}\right)\widetilde{\M}^{\nu}\nonumber\\
& = &  -e\, \bar{u}^{s'}(p')\, \gamma_{\mu}\, u^{s}(p)\, \frac{1}{q^{2}+i\varepsilon}\, \widetilde{\M}^{\mu}\label{eq:mepa}
\end{eqnarray}
with $q\,=\,p-p'$.

\begin{fmffile}{fd3}
\end{fmffile}
%
$\widetilde{\M}$ denotes the matrix element of the process with the photon as an incoming particle; it is arbitrary and does not have any influence on the derivation of the EPA spectrum.

Squaring the expression given by (\ref{eq:mepa}) and averaging over incoming and summing over outgoing particle helicities for unpolarized cross sections, we arrive at

\begin{eqnarray}
\left|\M\right|^{2}_{t} & = &\frac{1}{2} \frac{4\pi\alpha}{q^{4}+\varepsilon^{2}} \,Tr \Big( \left( p \mkern-8mu / +m \right) \gamma_{\nu} \left( p' \mkern-12mu / +m \right) \gamma_{\mu} \Big)  \,\widetilde{\M}^{\mu}\widetilde{\M}^{*\,\nu}\nonumber\\
 & = & \frac{4\pi\alpha}{-q^{2}}\,\rho_{\mu\nu}\, \widetilde{\M}^{\mu}\widetilde{\M}^{*\,\nu} 
\end{eqnarray}
with

\begin{eqnarray}
\rho^{\mu\nu} & = & \frac{1}{2\,(-q^{2})}\:Tr \Big( \left( p \mkern-8mu / +m \right) \gamma_{\nu} \left( p' \mkern-12mu / +m \right) \gamma_{\mu} \Big)\nonumber \\
& = & - \left( g^{\mu\nu}-\frac{q^{\mu}q^{\nu}}{q^{2}}\right)\,-\frac{\left(2p-q\right)^{\mu}\,(2p-q)^{\nu}}{q^{2}}\;. \label{eq:rhoepa}
\end{eqnarray} 
In this section, $m\,\equiv\,m_{e}$ will denote the electron/ positron mass\footnote{In general, the EPA can be derived for any charged particle; see \cite{Budnev:1975} for details.}.
As we are interested in total or differential cross sections, we now need to plug this expression into the formula for a differential cross section for a $2\,\rightarrow\,n$ particle reaction given by (\ref{eq:dsig1}). We obtain

\begin{eqnarray}
d\sigma & = & \frac{4\pi\alpha}{(-q^{2})}\,\widetilde{\M}^{*\,\mu}\,\widetilde{\M}^{\nu}\,\rho_{\mu\nu}\,\frac{(2\,\pi)^{4}\,\delta(p+P-p'-k)\,d\Gamma}{2\,w}\,\frac{d^{3}p'}{2\,E'\,(2\pi)^{3}}\nonumber\\
& & \label{eq:sigepa}
\end{eqnarray}
where $d\Gamma$ denotes the differential phase space volume of all products $k_{i}$ of the inner reaction:

\begin{eqnarray*}
d\Gamma&=&\prod_{i}\,\frac{d^{3}k_{i}}{2\,E_{i}\,(2\pi)^{3}}\,.
\end{eqnarray*}
$w$ is given according to (\ref{eq:dsig1}).

Integrating (\ref{eq:sigepa}) over $d\Gamma$ gives

\begin{eqnarray}
d\sigma & = & \frac{4\pi\alpha}{(-q^{2})}\,\rho_{\mu\nu}\,\frac{W^{\mu\nu}}{w}\,\frac{d^{3}p'}{2\,E'\,(2\pi)^{3}}\nonumber\\
& & \label{eq:sigepa2}
\end{eqnarray}
with

\begin{eqnarray*}
W^{\mu\nu}&=&\frac{1}{2}\,\int\M^{*\,\mu}\M^{\nu}\,(2\pi)^{4}\,\delta(q+P-\sum k_{i})\,d\Gamma\, ,\\
q^{\mu}&=&p^{\mu}-p^{'\,\mu}\,.
\end{eqnarray*}
$W^{\mu\nu}$ is connected to the total cross section of the inner reaction:

\begin{equation}
\widetilde{\sigma}\;=\;\frac{1}{\tilde{\wg}}\,\epsilon^{*}_{\mu}\,W^{\mu\nu}\epsilon_{\nu}\;.
\end{equation}
It can be split into two parts corresponding to  the contributions from scalar and transverse polarizations of the photon respectively. Transferring (\ref{eq:rhoepa}) into the helicity basis given by (\ref{eq:helba}) and plugging this into (\ref{eq:sigepa2}) gives

\begin{eqnarray} 
d\sigma &=& \frac{\alpha}{4\pi^{2}\left|q^{2}\right|}\,\frac{\tilde{w}}{w}\,\left(2\rho^{+\,+}\widetilde{\sigma}_{T}\,+\rho^{0\,0}\,\widetilde{\sigma}_{S}\right)\,\frac{d^{3}p'}{E'}\,.\label{eq:sigepa3}
\end{eqnarray}
$\tilde{w},\widetilde{\sigma}_{T}\, $ and $\widetilde{\sigma}_{S}\,$ refer to the inner reaction; $\rho^{+\,+}\,$ and $\rho^{0\,0}\,$ denote the transverse and scalar parts of (\ref{eq:rhoepa}).

\subsection{Approximation and photon spectra}

\begin{figure}
\centering
\psfrag{E}{$E,\vec{p}$}
\psfrag{E'}{$E',\vec{p}'$}
\psfrag{q}{$q$}
\psfrag{t}{$\theta$}
\includegraphics[width=0.5\textwidth]{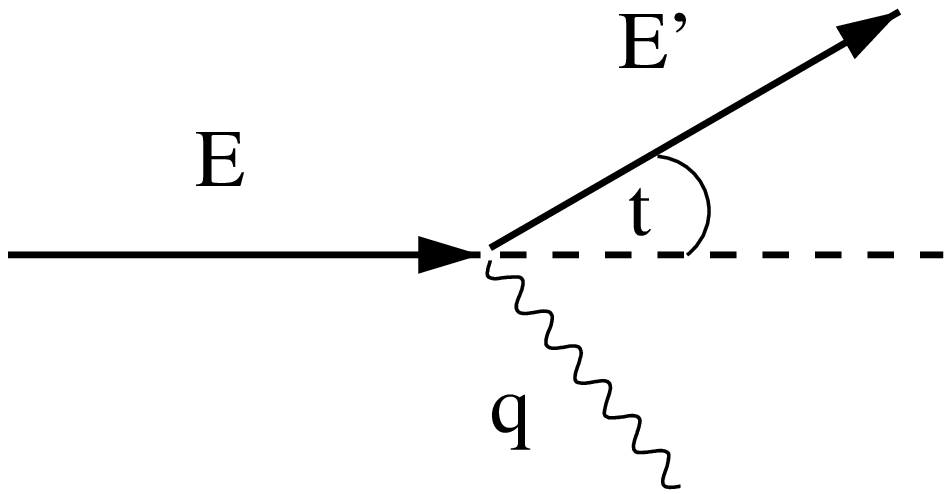}
\caption{kinematics in lab frame } 
\label{fig:depa}                                      
\end{figure}

From general kinematics (see figure \ref{fig:depa}), we get

\begin{eqnarray*}
-q^{2}&=&2\,(E\,E'-m^{2}-\left|\vec{p}\right|\left|\vec{p}\,'\right| \cos\theta),
\end{eqnarray*}
with $p^{\mu}\,$ and $p^{'\,\mu}\,$ being the fourvectors of the in- and outgoing electron and $\vartheta\,$ its scattering angle.

We obtain the following limits for $-q^{2}\,$:

\begin{eqnarray}
q^{2}_{min}&=&2\,\left(E E'-m^{2}-\left|\vec{p}\right|\left|\vec{p}\,'\right|\right)\,,\nonumber\\
q^{2}_{max}&=&2\,\left(E E'-m^{2}+\left|\vec{p}\right|\left|\vec{p}\,'\right|\right)\,,\label{eq:qminmax}
\end{eqnarray}
such that $q^{2}_{min}\,\leq\,-q^{2}_{i}\,\leq\,q^{2}_{max}$. For $E-\wg\,\gg\,m^{2}$

\begin{eqnarray}
q^{2}_{min}&=&\frac{m^{2}\,\wg^{2}}{EE'}\,\left(1+{\cal{O}}\left(\frac{m^{2}}{E'\,^{2}}\right)\right)\,,\nonumber\\
q^{2}_{max}&=&4E\,(E-\wg)\,.\label{eq:qminmax2}
\end{eqnarray}
In terms of the variables introduced in (\ref{eq:xval}), this becomes

\begin{eqnarray}
q^{2}_{min}\,=\,\frac{m^{2}\,x^{2}}{\left(1-x\right)}&,&
q^{2}_{max}\,=\,S\,\left(1-x\right)\,.\label{eq:qminmaxx}
\end{eqnarray}
It can be shown that

\begin{eqnarray}
\widetilde{\sigma}_{T}&=&\widetilde{\sigma}_{\gamma}\,(1+\mathcal{O}(q^{2}))\,,\nonumber\\
\frac{\widetilde{\sigma}_{S}}{\widetilde{\sigma}_{T}}&=&\mathcal{O}(q^{2})\,;
\end{eqnarray}
$\widetilde{\sigma}_{\gamma}$ is the total cross section for onshell photons. Using further that

\begin{displaymath}
\frac{d^{3}p'}{E'}\,=\, \frac{d\wg\,d(-q^{2})\,d\varphi}{2\,\sqrt{E^{2}-m^{2}}}\,\longrightarrow\,\pi\,\frac{d\wg\,d(-q^{2})}{\sqrt{E^{2}-m^{2}}}\,,
\end{displaymath}
we obtain in the limit of small $|q^{2}|$ for (\ref{eq:sigepa3})

\begin{eqnarray}
d\sigma&=&\widetilde{\sigma}_{\gamma}\left(\wg\right)\,d\tilde{n}\left(\wg,q^{2}\right) \label{eq:dsfluxepa}
\end{eqnarray}
with

\begin{eqnarray*}
d\tilde{n}(\wg,q^{2})&=&\frac{\alpha}{2\pi}\,\rho^{{  +\,+}}\,\frac{\wg\,d\wg\,d(-q^{2})}{E^{2}\,\left|q^{2}\right|}\,.
\end{eqnarray*}
Integration over $d(-q^{2})\,$ in the limits given by (\ref{eq:qminmax2}) gives

\begin{eqnarray}
dn(\wg)&=& \int^{q^{2}_{max}}_{q^{2}_{min}}d\tilde{n}(\wg,q^{2})\,=\,N(\wg)\,\,\frac{d\wg}{\wg}\,,\nonumber\\
N(\wg)&=&\frac{\alpha}{\pi}\left[\left(1-\frac{\wg}{E}+\frac{\wg^{2}}{2\,E^{2}}\right)\,\ln\frac{q^{2}_{max}}{q^{2}_{min}}\,-\,\left(1-\frac{\wg}{E}\right)\,\left(1-\frac{q^{2}_{min}}{q^{2}_{max}}\right)\right]\,.\nonumber\\
& & \label{eq:efluxepa}
\end{eqnarray}

Putting together (\ref{eq:dsfluxepa}) and (\ref{eq:efluxepa}) and changing to $x\,$ and $S\,$ as variables, we obtain

\begin{eqnarray}
d\sigma(x)&=&d\widetilde{\sigma}_{\gamma}(x)\,N(x)\,\frac{dx}{x}
\end{eqnarray}
with

\begin{equation}
N(x)\,=\,\frac{\alpha}{\pi}\left[\left(1-x+\frac{x^{2}}{2}\right)\,\ln\frac{q^{2}_{max}}{q^{2}_{min}}\,-\,\left(1-x\right)\,\left(1-\frac{q^{2}_{min}}{q^{2}_{max}}\right)\right]\,.\label{eq:Nepa}
\end{equation}
$q^{2}_{min}\,$ and $q^{2}_{max}\,$ are given by (\ref{eq:qminmaxx}).

\vspace{2mm}
For the Double Equivalent Photon Approximation, we obtain in a similar manner \cite{Budnev:1975}

\begin{eqnarray}
d\sigma(x_{1},x_{2})&=&\widetilde{\sigma}_{\gamma\gamma}(x_{1},x_{2})\,N(x_{1})\,N(x_{2})\,\frac{dx_{1}}{x_{1}}\,\frac{dx_{2}}{x_{2}}
\end{eqnarray}
with $N(x_{i})\,$ given by (\ref{eq:Nepa}).

\section{The $\gamma\,\gamma\,$-collider option }\label{sec:gammacoll}

\subsection{General idea}

\begin{figure}
\centering
\includegraphics[ width=0.7\textwidth]{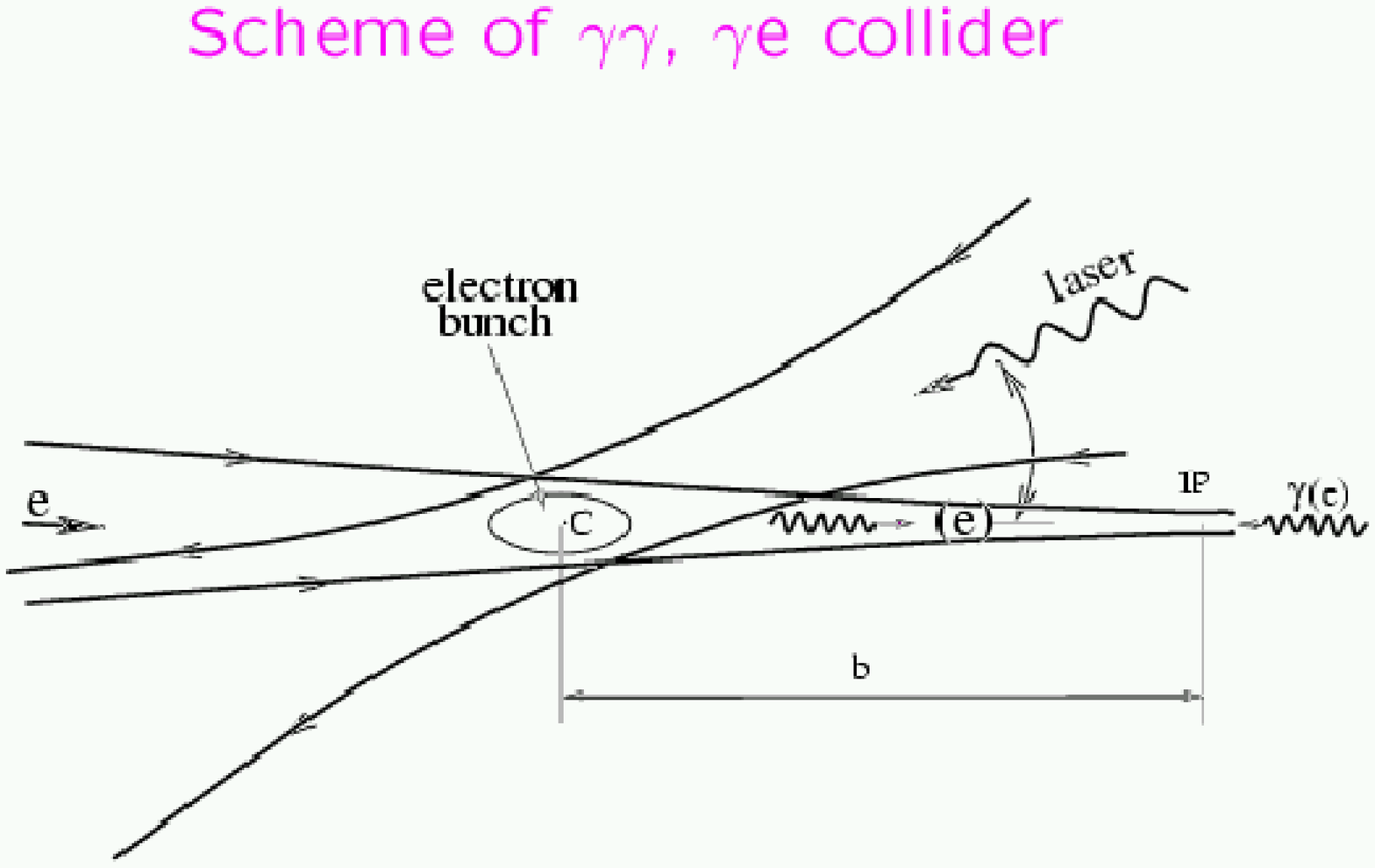}
\caption{principle of a photon collider; \cite{Telnov:2001}}
\label{fig:princphot} 
\end{figure}

Considerations for realization of a $\gamma\gamma\,$ collider for onshell photons go back to the 1980s \cite{Ginzburg:1983vm}, \cite{Ginzburg:1984yr}. The general idea is the production of photons by Compton-scattering; this way, the produced photons can have energies close to the initial electron/ positron energy.
Photon colliders are based on the collision of $e^{-}e^{-}$ beams. The Compton scattering takes place in the so-called conversion region; from this region, the produced photons travel a short distance to the interaction point where they collide with photons or electrons from the second $e^{-}$ beam. Photon colliders provide an environment for $\gamma\gamma,\,\gamma\,e,\,e^{+}e^{-}$ and $e^{-}e^{-}$ collisions.
A scheme of a photon collider is given in figure \ref{fig:princphot}.

\subsection{Compton scattering}
 
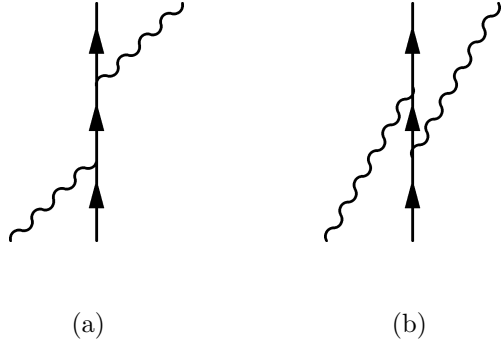
\begin{figure}[!h]
\centering
\subfigure[]{
\begin{fmffile}{fd41}
\fmfstraight
\fmfframe(15,20)(20,10){
\begin{fmfgraph*}(65,90)
\fmfbottomn{i}{2}
\fmftopn{o}{2}
\fmf{photon}{i1,v1}
\fmf{photon}{v2,o2}
\fmf{fermion}{i2,v1}
\fmf{fermion}{v1,v2}
\fmf{fermion}{v2,o1}
\fmfforce{(.5w,0.h)}{i2}
\fmfforce{(.5w,0.35h)}{v1}
\fmfforce{(.5w,0.65h)}{v2}
\fmfforce{(.5w,1.h)}{o1}
\end{fmfgraph*}}
\end{fmffile}}
\subfigure[]{
\begin{fmffile}{fd42}
\fmfstraight
\fmfframe(15,20)(20,10){
\begin{fmfgraph*}(65,90)
\fmfbottomn{i}{2}
\fmftopn{o}{2}
\fmf{photon}{i1,v2}
\fmf{photon}{v1,o2}
\fmf{fermion}{i2,v1}
\fmf{fermion}{v1,v2}
\fmf{fermion}{v2,o1}
\fmfforce{(.5w,0.h)}{i2}
\fmfforce{(.5w,0.35h)}{v1}
\fmfforce{(.5w,0.65h)}{v2}
\fmfforce{(.5w,1.h)}{o1}
\end{fmfgraph*}}
\end{fmffile}
}
\caption{Feynman diagrams contributing to Compton scattering}
\label{fig:Compscat}
\end{figure}

The basic Feynman diagrams describing Compton scattering are given by figure \ref{fig:Compscat}.

Calculations leading to differential cross sections for Compton scattering can be found in standard textbooks; therefore, we will only list the results and refer to the literature for further discussion (see e.g. \cite{Berestetzki:1980}).

\begin{figure}[b]
\centering
\psfrag{E0}{$E_{0}$}
\psfrag{w0}{$\wg_{0}$}
\psfrag{w}{$\wg$}
\psfrag{a}{$\al_{0}$}
\psfrag{t}{$\vartheta$}
\includegraphics[ width=0.4\textwidth]{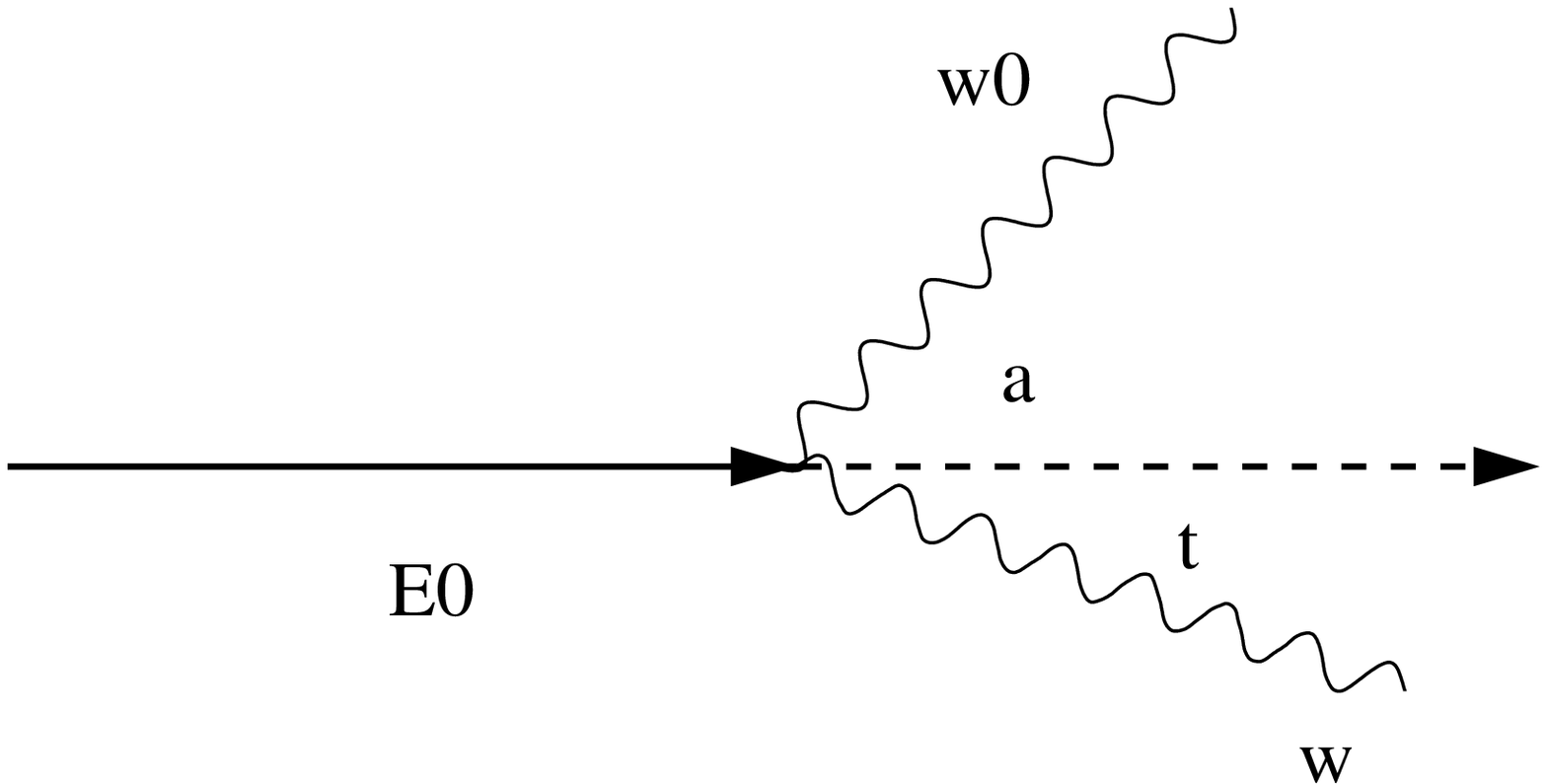}
\caption{Compton scattering in lab frame } 
\label{fig:compscatt}                                      
\end{figure}
The normalized differential cross section for Compton scattering is given by \cite{Ginzburg:1984yr},\cite{Telnov:1995hc}:

\begin{eqnarray}
\frac{1}{\sigma_{c}}\,\frac{d\sigma_{c}}{dx}&=&\frac{2\sigma_{0}}{x_{c}\,\sigma_{c}}\,\left[\frac{1}{1-x}+1-x-4\,r(1-r)+2\,\lambda_{e}P_{l}\,r\,x_{c}(1-2r)\,(2-x)\right]\nonumber\\ \label{eq:sigcomp}
\end{eqnarray}
with

\begin{eqnarray*}
x_{c}\,=\,\frac{4\,E_{0}\,\wg_{0}\,\cos^{2}(\al_{0}/2)}{m^{2}_{e}}\; &, & x\,=\,\frac{\wg}{E_{0}},\\
r\,=\,\frac{x}{x_{c}\,(1-x)}&, &\sigma_{0}\,=\,\pi\,\frac{e^{4}}{m^{2}}\\
\lambda_{e}\,:\;\mbox{mean electron/ positron helicity}&,&P_{l}\,:\;\mbox{mean photon helicity}\;. 
\end{eqnarray*}
$E_{0}\,$ and $\wg_{0}\,$ are the initial electron and photon energy respectively, $\al_{0}\,$ denotes the angle between them in the lab system (see figure \ref{fig:compscatt}).

In  case of nonzero electron and photon helicity, the total cross section is split into a non-polarized and a polarized part according to

\begin{eqnarray}
\sigma_{c}&=&\sigma^{np}_{C}\,+\,2\lambda_{e}\,P_{l}\,\sigma_{1}\label{eq:sigcomp2}
\end{eqnarray}
with

\begin{eqnarray*}
\sigma^{np}_{C}&=&\frac{2\sigma_{0}}{x_{c}}\left[\left(1-\frac{4}{x_{c}}-\frac{8}{x_{c}^{2}}\right)\ln(x_{c}+1)\,+\frac{1}{2}+\frac{8}{x_{c}}-\frac{1}{2(x_{c}+1)^{2}}\right]\;,\nonumber\\
\sigma_{1}&=&\frac{2\sigma_{0}}{x_{c}}\left[\left(1+\frac{2}{x_{c}}\right)\ln(x_{c}+1)\,-\frac{5}{2}+\frac{1}{x_{c}+1}-\frac{1}{2(x_{c}+1)^{2}}\right]\;.
\end{eqnarray*}
The number of photons with $\wg\,\in\,[E\,x;E(x+dx)]$ is given by

\begin{equation}
dN_{\gamma}\;=\;N_{e}\,\frac{k}{\sigma_{c}}\,\frac{d\sigma_{c}}{dx}\,dx
\end{equation}
with $N_{e}\,$ being the number of electrons and $k$ the conversion coefficient depending on the experimental setup; in short, $k$ is the number of photons produced per initial electron:

\begin{displaymath}
k\;=\;\frac{N_{\gamma}}{N_{e}}\;.
\end{displaymath}
Therefore, we see that we can relate an equivalent photon flux $N(x)\,$ to the cross section: 

\begin{equation}
N(x)\;\equiv\;\frac{1}{N_{e}}\,\frac{dN_{\gamma}}{dx}\;=\;k\frac{1}{\sigma_{c}}\,\frac{d\sigma_{c}}{dx}\label{eq:Ngg}
\end{equation}
such that 
\begin{displaymath}
d{\cal L}_{\gamma\gamma}\,=\, {\cal L}_{e_{+}e_{-}}\,N(x_{1})\,N(x_{2})\,dx_{1}\,dx_{2}
\end{displaymath}
according to (\ref{eq:Lrel2}).

\subsection{Additional effects in the conversion and interaction region}\label{sec:addeff}

Besides the part of the photon spectrum arising from Compton scattering, several additional effects in the conversion region lead to smearing of the original spectrum \cite{Telnov:1995hc},\cite{Badelek:2001xb}:

\begin{itemize}
\item{Nonlinear QED effects in Compton scattering}\\
In addition to the Feynman diagrams describing the scattering of one photon off an electron as described by Compton scattering, simultaneous interactions of one electron with several photons can take place as well. These are characterized by the parameter $\xi^{2}\,$. For $\xi^{2}\,\ll\,1\,$, the electron interacts with one photon only, for  $\xi^{2}\,\gg\,1\,$, simultaneous scattering with several photons takes place. The nonlinear QED effects lead to a shift in the effective electron mass and therefore a decrease in the maximal energy according to \cite{Badelek:2001xb}

\begin{equation}
\frac{\Delta\wg_{max}}{\wg_{max}}\;\propto\;\xi^{2}
\end{equation}
\item{Linear and nonlinear $e^{+}e^{-}\,$ creation in the conversion region}\\
Another important effect changing the photon spectra from an unsmeared Compton spectrum is the $e^{+}e^{-}\,$ pair creation by single (corresponding to linear creation) or multiple (corresponding to nonlinear creation) photon-collisions in the conversion region. It can be shown (\cite{Badelek:2001xb} and references therein) that the cross section for these processes highly depend on $x_{c}\,$ introduced in the previous section (see (\ref{eq:sigcomp})); for suppression of linear pair-creation, $x_{c}\,$ has to be smaller than 4.8. For non-linear effects, this limit is modified according to $x'_{c}\,=\,(1+\xi^{2})\,x_{c}$.
\item{Depolarization of initial electrons and photons}\\
Depolarization of initial photons and electrons can take place in the conversion region as well as the interaction region. The depolarization of the photons results from interaction with the polarized laser beam in the conversion region and  from the beam field of the $e^{-}e^{-}\,$ beams in the interaction region. Depolarization of the electrons is due to Compton scattering and interactions with the beam field. However, all effects have been found to be negligible \cite{Badelek:2001xb}.
\item{Coherent and incoherent $e^{+}e^{-}\,$ creation in the interaction region}\\
In interaction with the magnetic field arising from the beam, coherent as well as incoherent $e^{+}e^{-}\,$ creation can take place in the interaction region, resulting from single photons and $\gamma \gamma\,,\gamma e\,,e^{+}e^{-}\,$ interactions respectively. For the TESLA environment, these effects have been shown to be very small (see \cite{Badelek:2001xb}).
\item{beamstrahlung}\\
The beamstrahlung at $e^{+}e^{-}\,$ colliders contributes to the total luminosity as well as the photon-spectrum; for he latter, it mainly causes a peak in the low-energy regime (see also section \ref{sec:compspectra}). For further investigation, see \cite{Telnov:1995hc},\cite{Noble:1987yz}.
\item{Deflection by magnetic fields; synchrotron radiation}\\
These processes mainly take place in the region between the conversion and the interaction point; they are due to the field generated by possible extra magnets and the detectors.
\end{itemize}
The items given above describe the main additional effects taking place at a possible photon collider; they are partly included in analytical and numerical descriptions of the spectra given in the next section.

\subsection{Parameters for TESLA; numerical and analytical spectra}

The parameter for the TESLA photon collider are given by\footnote{Unfortunately, there are different notations concerning the electron polarization in the literature; in (\ref{eq:sigcomp2}), $|\lambda_{e}|\,\leq\,0.5\,$ denotes the electron helicity while in the TDR simulation code  $|\lambda_{e}|\,\leq\,1\,$ refers to electron polarization.} \cite{Badelek:2001xb}:

\begin{equation}
k  =  0.64,\, 
P_{c}  =  -1,\,
\lambda_{e}\,=\,0.85,\;
x_{c}  =  4.8,\,
\xi^{2}  =  0.3\,. \label{eq:tesval}
\end{equation}
For our calculation, we used spectra from a simulation done by V.Telnov \cite{Telnov:2001}; an upgraded version is available using \cite{Ohl:1997fi}. The simulation includes the effects of linear and non-linear Compton scattering, $e^{+}e^{-}\,$ pair creation, effects of additional magnetic fields, and beamstrahlung.

An analytic description \cite{Galynskii:2000fk} respects linear and non-linear Compton scattering and gives a good description of the high-energy peak of the Compton spectrum; however, the low-energy description is not accurate. Therefore, the analytic description has not been used in this work; a short description is given in Appendix \ref{chap:theophot}.

The photon spectrum taking from simulation files \cite{Telnov:2001} has to be modified \cite{Badelek:2001xb} according to:
\begin{equation}
{\cal L}_{\gamma\gamma}(z\,\ge\,0.64)\,=\,\frac{1}{3}\,{\cal L}_{e^{+}\,e^{-}} \label{eq:Lrel}
\end{equation}
with \cite{Ginzburg:1983vm}
\begin{eqnarray}
z&=&\sqrt{x_{1}\,x_{2}}\,,\nonumber\\ 
\frac{d\Lc(z)}{dz}&=&2\,z\,\int^{x_{max}}_{z^{2}/x_{max}} N(z^{2}/x)\,N(x)\,\frac{dx}{x}\,,
\end{eqnarray}
and  $x_{max}\;=\;\frac{x_{c}}{x_{c}+1}$.

\section{Comparison of spectra from DEPA and photon-collider}\label{sec:compspectra} 

We just list a short comparison of the spectra obtained for the DEPA (see (\ref{eq:Nepa})) and the photon collider at TESLA (\ref{eq:sigcomp}) using the spectra from simulations.

\begin{figure}
\centering
\includegraphics[angle=-90, width=0.95\textwidth]{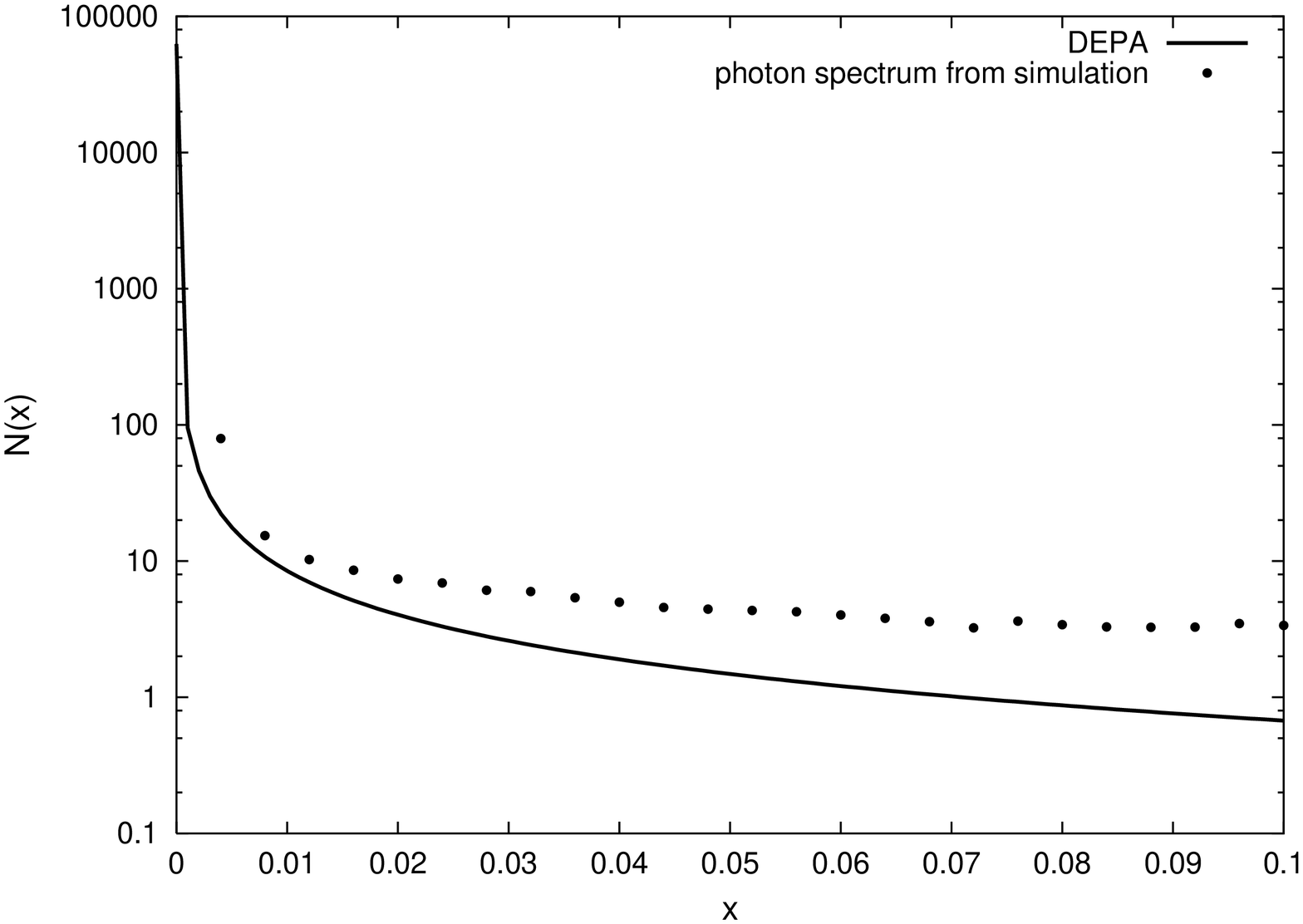}
\caption{comparison for $x\,\in\,[0;0.1]$ } 
\label{fig:app1}                                      
\end{figure}

Comparing 

\begin{equation}
\Nc_{x'}\;=\;\int^{x'}_{x_{min}} N(x)\,dx\,,
\end{equation}
we obtain for the photon spectra for the spectrum given by the DEPA and the $\gamma\gamma$-collider option

\begin{displaymath}
\centering
\begin{array}{|c|c|c|} \hline
 & \mbox{DEPA} & \gamma \mbox{-collider} \\ \hline \hline
\Nc_{1}& 1.34 & 1.94  \\ \hline
\Nc_{0.1}& 1.24 & 0.82 \\ \hline
\end{array}
\end{displaymath}
Here, the lower limit $x_{1\,\min}\,\equiv\,x_{min}\,=\,2.91\cdotp 10^{-7}\,$ is taken from (\ref{eq:limkt}) by setting $x_{2}\,=\,1$:

\begin{equation}
S\,x_{1}\,x_{2}\;\geq\,4\,m^{2}_{\pi}\,.
\end{equation}
Actually, from figure \ref{fig:app1}, we expect for the low energy regime with $x\,\leq\,0.1$, $\frac{N_{DEPA}}{N_{\gamma\,-coll}}\,\lessapprox\,1$; however, the file used for the modified Compton photon spectrum is limited in accuracy for very low x behavior; a better description should be achieved using \cite{Ohl:1997fi}. For high energies the Compton peak around $x\,=\,0.8$ gives a higher contribution form the photon collider spectrum. As the total expressions for the differential cross sections are peaked around low values of $x_{i}$, we expect higher cross sections for the direct $e^{+}e^{-}$ option.

For interpretation of the results using the spectrum from the DEPA and the simulation for the photon collider, we have to keep in mind that the former corresponds to a purely theoretical spectrum leaving out any smearing effects, while the latter includes smearing effects as described in section \ref{sec:addeff}. 

\begin{figure}
\centering
\includegraphics[angle=-90, width=0.95\textwidth]{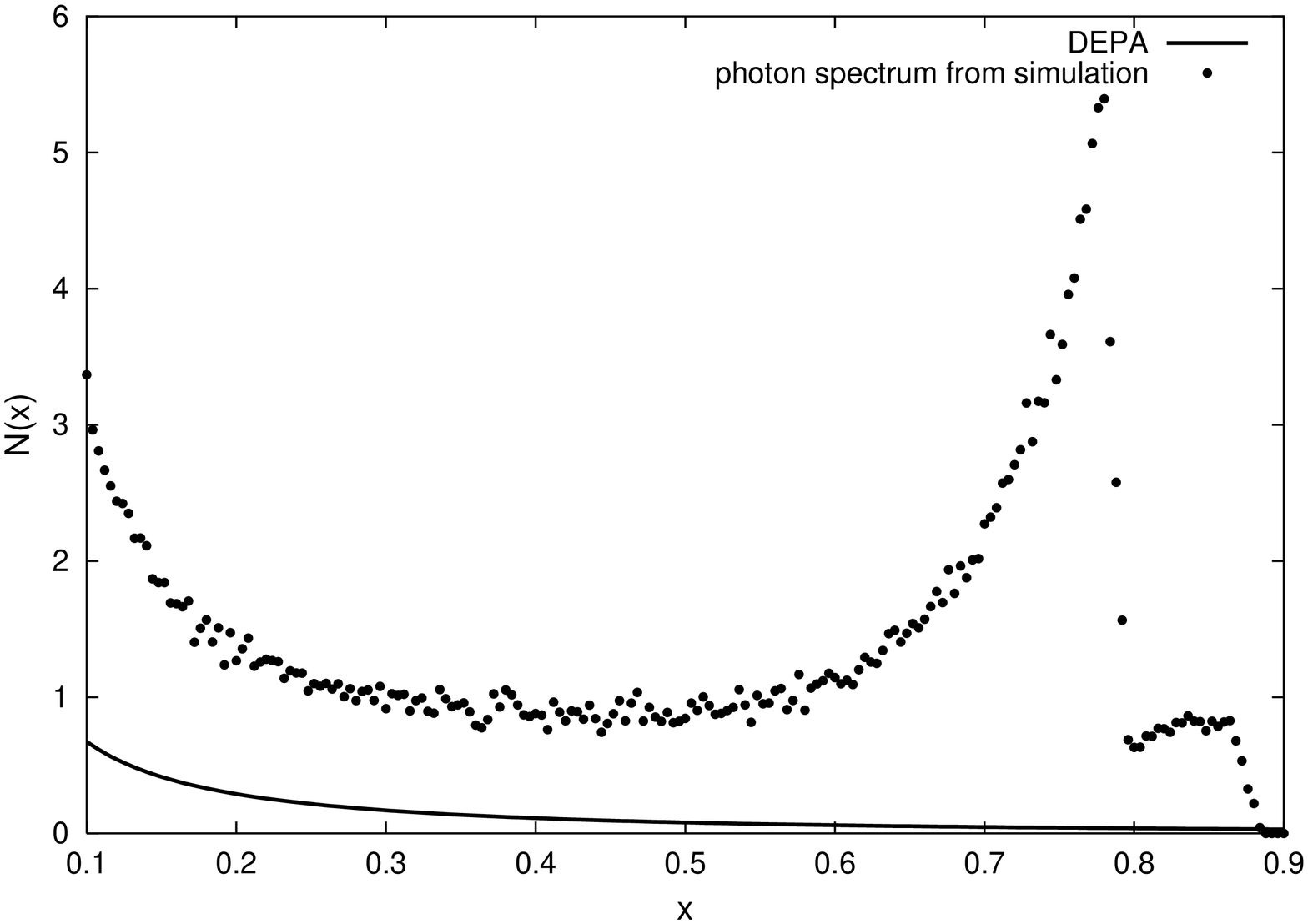}
\caption{comparison for $x\,\in\,[0.1;0.95]$}
\label{fig:app2}
\end{figure}

\section{Final expressions for {\large $\frac{d\sigma}{dk_{t}}\,, \frac{d\sigma}{dk_{l}}$} and {\large$\frac{d\sigma}{d\left|k_{t}\right|^{2}}$}}

Combining now the results of (\ref{eq:exprdsx}),(\ref{eq:Lrel2}), and (\ref{eq:efluxepa}), we obtain the following final expressions for the differential cross sections:

\begin{eqnarray}
\frac{d\sigma}{dk_{t}} & = &c'\, \frac{1}{8\,S\,\pi\,x_{1}\,x_{2}}\,\left|k_{t}\right|\, N(x_{1})\,N(x_{2})\,dx_{1}dx_{2}\times\nonumber\\ 
 & &  \left(\frac{1}{k^{0}_{1}{\left(k_{l+}\right)}}\,\frac{1}{\left|f'(k_{l+})\right|}\,\left|\cal M\right|^{2}_{t}{ (k_{l+})}\,+\,\frac{1}{k^{0}_{1}(k_{l-})}\,\frac{1}{\left|f'{ (k_{l-})}\right|}\,\left|\cal M\right|^{2}_{t}{ (k_{l-})}\right)\;, \nonumber \\ \nonumber \\
\frac{d\sigma}{dk_{l}} & = &c'\, \frac{1}{8\,S\,\pi\,x_{1}\,x_{2}}\,\left|k_{t}{ (k_{l})}\right|\,\frac{1}{k^{0}_{1}{ (k_{t})}}\,\frac{1}{\left|f'{ (k_{t})}\right|}\,\left|\cal M\right|^{2}_{t}{ (k_{t})}\,N(x_{1})\,N(x_{2})\,dx_{1}dx_{2}\;, \nonumber\\ \nonumber \\
\frac{d\sigma}{d\left|k_{t}\right|^{2}} & = &c'\, \frac{1}{16\,S\,\pi\,x_{1}\,x_{2}}\,N(x_{1})\,N(x_{2})\,dx_{1}dx_{2}\,\times\nonumber\\
 & & \left(\frac{1}{k^{0}_{1}{\left(k_{l+}\right)}}\,\frac{1}{\left|f'(k_{l+})\right|}\,\left|\cal M\right|^{2}_{t}{ (k_{l+})}\,+\,\frac{1}{k^{0}_{1}(k_{l-})}\,\frac{1}{\left|f'{ (k_{l-})}\right|}\,\left|\cal M\right|^{2}_{t}{ (k_{l-})}\right)\;. \nonumber\\ \label{eq:exprdsf}
\end{eqnarray}
Multiplication of any of these cross sections with the luminosity $\Lc_{e^{+}e{-}}$ for the $e^{+}e^{-}$ beam of the linear collider will give the expected counting rates as

\begin{equation}
\mathcal{N}\;=\;\sigma\,\Lc_{e^{+}e^{-}}.
\end{equation}
We already introduced the relation between $\Lc_{e{+}e^{-}}$ and $\Lc_{\gamma\gamma}$ via (\ref{eq:Lrel2}).
 
Here, $c'$ is a constant respecting (\ref{eq:Lrel}) for the photon collider option; it depends on the used file from \cite{Telnov:2001}. When we consider the direct $e^{+}e^{-}$ option, $c'\,=\,1$.  
The file used in our calculation corresponds to the following parameters for the photon collider:

\begin{equation}
k  =  0.632,\, 
P_{c}  =  -1,\,
\lambda_{e}\,=\,0.85,\;
x_{c}  =  4.6,\,
\xi^{2}  =  0.3\,
\end{equation}
with photon production by Compton scattering for the electron as well as the positron beam; here, $c'\,\approx\,9.6\,\times\,10^{-12}$.

\section{Detector cuts}\label{sec:detcut}

\begin{figure}
\centering
\psfrag{e+}{$e^{+}$}
\psfrag{e-}{$e^{-}$}
\psfrag{kl}{${\scripts \vec{k_{l}}}$}
\psfrag{kt}{${\scripts \vec{k_{t}}}$}
\psfrag{t}{${\scripts \theta}$}
\psfrag{k}{${\scripts \vec{k}}$}
\psfrag{ez}{$\hat{e}_{z}$}
\psfrag{D}{$\mbox{detector}$}
\includegraphics[width=0.5\textwidth]{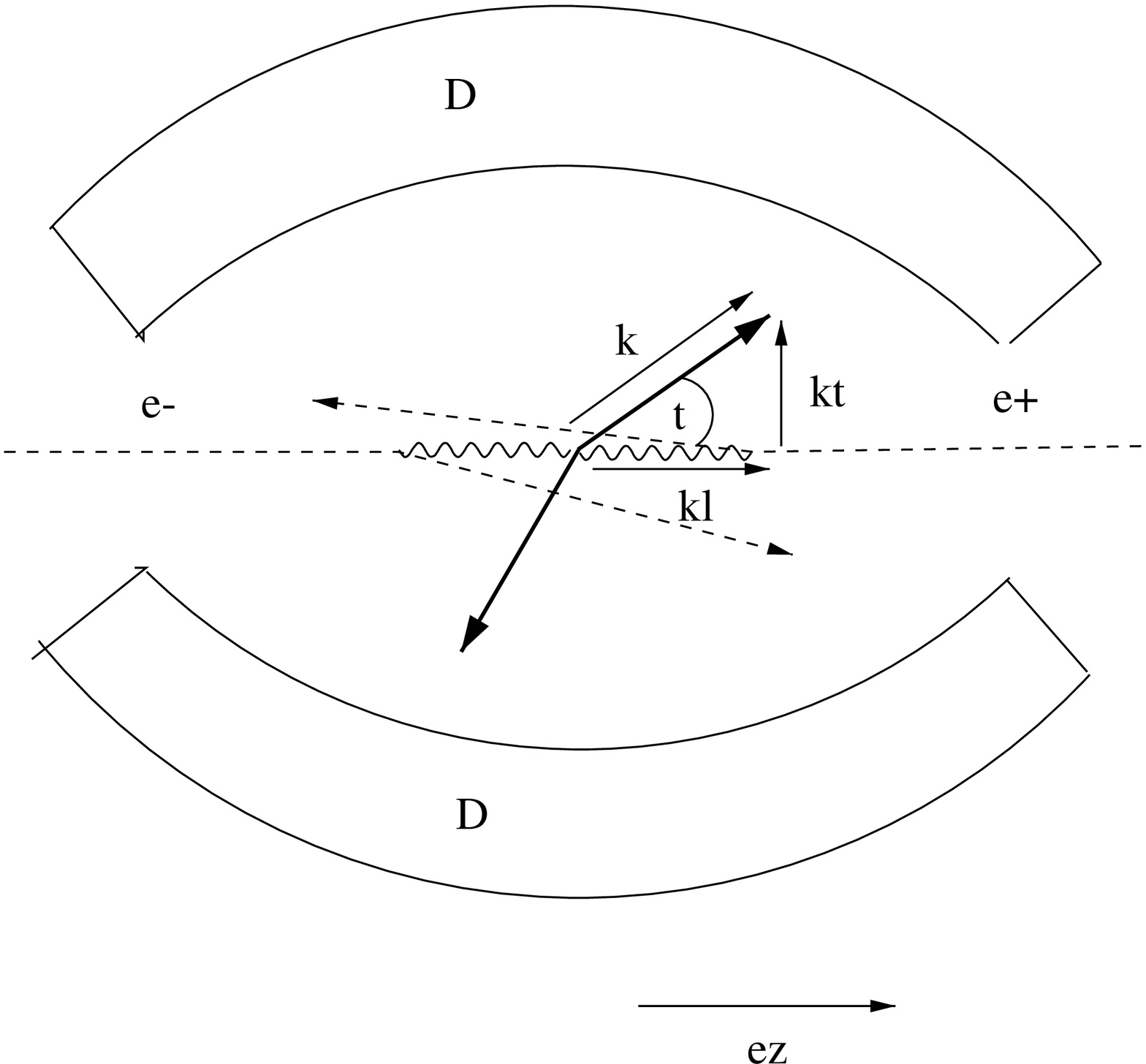}
\caption{kinematics in lab frame } 
\label{fig:cuts}                                      
\end{figure}

In experimental data taking, the accessible kinematic area is limited by the detector geometry as well as energy solution. We take this into account by introducing detector cuts, which are given by a minimal angle required for outgoing particles as well as a minimal energy for each of them. The limitation of the angle is done by introducing $|\cos \theta|_{\max}\,$.

Taking $|\cos \theta|_{\max}\,$ and $E_{\min}\,$ to be given, we can easily determine the corresponding kinematic restrictions;
from

\begin{eqnarray}
|\cos \theta|&\leq&|\cos \theta_{max}|\,,\nonumber \\
|\cos \theta|&=&\frac{|k_{l}|}{|\vec{k}|}
\end{eqnarray}
with $|\vec{k}|\,=\,\sqrt{k^{2}_{l}+k^{2}_{t}}\,$, we arrive at
\begin{eqnarray}
|k_{l}| & \leq &\frac{k_{t}}{\sqrt{c_{1}}}\label{eq:limkld}
\end{eqnarray}
with

\begin{displaymath}
 c_{1}\;=\;\frac{1}{(|cos\theta|_{max})^{2}}-1\,.
\end{displaymath}
From (\ref{eq:zvar}) and (\ref{eq:xval}), we know that

\begin{displaymath}
k^{0}_{2}\;=\;\frac{\sqrt{S}}{2}\,(x_{1}+x_{2})-k^{0}_{1}\;,
\end{displaymath} 
so we can test 

\begin{equation}
k^{0}_{i}\;\leq\;E_{\min}
\end{equation} 
for both outgoing particles.
\vspace{3mm}

The values used for the single experiments are

\vspace{3mm}
\begin{displaymath}
{\scriptstyle
\begin{array}{|l |c|c|c|} \cline{1-4}
\rule[-2mm]{0mm}{7mm}  Experiment &  |\cos\theta |_{max} & E_{min} \;[\mbox{GeV}] & \sqrt{S_{lab}}\;[\mbox{GeV}]  \\  \hline \hline
TESLA(e^{+}e^{-})\;\footnotemark  & 0.98  & 0.5& 500   \\ \cline{1-4}
TESLA(\gamma\gamma)\;\footnotemark &0.98 &0.5 & 500  \\  \cline{1-4}
OPAL(e^{+}e^{-}) {\scriptstyle(LEPI)}\;\footnotemark & 0.81 & 0.4 & 92   \\ \cline{1-4}
OPAL (e^{+}e^{-}){\scriptstyle(LEPII)}\;\addtocounter{footnote}{-1} \footnotemark  & 0.81 & 0.4 & 200  \\ \cline{1-4}
BaBar(e^{+}e^{-})\;\footnotemark  &0.96; 0.77 \footnotemark &0.08 & 10.58  \\ \cline{1-4}
\end{array}
}
\addtocounter{footnote}{-4}\footnotetext{\cite{Desch:2002}}
\stepcounter{footnote}\footnotetext{\cite{Telnov:2002}}
\stepcounter{footnote}\footnotetext{\cite{Lillich:2002}}
\stepcounter{footnote}\footnotetext{\cite{Aubert:2001tu}, \cite{Schieck:2002}}
\stepcounter{footnote}\footnotetext{Backward scattering}
\end{displaymath}

For comparison, we are also investigating the effects of cuts on $k_{t}\,$ instead of $k^{0}_{i}\,$. 

Notice that the values for  $ E_{min}\,$  and $|\cos\theta|_{max}\,$  are given in the lab-frames of the corresponding experiment. In cases where this does not coincide with the center of mass frame of the $e^{+}e^{-}$ system, we have to convert these values into the $e^{+}e^{-}$ - lab-frame using

\begin{eqnarray}
\binom{ k^{0}_{i}}{k^{3}_{i}}_{cm} & = & \left(\begin{array}{cc} \cosh\eta & -\sinh\eta \\ -\sinh \eta & \cosh \eta \\ \end{array}\right)\,\binom{ k^{0}_{i}}{k^{3}_{i}}_{lab}\,.
\end{eqnarray}
For BaBar, $\eta\, \approx\,0.533\,$.

In the same environment, there are two different angular limitations corresponding to forward and backward scattering in the lab frame. The values for BaBar are given by 

\begin{displaymath}
-0.77\,\leq\,\cos\theta_{lab}\,\leq\,0.96\,.
\end{displaymath}
Of course this equally has to be taken into account when applying the cuts in cross section calculations.

In contrast to experiments at electron-proton colliders, the outgoing electrons are not detected at the linear colliders mentioned above; therefore, there are no cuts on the incoming photon energies $\wg_{i}$ (for a comparison, see for example \cite{Kilian:1998ew},\cite{Tapprogge:1996}).

\newpage 

\chapter[{\normalsize{Results for various sets of parameters and different colliders}}]{\LARGE{Results for various sets of parameters and different colliders}}

\section{General expectation; $e^{+}e^{-}$-option}\label{sec:genexp}

First, we consider the differential cross sections in the environment of a linear $e^{+}e^{-}$ collider without any detector cuts for center of mass energy $\sqrt{S}\,=\,500\,$GeV like the TESLA environment. The results can easily be transferred to results for lower-energy machines such as LEP or BaBar. The photon collider option at TESLA will be treated separately.

\vspace{3mm}

\begin{figure*}
\begin{itemize}
\item{ for $\frac{d\sigma}{d|k_{t}|^{2}}$:}
\vspace{-13mm}
\begin{center}
\includegraphics[angle=-90, width=0.95\textwidth]{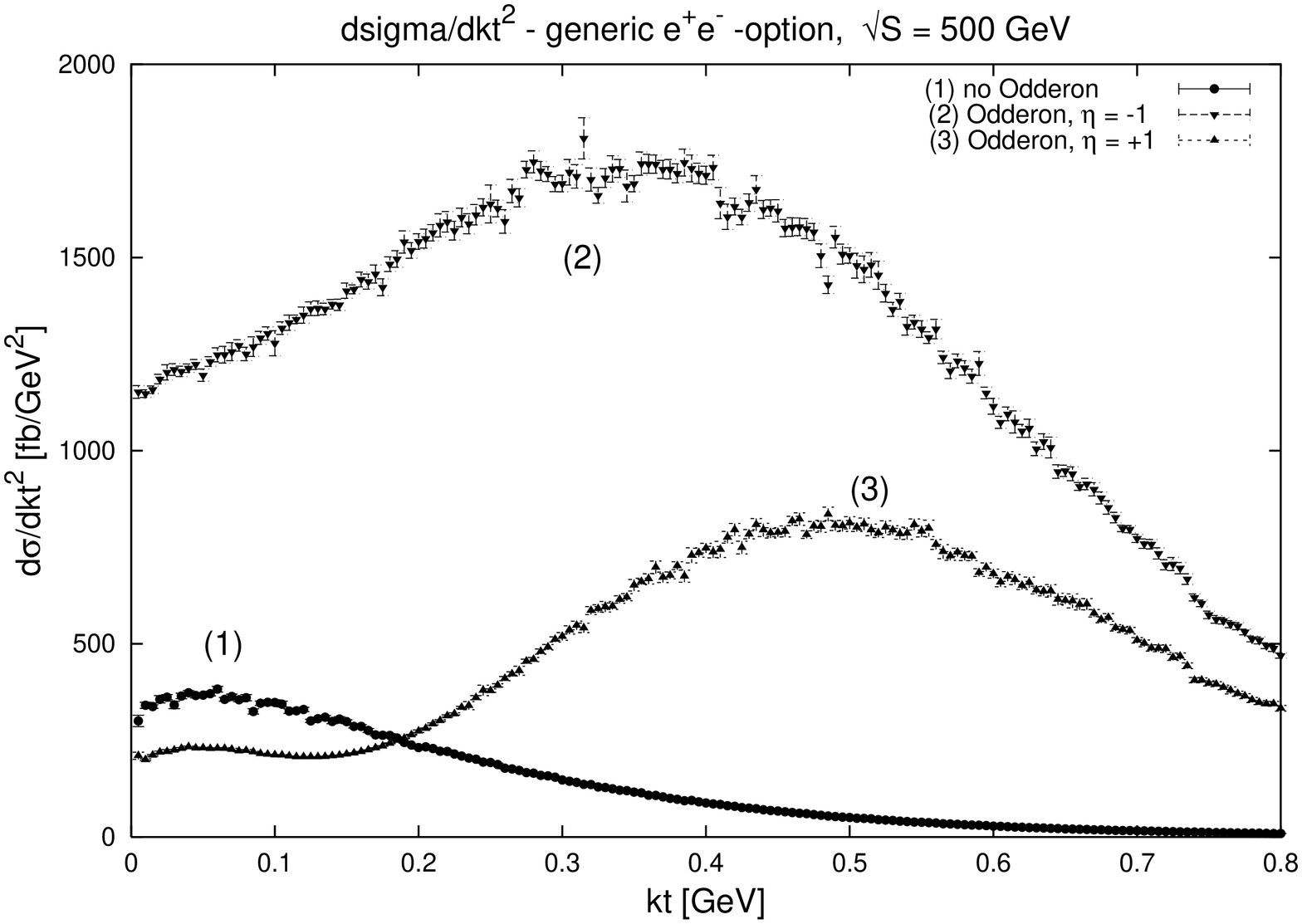}
\caption{different values for $\eta_{\odd}$}
\label{fig:1}
\end{center}
\vspace{-5mm}
\item{ for $\frac{d\sigma}{dk_{l}}$:}
\vspace{-13mm}
\begin{center}
\includegraphics[angle=-90, width=0.95\textwidth]{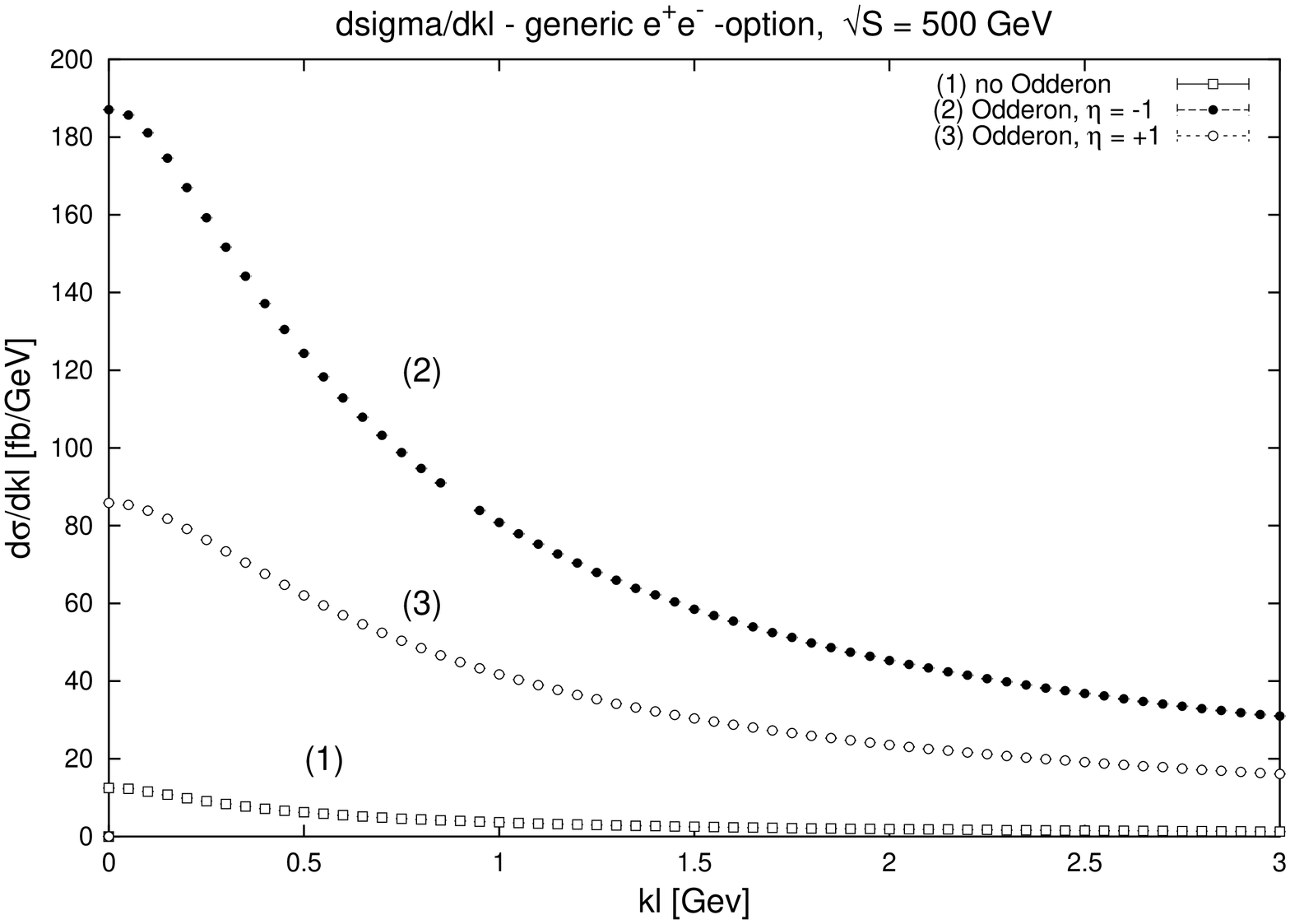}
\caption{different values for $\eta_{\odd}$}
\label{fig:2}
\end{center}
\end{itemize}
\end{figure*}

With the expressions for $\frac{d\sigma}{d|k_{t}|^{2}}\,$ and $\frac{d\sigma}{dk_{l}}\,$ given by (\ref{eq:exprdsf}), we obtain the differential cross sections shown in figures \ref{fig:1} and \ref{fig:2}.

We can explain the shapes of the differential cross sections by remembering the modification of the matrix element through the Odderon contribution as described in section \ref{sec:incluodd}. We will here focus in the discussion of $\frac{d\sigma}{d|k_{t}|^2}\,$, as effects of phase and parameter variation lead to more obvious modifications of this cross section.

From (\ref{eq:totmgodd}), we see that the inclusion of the Odderon contribution lead to modifications depending on constants $C^{\gamma+\odd}_{i}$ given by

\begin{equation}
C^{\gamma+\odd}_{i}\;=\;1+\kappa\,\lambda_{i}(s/s_{o})^{\al_{\odd}(t)-1}\,\left[2\cos\left(\frac{\pi}{2}(\al_{\odd}(t)-1)\right)+\kappa\,\lambda_{i}(s/s_{0})^{\al_{\odd}(t)-1}\right]
\end{equation}\label{eq:Cagain}
with

\begin{displaymath}
\lambda_{i}\;=\;\left\{ \begin{array}{c} t\;\mbox{for $i\,=\,1$} \\ u\;\mbox{for $i\,=\,2$} \end{array} \right.
\end{displaymath}
and $C^{\gamma+\odd}_{3}$ defined similarly but depending on $t$ as well as on $u$. $C^{\gamma+\odd}_{1/2}$ modify the exchange in the $t$- and $u$-channel respectively while $C^{\gamma+\odd}_{3}$ modifies the matrix elements corresponding to mixtures of $t$ and $u$ channel exchange. Looking at $C^{\gamma+\odd}_{1}\,$ as given by (\ref{eq:Cagain}), we see that we can distinguish three contributions. In (\ref{eq:Mmodbyodd}), we calculated the Odderon contribution by modifying the propagator:

\begin{equation}
\frac{1}{q^{2}_{i}}\;\rightarrow\;\frac{1}{q^{2}_{i}}+\frac{1}{q^{2}_{i}}\,\kappa q^{2}_{i}(-is/s_{0})^{\al_{\odd}(t)-1}\,,\label{eq:oddsub2}
\end{equation}

\begin{figure}[h!]
\centering
\subfigure{
\begin{fmffile}{fd10a}
\fmfstraight
\begin{fmfgraph*}(45,5)
\fmfleft{i1}
\fmfright{o1}
\fmf{photon}{i1,o1}
\end{fmfgraph*}
\end{fmffile}}
$\longrightarrow$
\subfigure{
\begin{fmffile}{fd10b}
\fmfstraight
\begin{fmfgraph*}(45,5)
\fmfleft{i1}
\fmfright{o1}
\fmf{photon}{i1,o1}
\end{fmfgraph*}
\end{fmffile}}
+
\subfigure{
\begin{fmffile}{fd10c}
\fmfstraight
\begin{fmfgraph*}(45,5)
\fmfleft{i1}
\fmfright{o1}
\fmf{zigzag}{i1,o1}
\end{fmfgraph*}
\end{fmffile}}
$\,.$
\end{figure}
Considering for example $|\M^{\gamma+\odd}_{1}|^{2}$, we can write

\begin{figure}[h!]
\centering
$|\M^{\gamma+\odd}_{1}|^{2}\;\propto\;\Big($
\subfigure{
$|$
\begin{fmffile}{fd11a}
\begin{fmfgraph*}(15,5)
\fmfleft{i1}
\fmfright{o1}
\fmf{photon}{i1,o1}
\end{fmfgraph*}
\end{fmffile}
$|^{2}$
}
+
\subfigure{
$|$
\begin{fmffile}{fd11b}
\fmfstraight
\begin{fmfgraph*}(15,5)
\fmfleft{i1}
\fmfright{o1}
\fmf{zigzag}{i1,o1}
\end{fmfgraph*}
\end{fmffile}
$|^{2}$
}
+
\subfigure{
\subfigure{
\begin{fmffile}{fd11c}
\fmfstraight
(
\begin{fmfgraph*}(15,5)
\fmfleft{i1}
\fmfright{o1}
\fmf{photon}{i1,o1}
\end{fmfgraph*}
)
(
\begin{fmfgraph*}(15,5)
\fmfleft{i1}
\fmfright{o1}
\fmf{zigzag}{i1,o1}
\end{fmfgraph*}
)$^{*}$
\end{fmffile}}
+
\subfigure{
\begin{fmffile}{fd11d}
(
\begin{fmfgraph*}(15,5)
\fmfleft{i1}
\fmfright{o1}
\fmf{zigzag}{i1,o1}
\end{fmfgraph*}
)
(
\begin{fmfgraph*}(15,5)
\fmfleft{i1}
\fmfright{o1}
\fmf{photon}{i1,o1}
\end{fmfgraph*}
)$^{*}\,\Big)\;\times\,|\M^{\gamma}_{1}|^{2}$
\end{fmffile}}
}
\caption{}
\label{fig:mprop}
\end{figure}
\begin{figure}[t]
\centering
\includegraphics[angle=-90, width=0.5\textwidth]{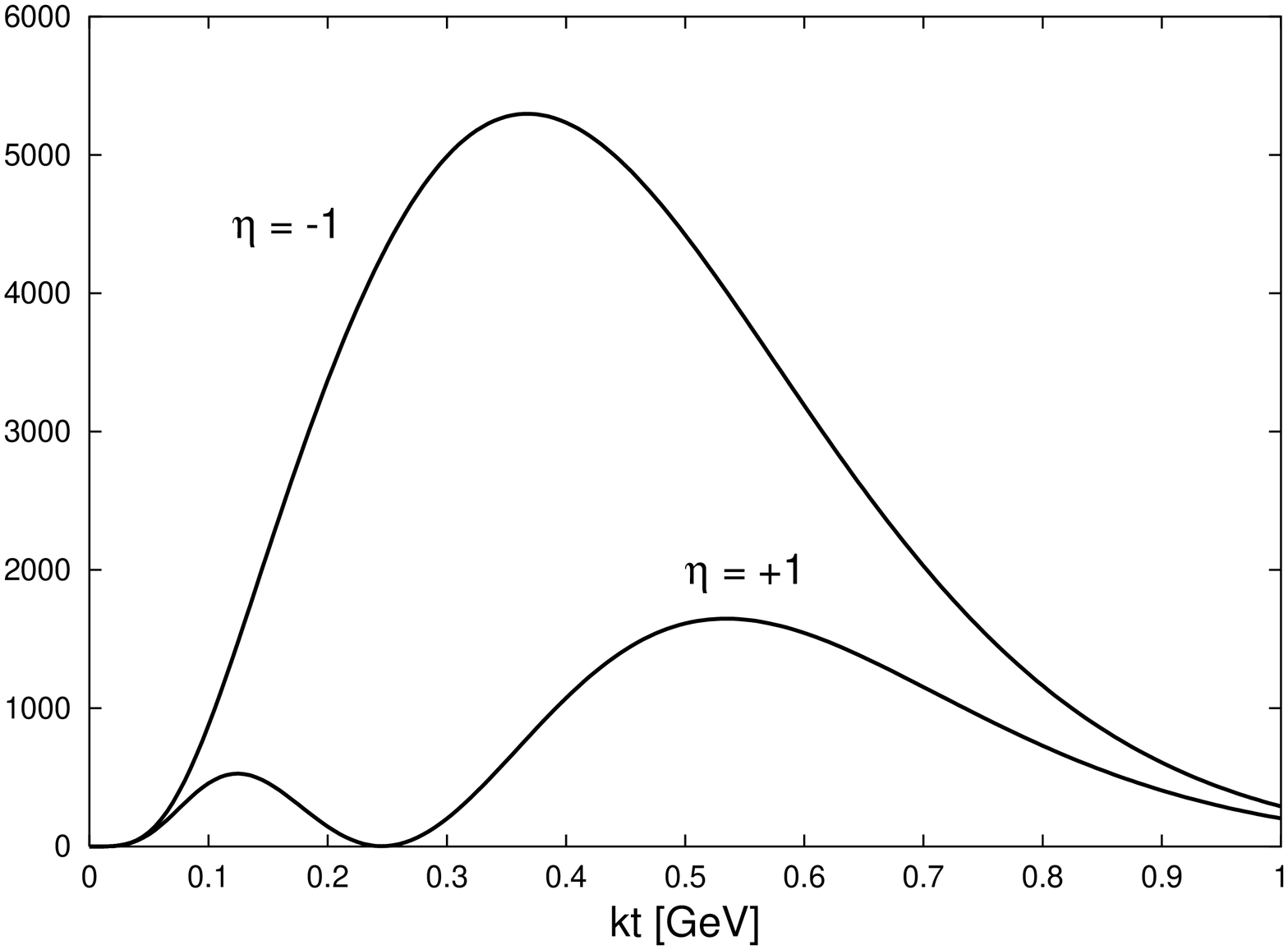}
\caption{results for $\eta_{\odd}\,=\pm1$ from simple estimation}
\label{fig:bothetas}
\end{figure}
\begin{figure}[ht]
\centering
\includegraphics[angle=-90, width=0.5\textwidth]{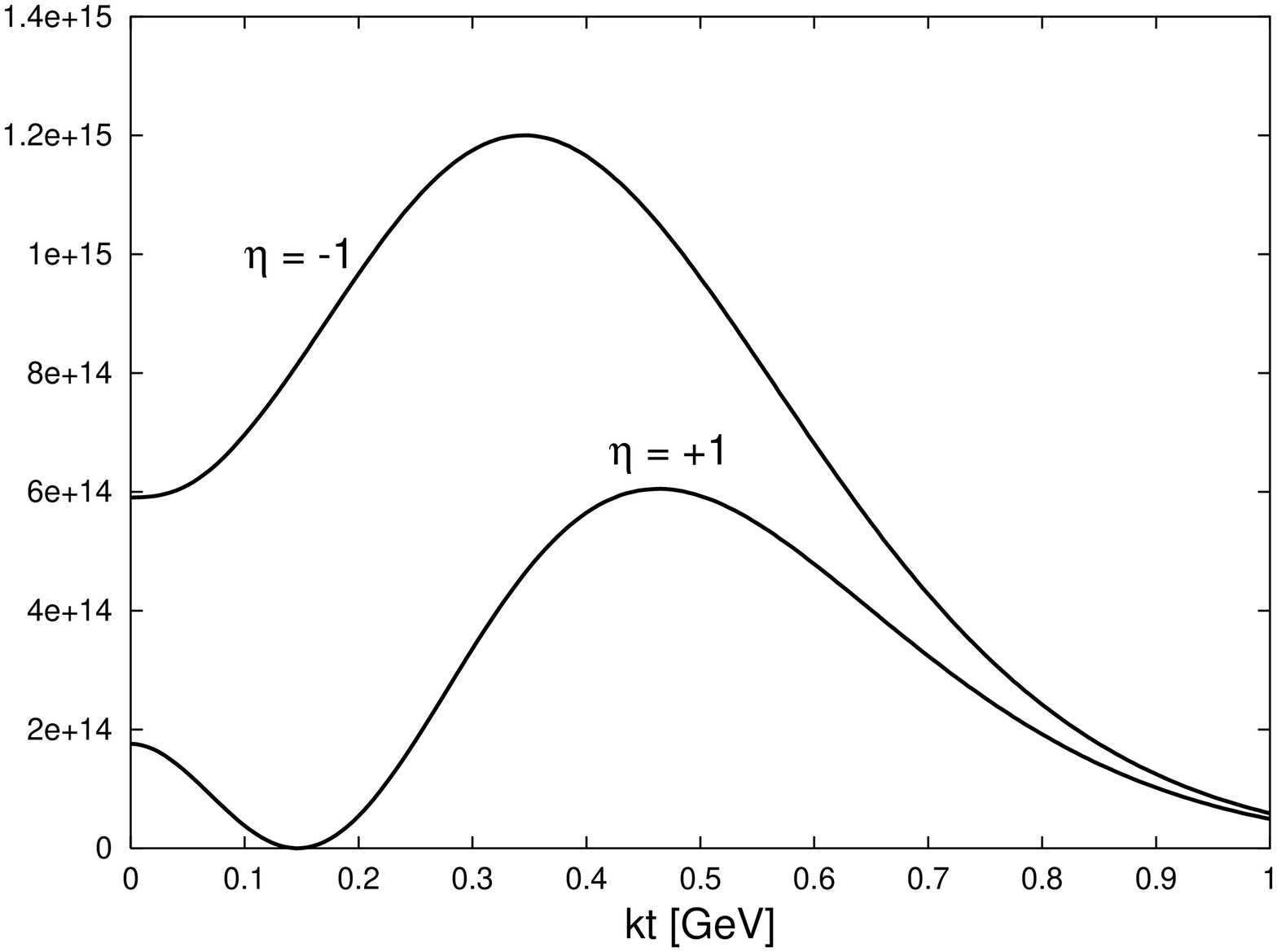}
\caption{results for $\eta_{\odd}\,=\pm1$ from simple estimation, higher order mass terms included}
\label{fig:bothetas2}
\end{figure}

\noindent
(see (\ref{eq:m1goddtot})).
We can now associate the single terms of $C^{\gamma+\odd}_{1}$ with the contributions in figure \ref{fig:mprop}: 
\begin{displaymath}
\begin{array}{l c l c}
1&:&\mbox{photon contribution} & (1)\;, \\
(\kappa\,t\,(s/s_{0})^{\al_{\odd}(t)-1})^{2}&:&\mbox{Odderon contribution}& (2)\;,\\
\kappa\,t\,(s/s_{0})^{\al_{\odd}(t)-1}\,2\cos(\frac{\pi}{2}\,(\al_{\odd}(t)-1))&:&\mbox{mixed contributions}& (3)\;. 
\end{array}
\end{displaymath}
We investigate the contributions $(2)$ and $(3)$ separately in order to determine the influence of the Odderon contribution on $\frac{d\sigma}{d|k_{t}|^{2}}$. Taking the parameters for the Odderon propagator and coupling from (\ref{eq:oddval}), we see that (2) dominates for $k_{t}\,\leq\,0.35\;\mbox{GeV}$.
We will now make an approximation for $\frac{d\sigma}{d|k_{t}|^{2}}$ by only considering terms with $x_{1}\,=\,x_{2}\,\equiv\,x_{\min}$ with $x_{min}$ given by

\begin{equation}
x_{min}\;=\;2\sqrt{\frac{m^{2}+k^{2}_{t}+\vare}{S}}
\end{equation}
(see (\ref{eq:limkt})). This approximation only takes the contribution resulting from the Jacobian peak into account; it already provides a rough estimate of the $k_{t}$-dependence of the differential cross section. 
In this limit, the quantities describing the scattering process and the kinematics given in chapters \ref{chap:kin} and \ref{chap:Mel} are given by:

\begin{eqnarray*}
k^{\pm}_{l}\;=\;\pm\sqrt{\vare}&,&
|f'(k_{t})|\;=\;4\sqrt{\vare}\;,\\
k^{0}_{1}\; = \; \sqrt{k_{t}^{2}+m^{2}}&,&
s\;=\;4\,(k^{2}_{t}+m^{2})\;,
\end{eqnarray*}
\begin{eqnarray}
t&=&-2\,k^{2}_{t}-m^{2}\,\pm\sqrt{\vare}\,\sqrt{m^{2}+k^{2}_{t}}\;, \nonumber\\
u&=&-2\,k^{2}_{t}-m^{2}\,\mp\sqrt{\vare}\,\sqrt{m^{2}+k^{2}_{t}}\;.\label{eq:approxval}
\end{eqnarray}
The expression above are valid up to $\mathcal{O}(\vare)$. Taking further
\begin{displaymath}
t\,\approx\,u\,\approx\,-2\,k^{2}_{t}-m^{2}
\end{displaymath}
 and leaving out higher order terms in $m^{2}$ in $|\M|^{2}$, we obtain
\begin{equation}
|\M|^{2}\;\approx\;t^{2}\,T_{t}^{4}\,(2+\frac{s^{4}}{t^{4}})\;.\label{eq:Mapprox}
\end{equation}
Here, we assumed $m^{2}\,\ll\,k^{2}_{t}$. Of course, this condition does not hold for small values of $k_{t}$. However, we will see that this approximation still gives a good estimation for the behavior of the differential cross section. A more thorough investigation should include higher order mass terms in $|\M|^{2}$.
Equally, we obtain in this limit
\begin{equation}
C^{\gamma+\odd}_{1}\;=\;C^{\gamma+\odd}_{2}\;=\;C^{\gamma+\odd}_{3}\;\equiv\;C^{\gamma+\odd}\;.
\end{equation}
In order to understand the behavior of $\frac{d\sigma}{d|k_{t}|^{2}}$, we therefore have to distinguish the behavior of $|\M|^{2}$ depending on $k_{t}$, the behavior of $C^{\gamma+\odd}$ depending on $k_{t}$, kinematic effects, and the influence of the photon spectra.
\begin{itemize}
\item{$|\M|^{2}$}\\
From (\ref{eq:approxval}), we see that $t^{2}\sim k^{4}_{t}$ while $T^{4}\propto (\frac{1}{1+k^{2}_{t}/c})^{4}$. Using the approximations given above, we obtain $|\M|^{2}\,(k_{t})$ given by figure \ref{fig:Msq} with a maximum at approximately $k_{t}\approx\,0.54$ GeV.

\begin{figure}[th]
\centering
\includegraphics[angle=-90, width=0.5\textwidth]{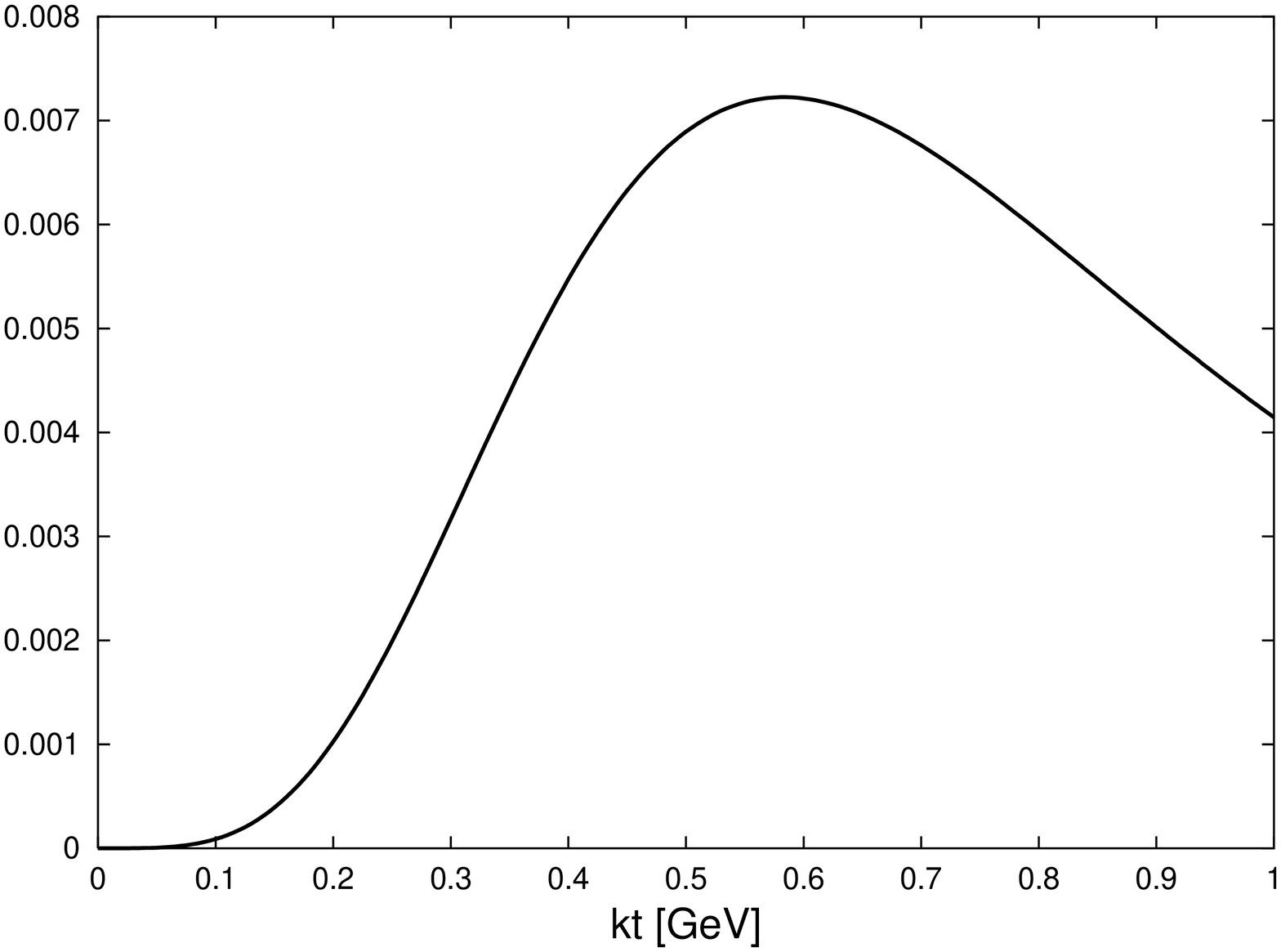}
\caption{$|\M|^{2}(k_{t})$, no Odderon; not to scale}
\label{fig:Msq}
\end{figure}

\item{Odderon contribution}\\
For the two terms in $C^{\gamma+\odd}$ describing the pure and mixed Odderon contributions, we obtain
\begin{eqnarray}
\mbox{(2)}&\propto&|\kappa|^{2}\,k^{4}_{t}\,s^{2(\al(0)-2\al'\,k^{2}_{t}-1)}\,,\nonumber\\
\mbox{(3)}&\propto & -2\,\eta_{\odd}\,|\kappa|\, k^{2}_{t}\cos(\frac{\pi}{2}(\al(0)-2\al'\,k^{2}_{t}-1))\,s^{\al(0)-2\al'\,k^{2}_{t}-1}\,.\nonumber\\ \label{eq:propprop}
\end{eqnarray}
The single contributions can be seen in figure \ref{fig:oddconnomat}. Multiplication with $|\M|^{2}$ leads to the form of the cross section displayed in figure \ref{fig:oddconmat}.
\begin{figure}[th]
\centering
\subfigure[$C^{\gamma+\odd}$; not to scale]{\label{fig:oddconnomat}
\includegraphics[angle=-90, width=0.45\textwidth]{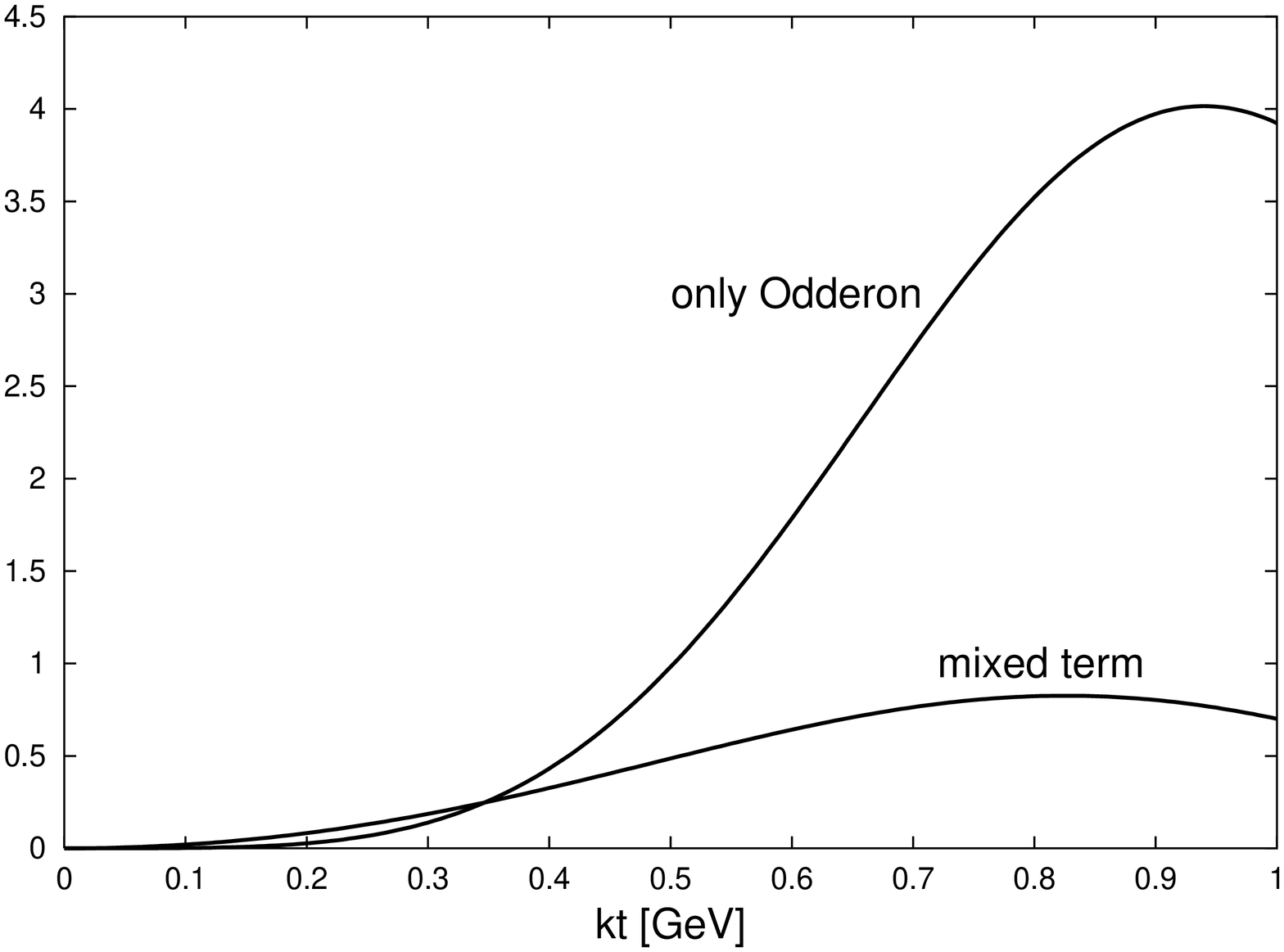}
}
\subfigure[$C^{\gamma+\odd}\,|\M|^{2}$; not to scale]{\label{fig:oddconmat}
\includegraphics[angle=-90, width=0.45\textwidth]{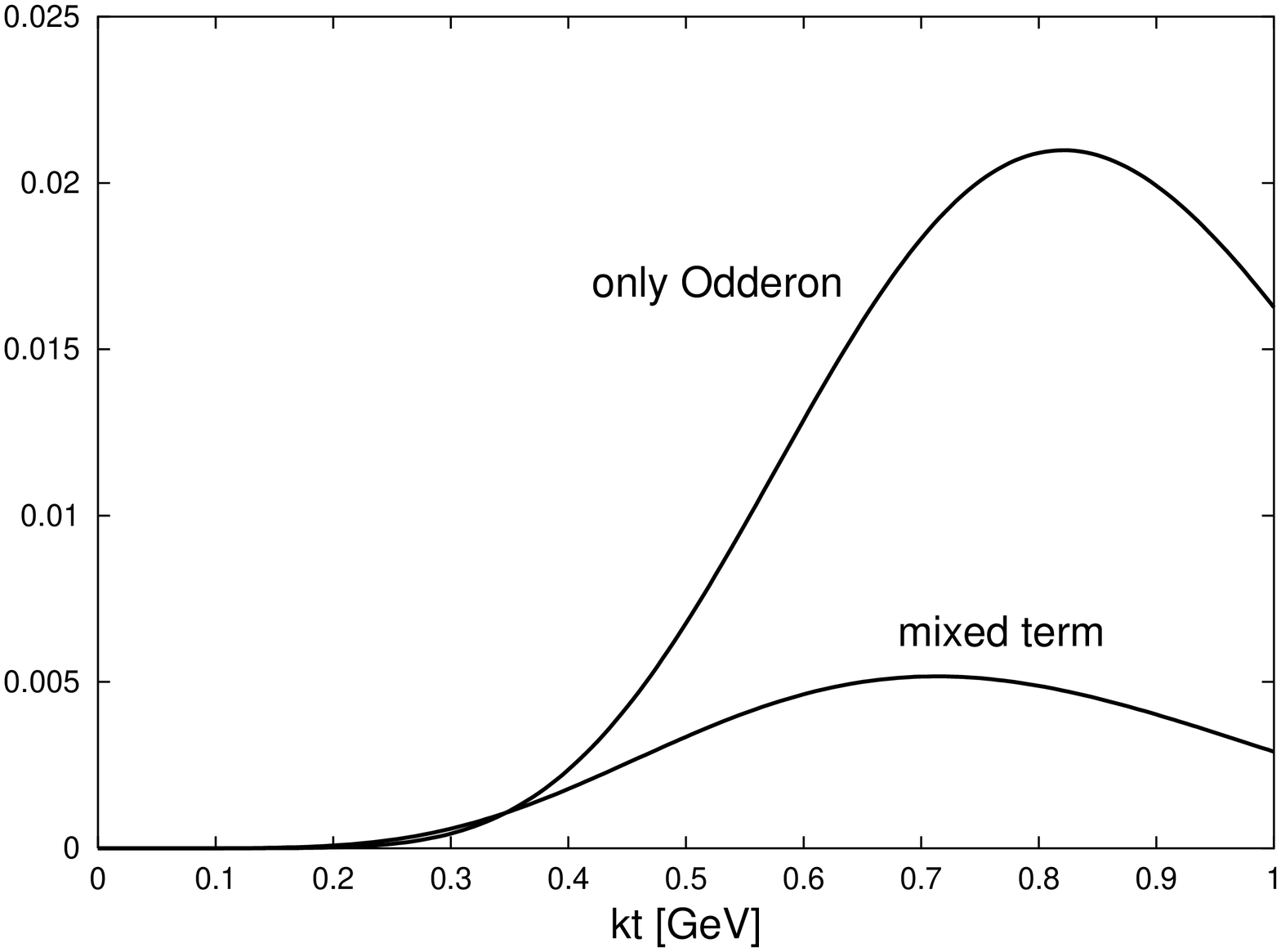}
}
\caption{}
\label{}
\end{figure}

\item{kinematics}\\
From the expressions for the differential cross sections (\ref{eq:exprdsx}), we know that $ d\sigma\,\propto\,\frac{1}{s\,k^{0}_{1}}$. In our approximation, 

\begin{displaymath}
d\sigma\,\propto\,\frac{1}{(k^{2}_{t}+m^{2})^{3/2}}\,.
\end{displaymath}
Including this $k_{t}$ dependence leads to an additional modification shown in figure \ref{fig:dsig}.
\begin{figure}[ht]
\centering
\includegraphics[angle=-90, width=0.5\textwidth]{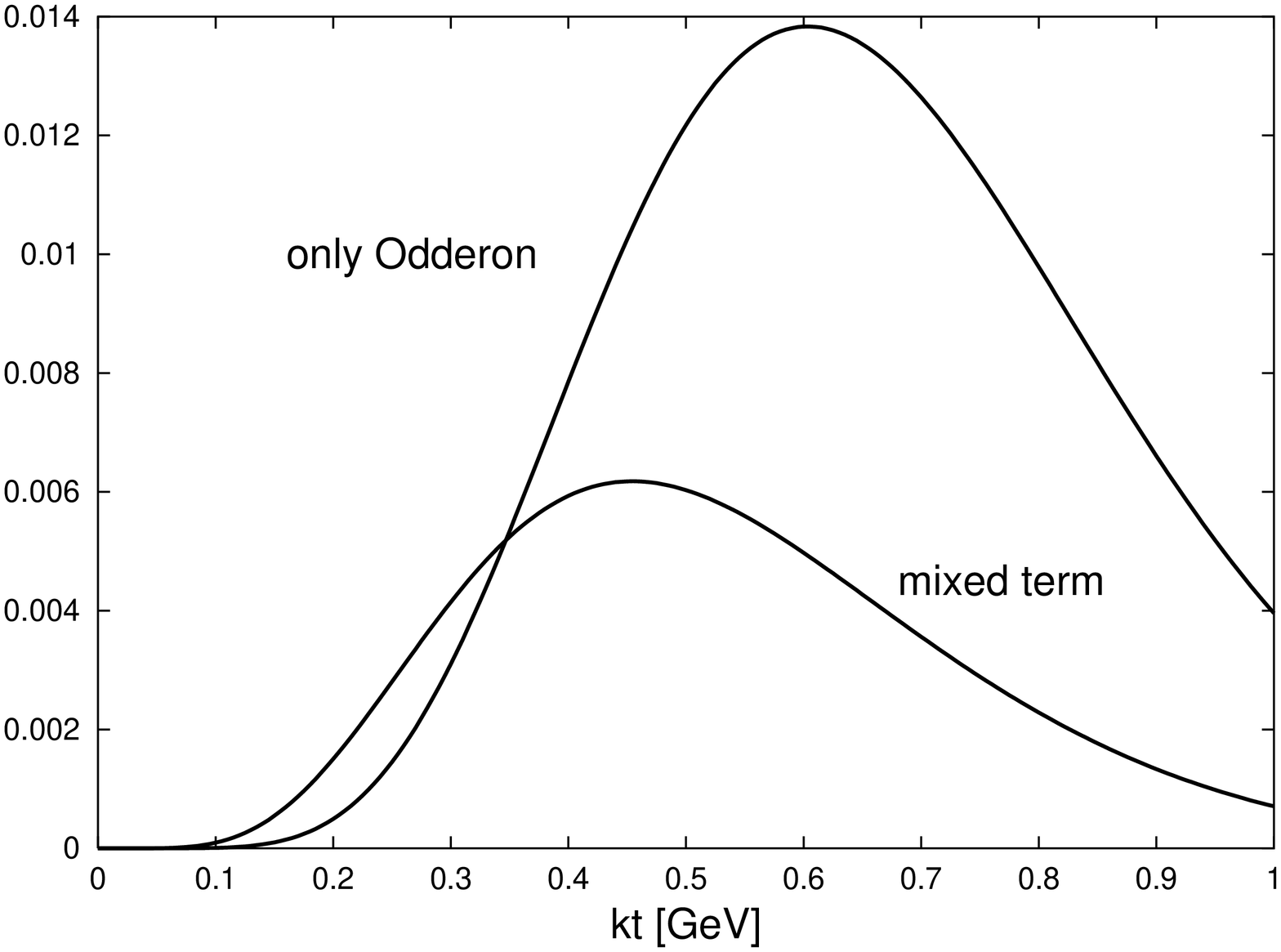}
\caption{$d\sigma$; not to scale}
\label{fig:dsig}
\end{figure}
\item{photon spectrum}\\
As a final step, we have to include the effects of the photon spectrum. We can approximate the spectra from the DEPA by
\begin{equation}
N(x)\;\propto\;\frac{1}{x}\,\ln\left(\frac{S}{(mx)^{2}}\right)\;\propto\;\frac{1}{x}\;.
\end{equation}
Including this, we obtain the final result for $d\sigma\,N(x)\,N(x)$ given by figure \ref{fig:specsig}.

\begin{figure}[ht]
\centering
\includegraphics[angle=-90, width=0.5\textwidth]{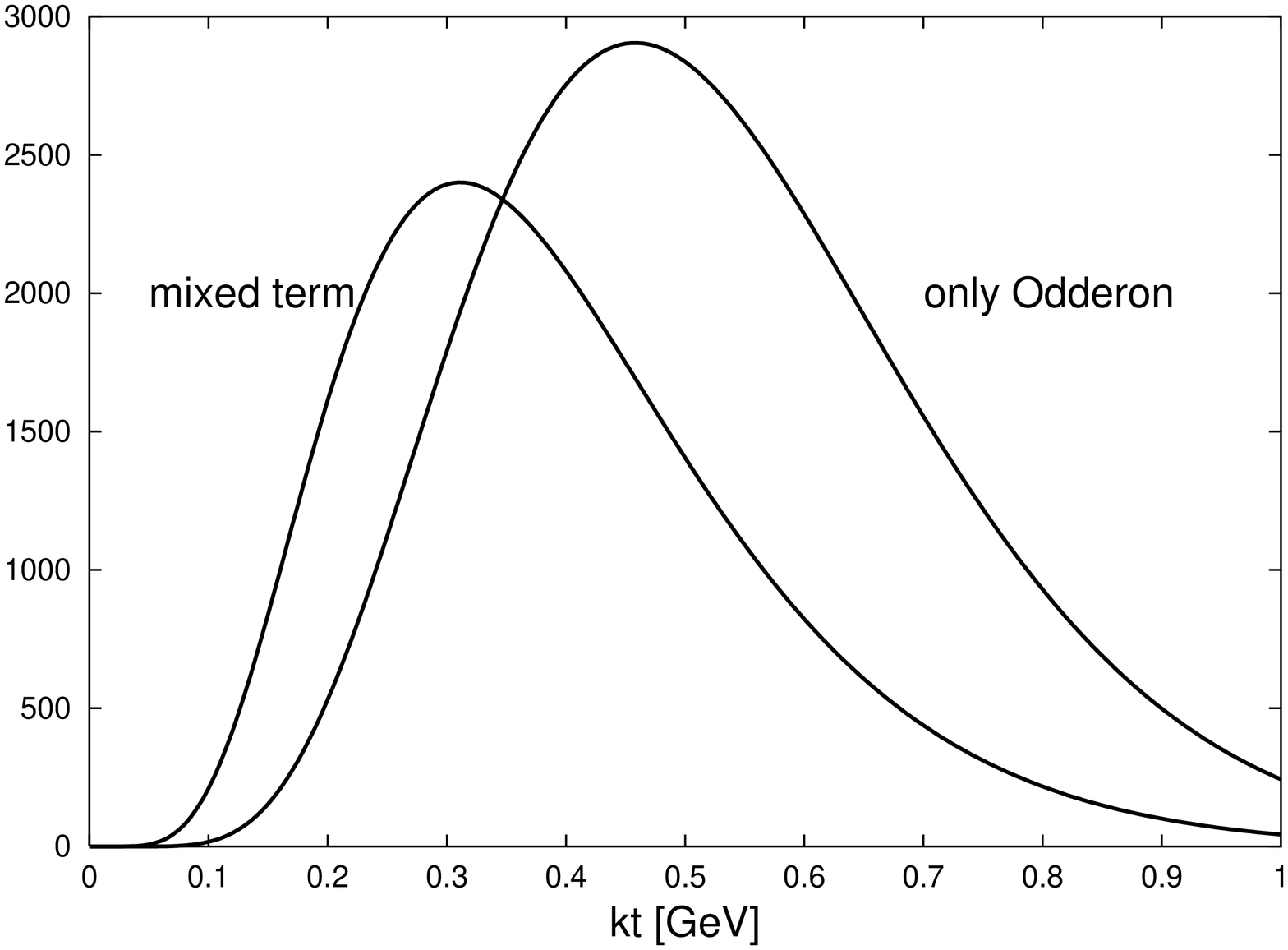}
\caption{$d\sigma$, including N(x); not to scale}
\label{fig:specsig}
\end{figure}
\end{itemize}
We see that the shape of the differential cross section depends the matrix element without Odderon contribution, the Odderon propagator and coupling, the kinematics, and the photon spectrum. Both $(2)$ and $(3)$ contain terms increasing as well as decreasing with $k_{t}$. Summarizing, we obtain
\begin{itemize}
\item{for (2) (Odderon contribution only):}\\
\begin{eqnarray}
\mbox{increase}&:&\left(k^{4}_{t}\right)_{|\M|^{2}}\times\left(k^{4}_{t}\right)_{prop}\;,\nonumber\\
\mbox{decrease}&:&\left(\frac{1}{s^{4\al'\,k^{2}_{t}}}\right)_{prop}\times\left(\frac{1}{(k^{2}_{t}+m_{\pi}^{2})^{3/2}}\right)_{kin}\times\left(\frac{1}{(1+\frac{(2k^{2}_{t}+m_{\pi}^{2})}{0.68})^{4}}\right)_{coup}\nonumber\\
 & & \times \left(\frac{1}{k^{2}_{t}+m_{\pi}^{2}}\right)_{spec}\;. \label{eq:updown}
\end{eqnarray}
\item{for (3) (mixed terms):}\\
\begin{eqnarray}
\mbox{increase}&:&\left(k^{4}_{t}\right)_{|\M|^{2}}\times \left(k^{2}_{t}\,\cos(\pi\,\al'\,k^{2}_{t})\right)_{prop}\;, \nonumber\\
\mbox{decrease}&:&\left(\frac{1}{s^{2\al'\,k^{2}_{t}}}\right)_{prop}\,\times\,\left(\frac{1}{(k^{2}_{t}+m_{\pi}^{2})^{3/2}}\right)_{kin}\,\times\,\left(\frac{1}{(1+\frac{(2k^{2}_{t}+m_{\pi}^{2})}{0.68})^{4}}\right)_{coup} \nonumber\\ 
 & & \times \left(\frac{1}{k^{2}_{t}+m_{\pi}^{2}}\right)_{spec}\,.
\end{eqnarray}
\end{itemize}
We used $\vare'\,=\,0$ and $\al'\,=\,0.25\;\mbox{GeV}^{2}$.
The resulting functions are displayed in figure \ref{fig:specsig}. Final results for $\eta_{\odd}\,=\pm\,1$ are given in figure \ref{fig:bothetas}; we clearly see the negative interference for $\eta_{\odd}\,=\,+1$ (see (\ref{eq:propprop}); for the parameters given above, $\cos(\frac{\pi}{2}\,\al(t))\,\geq\,0$). A comparison to the actual results of the numerical integration in figure \ref{fig:1} show that the naive approximation done above already gives a rough estimate of the $k_{t}$-dependence of $\frac{d\sigma}{d|k_{t}|^{2}}$. 
$\eta_{\odd}$ can easily be determined by measuring  $\frac{d\sigma}{d|k_{t}|^{2}}\,$.
Figure \ref{fig:bothetas2} shows results for the same approximation including higher order mass terms; as expected, the cross section differs for small values of $k_{t}$ where the approximation $k^{2}_{t}\,\gg\,m^{2}$ is no longer valid.

\section{Effects of parameter variation}\label{sec:parvar}
\indent
Next, we study the effects of parameter variation for the Odderon propagator and coupling. From section \ref{sec:effpropsec}, we know that there are basically three free parameters which need fitting by experiment. Remember that the Regge-trajectory of a particle is defined by

\begin{equation}
\al(t)\;=\;\al(0)+\al't\,.
\end{equation}
Furthermore, $\frac{\be_{\odd}}{\be_{\pom}}$ is as well unknown a priori (see section(\ref{sec:effprop})). As standard values, we choose (\ref{eq:oddval}):

\begin{equation}
\al(0)\,=\,1\;,\al'\,=\,0.25\;\mbox{GeV}^{-2}\;,\be^{2}_{\odd}\,=\,0.05\,\be^{2}_{\pom}\,.\label{eq:oddval2}
\end{equation}
In varying the parameters describing the Odderon contribution, we have to take experimental limits into account. From section \ref{sec:effprop}, we remember that the following relation holds for $\rho\,$ in $p\,p\,$ and $\pb\,p\,$-scattering for $t\,\rightarrow\,0$: 

\begin{eqnarray}
\rho^{(pp)}(s)-\rho^{(\pb p)}(s) & \longrightarrow & -\;2\,\eta_{\odd}\,(\frac{\be_{\odd}}{\be_{\pom}})^{2}\,(\frac{s}{s_{0}})^{\vare'-\vare}\,\frac{ \cos(\frac{\vare'}{2}\pi)}{\cos(\frac{\vare}{2}\pi)} \label{eq:rholim3}
\end{eqnarray}
Closely following \cite{Kilian:1998ew},\cite{Nachtmann:1991ua}, and \cite{Armesto:1997gg}, we assume a limit for (\ref{eq:rholim3}):

\begin{eqnarray}
|\rho^{(pp)}(s)-\rho^{(\pb p)}(s)| & \lesssim & 0.05\label{eq:limrho}
\end{eqnarray}
for $\sqrt{s}\,\geq\,100\,$GeV.

\begin{figure}
\begin{itemize}
\item{ for $\frac{d\sigma}{d|k_{t}|^{2}}$:}
\vspace{-13mm}
\begin{center}
\includegraphics[angle=-90, width=0.95\textwidth]{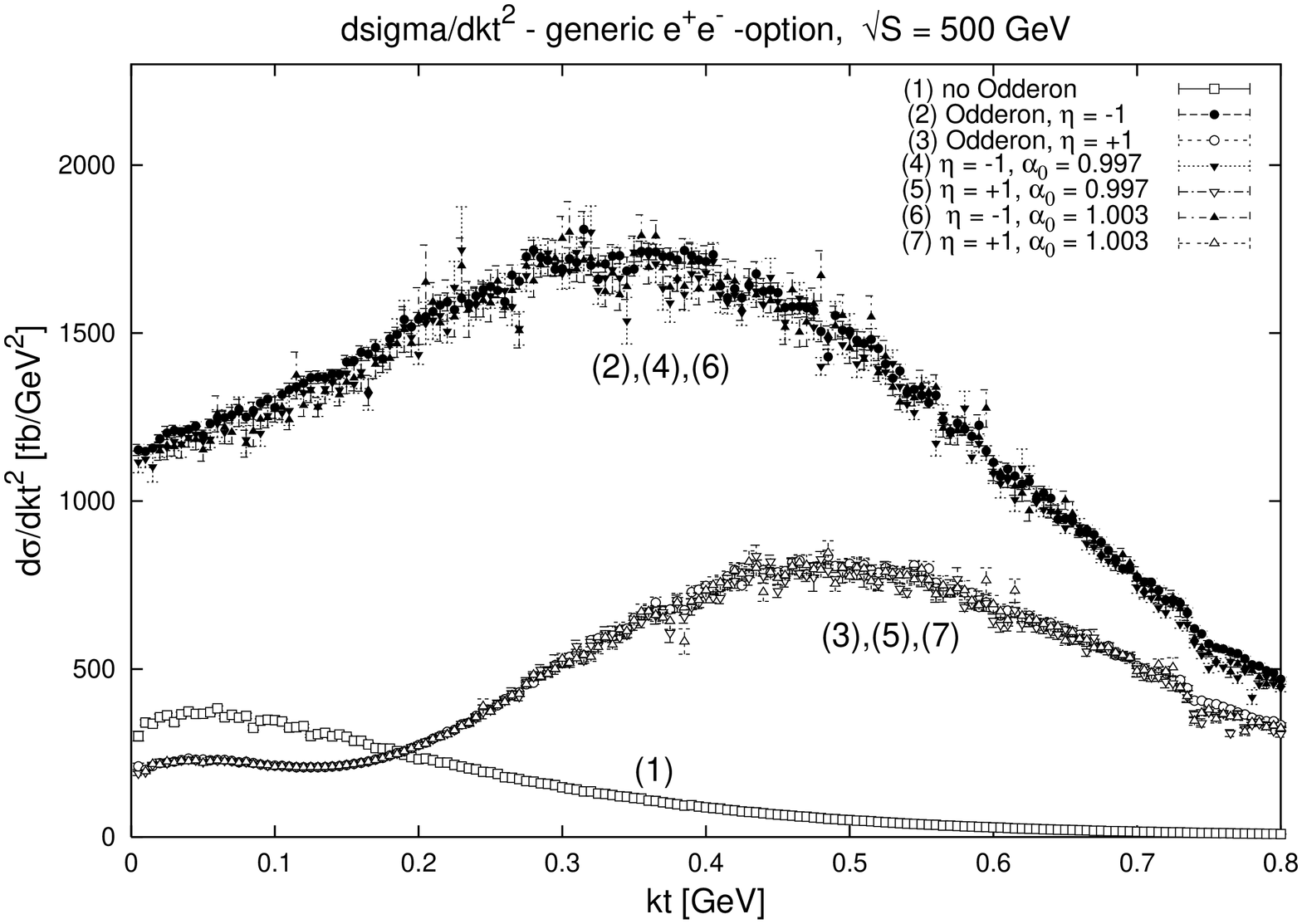}
\caption{Variation of $\vare'$ at $(\frac{\be_{\odd}}{\be{\pom}})^{2}\,=\,0.05$}
\label{fig:3}
\end{center}
\vspace{-5mm}
\item{ for $\frac{d\sigma}{dk_{l}}$:}
\vspace{-13mm}
\begin{center}
\includegraphics[angle=-90, width=0.95\textwidth]{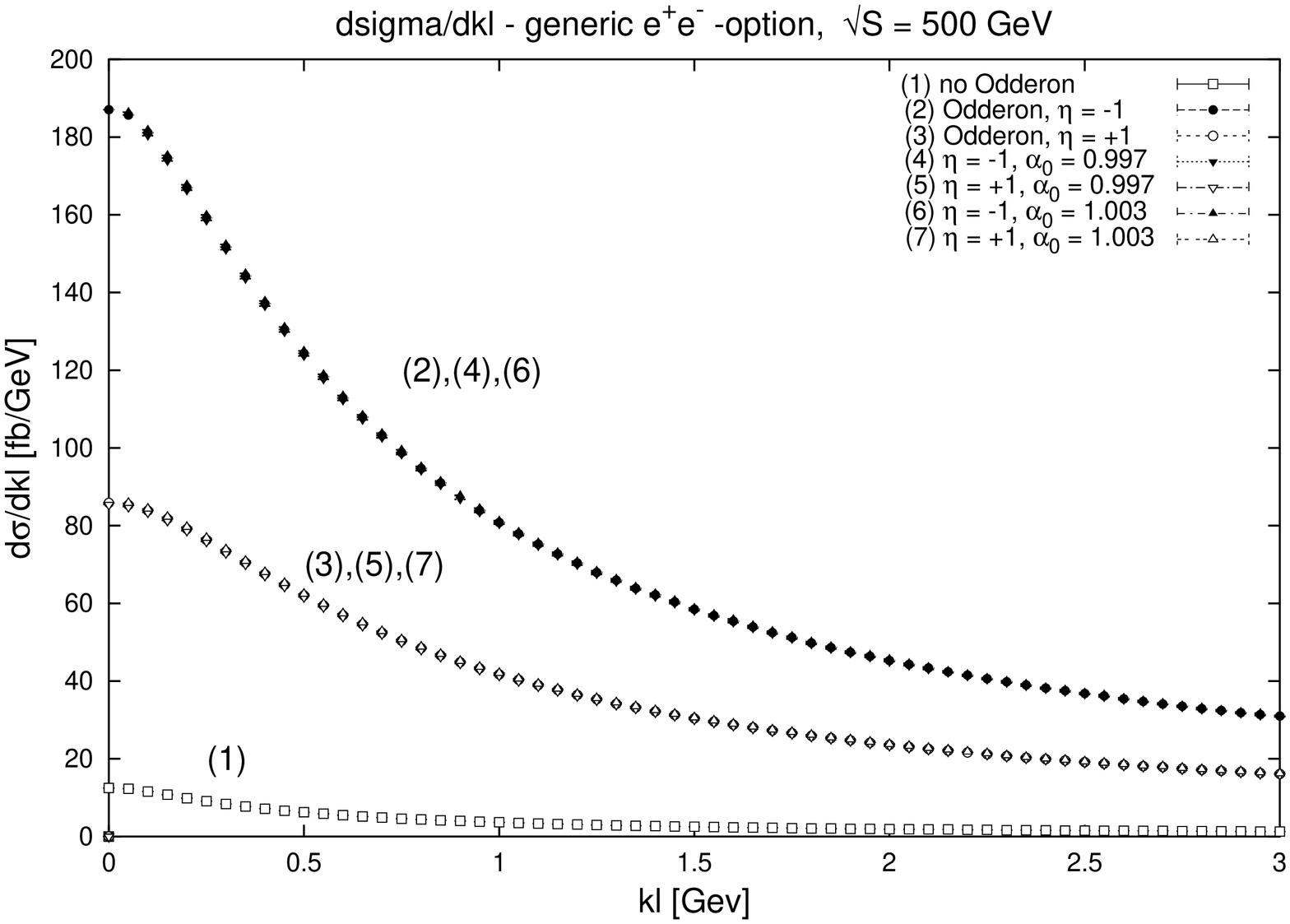}
\caption{Variation of $\vare'$ at $(\frac{\be_{\odd}}{\be{\pom}})^{2}\,=\,0.05$}
\label{fig:4}
\end{center}
\end{itemize}
\end{figure}

\begin{figure}
\begin{center}
\includegraphics[angle=-90, width=0.95\textwidth]{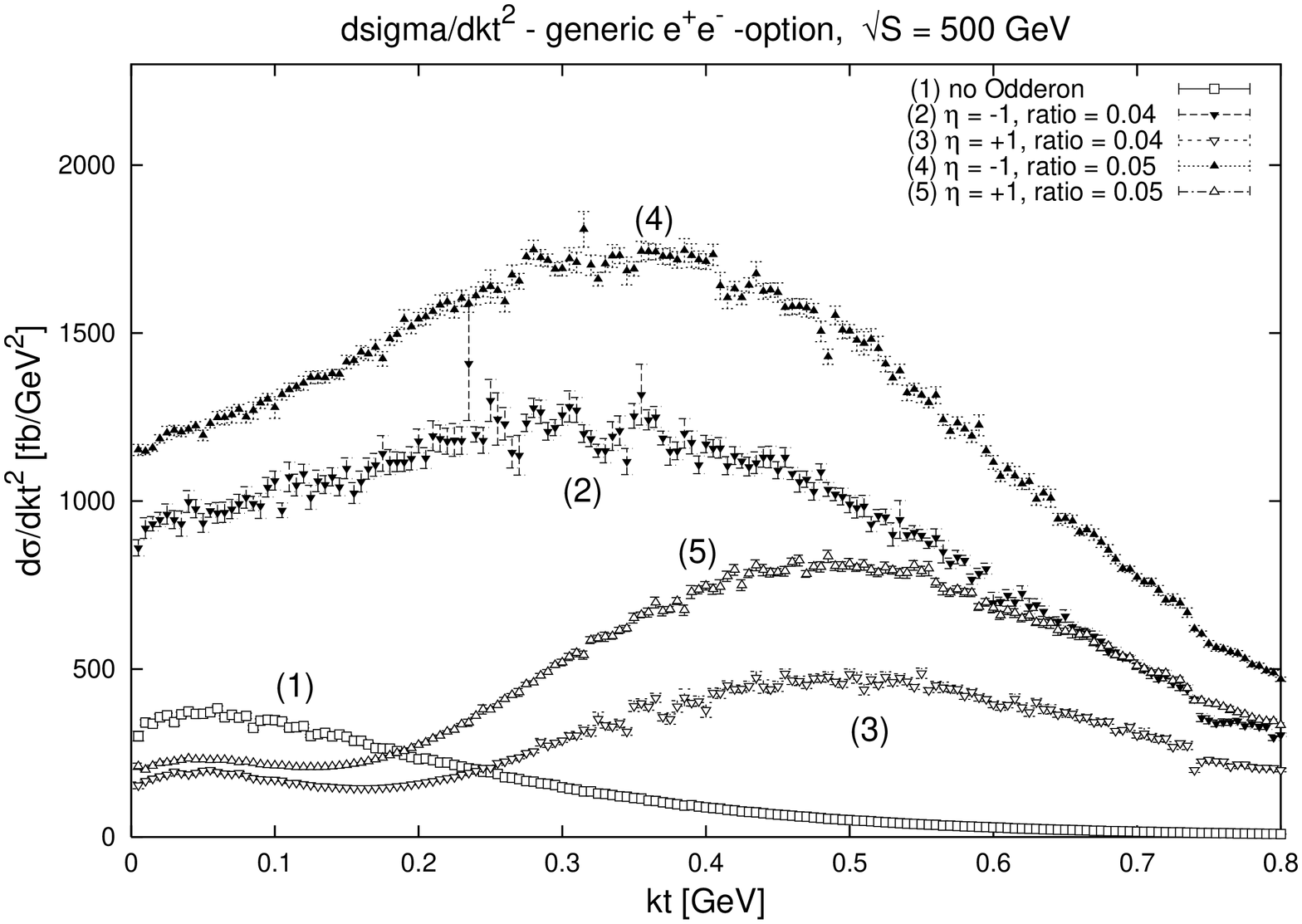}
\caption{Variation of $(\frac{\be_{\odd}}{\be_{\pom}})^{2}$ at $\vare'=0$}
\label{fig:5}
\end{center}
\begin{center}
\includegraphics[angle=-90, width=0.95\textwidth]{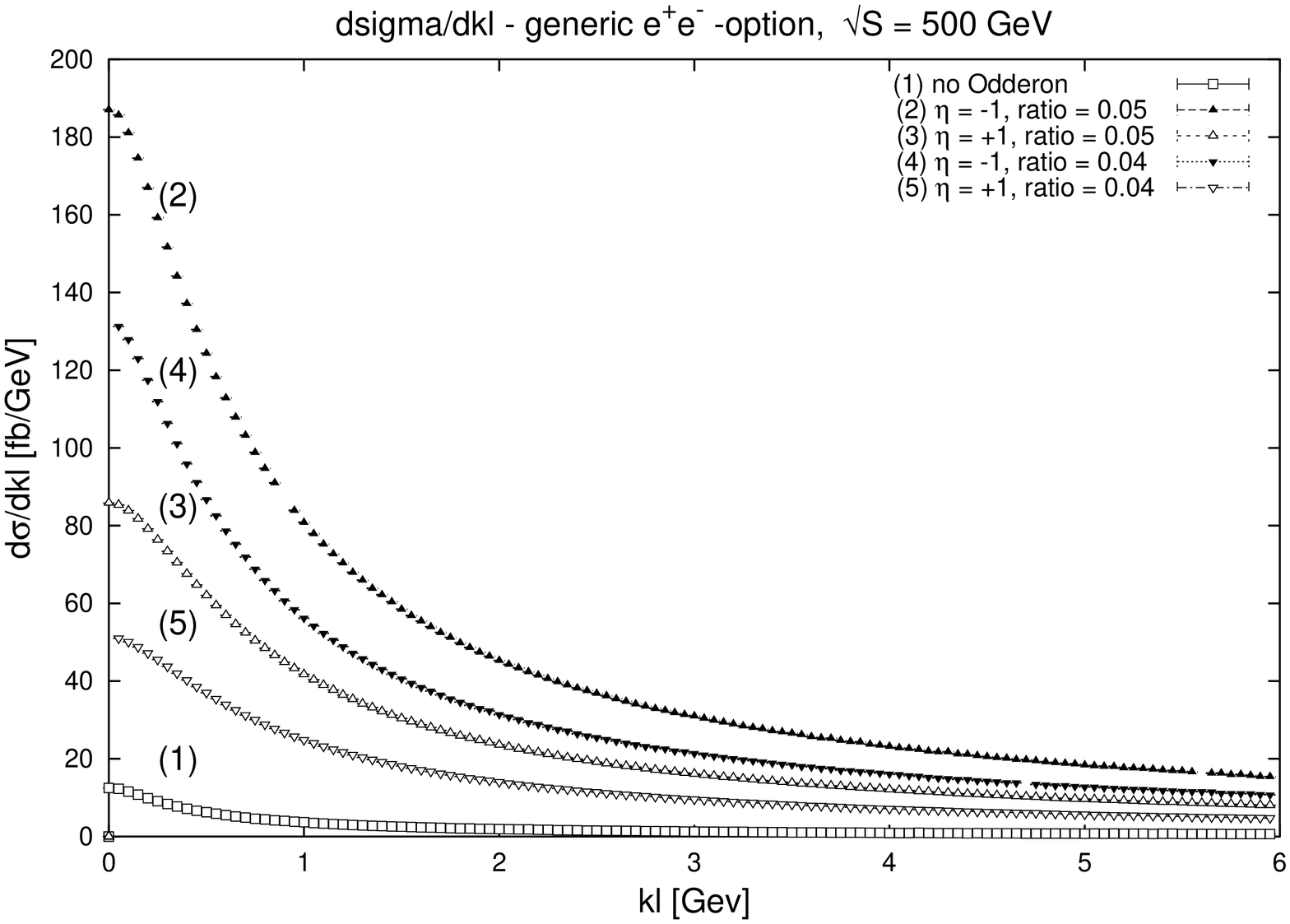}
\caption{Variation of $(\frac{\be_{\odd}}{\be_{\pom}})^{2}$ at $\vare'=0$}
\label{fig:6}
\end{center}
\end{figure}

\begin{figure}
\centering
\includegraphics[angle=-90, width=0.5\textwidth]{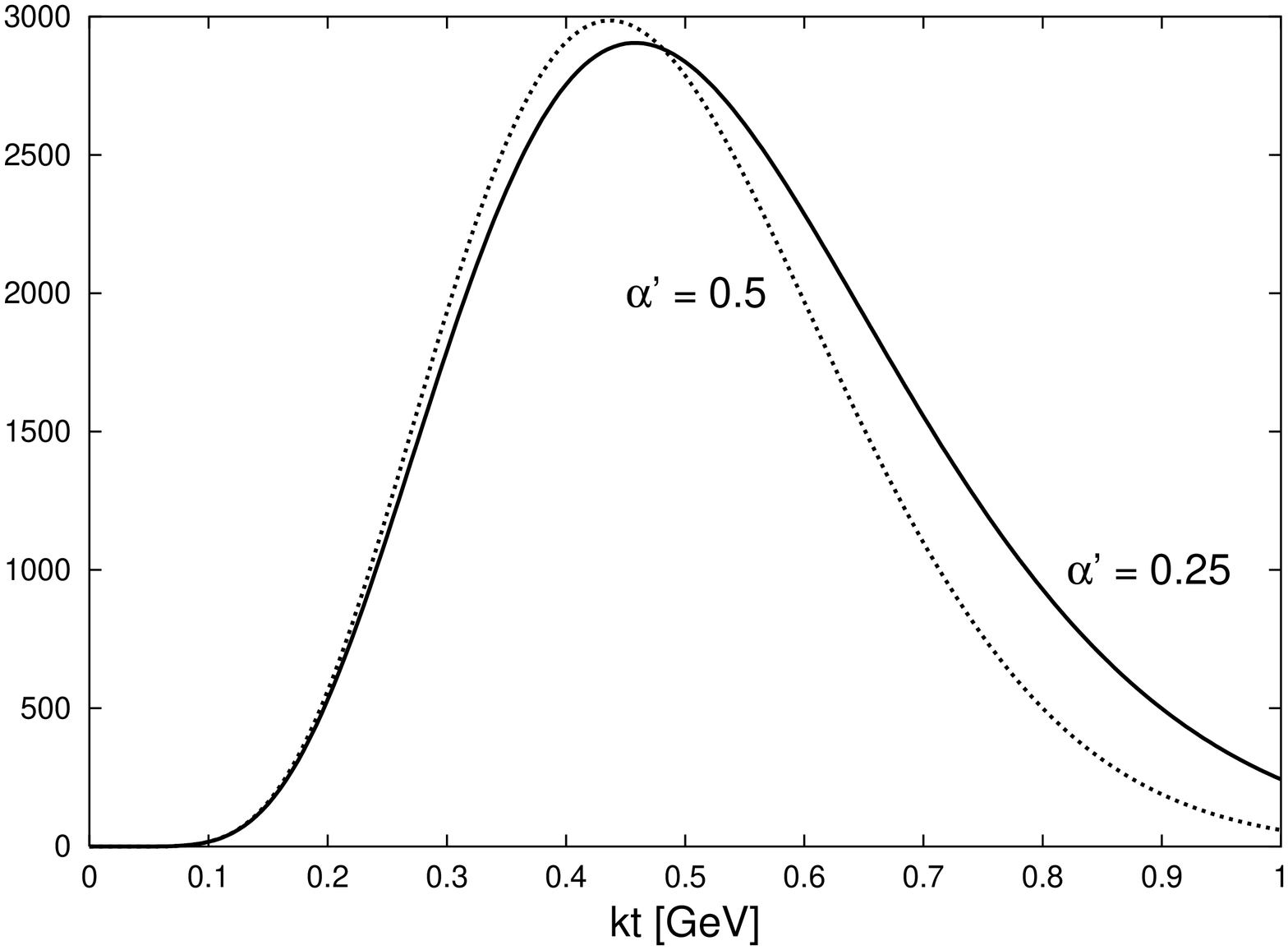}
\caption{$d\sigma$, including $N(x)$, different values for $\al'$, dominant contribution only }
\label{fig:alsl}
\end{figure}

\begin{figure}
\centering
\includegraphics[angle=-90, width=0.95\textwidth]{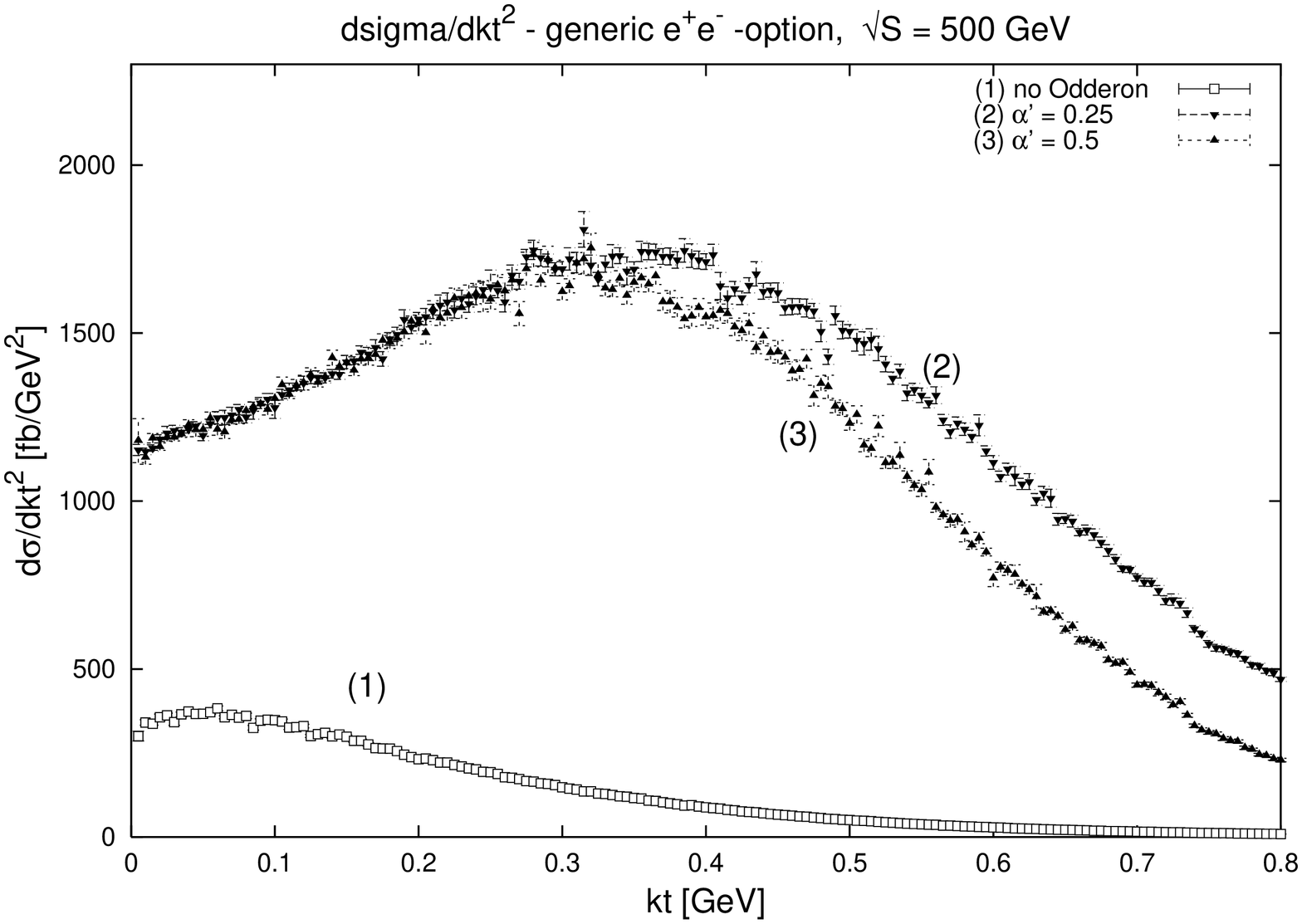}
\caption{variation of $\al'_{\odd},\,\eta_{\odd}=-1$ } 
\label{fig:7}                                      
\end{figure}

\begin{figure}
\centering
\includegraphics[angle=-90, width=0.95\textwidth]{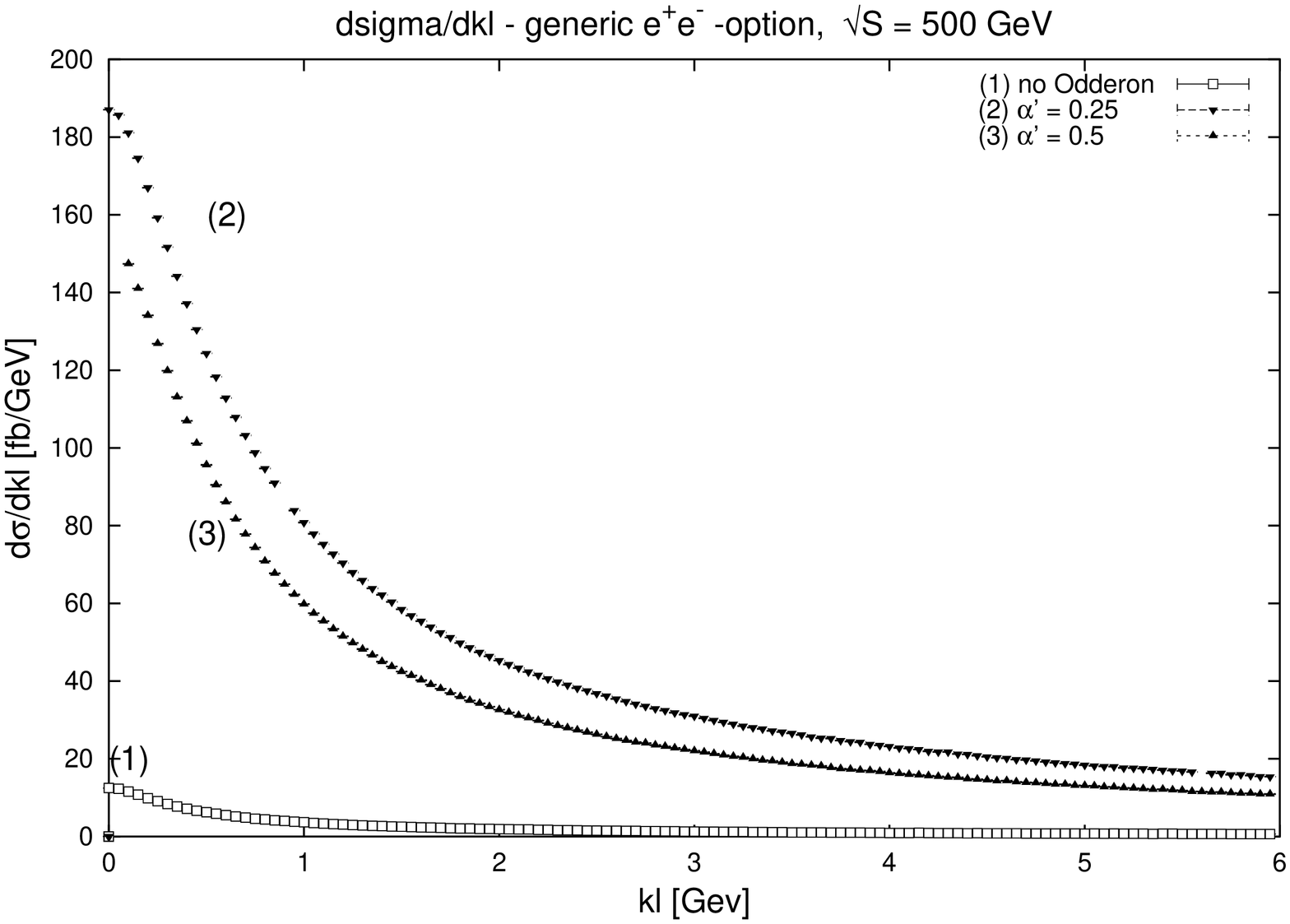}
\caption{variation of $\al'_{\odd},\,\eta_{\odd}=-1$}
\label{fig:8}
\end{figure}

\begin{figure}
\centering
\includegraphics[angle=-90, width=0.95\textwidth]{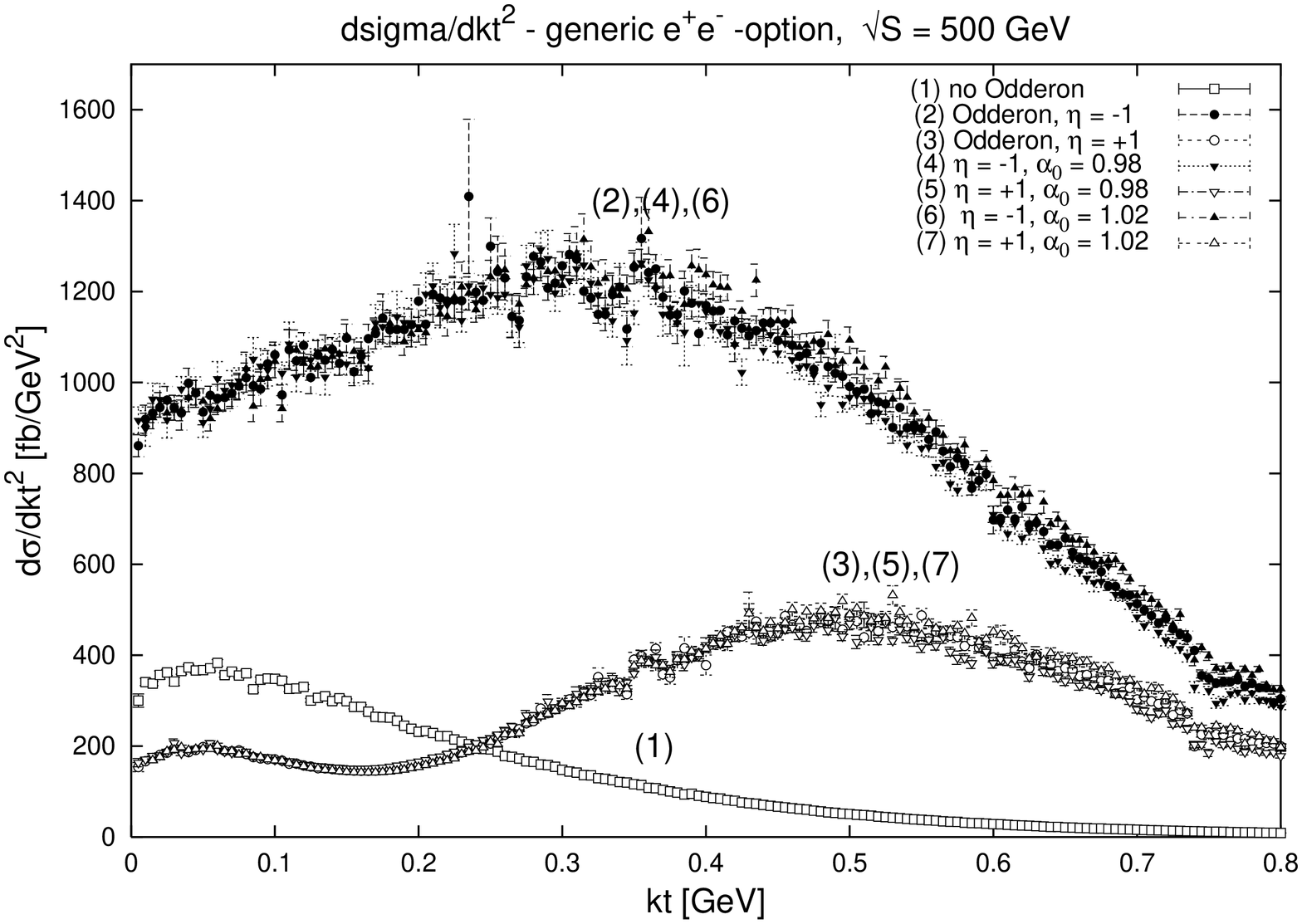}
\caption{variation of $\vare'$ at $(\frac{\be_{\odd}}{\be_{\pom}})^{2}=0.04$ } 
\label{fig:8a}                                      
\end{figure}

\begin{figure}
\centering
\includegraphics[angle=-90, width=0.95\textwidth]{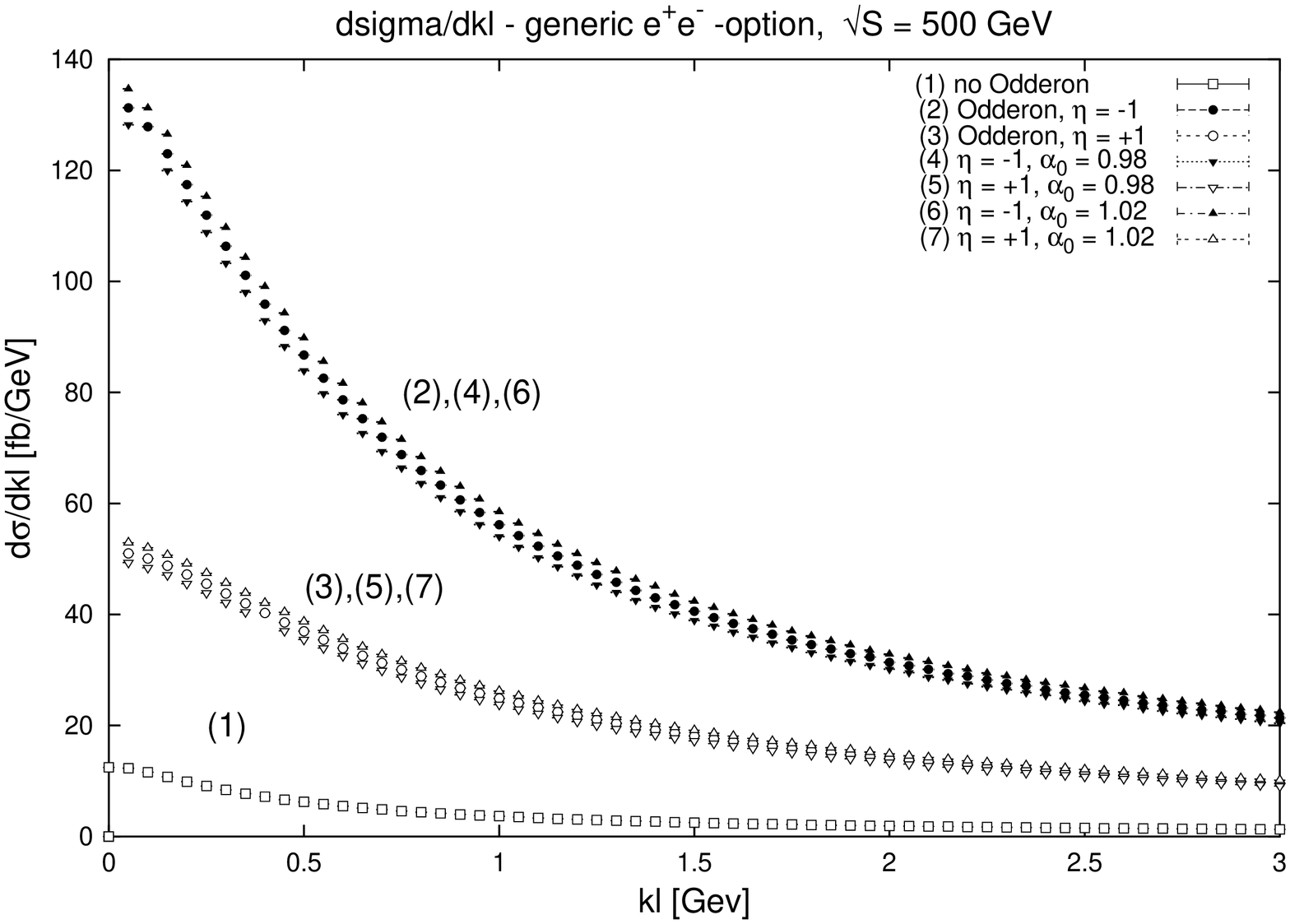}
\caption{variation of $\vare'$ at $(\frac{\be_{\odd}}{\be_{\pom}})^{2}=0.04$ }
\label{fig:8b}
\end{figure}

\begin{figure}
\centering
\includegraphics[angle=-90, width=0.95\textwidth]{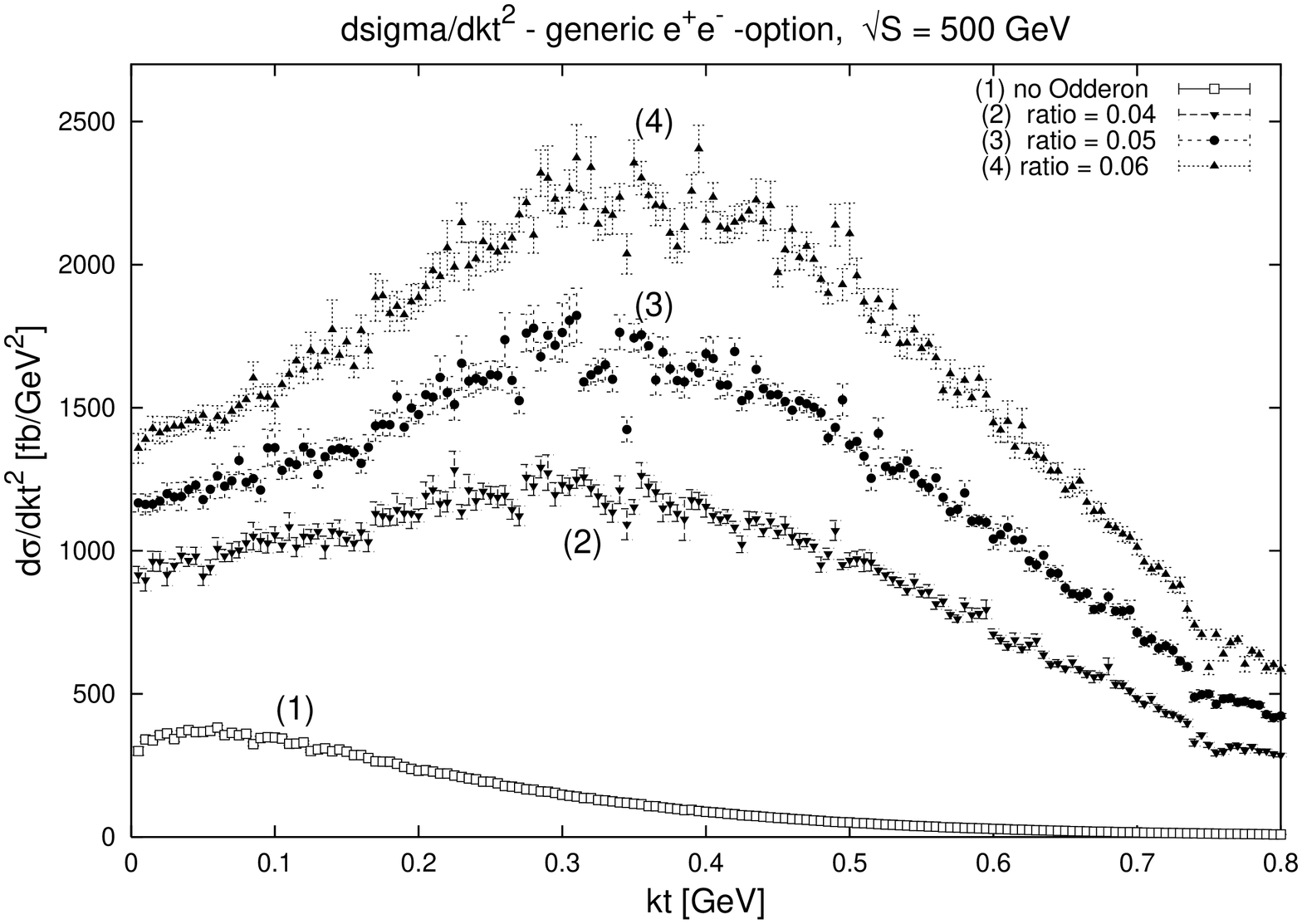}
\caption{variation of $(\frac{\be_{\odd}}{\be_{\pom}})^{2}$ at $\vare'\,=\,-0.02,\,\eta_{\odd}\,=\,-1$ } 
\label{fig:8e}                                      
\end{figure}

\begin{figure}
\centering
\includegraphics[angle=-90, width=0.95\textwidth]{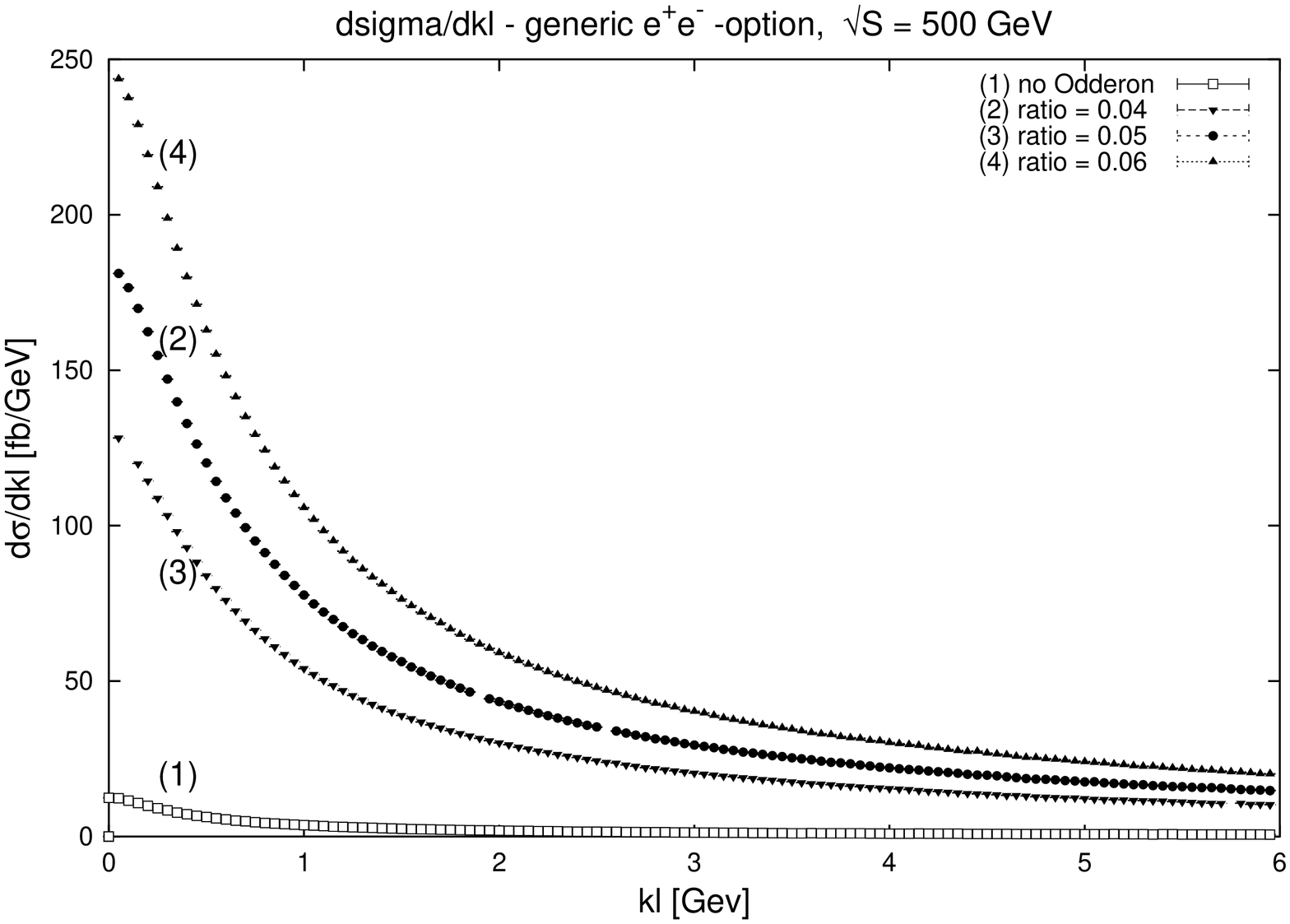}
\caption{variation of $(\frac{\be_{\odd}}{\be_{\pom}})^{2}$ at $\vare'\,=\,-0.02,\,\eta_{\odd}\,=\,-1$ }
\label{fig:8f}
\end{figure}

\begin{figure}
\centering
\includegraphics[angle=-90, width=0.95\textwidth]{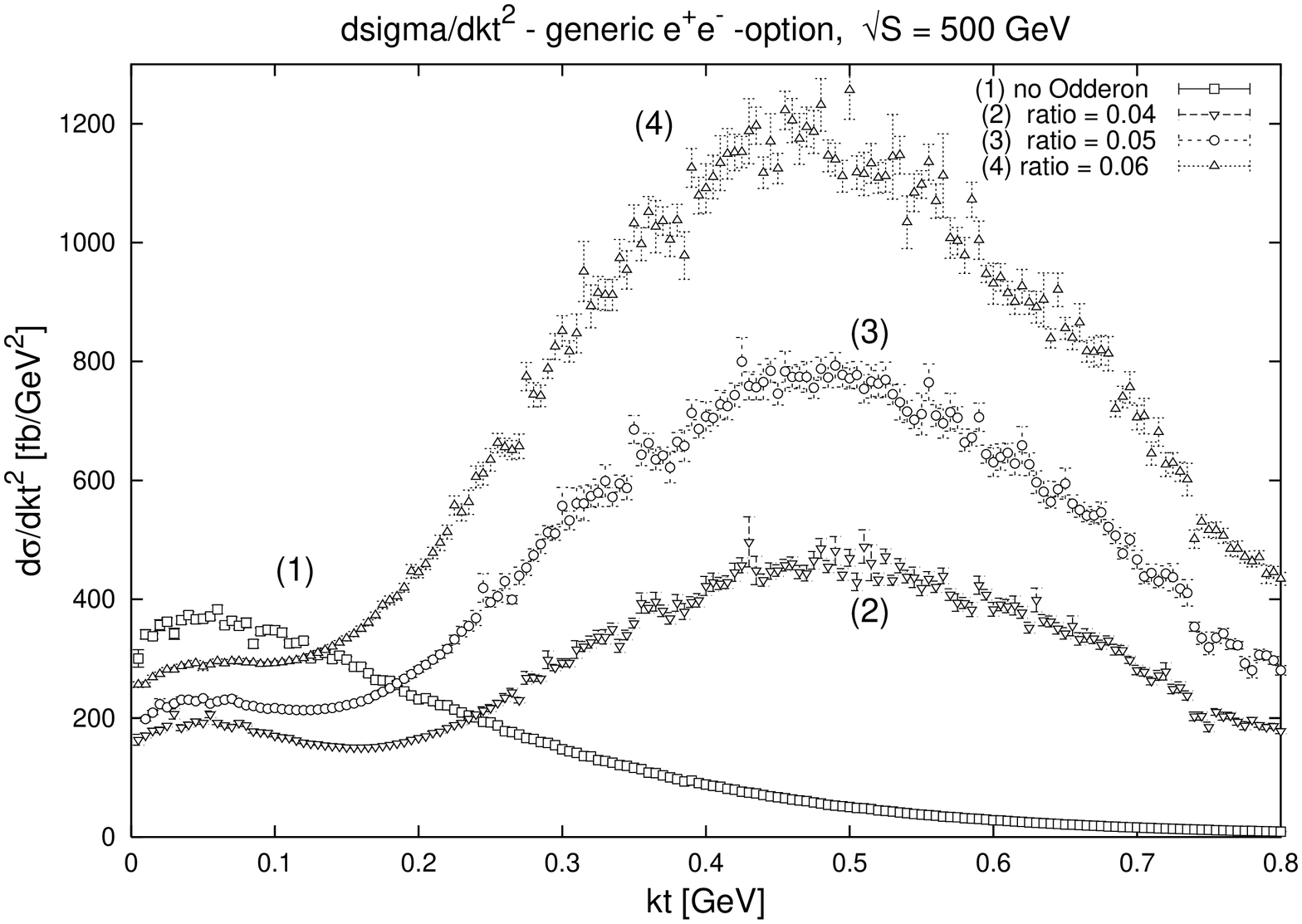}
\caption{variation of $(\frac{\be_{\odd}}{\be_{\pom}})^{2}$ at $\vare'\,=\,-0.02,\,\eta_{\odd}\,=\,+1$ } 
\label{fig:8g}                                      
\end{figure}

\begin{figure}
\centering
\includegraphics[angle=-90, width=0.95\textwidth]{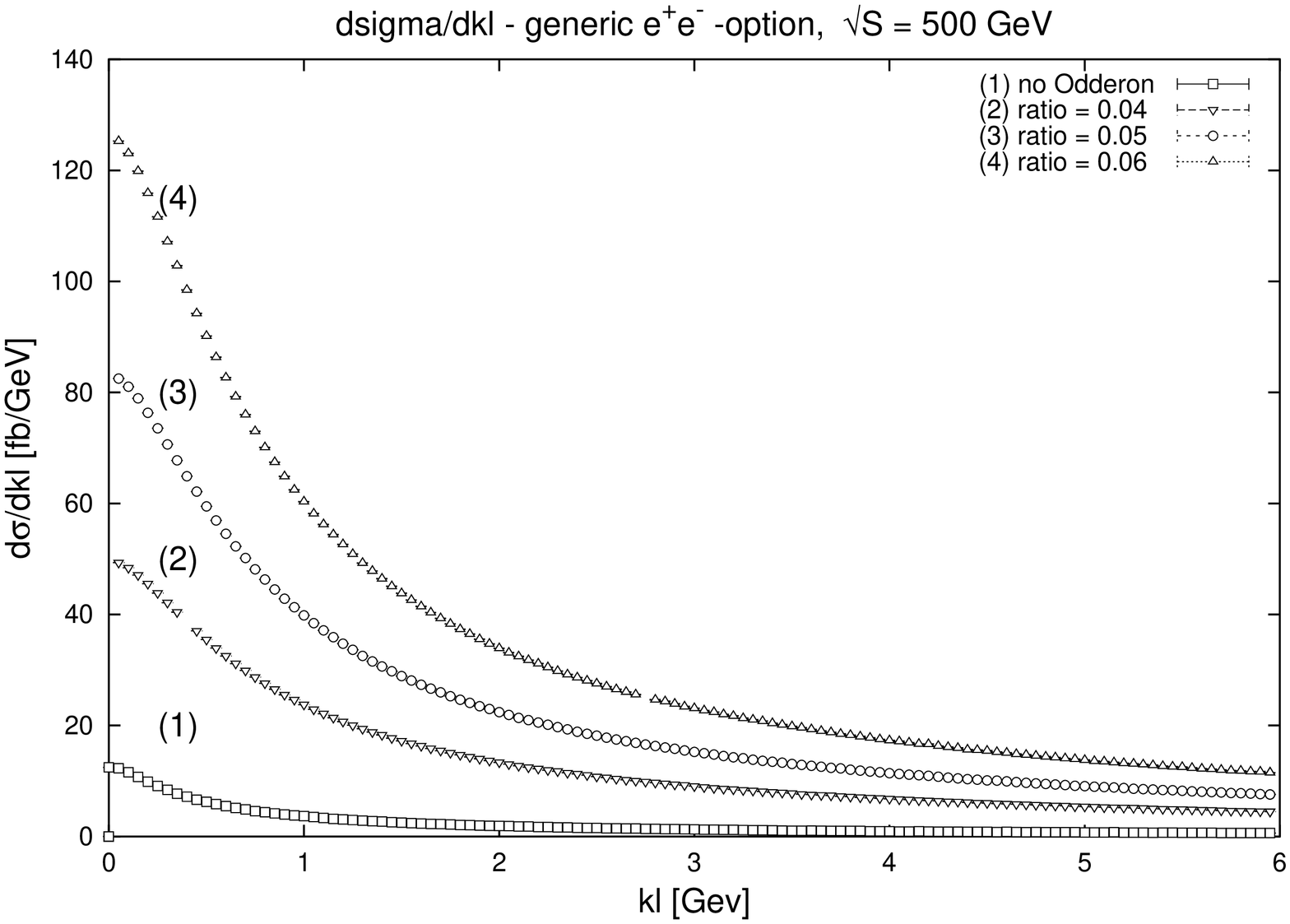}
\caption{variation of $(\frac{\be_{\odd}}{\be_{\pom}})^{2}$ at $\vare'\,=\,-0.02,\,\eta_{\odd}\,=\,+1$ }
\label{fig:8h}
\end{figure}

We immediately see that this implies limits for possible parameter variations.
In detail, we will consider 
\begin{itemize}
\item{Variation of $\vare'\,$}\\
The limit given above provides a correlation between the variation of $\left(\frac{\be_{\odd}}{\be_{\pom}}\right)^{2}\,$ and $\vare'\,$. 
In varying $\vare'\,$  by keeping $\left(\frac{\be_{\odd}}{\be_{\pom}}\right)^{2}\,=\,0.05$, we are restricted to $|\vare'|\,\leq\,0.003$.

\item{Variation of $\al'_{\odd}$}\\
The slope of the Odderon Regge trajectory, $\al'_{\odd}\,$, does not appear in (\ref{eq:rholim3}). Therefore, we can freely vary this parameter. We will consider the value $\al'_{\odd}\,=\,0.5\;\mbox{GeV}^{-2}$.
\item{Variation of $(\frac{\be_{\odd}}{\be_{\pom}})^{2}\,$}\\
The parameter $(\frac{\be_{\odd}}{\be_{\pom}})^{2}\,$ is strongly limited by the restriction given above. Actually, (\ref{eq:limrho}) implies
\begin{displaymath}
(\frac{\be_{\odd}}{\be_{\pom}})^{2}\,\lesssim\,0.05182
\end{displaymath}
with all other parameters taking standard values (\ref{eq:oddval2}) and $\sqrt{s}\,=\,100\,$GeV. For $\vare'\,=\,0$, we will only consider $(\frac{\be_{\odd}}{\be_{\pom}})^{2}\,=\,0.04$.
\item{Combined variations}\\
For a better investigation of Odderon coupling strength and intercept variations, we will consider the following combinations:

\begin{eqnarray*}
\vare'&\longrightarrow&\pm 0.02 \;\;\mbox{for}\;\left(\frac{\be_{\odd}}{\be_{\pom}}\right)^{2}=0.04\;,\\
\left(\frac{\be_{\odd}}{\be_{\pom}}\right)^{2}&\longrightarrow&0.05\pm 0.01\;\;\mbox{for}\;\vare'=-0.02\;.
\end{eqnarray*}

\end{itemize}
Results for the variations given above are shown in figures \ref{fig:3} to \ref{fig:8h}.

From  figures \ref{fig:3} and \ref{fig:4} (variation for $(\frac{\be_{\odd}}{\be_{\pom}})^{2}\,=\,0.05$) as well as from figures \ref{fig:8e} and \ref{fig:8f} (variation for $(\frac{\be_{\odd}}{\be_{\pom}})^{2}\,=\,0.04$), we immediately see that a variation of $\vare'\,$  does not have a significant effect on the cross sections. Going back to the considerations in the previous section, we see that a variation of $\al(0)$ concerns the Odderon-propagator and therefore modifies $|\M|^{2}$:

\begin{eqnarray}
s^{2\,\al't}&\longrightarrow&s^{2\,(\vare'+\al't)}\;,\nonumber\\
s^{\al't}\,\cos(\frac{\pi}{2}\al't)&\longrightarrow& s^{\vare'+\al't}\,\cos\left(\frac{\pi}{2}(\vare'+\al't)\right)\nonumber \\
 & & =  s^{\vare'+\al't}\,\left(\cos(\frac{\pi}{2}\vare')\,\cos(\frac{\pi}{2}\al't)-\tan(\frac{\pi}{2}\vare')\,\sin(\frac{\pi}{2}\al't)\right)\;.\nonumber\\
\end{eqnarray}
Taylor expansions of the varied terms give
\begin{eqnarray}
s^{2\,\al't}&\longrightarrow&s^{2\,\al't}\,\left(1+2\vare'\,\ln(s)+\mathcal{O}(\vare'^{2})\right)\nonumber\\
& & \mbox{(modification of $(2)$)}\;,\nonumber\\
& & \nonumber \\
s^{\al't}\,\cos(\frac{\pi}{2}\al't)&\longrightarrow&s^{\al't}\,\cos(\frac{\pi}{2}\al't)\,\left(1+\vare'\left(\ln s-\frac{\pi}{2}\sin(\frac{\pi}{2}\al't)\right)+\mathcal{O}(\vare'^{2})\right) \nonumber\\
& & \mbox{(modification of $(3)$)}\;. \label{eq:resmodeps}
\end{eqnarray}
We can therefore estimate the effects of $\vare'\,\neq\,0$:
\begin{itemize}
\item{pure Odderon term}\\
As $s$ is limited by (\ref{eq:limkt}) to $s_{min}\,=\,4\,m^{2}_{\pi}$, $|\ln s|$ takes its maximal value at $s\,\rightarrow\,S$. However, taking the form of the photon spectra into account, we rather use $x\,=\,0.06$ leading to $s\,=\,900\;\mbox{GeV}^{2}$. For this value, $\ln s\,=\,6.8$. However, we can also follow the method of approximation given in the last section; if we consider the maxima of the cross sections, i.e. $k_{t}\,\approx\,0.35\;\mbox{GeV}$ for $\eta_{\odd}\,=\,-1$ and $k_{t}\,\approx\,0.5\;\mbox{GeV}$ for $\eta_{\odd}\,=\,1$, we obtain $s_{-}\,=\,0.56\;\mbox{GeV}$ and $s_{+}\,=\,1.07\;\mbox{GeV}$ leading to $|\ln s_{-}|\,=\,0.68$ and $\ln s_{+}\,=\,0.07$.
Applying this to the modifications, we obtain:
\begin{eqnarray}
\lefteqn{\mbox{for}\;\vare'\,=\,\pm\,0.003\;:\;} & & \nonumber\\
 \Delta_{max}\,\approx\,0.04\,&
 \Delta_{+}\,\approx\,0.0004,&
 \Delta_{-}\,\approx\,0.004\;,\nonumber\\
\lefteqn{\mbox{for}\;\vare'\,=\,\pm\,0.02\;:\;} & & \nonumber\\
\Delta_{max}\,\approx\,0.26,&  
  \Delta_{+}\,\approx\,0.003,&
 \Delta_{-}\,\approx\,0.03\nonumber\\
\end{eqnarray}
with $s^{2\,\al't}\,\longrightarrow\,s^{2\,\al't}\,(1+\Delta(\vare')+\mathcal{O}(\vare'^{2}))$.
\item{mixed terms}\\
Taking also the mixed terms into account, we obtain
\begin{eqnarray}
\lefteqn{\mbox{for}\;\vare'\,=\,\pm\,0.003\;:\;} & & \nonumber\\
 \Delta'_{max}\,\approx\,0.022\;&
 \Delta'_{+}\,\approx\,0.0001;&
 \Delta'_{-}\,\approx\,0.0023\nonumber\\
\lefteqn{\mbox{for}\;\vare'\,=\,\pm\,0.02\;:\;} & & \nonumber\\
\Delta'_{max}\,\approx\,0.15;&  
  \Delta'_{+}\,\approx\,0.0044;&
 \Delta'_{-}\,\approx\,0.2\nonumber\\
\end{eqnarray}
for $s^{\al't}\,\cos(\frac{\pi}{2}\al't)\;\longrightarrow\;s^{\al't}\,\cos(\frac{\pi}{2}\al't)\,(1+\Delta'(\vare')+\mathcal{O}(\vare'^{2}))$ following a similar argumentation. However, we have to keep in mind that the contributions from the mixed terms are small compared to the pure Odderon contributions. 
\end{itemize}
From the numerical calculations of the cross sections, we obtain
\begin{equation}
\Delta_{+}\,\approx\,0.09\;;\;\Delta_{-}\,\approx\,0.04\;
\end{equation}
for $\vare'\,=\,\pm 0.02$ which is well within the limits given by $\Delta_{max}$.

Effects of varying $(\frac{\be_{\odd}}{\be_{\pom}})^{2}\,$ are displayed in figures \ref{fig:5} and \ref{fig:6} as well as figures \ref{fig:8e} to \ref{fig:8h}. Neglecting the contribution from the mixed terms, we obtain

\begin{equation}
d\sigma\;\propto\;\kappa^{2}\;\propto\;\left(\frac{\be_{\odd}}{\be_{\pom}}\right)^{4};
\end{equation}
we therefore expect 
\begin{eqnarray}
\frac{d\sigma_{0.04}}{d\sigma_{0.05}}&\propto&\left(\frac{0.04}{0.05}\right)^{2}\,=\,0.64\,, \nonumber \\
\frac{d\sigma_{0.04}}{d\sigma_{0.06}}&\propto&\left(\frac{0.04}{0.06}\right)^{2}\,=\,0.44\,, \nonumber \\
\frac{d\sigma_{0.05}}{d\sigma_{0.06}}&\propto&\left(\frac{0.05}{0.06}\right)^{2}\,=\,0.69\,. \nonumber\\
& & \label{eq:exprat}  
\end{eqnarray}
An additional inclusion of the mixed terms in our approximation from the last section gives:

\begin{equation}
d\sigma_{\kappa}\;\propto\;\kappa\,(k_{t}^{2}/s^{2\,\al'\,k_{t}^{2}\,} \kappa -2\,\cos(\pi\al'\,k_{t}^{2}))\,.\label{eq:expkt2fac}
\end{equation}
We see that the terms corresponding to pure Odderon exchange are proportional to $\kappa^{2}$, the mixed terms to $\kappa$. This leads to an expected shift of the position of the maximum to wards lower values of $k_{t}$ for $\eta_{\odd}\,=\,-1$ and towards higher values of $k_{t}$ for $\eta_{\odd}\,=\,+1$; compare to figures \ref{fig:5}, \ref{fig:8e}, and \ref{fig:8g}.

From figures  \ref{fig:5}, \ref{fig:6}, and \ref{fig:8e} to \ref{fig:8h}, we obtain

\begin{displaymath}
\begin{array}{|c| c| c| c|}\hline
\mbox{ratio}& \eta_{\odd} = -1 & \eta_{\odd}= +1 & \vare' \\ \hline\hline
0.04 : 0.05 & 0.72\, /\, 0.68 & 0.55\, /\, 0.6 & 0\\ \hline
& 0.69\, /\, 0.70 & 0.6\, /\, 0.6 & -0.02 \\ \hline
0.04 : 0.06 & 0.56\, /\, 0.53 & 0.42\, /\, 0.38 & -0.02\\ \hline
0.05 : 0.06 & 0.79\, /\, 0.76 & 0.69\, /\, 0.66 & -0.02 \\ \hline
\end{array}
\end{displaymath} 
where the first / second number corresponds to the values for $\frac{d\sigma}{d|k_{t}|^{2}}/ \frac{d\sigma}{dk_{l}}$  taken at the respective maxima. We see that the values from from the calculations are in the order of magnitude of the expected ratios from (\ref{eq:exprat}).

Finally, we are interested in varying $\al'\,$ while keeping $\al(0)\,$ fixed. The corresponding cross sections are given in figures \ref{fig:7} and \ref{fig:8}. 

From the terms determining the shape of $\frac{d\sigma}{d|k_{t}|^{2}}$ as given by (\ref{eq:updown}) and again only considering the dominant pure Odderon contribution, we see that 

\begin{equation}
d\sigma\;\propto\;\frac{1}{s^{4\al'\,kt^{2}}}
\end{equation}
for $\al(0)\,=\,1$. This leads to a shift of the maximum as well as a faster decrease for $\frac{d\sigma}{d|k_{t}|^{2}}$. We displayed the approximation in figure \ref{fig:alsl}. A comparison with figure \ref{fig:7} shows again that this gives a good description of the behavior of the cross section. We see that higher values of $\al'\,$ lead to a faster decrease after the maximum when considering the transverse momenta; for longitudinal ones, the distinguishability is less visible, especially in combination with simultaneous variations of $(\frac{\be_{\odd}}{\be_{\pom}})^{2}$ (see figure \ref{fig:6} for comparison). 

\section{Effects of detector cuts}\label{sec:effdetcut}

\begin{figure}
\centering
\includegraphics[angle=-90, width=0.95\textwidth]{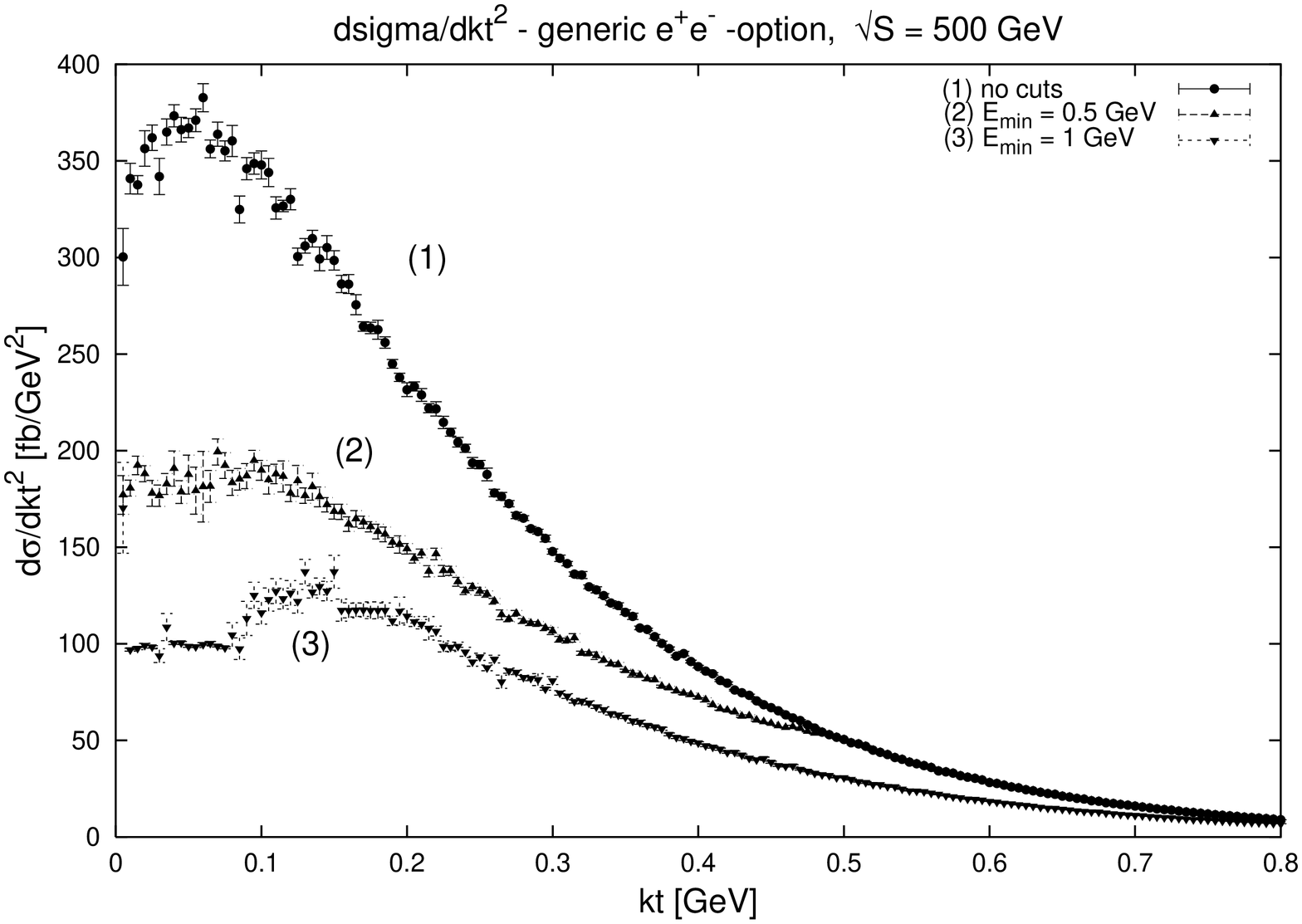}
\caption{effects of energy cuts; no Odderon } 
\label{fig:9}                                      
\end{figure}

\begin{figure}
\centering
\includegraphics[angle=-90, width=0.95\textwidth]{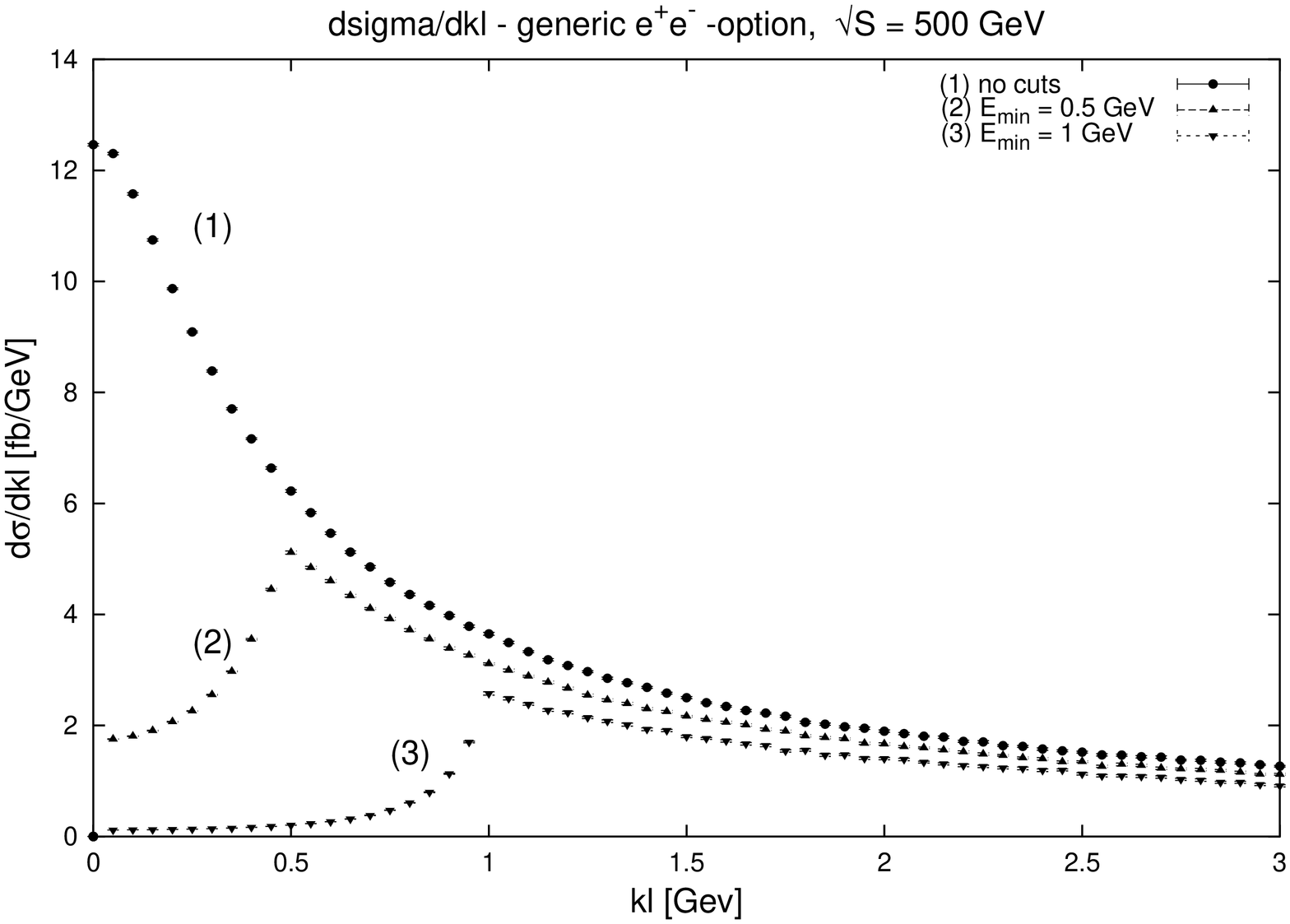}
\caption{effects of energy cuts; no Odderon } 
\label{fig:11}                                      
\end{figure}

\begin{figure}
\centering
\includegraphics[angle=-90, width=0.95\textwidth]{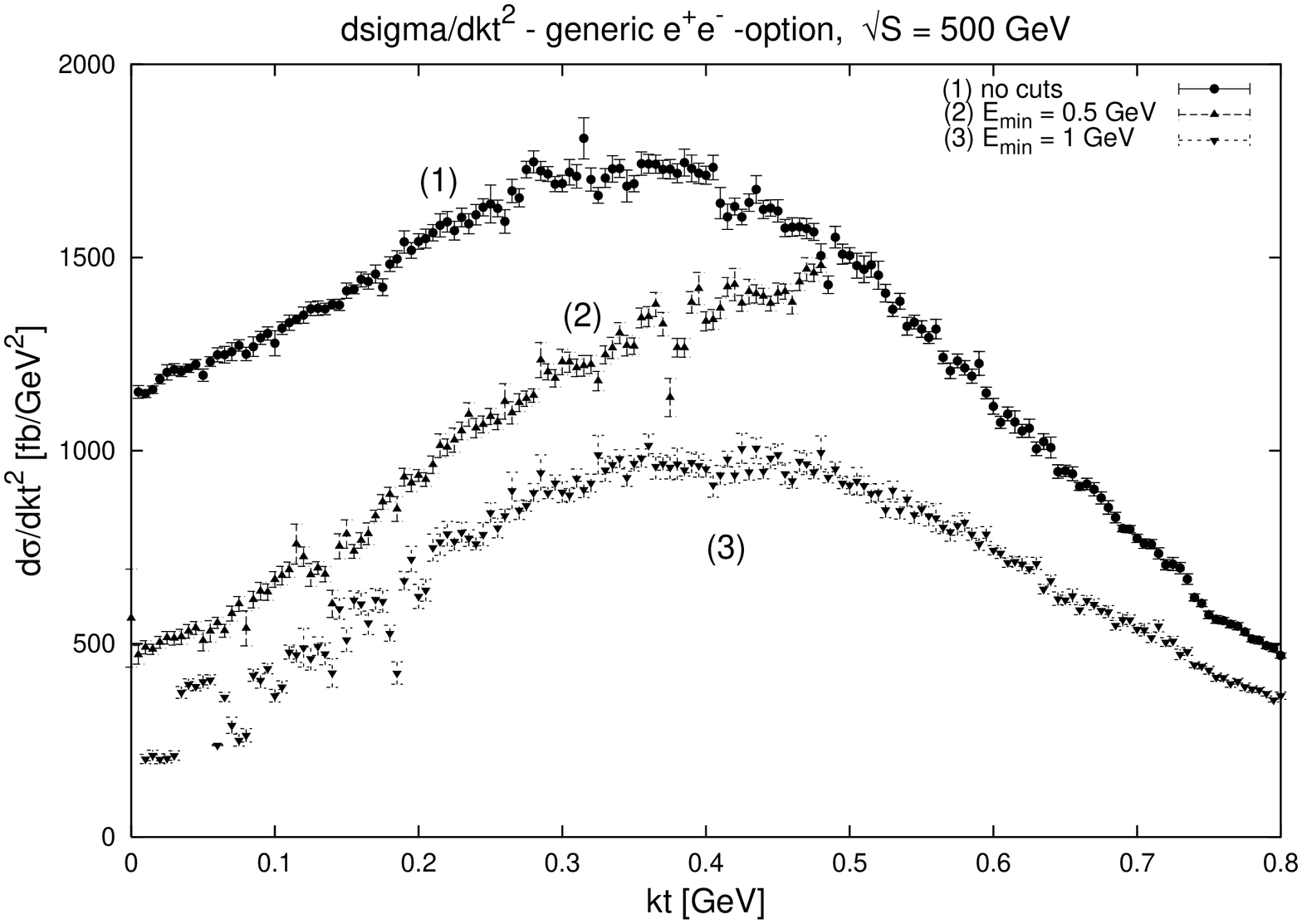}
\caption{effects of energy cuts; Odderon, $\eta_{\odd}=-1$}
\label{fig:10}
\end{figure}

\begin{figure}
\centering
\includegraphics[angle=-90, width=0.95\textwidth]{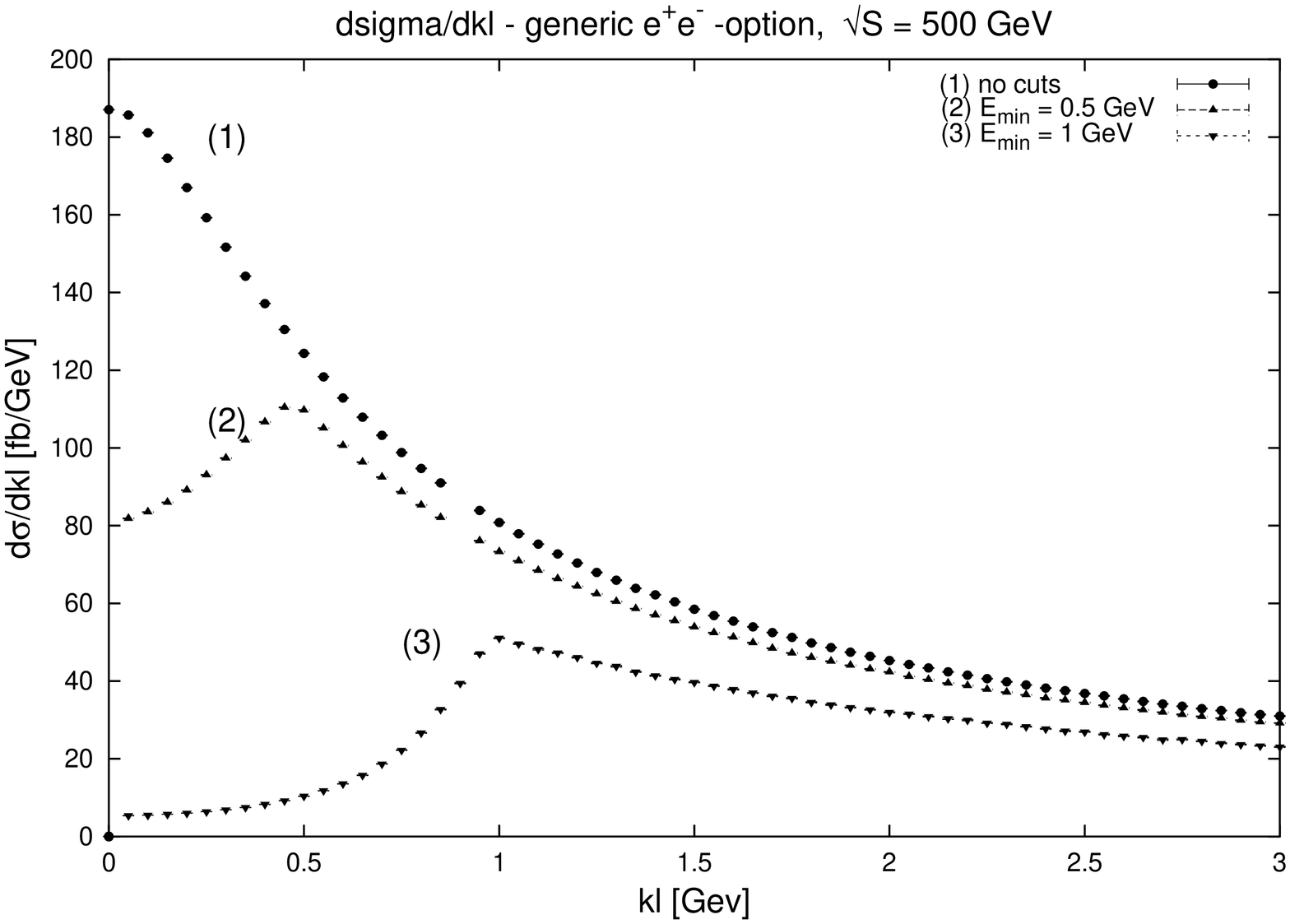}
\caption{effects of energy cuts; Odderon, $\eta_{\odd}=-1$}
\label{fig:12}
\end{figure}

We will now study the effects of detector cuts while keeping the values for $\al(0)\,$ and $\al'(t)\,$ fixed. We refrain from an in-depth analysis. The behavior resulting from energy cuts is displayed in figures \ref{fig:9} to \ref{fig:12}. $k^{0}_{1}\,\leq\,E_{min}$ implies
\begin{eqnarray}
k^{2}_{t}&\geq& E^{2}_{\min}-k^{2}_{l}-m^{2}_{\pi}\;\;\mbox{for $\frac{d\sigma}{d|k_{t}|^{2}}$}\,,\nonumber \\
k^{2}_{l}&\geq& E^{2}_{\min}-k^{2}_{t}-m^{2}_{\pi}\;\;\mbox{for $\frac{d\sigma}{dk_{l}}$}\,.
\end{eqnarray}
As can be seen from figures \ref{fig:9} to \ref{fig:12}, the results coincide with the cross sections without energy cuts for $k^{2}_{t}\,\geq\,E^{2}_{min}-m^{2}_{\pi}$. Below this boundary, the area of integration over $x_{i}$ is limited in the calculations of the differential cross sections leading to lower values for $\frac{d\sigma}{d|k_{t}|^{2}}$. For $\frac{d\sigma}{dk_{l}}$, we observe an approximate equality for $k^{2}_{l}\,\geq\,E^{2}_{min}-m^{2}_{\pi}$.

\begin{figure}
\centering
\includegraphics[width=0.75\textwidth]{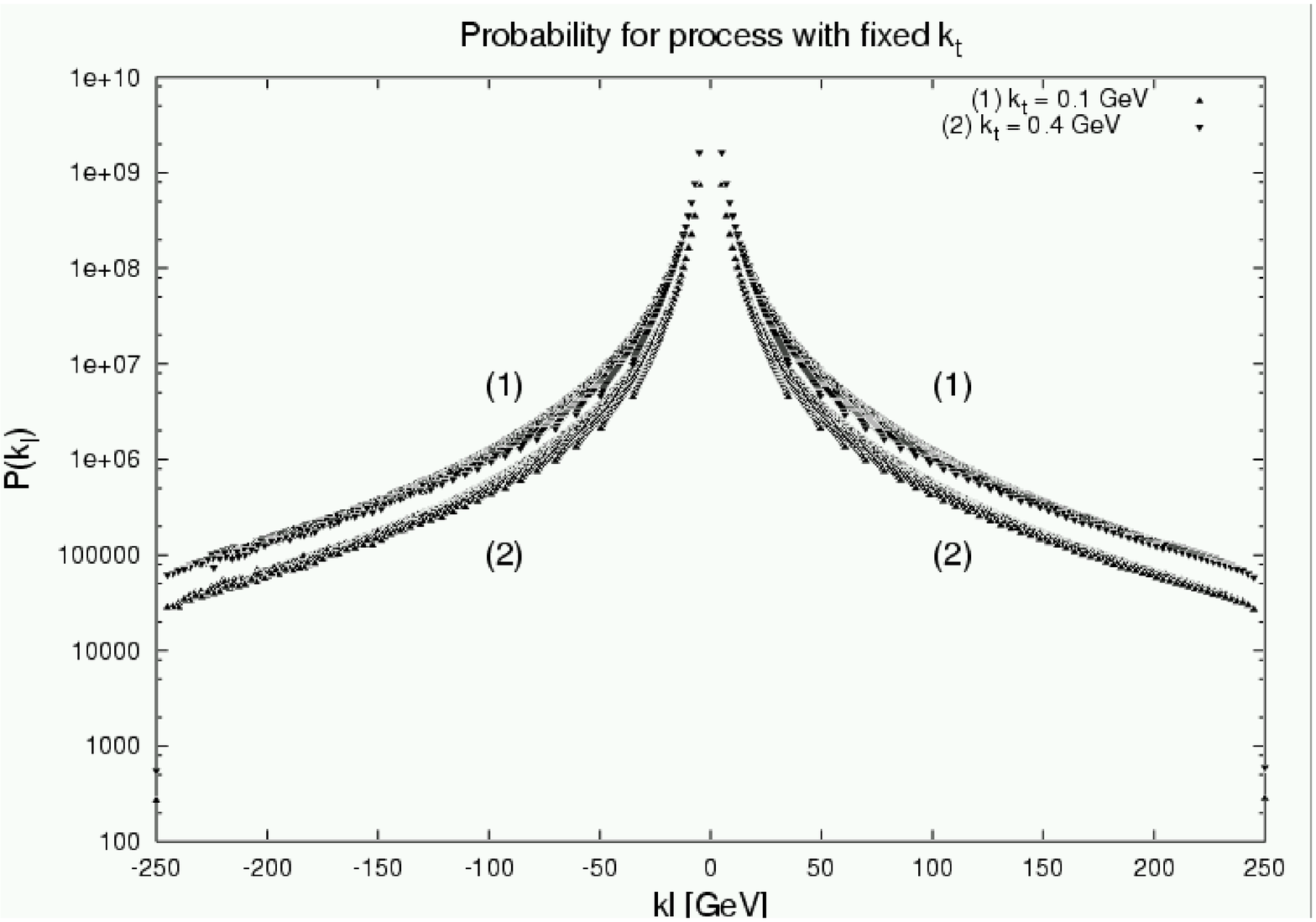}
\caption{$d\sigma(k_{l})\,$for fixed values of $k_{t}$; not to scale.}
\label{fig:probkl}
\end{figure}

\begin{figure}
\centering
\includegraphics[angle=-90, width=0.95\textwidth]{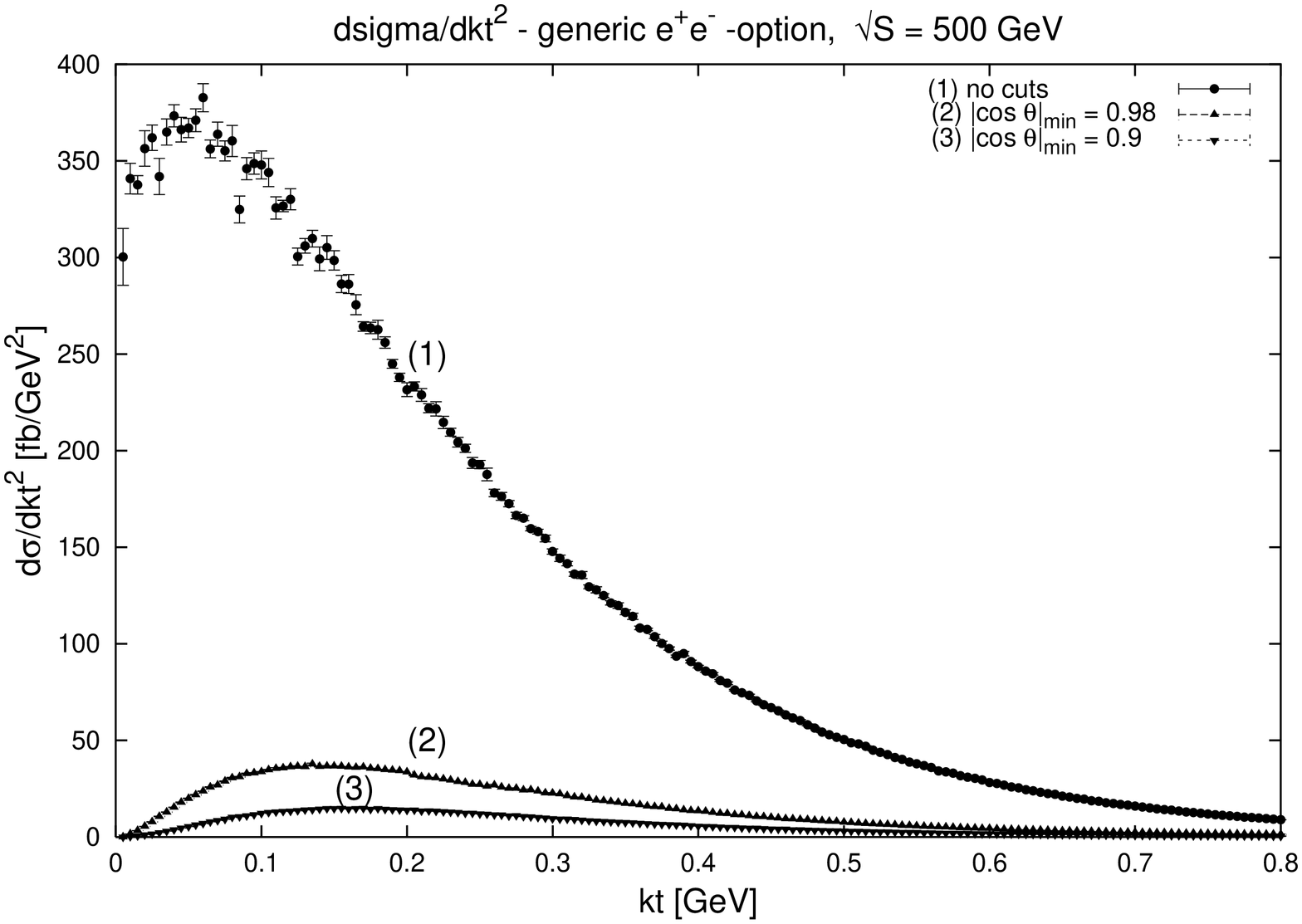}
\caption{effects of angular cuts; no Odderon } 
\label{fig:13}                                      
\end{figure}

\begin{figure}
\centering
\includegraphics[angle=-90, width=0.95\textwidth]{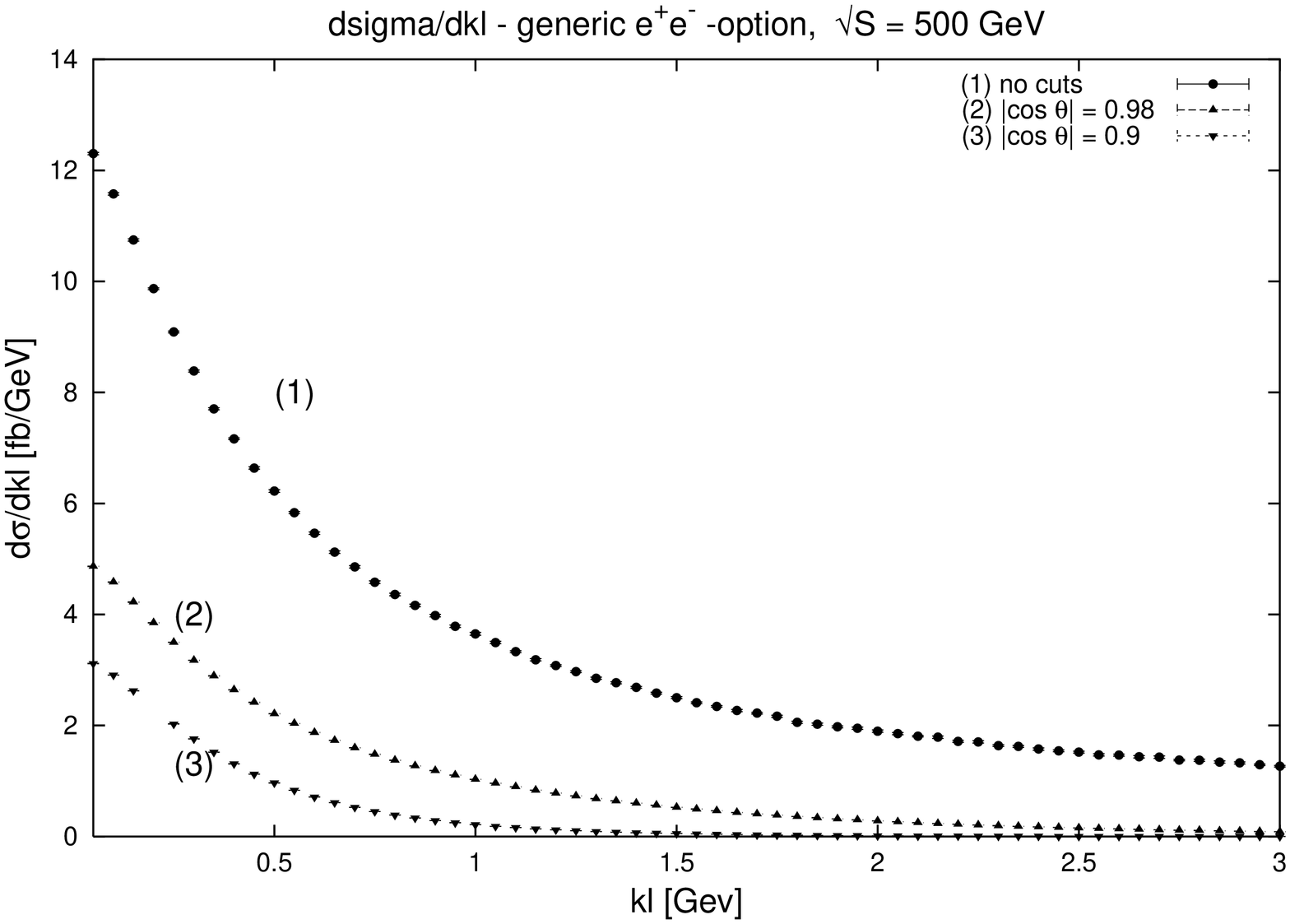}
\caption{effects of angular cuts; no Odderon } 
\label{fig:15}                                      
\end{figure}

\begin{figure}
\centering
\includegraphics[angle=-90, width=0.95\textwidth]{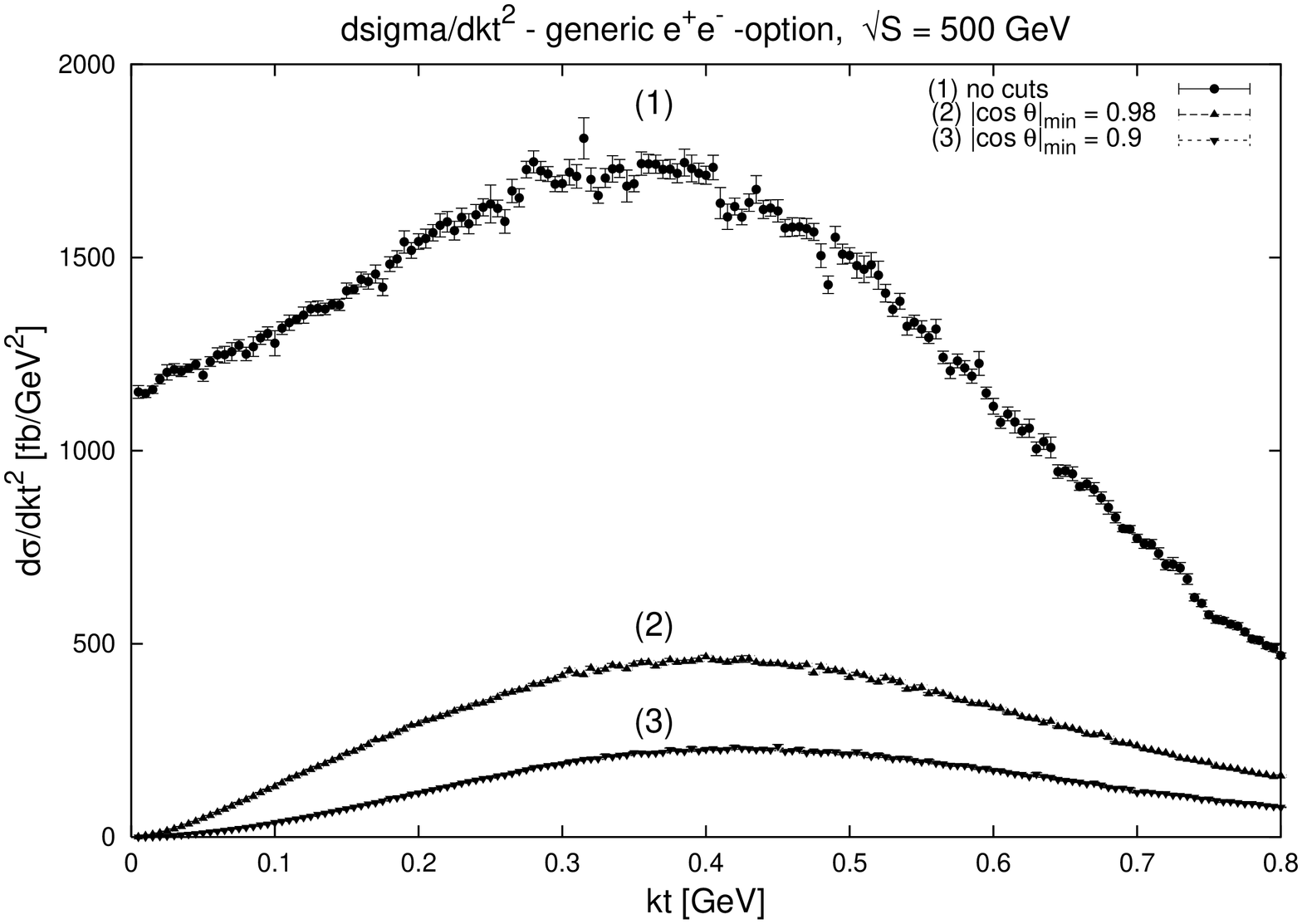}
\caption{effects of angular cuts, Odderon, $\eta_{\odd}\,=\,-1$}
\label{fig:16}
\end{figure}

\begin{figure}
\centering
\includegraphics[angle=-90, width=0.95\textwidth]{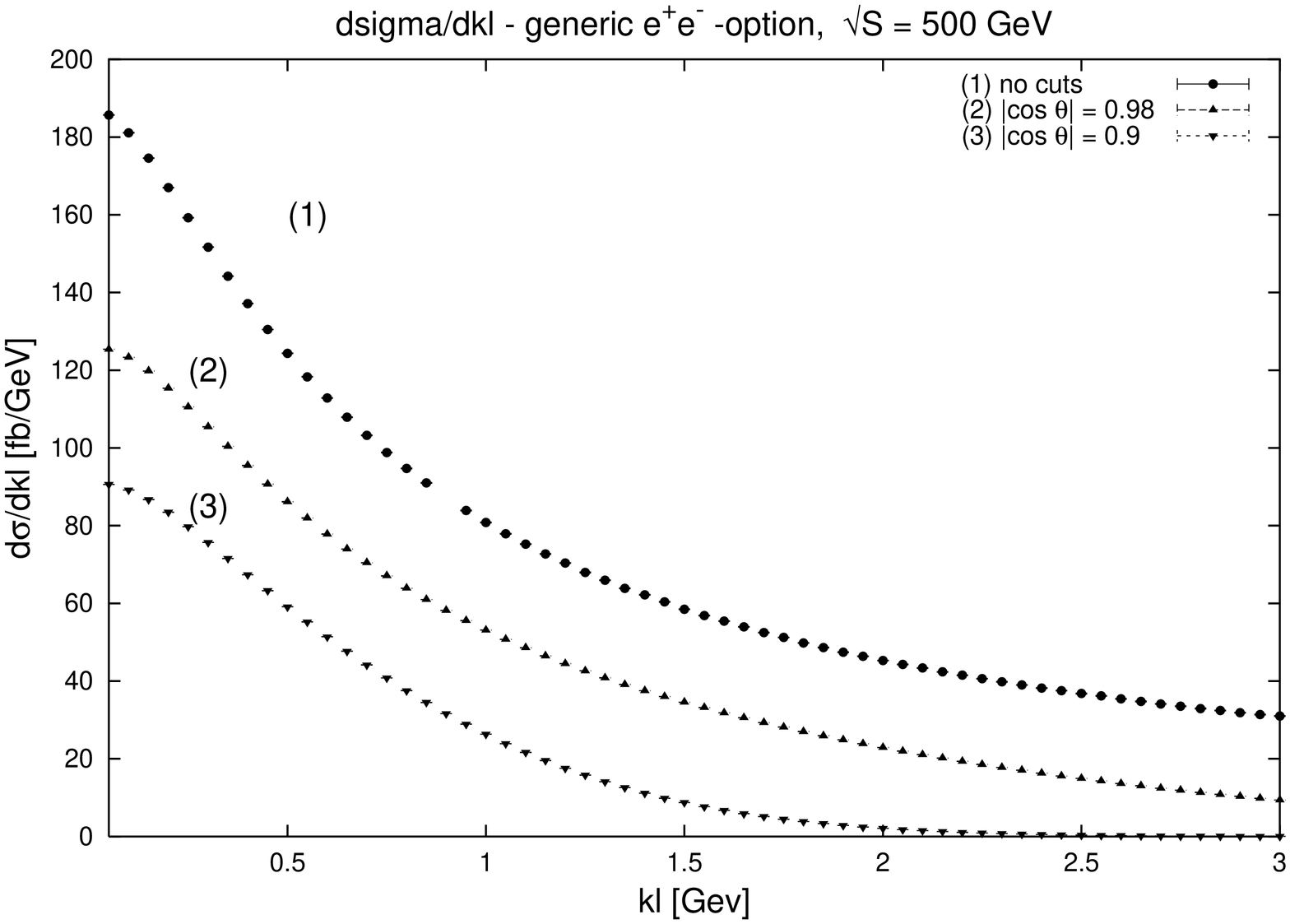}
\caption{effects of angular cuts, Odderon, $\eta_{\odd}\,=\,-1$  }
\label{fig:17}
\end{figure}

\begin{figure}
\centering
\includegraphics[angle=-90, width=0.95\textwidth]{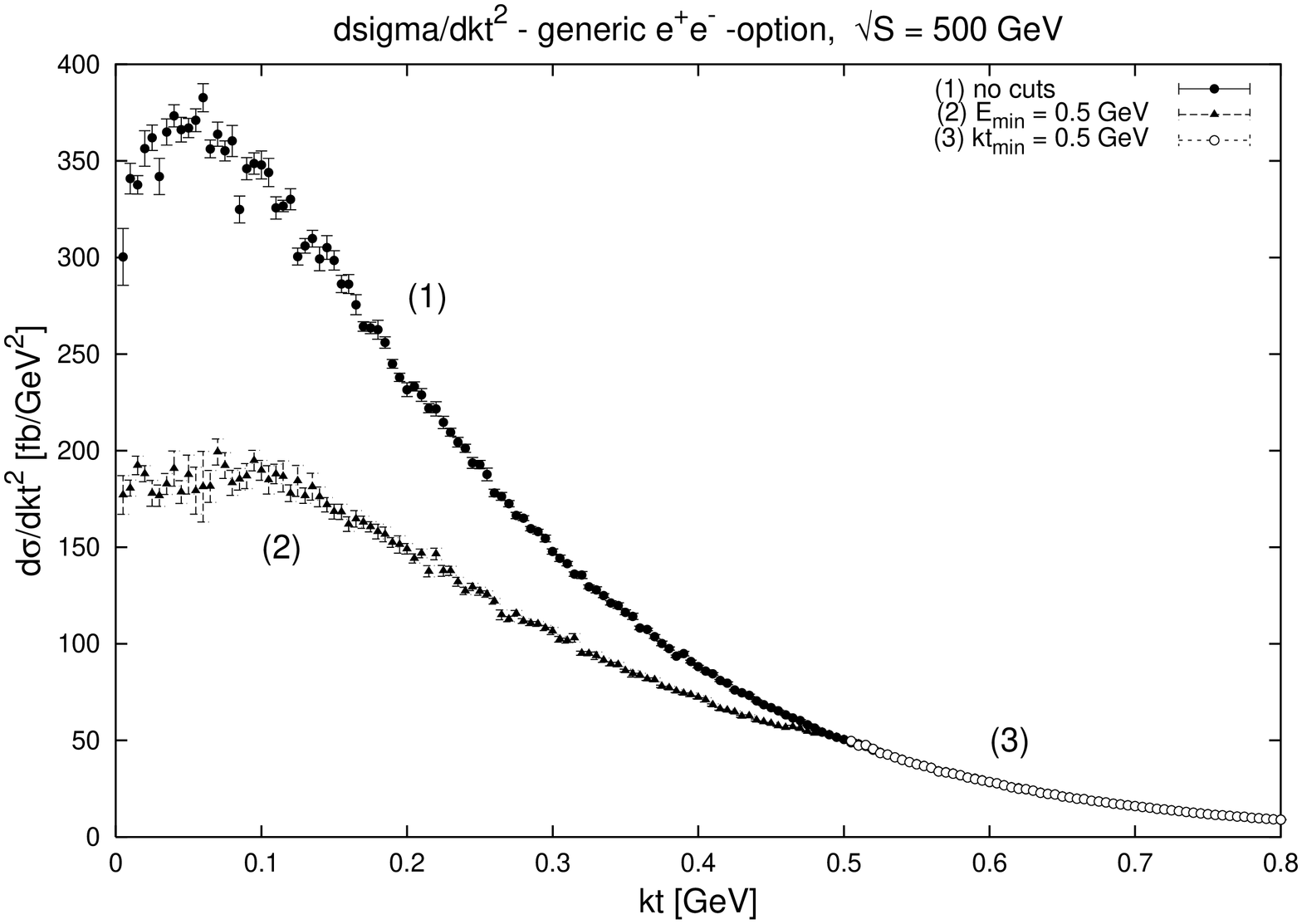}
\caption{comparison between energy and transverse momentum cuts; no Odderon } 
\label{fig:18}                                      
\end{figure}

\begin{figure}
\centering
\includegraphics[angle=-90, width=0.95\textwidth]{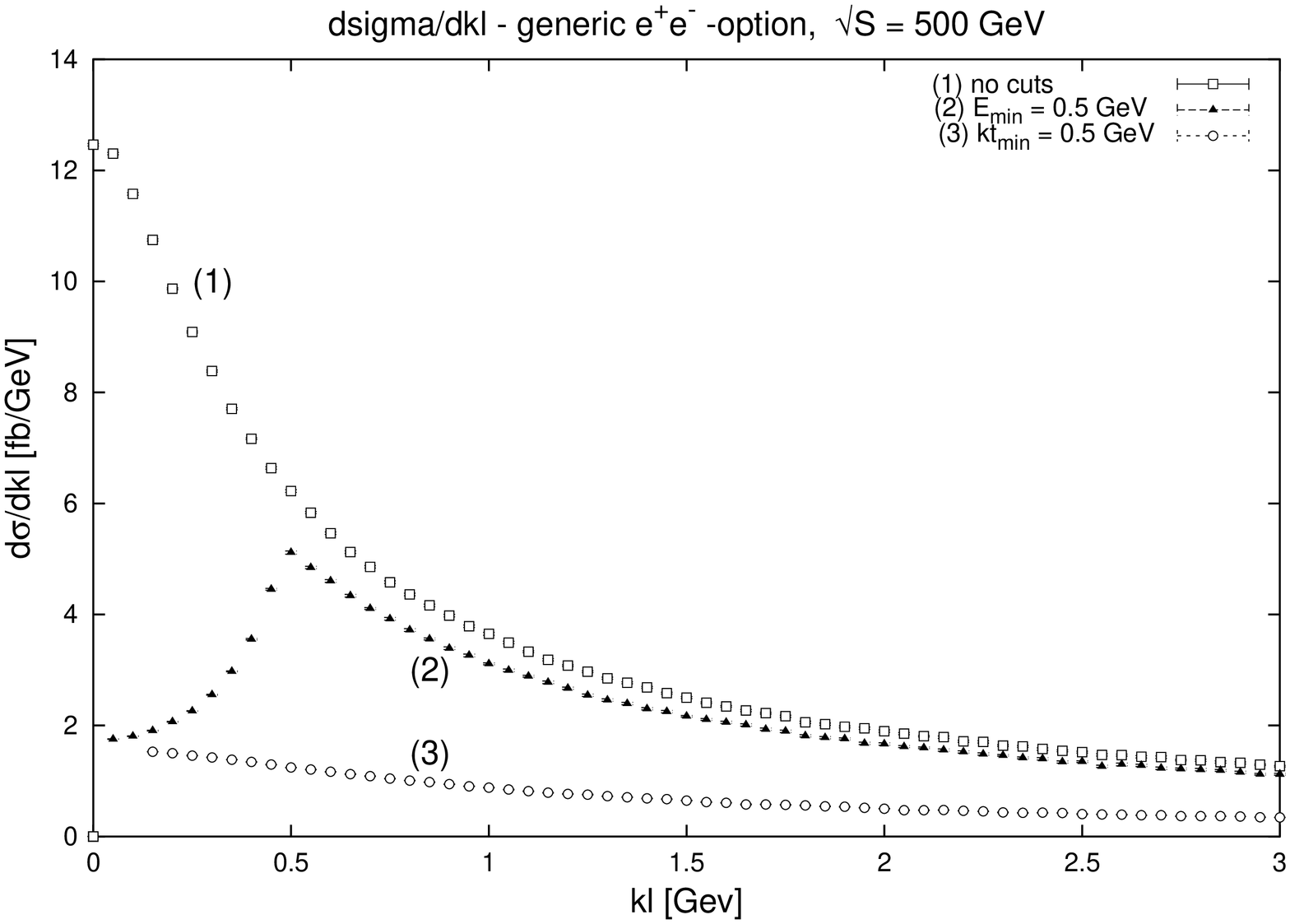}
\caption{comparison between energy and transverse momentum cuts; no Odderon}
\label{fig:19}
\end{figure}

\begin{figure}
\centering
\includegraphics[angle=-90, width=0.95\textwidth]{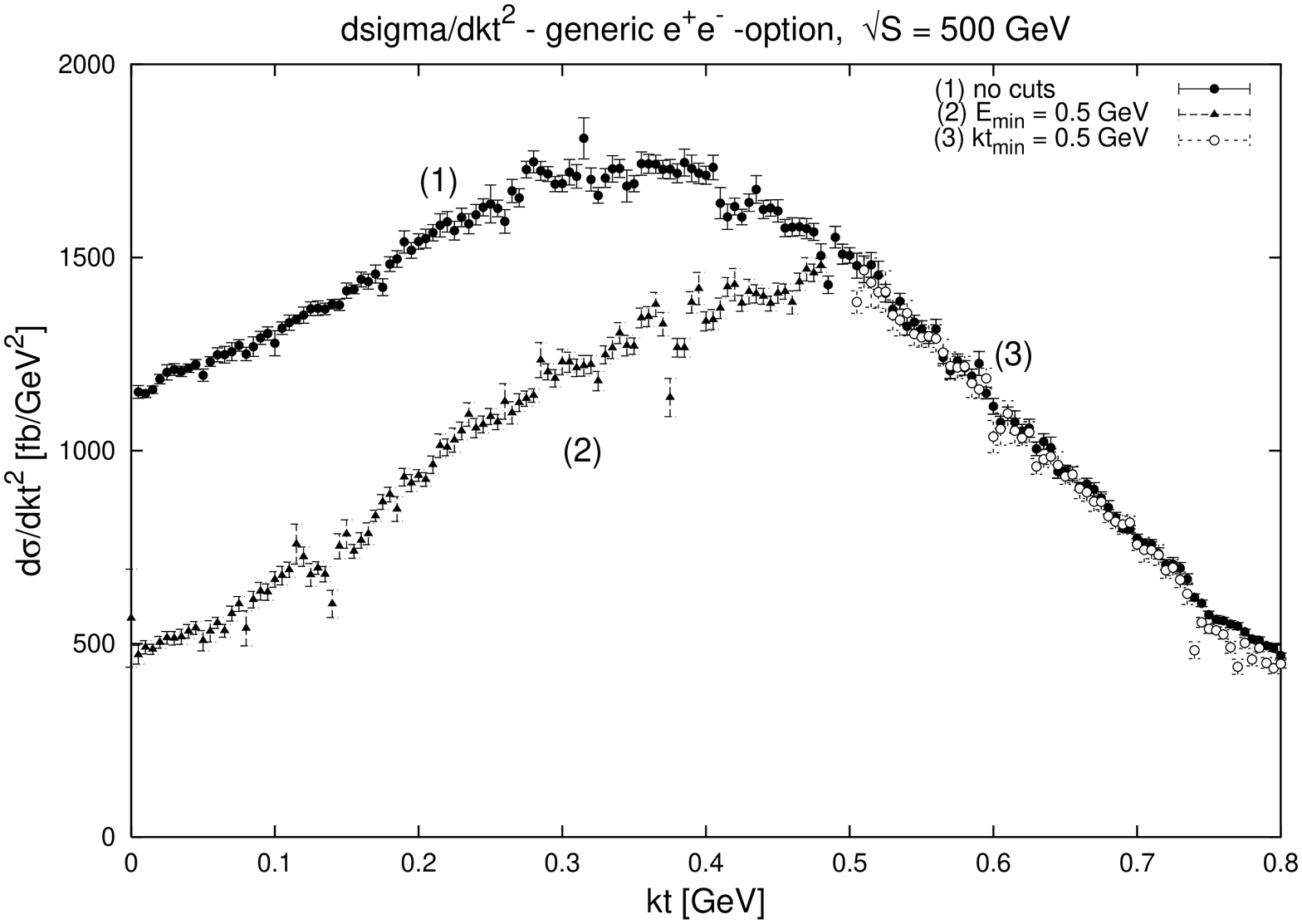}
\caption{comparison between energy and transverse momentum cuts; Odderon, $\eta_{\odd}\,=\,-1$ } 
\label{fig:20}                                      
\end{figure}

\begin{figure}
\centering
\includegraphics[angle=-90, width=0.95\textwidth]{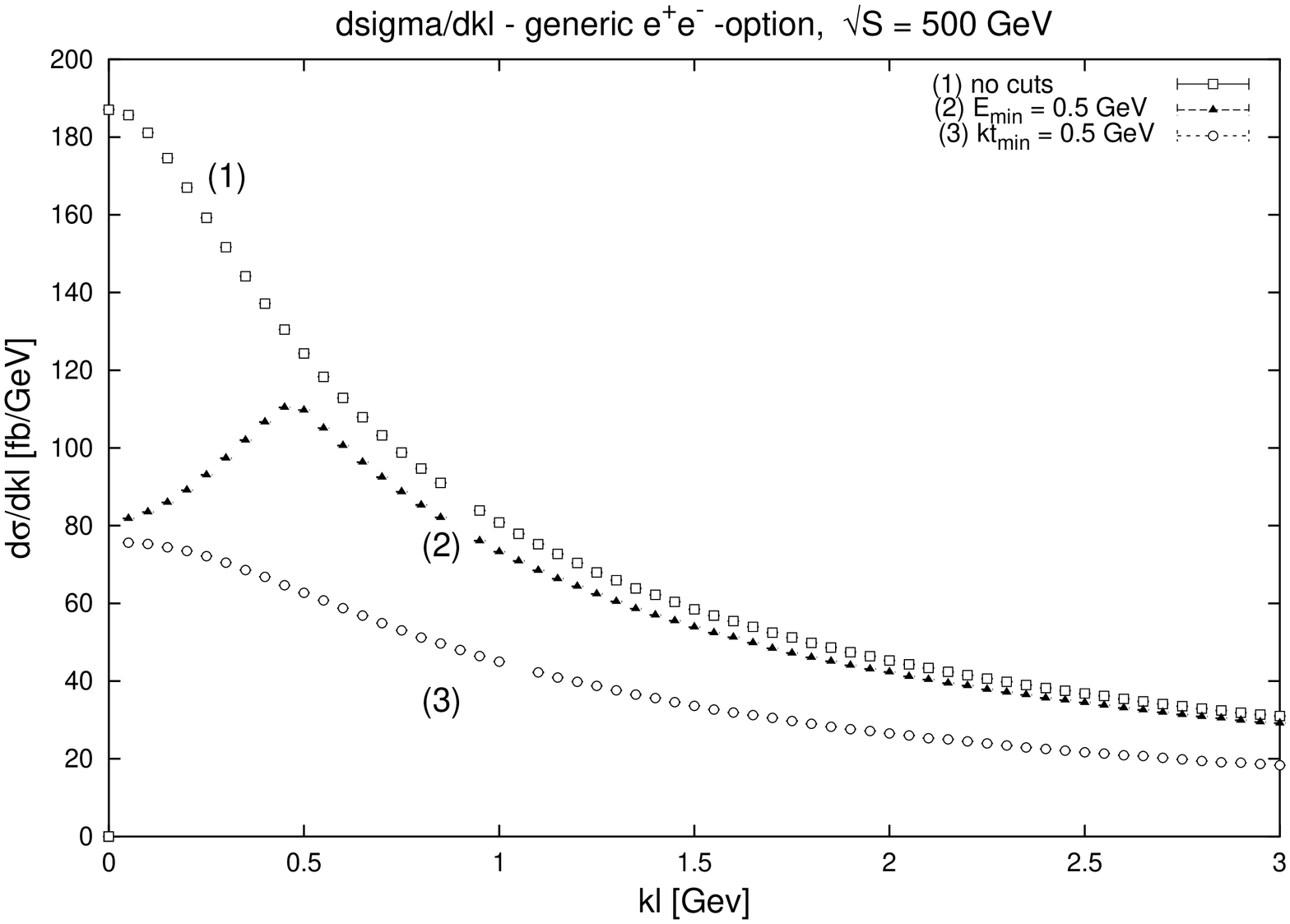}
\caption{comparison between energy and transverse momentum cuts; Odderon, $\eta_{\odd}\,=\,-1$}
\label{fig:21}
\end{figure}

The results of angular cuts are displayed in figures \ref{fig:13} to \ref{fig:17}. An angular cut implies a cut on the ratio of $k_{t}$ and $k_{l}$ according to (\ref{eq:limkld}): $k_{l}\;\leq\;k_{t}/\sqrt{c_{1}}$, i.e. it acts like a filter for high-$k_{l}$ outgoing particles for $k_{t}$ fixed or low-$k_{t}$ outgoing particles for $k_{l}$ fixed. We obtain

\begin{equation}
\frac{1}{\sqrt{c_{1}}}\;=\; \left\{ \begin{array}{c} 4.92 \;\;\; \mbox{for}\, |\cos\theta|_{max}\,=\,0.98\,,\\
 2.06 \;\;\; \mbox{for}\, |\cos\theta|_{max}\,=\,0.9\,. \end{array} \right.
\end{equation}
Plugging this into (\ref{eq:limkld}) in the calculation of $\frac{d\sigma}{d|k_{t}|^{2}}$ limits $k_{l}$ according to

\begin{equation}
k_{l}\,\leq\,4.92\;\mbox{GeV}\;;\;  
k_{l}\,\leq\,2.06\;\mbox{GeV}
\end{equation}
for $k_{t}\,=\,1$ GeV. Figure \ref{fig:probkl} gives $d\sigma (k_{l})$ for fixed values of $k_{t}$; we see that, although $d\sigma$ is peaked at $k_{l}\,=\,0$, $\frac{d\sigma}{d|k_{t}|^{2}}$ is still severely limited by angular cuts after integration over all values for $k_{l}$. The latter can be roughly equated with the integration over $x_{i}$.
Finally, we will compare the given detector cuts with "naive" cuts; in figures \ref{fig:18} to \ref{fig:21}, we display the results of comparison between energy and transverse momentum cuts. We see that this substitution leads to negligence of terms with $k_{t}\,\le\,E_{cut}\,$ for $\frac{d\sigma}{d|k_{t}|^{2}}\,$; similarly,  $\frac{d\sigma}{dk_{l}}\,$ becomes smaller. We conclude that naive cuts on the transverse momentum qualitatively lead to an artificially smaller value for $\frac{d\sigma}{d|k_{t}|^{2}}\,$ for  $k_{t}\,\leq\,E_{cut}\,$ as well as $\frac{d\sigma}{dk_{l}}$ for all values of $k_{l}$.

\section{Results for individual experiments}

We will now display the results for the individual experiments; here, we use the cuts given in section \ref{sec:detcut}. For different cut parameters, we refer to the comparison of different energy and angular cuts in the previous section. We consider:

\begin{itemize}
\item{}
Comparison between cross sections with and without Odderon, taking $\eta_{\odd}\,=\pm1$\,, 
\item{}Variation of $\al'_{\odd}$ from $\al'_{\odd}\,=\,0.25\;\mbox{GeV}^{2}$ to $\al'_{\odd}\,=\,0.5\;\mbox{GeV}^{-2}$ for $\eta_{\odd}=-1$\,,
\item{}Variation of $(\frac{\be_{\odd}}{\be_{\pom}})^{2}$ for $\eta_{\odd}=-1\,, \vare'\,=\,-0.02$.
\end{itemize}
If not mentioned otherwise, the reference values are given by (\ref{eq:oddval}):

\begin{eqnarray}
\vare' \,=\,0;&\al'_{\odd}\,=\,0.25\;\;\mbox{GeV}^{-2};&\left(\frac{\be_{\odd}}{\be_{\pom}}\right)^{2}\,=\,0.05\;.
\end{eqnarray}
Variations other than the ones given above can easily be extrapolated from the considerations in section \ref{sec:parvar}\,.

\subsection{LEP I}\label{sec:lepi}

\begin{figure}
\centering
\includegraphics[angle=-90, width=0.95\textwidth]{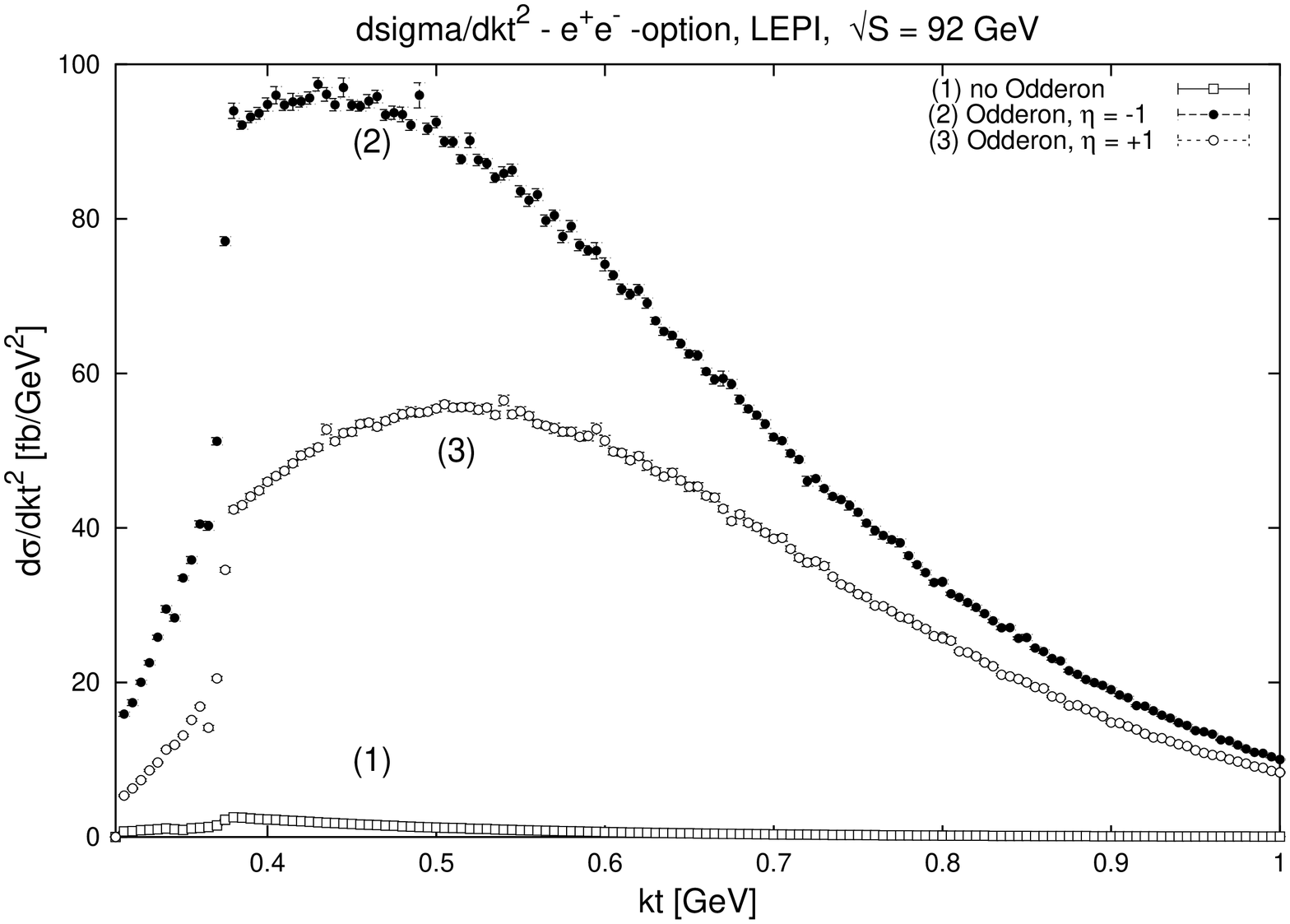}
\caption{different values for $\eta_{\odd}$ } 
\label{fig:22}                                      
\end{figure}

\begin{figure}
\centering
\includegraphics[angle=-90, width=0.95\textwidth]{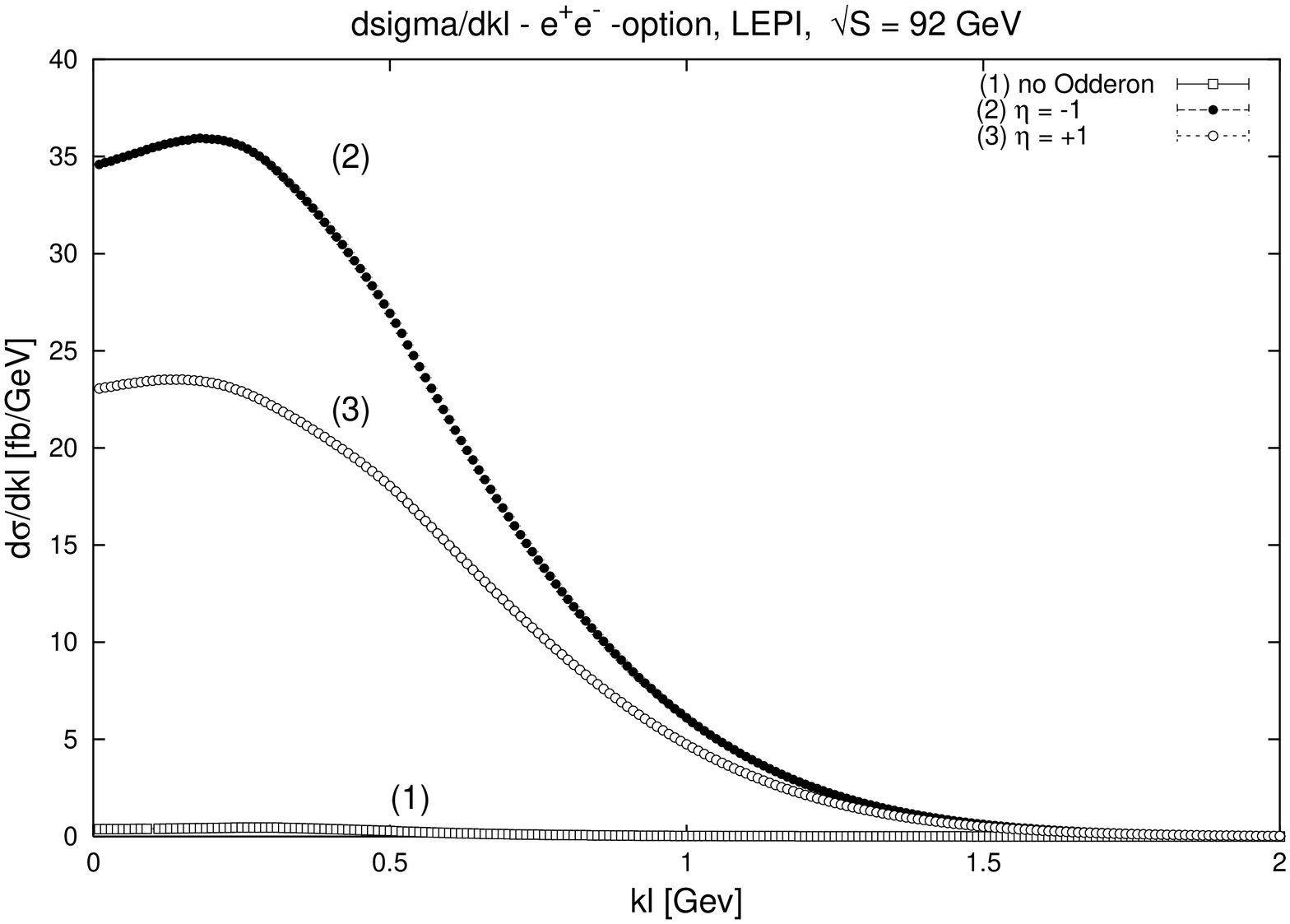}
\caption{different values for $\eta_{\odd}$}
\label{fig:23}
\end{figure}

\begin{figure}
\centering
\includegraphics[angle=-90, width=0.95\textwidth]{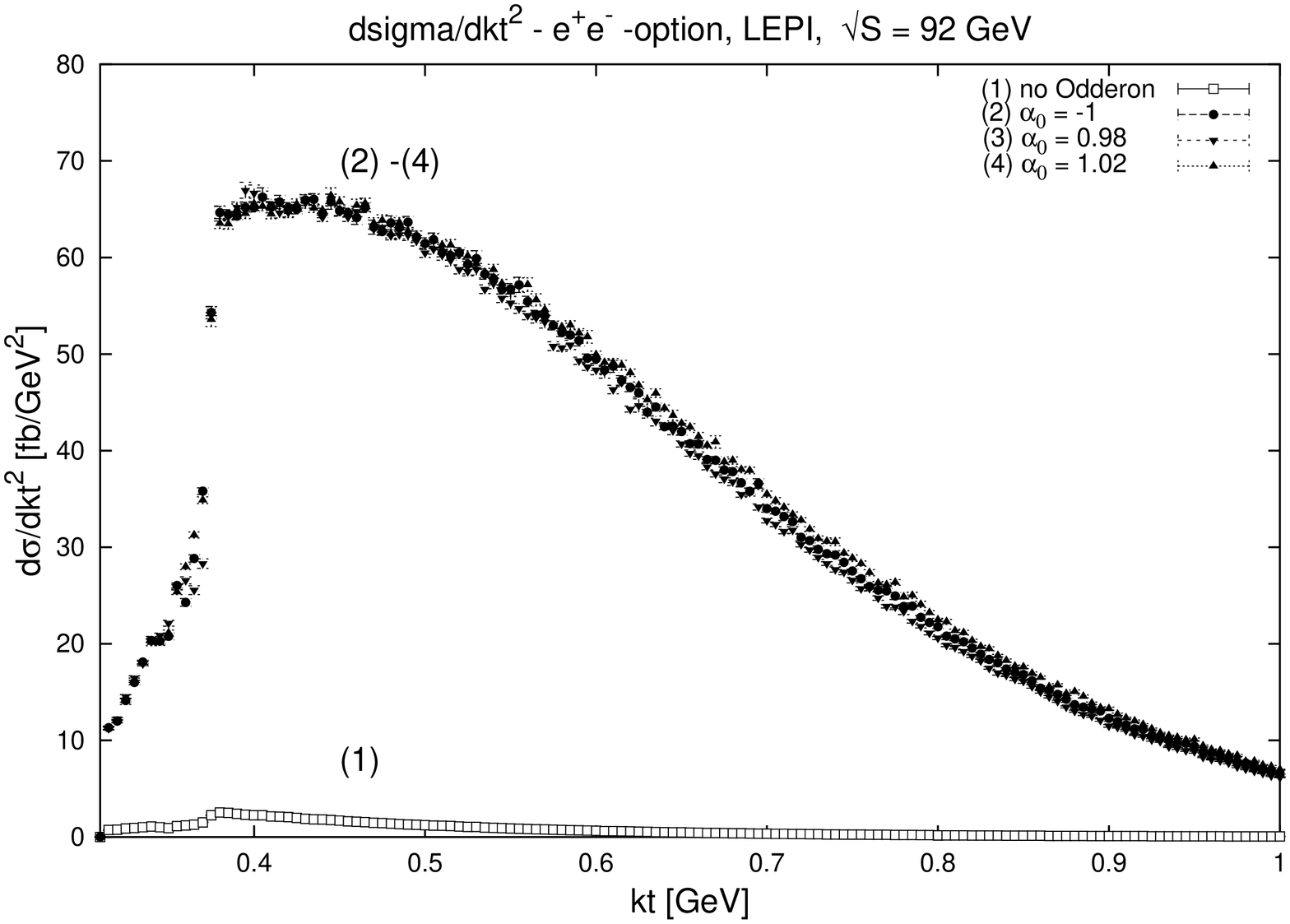}
\caption{comparison different values for $\al(0)\,$,$\eta_{\odd}$=-1} 
\label{fig:24}                                      
\end{figure}

\begin{figure}
\centering
\includegraphics[angle=-90, width=0.95\textwidth]{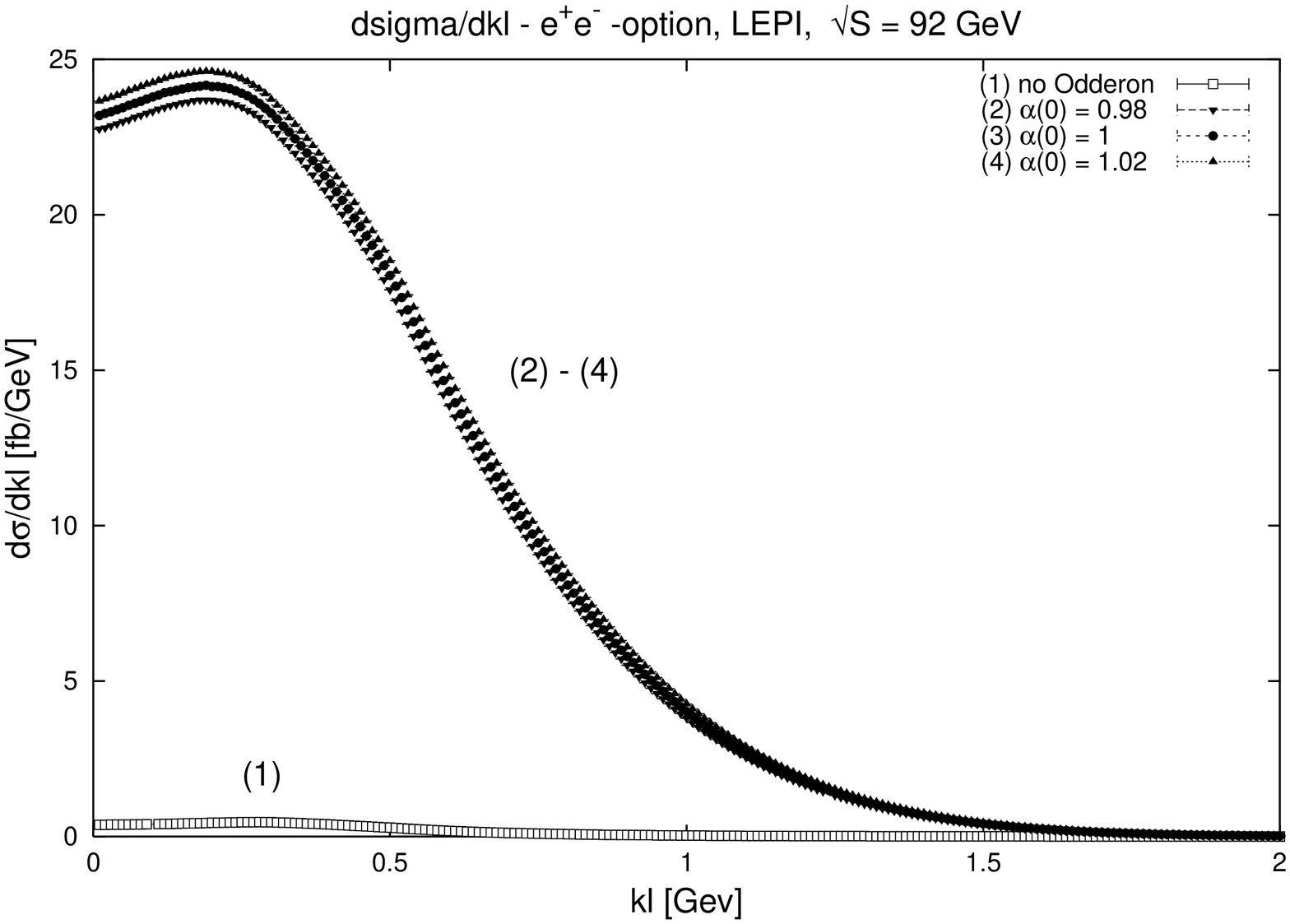}
\caption{comparison different values for $\al(0)\,$,$\eta_{\odd}$=-1} 
\label{fig:25}                                      
\end{figure}

\begin{figure}
\centering
\includegraphics[angle=-90, width=0.95\textwidth]{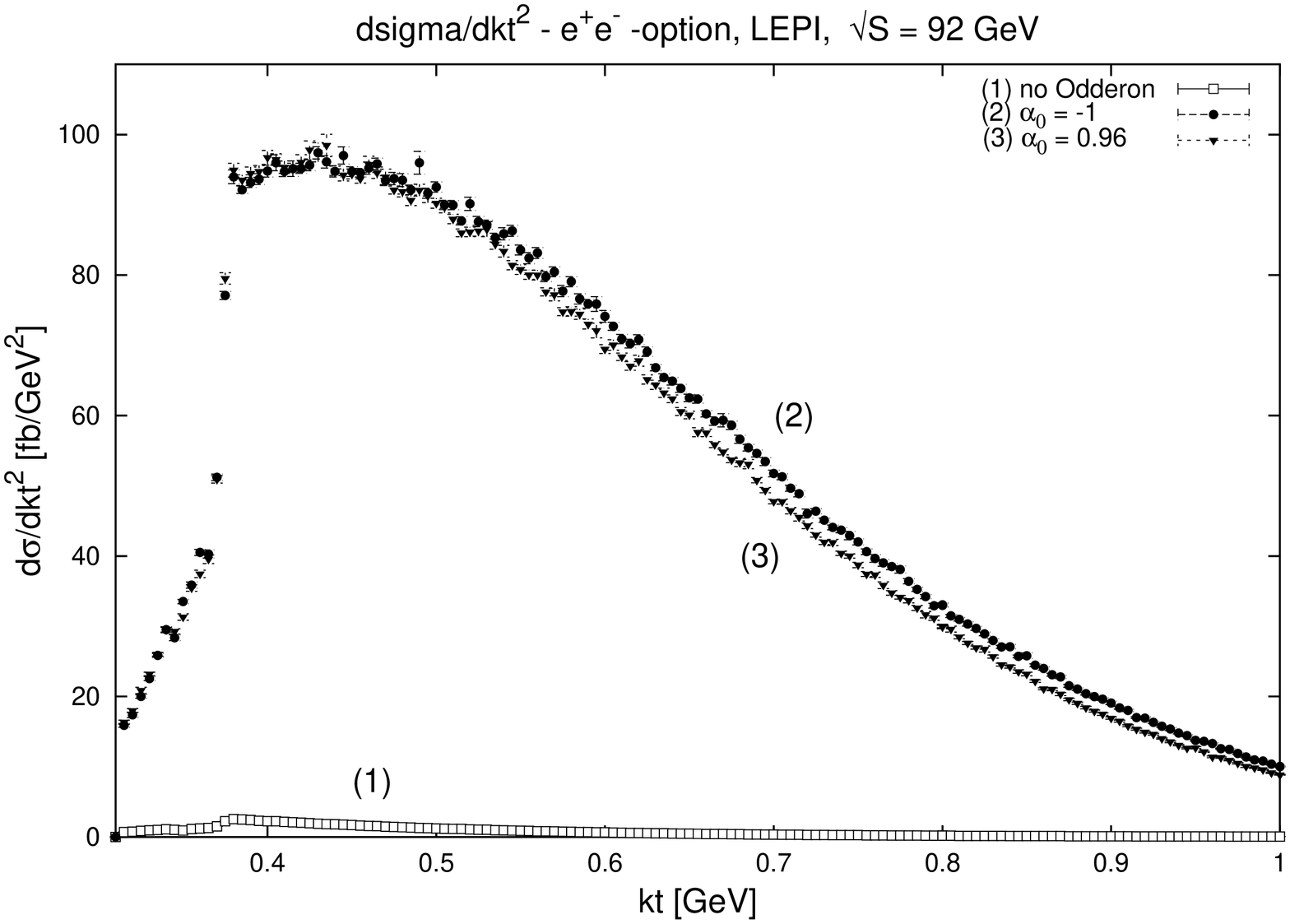}
\caption{comparison different values for $\al(0)\,$,$\eta_{\odd}$=-1} 
\label{fig:24-2}                                      
\end{figure}

\begin{figure}
\centering
\includegraphics[angle=-90, width=0.95\textwidth]{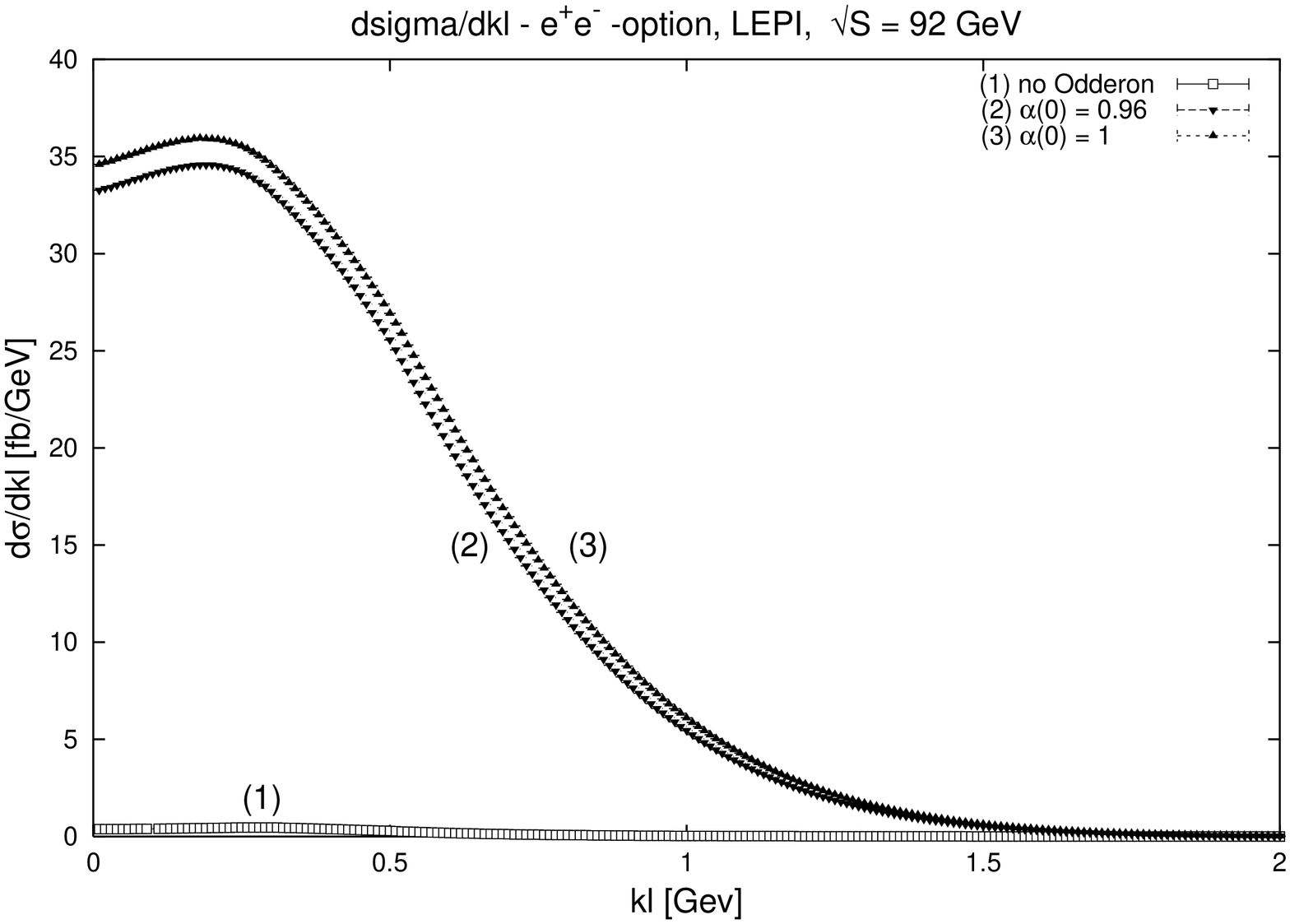}
\caption{comparison different values for $\al(0)\,$,$\eta_{\odd}$=-1} 
\label{fig:25-2}                                      
\end{figure}

\begin{figure}
\centering
\includegraphics[angle=-90, width=0.95\textwidth]{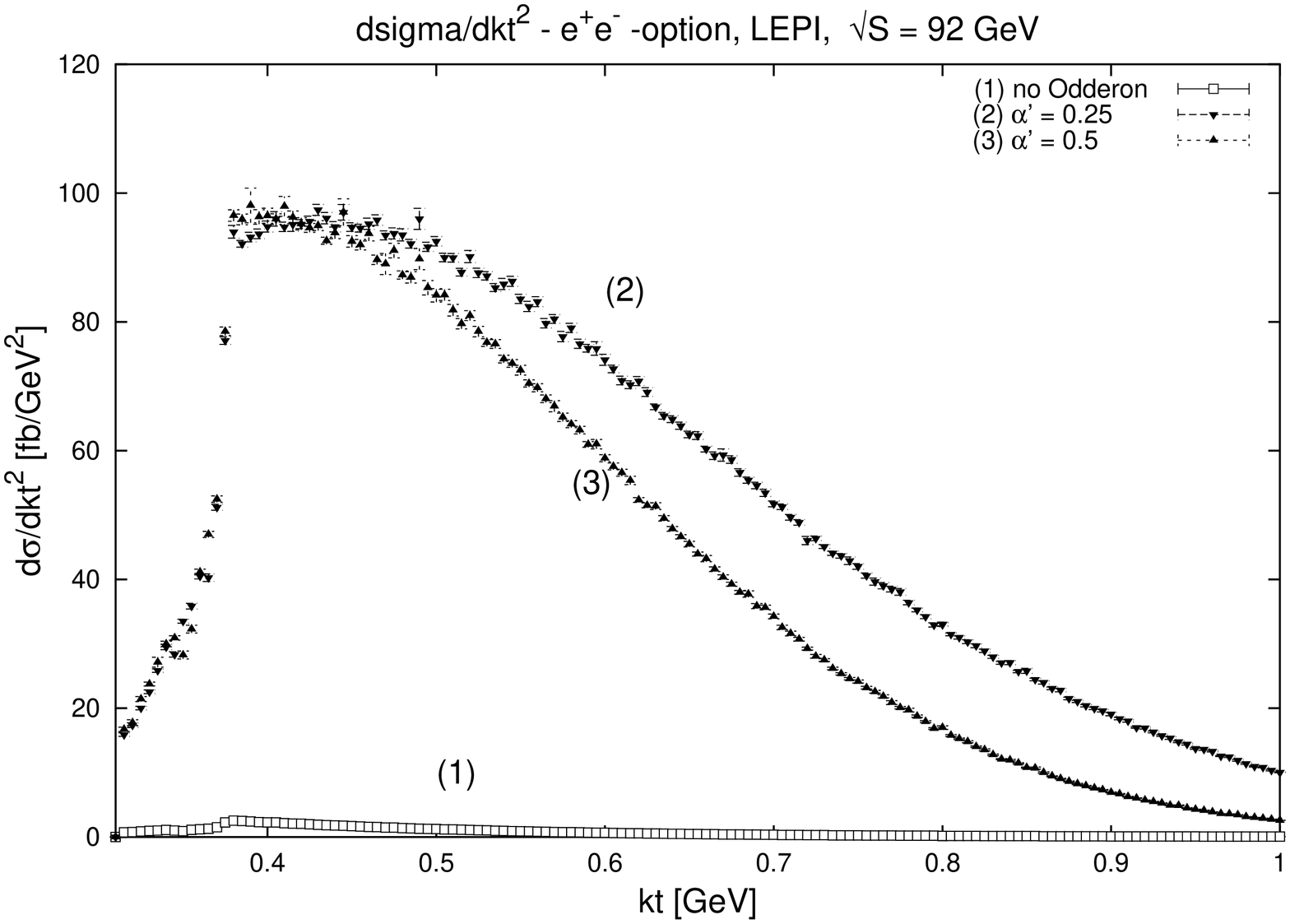}
\caption{comparison different values for $\al'\,$,$\eta_{\odd}$=-1} 
\label{fig:26}                                      
\end{figure}

\begin{figure}
\centering
\includegraphics[angle=-90, width=0.95\textwidth]{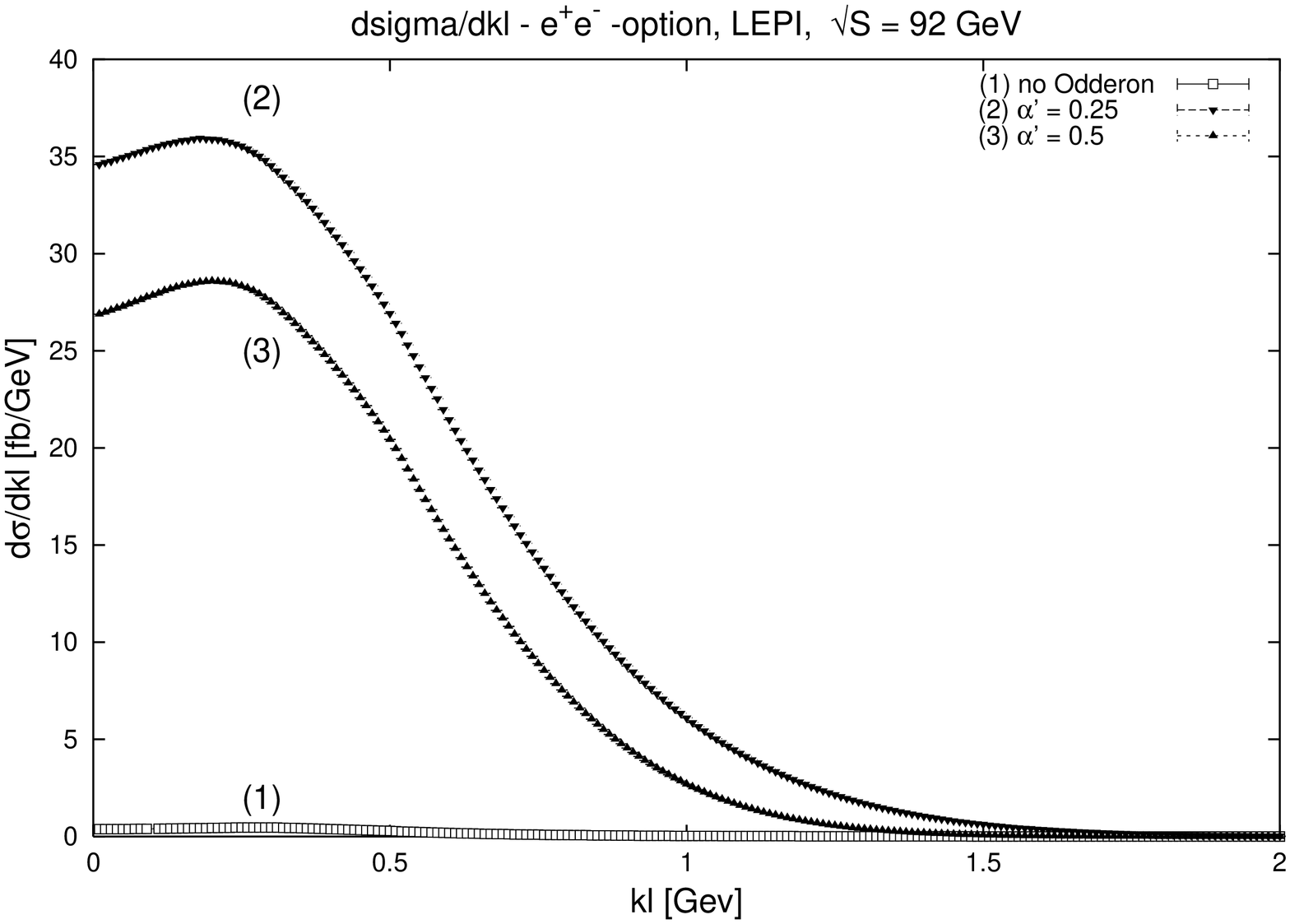}
\caption{comparison different values for $\al'\,$,$\eta_{\odd}$=-1}
\label{fig:27}
\end{figure}

\begin{figure}
\centering
\includegraphics[angle=-90, width=0.95\textwidth]{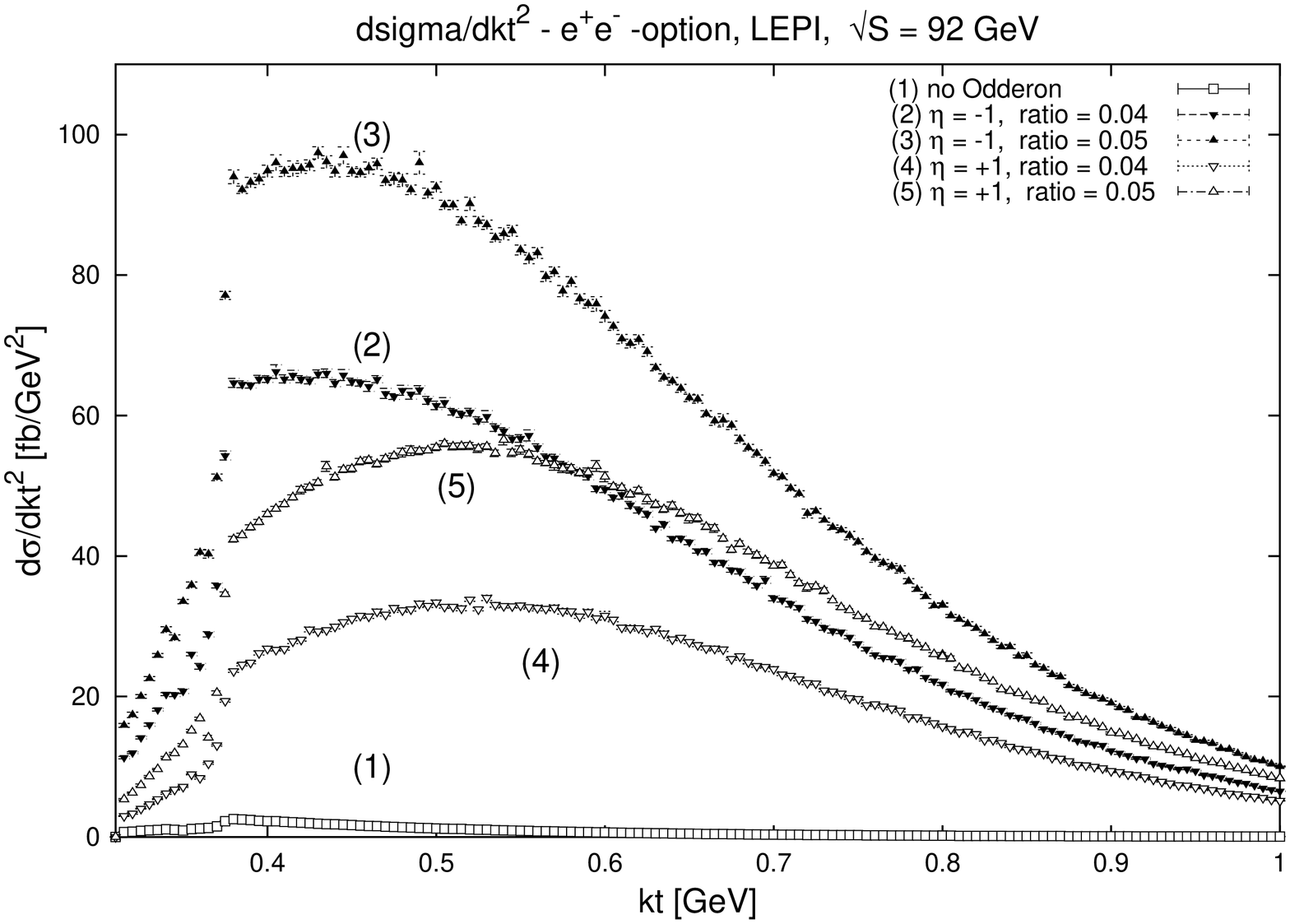}
\caption{comparison different values for $(\frac{\be_{\odd}}{\be_{\pom}})^{2}$ } 
\label{fig:27-1}                                      
\end{figure}

\begin{figure}
\centering
\includegraphics[angle=-90, width=0.95\textwidth]{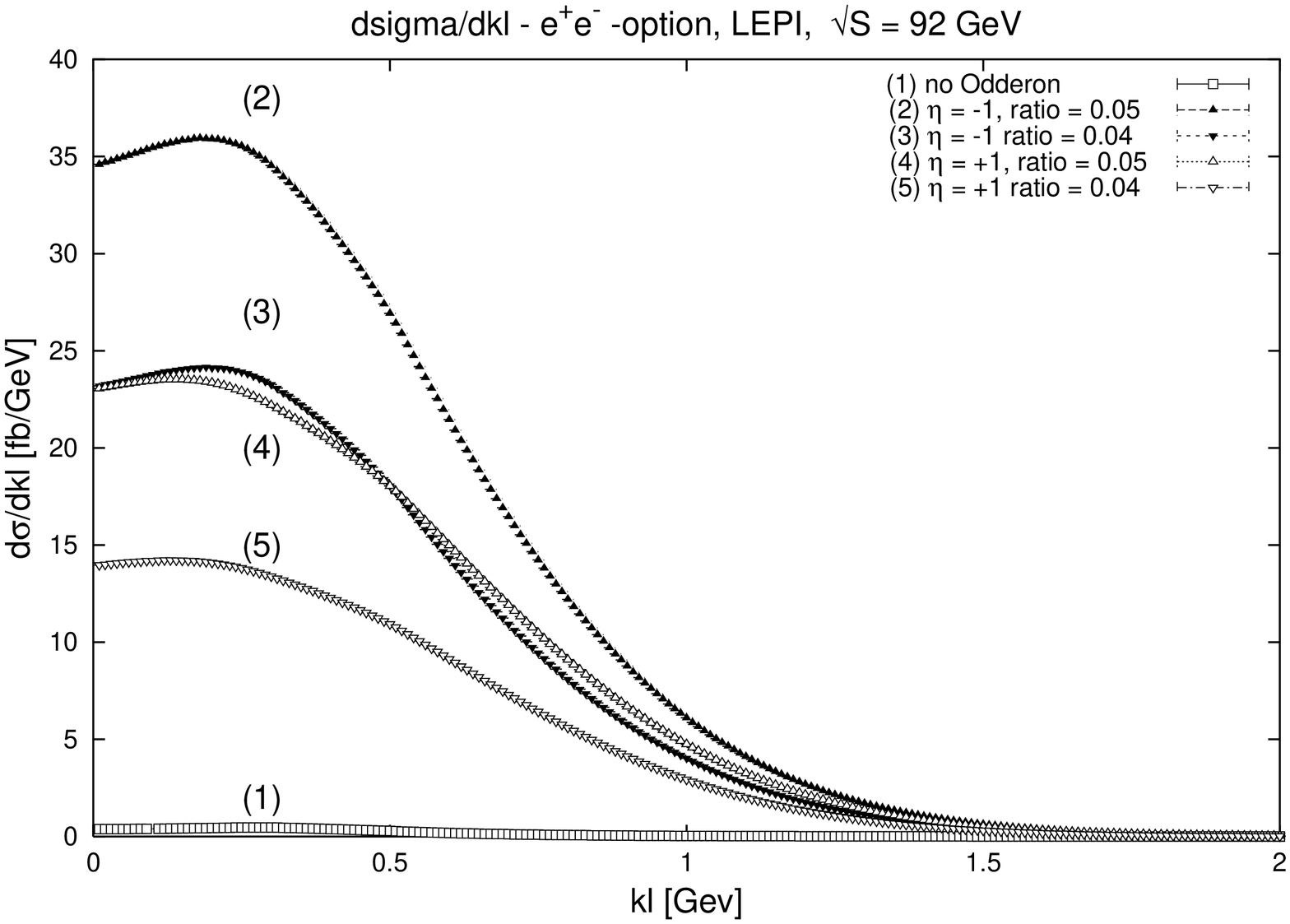}
\caption{comparison different values for $(\frac{\be_{\odd}}{\be_{\pom}})^{2}$}
\label{fig:27-2}
\end{figure}

For the OPAL detector at LEPI, we used the following values \cite{Lillich:2002}:

\begin{eqnarray*}
\sqrt{S_{lab}}\,=\,92\;\mbox{GeV};&|\cos\theta|_{max}\,=\,0.81;& E_{min}\,=\,0.4\;\mbox{GeV}\,.
\end{eqnarray*}
The results are displayed in figures \ref{fig:22} to \ref{fig:27-2}.

We obtain the following results:

\begin{itemize}
\item{Phase distinction}\\
As we can see from figures \ref{fig:22} and \ref{fig:23}, the Odderon phase can easily be distinguished from $\frac{d\sigma}{d|k_{t}|^{2}}\,$ as the differences between the result for $\eta_{\odd}\,=\pm 1$ are clearly visible. There is also a clear difference for the results for $\frac{d\sigma}{dk_{l}}\,$ depending on the sign of the Odderon phase. 
\item{Variation of $\vare'$}\\
In the generic case, the variation of $\vare'$ lead to an nearly unobservable changes in both cross sections for $|\vare'|\,\leq\,0.02$. The same holds for the real detector case including energy and momentum cuts: figures \ref{fig:24} and \ref{fig:25} show the results for $\vare'\,=\,\pm 0.02$. However, according to (\ref{eq:resmodeps}), larger variations should lead to bigger effects. The results for $\vare'\,=\,-0.04$ are shown in figures \ref{fig:24-2} and \ref{fig:25-2}; here, the intercept was motivated by a pQCD result \cite{Janik:1999ae}. The effects are still relatively small. The variation to intercepts $>\,1$ is limited by (\ref{eq:limrho}).
\item{variation of $\al_{\odd}'$}\\
For results for the variation of  $\al_{\odd}'$ are given in  figures \ref{fig:26} and \ref{fig:27}. We observe the same behavior with and without detector cuts, i.e. a faster decrease for $\frac{d\sigma}{d|k_{t}|^{2}}\,$ as well as lower values for $\frac{d\sigma}{dk_{l}}$; see figure \ref{fig:7} and \ref{fig:8} for comparison.
\item{variation of $(\frac{\be_{\odd}}{\be_{\pom}})^{2}$}\\
As expected, we obtain a decrease in the cross sections when varying $(\frac{\be_{\odd}}{\be_{\pom}})^{2}$; the results are displayed in figures \ref{fig:27-1} and \ref{fig:27-2}. See figures \ref{fig:5} and \ref{fig:6} for comparison. The ratios of the cross sections due to variation are approximately given by 0.68 for $\eta_{\odd}\,=\,-1$ and 0.6 for $\eta_{\odd}\,=\,+1$.
\end{itemize}

\subsection{LEP II}\label{sec:lepii}

\begin{figure}
\centering
\includegraphics[angle=-90, width=0.95\textwidth]{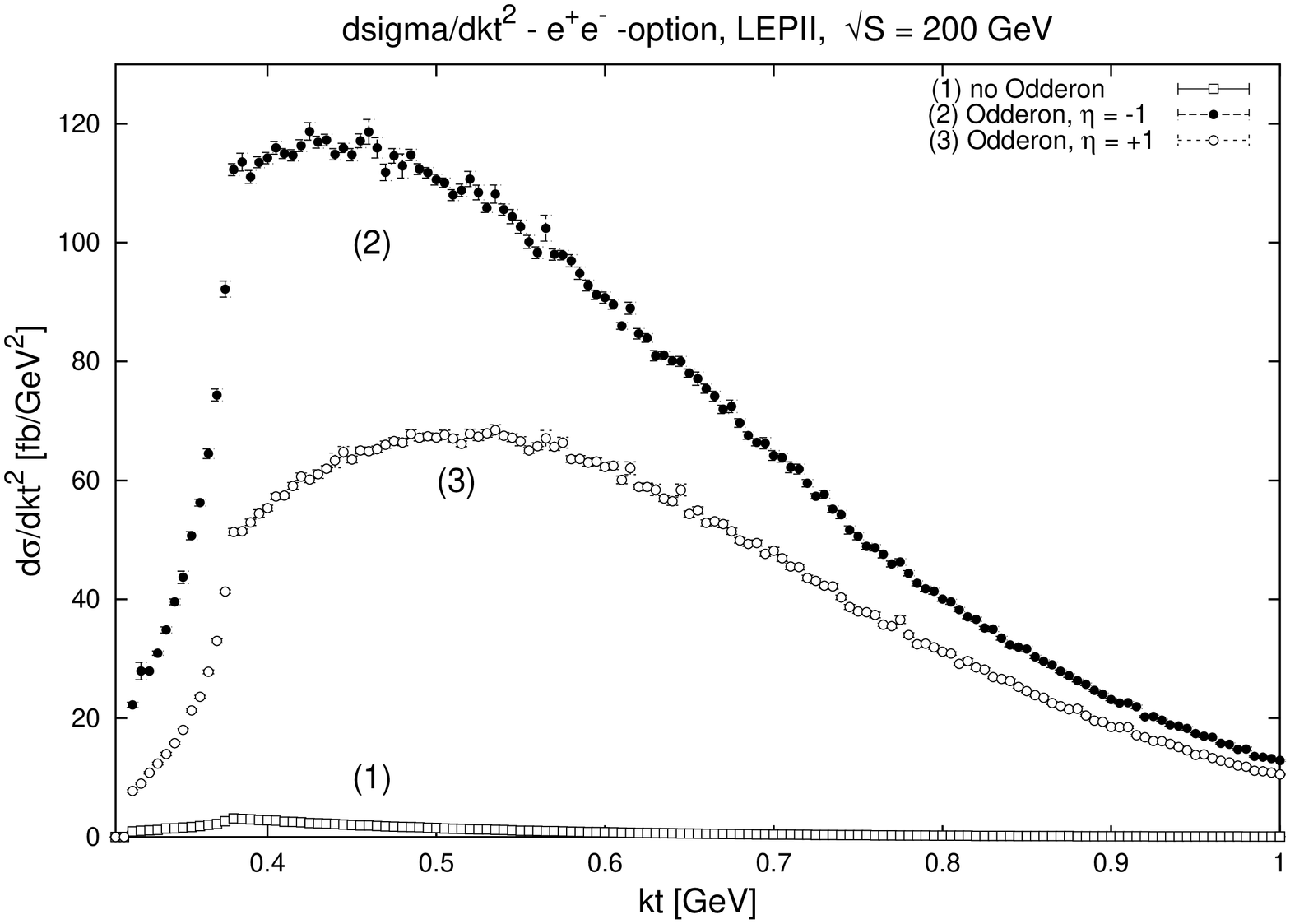}
\caption{different values for $\eta_{\odd}$ } 
\label{fig:28}                                      
\end{figure}

\begin{figure}
\centering
\includegraphics[angle=-90, width=0.95\textwidth]{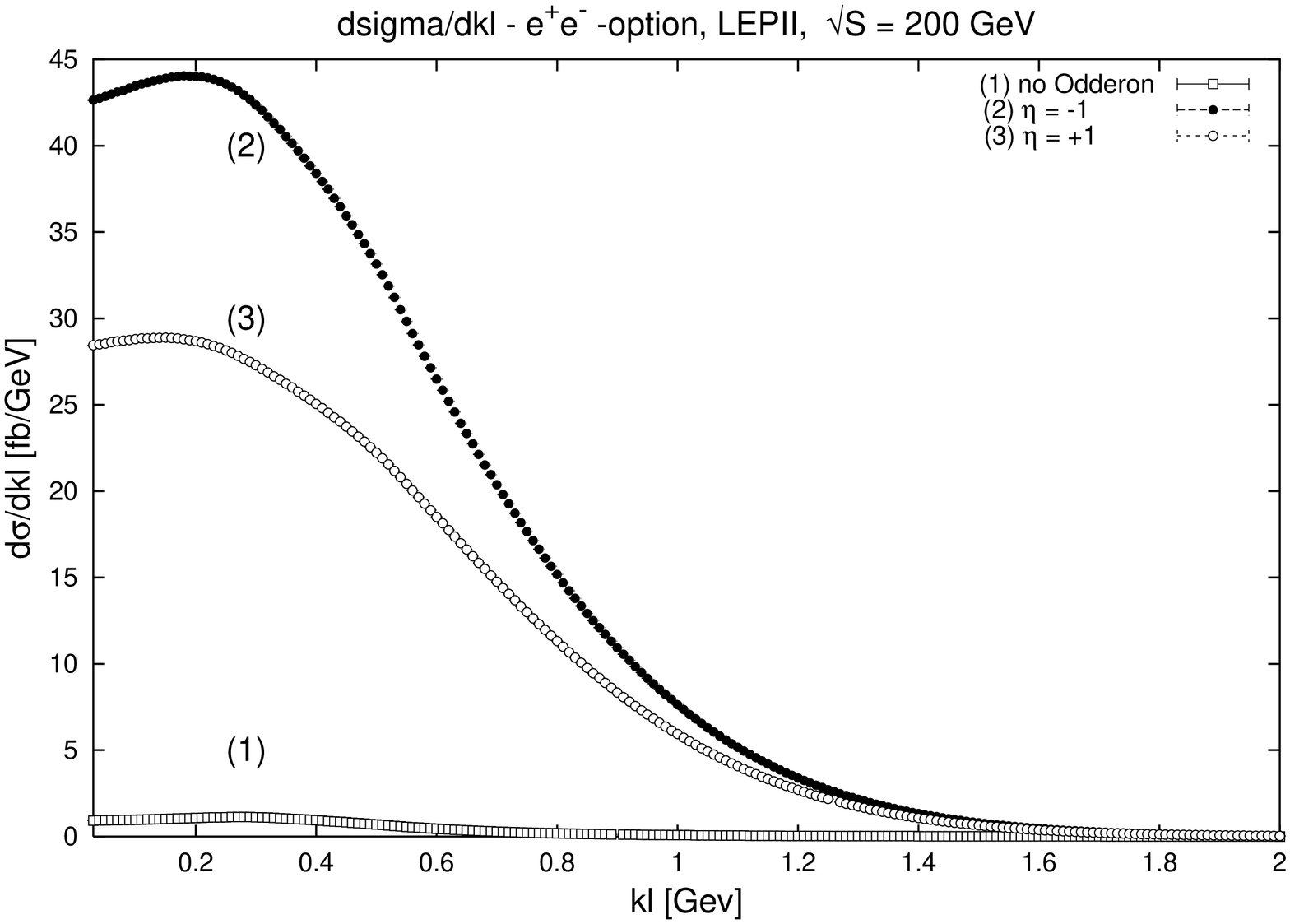}
\caption{different values for $\eta_{\odd}$ }
\label{fig:29}
\end{figure}

\begin{figure}
\centering
\includegraphics[angle=-90, width=0.95\textwidth]{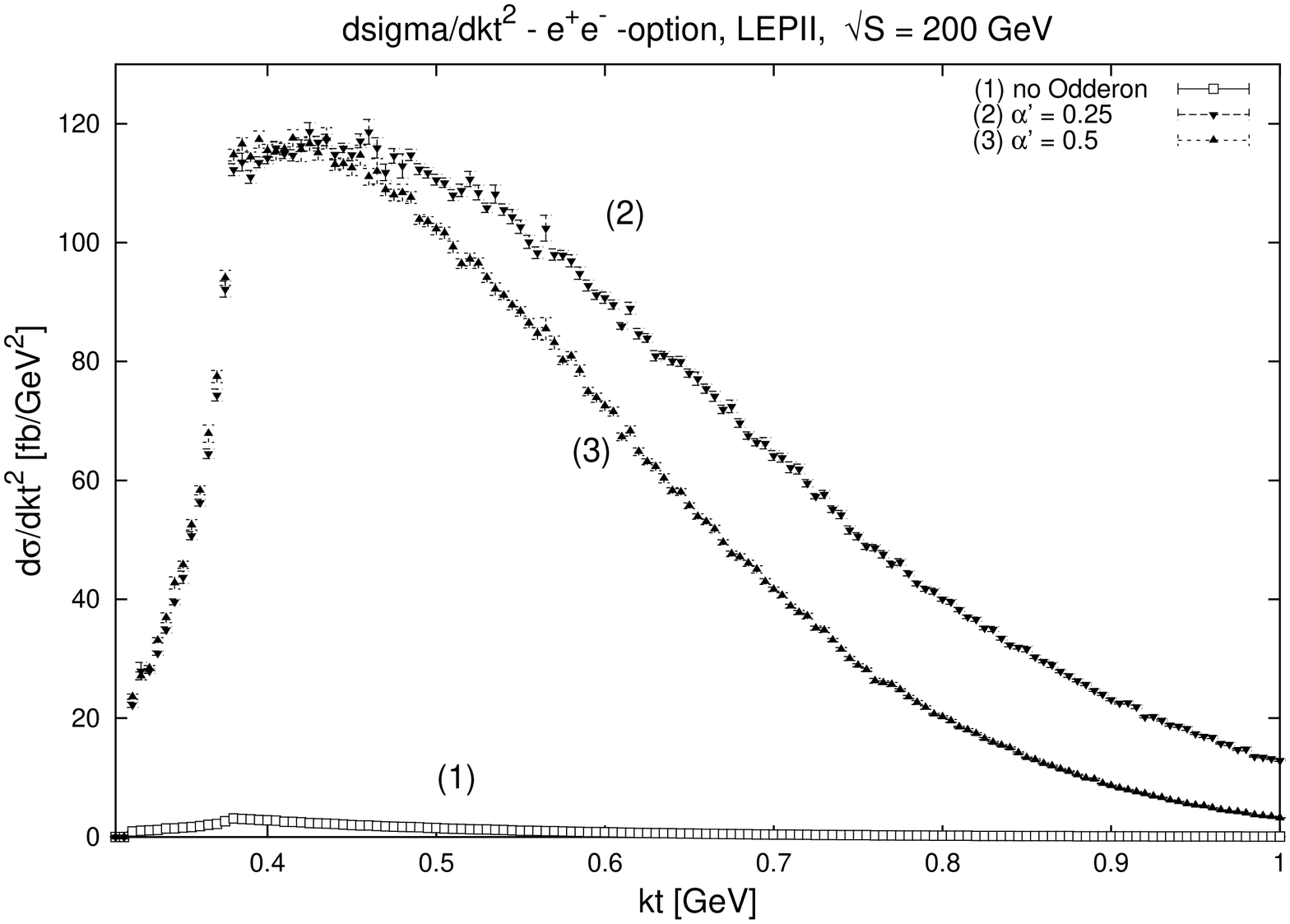}
\caption{comparison different values for $\al'$} 
\label{fig:32}                                      
\end{figure}

\begin{figure}
\centering
\includegraphics[angle=-90, width=0.95\textwidth]{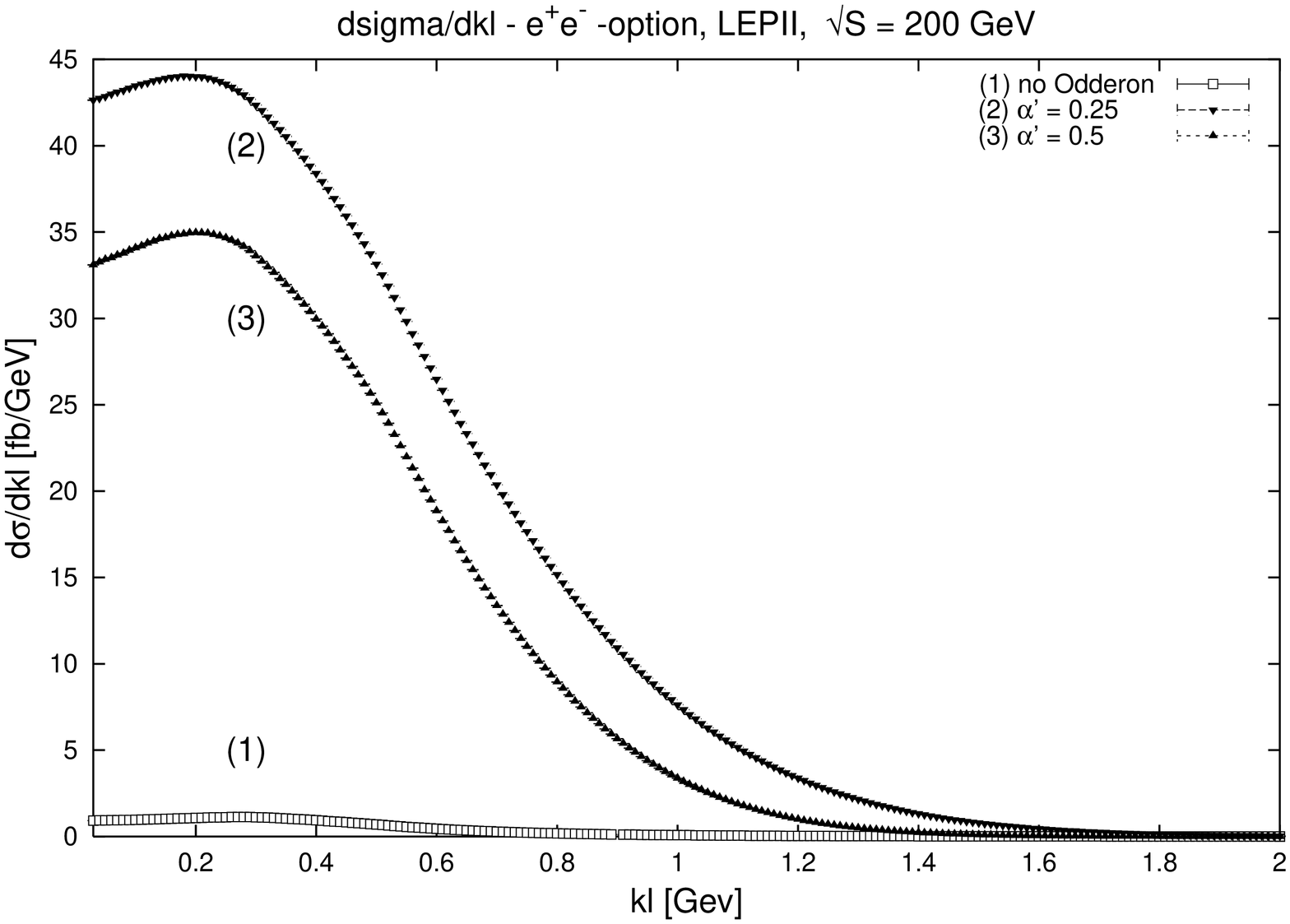}
\caption{comparison different values for $\al'$  }
\label{fig:33}
\end{figure}

\begin{figure}
\centering
\includegraphics[angle=-90, width=0.95\textwidth]{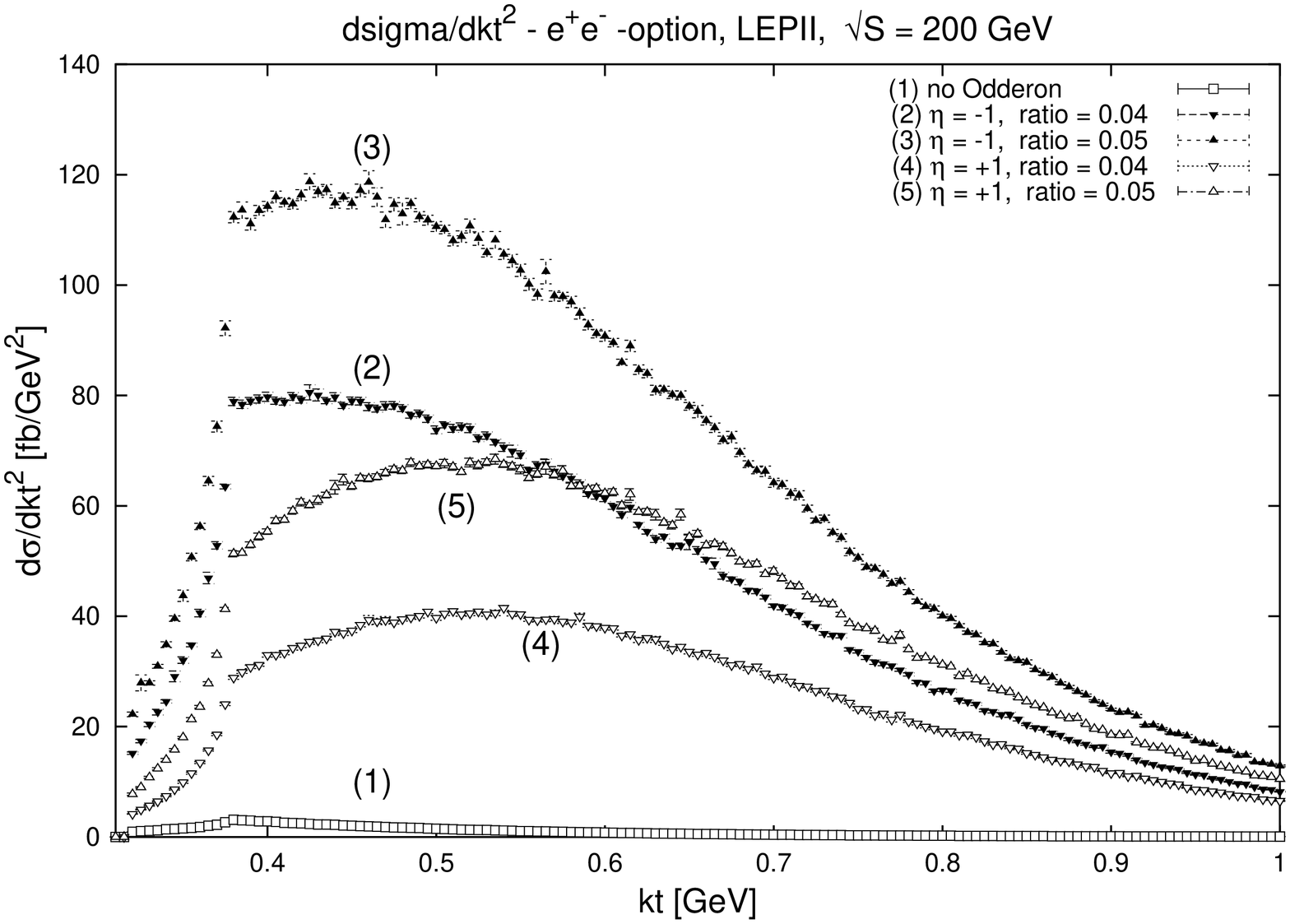}
\caption{comparison different values for $(\frac{\be_{\odd}}{\be_{\pom}})^{2}$ } 
\label{fig:33-1}                                      
\end{figure}

\begin{figure}
\centering
\includegraphics[angle=-90, width=0.95\textwidth]{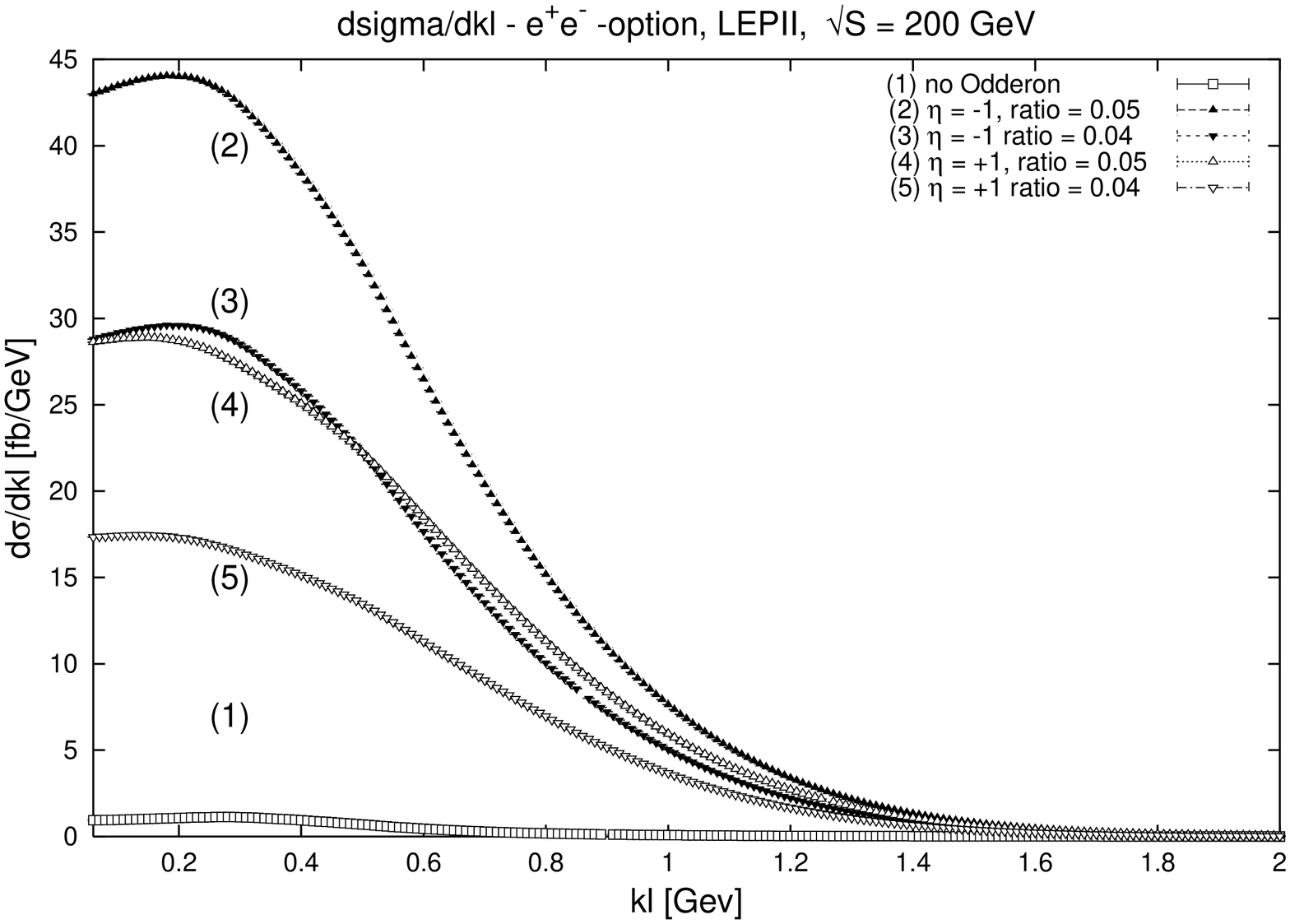}
\caption{comparison different values for $(\frac{\be_{\odd}}{\be_{\pom}})^{2}$}
\label{fig:33-2}
\end{figure}

For the OPAL detector at LEPII, we used the following values \cite{Lillich:2002}:

\begin{eqnarray*}
\sqrt{S_{lab}}\,=\,200\;\mbox{GeV};&|\cos\theta|_{max}\,=\,0.81;& E_{min}\,=\,0.4\;\mbox{GeV}\,.
\end{eqnarray*}
The results are displayed in figures \ref{fig:28} to \ref{fig:33-2}.

In section \ref{sec:effdetcut}, we saw that the magnitude of the differential cross section highly depends on the angular cuts; however, as we applied the same detector cuts for the LEPI and LEPII runs, we can now investigate the $\sqrt{S}$-dependence of the differential cross sections. Starting with the expressions given by (\ref{eq:klvalx}) to (\ref{eq:mandelsx}) for the kinematic variables, (\ref{eq:totmgodd}) for the matrix element, and (\ref{eq:exprdsf}) for the cross sections including the photon spectra, we obtain in a very rough estimate:
\begin{eqnarray}
s\,\propto\,S\,,&t\,\propto\,S\,,&u\,\propto\,S\,,\nonumber\\
|\M^{\gamma}|^{2}\,\propto\,S^{2}\,,&C_{i}\,\propto\,S^{2(\al_{\odd}(t)-1)}\,,&N(x)\,\propto\,\ln^{2}(S)\;.
\end{eqnarray}
\begin{itemize}
\item{for $\frac{d\sigma}{d|k_{t}|^{2}}$}\\
\begin{equation}
k_{l\,\pm}\,\propto\,\sqrt{S}\;,\; k^{0}_{1}\,\propto\,\sqrt{S}\;,\;
f'(k_{l})\,\propto\,\sqrt{S}\,.
\end{equation}
\item{for $\frac{d\sigma}{dk_{l}}$}\\
\begin{equation}
k_{t_{0}}\,\propto\,\sqrt{S}\;,\; k^{0}_{1}\,\propto\,\sqrt{S}\;,
f'(k_{t_{0}})\,\propto\,\sqrt{S}\,.
\end{equation}
\end{itemize}
For $\frac{d\sigma}{d|k_{t}|^{2}}$, this leads in our approximation above to
\begin{equation}
\frac{d\sigma(S)}{d\sigma(S')}\,\approx\,\left(\frac{\ln(S)}{\ln(S')}\right)^{2}\,\frac{S'^{\,k_{t}^{2}}}{S^{\,k_{t}^{2}}}
\end{equation}
with $\al_{\odd}(0)\,=\,1$ and $\al'_{\odd}\,=\,0.25\;\mbox{GeV}^{-2}$.
Comparing now the ratio of the differential cross sections taken at their maxima for $\eta_{\odd}\,=\,\pm 1$ at $k_{t}\,=\,0.5\;\mbox{GeV}$ and $k_{t}\,=\,0.4\;\mbox{GeV}$ respectively, we obtain

\begin{eqnarray}
\frac{d\sigma_{LEPI}}{d\sigma_{LEP II}}&\approx&0.83\;\;\mbox{for $\eta_{\odd}\,=\,-1$}\,,\nonumber\\
\frac{d\sigma_{LEPII}}{d\sigma_{LEPII}}&\approx&0.88\;\;\mbox{for $\eta_{\odd}\,=\,+1$}\,.
\end{eqnarray}
The actual values are given by 0.84 for $\eta_{\odd}\,=\,-1$ and 0.91 for $\eta_{\odd}\,=\,+1$; they are quite in agreement with the expectations.
For $\frac{d\sigma}{dk_{l}}$, we obtain a ratio of 0.79 and 0.82 for $\eta_{\odd}\,=\,\mp1$ respectively.
The behavior following from Odderon parameter variation is the same as for LEP I; we therefore refer to the discussion in the last section.

\subsection{TESLA - 500 GeV, $e^{+}e^{-}\,$-version}

\begin{figure}
\centering
\includegraphics[angle=-90, width=0.95\textwidth]{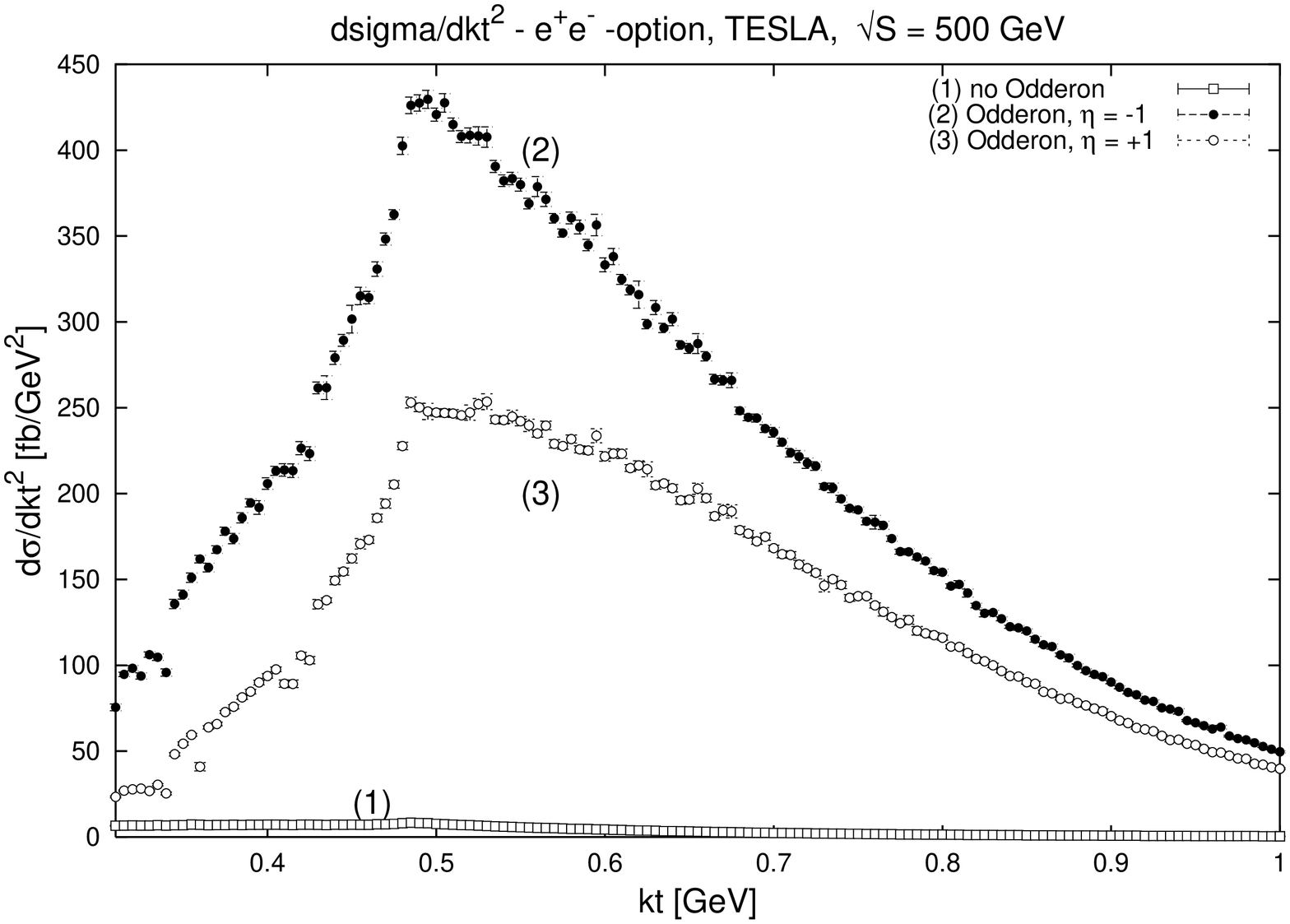}
\caption{comparison different values for $\eta_{\odd}$ } 
\label{fig:34}                                      
\end{figure}

\begin{figure}
\centering
\includegraphics[angle=-90, width=0.95\textwidth]{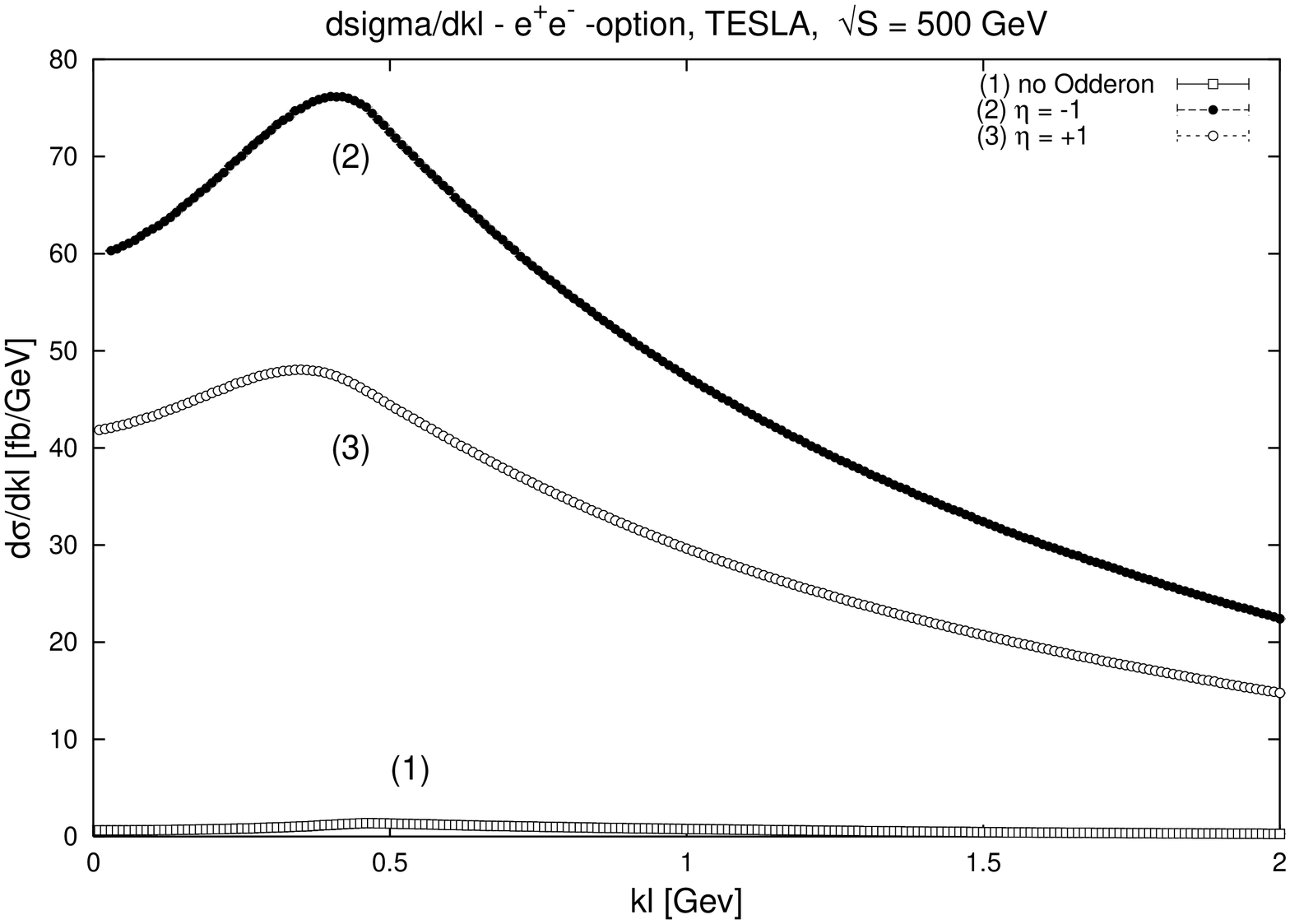}
\caption{comparison different values for $\eta_{\odd}$ }
\label{fig:35}
\end{figure}

\begin{figure}
\centering
\includegraphics[angle=-90, width=0.95\textwidth]{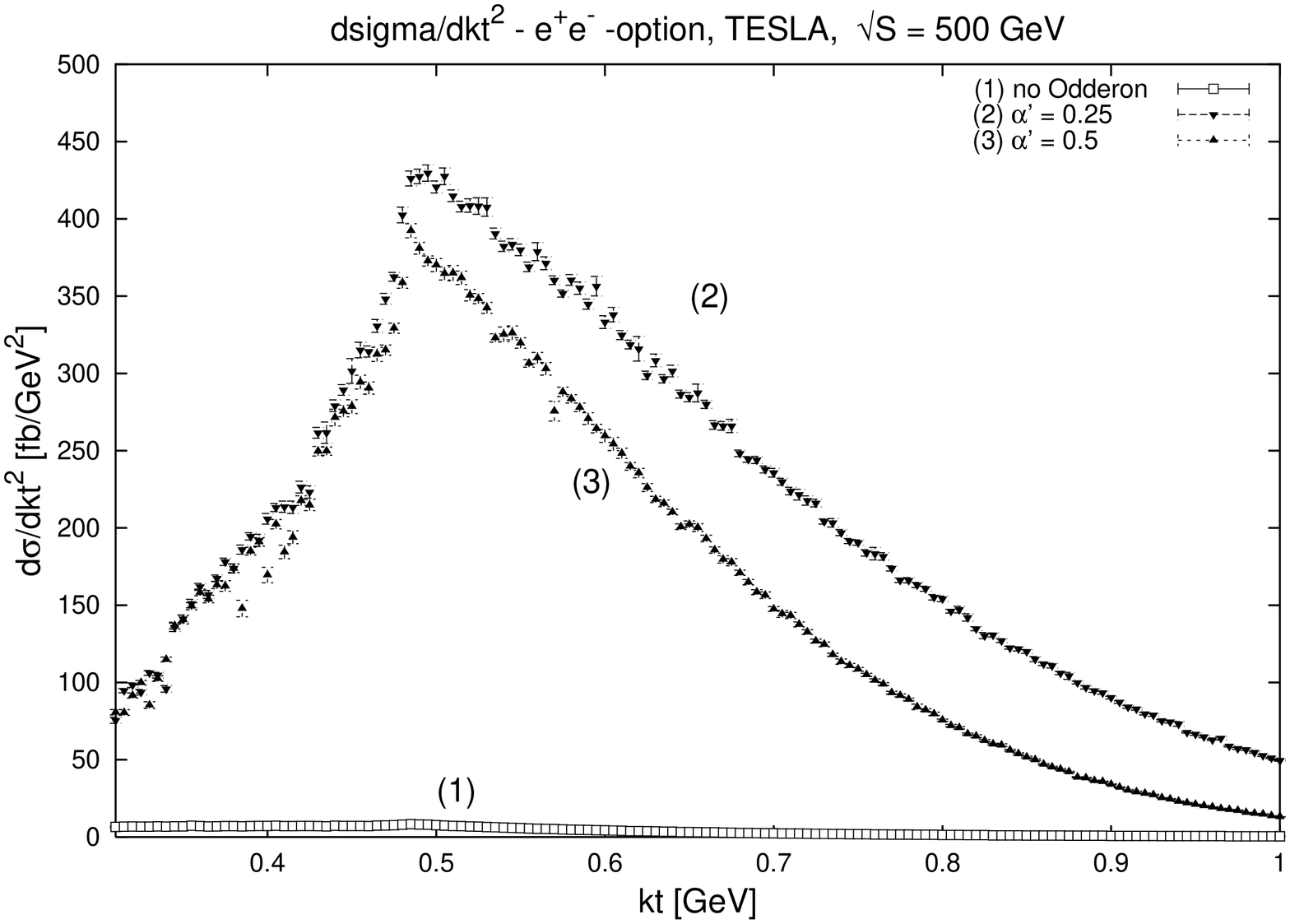}
\caption{comparison different values for $\al'$ } 
\label{fig:38}                                      
\end{figure}

\begin{figure}
\centering
\includegraphics[angle=-90, width=0.95\textwidth]{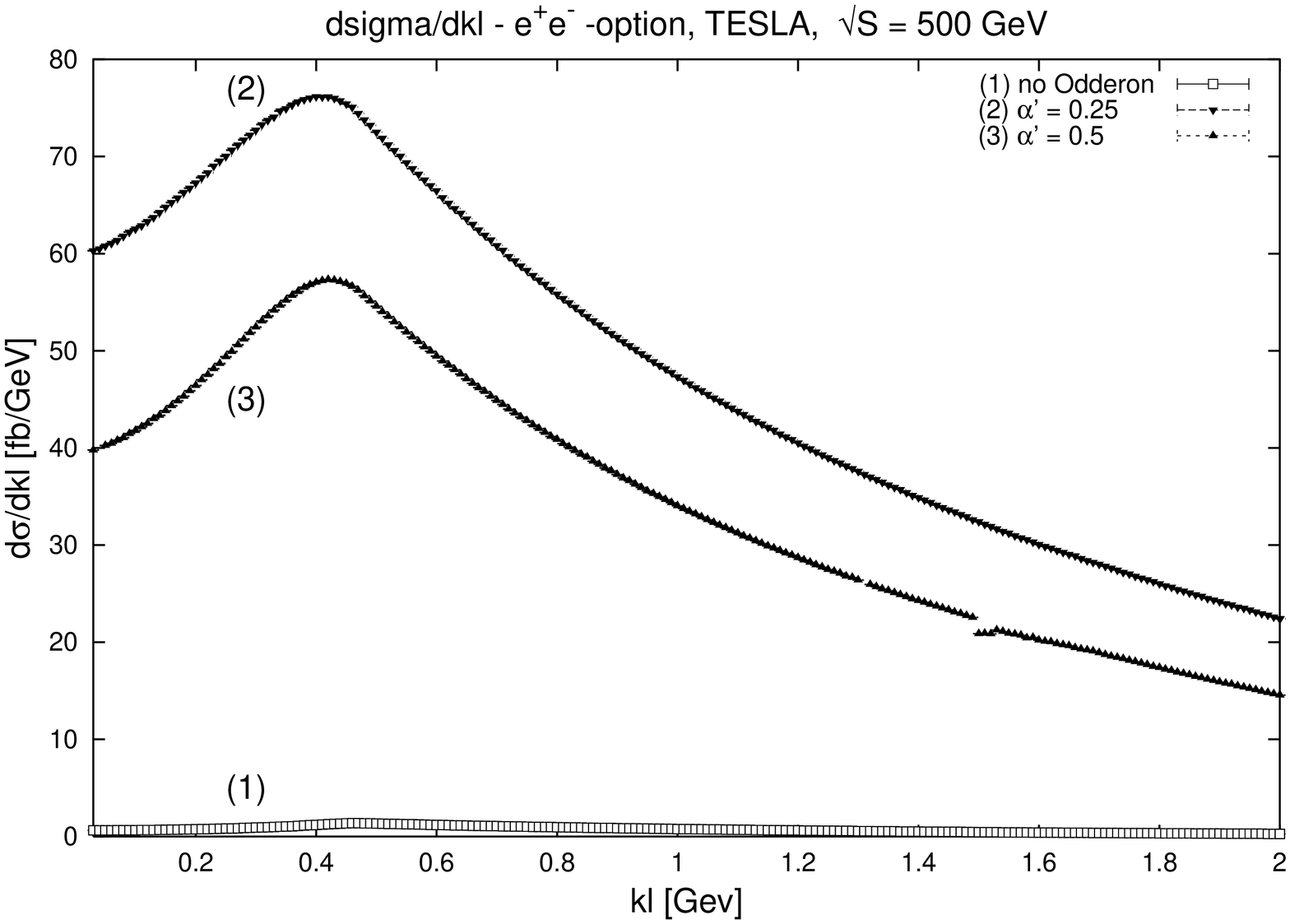}
\caption{comparison different values for $\al'$  }
\label{fig:39}
\end{figure}

\begin{figure}
\centering
\includegraphics[angle=-90, width=0.95\textwidth]{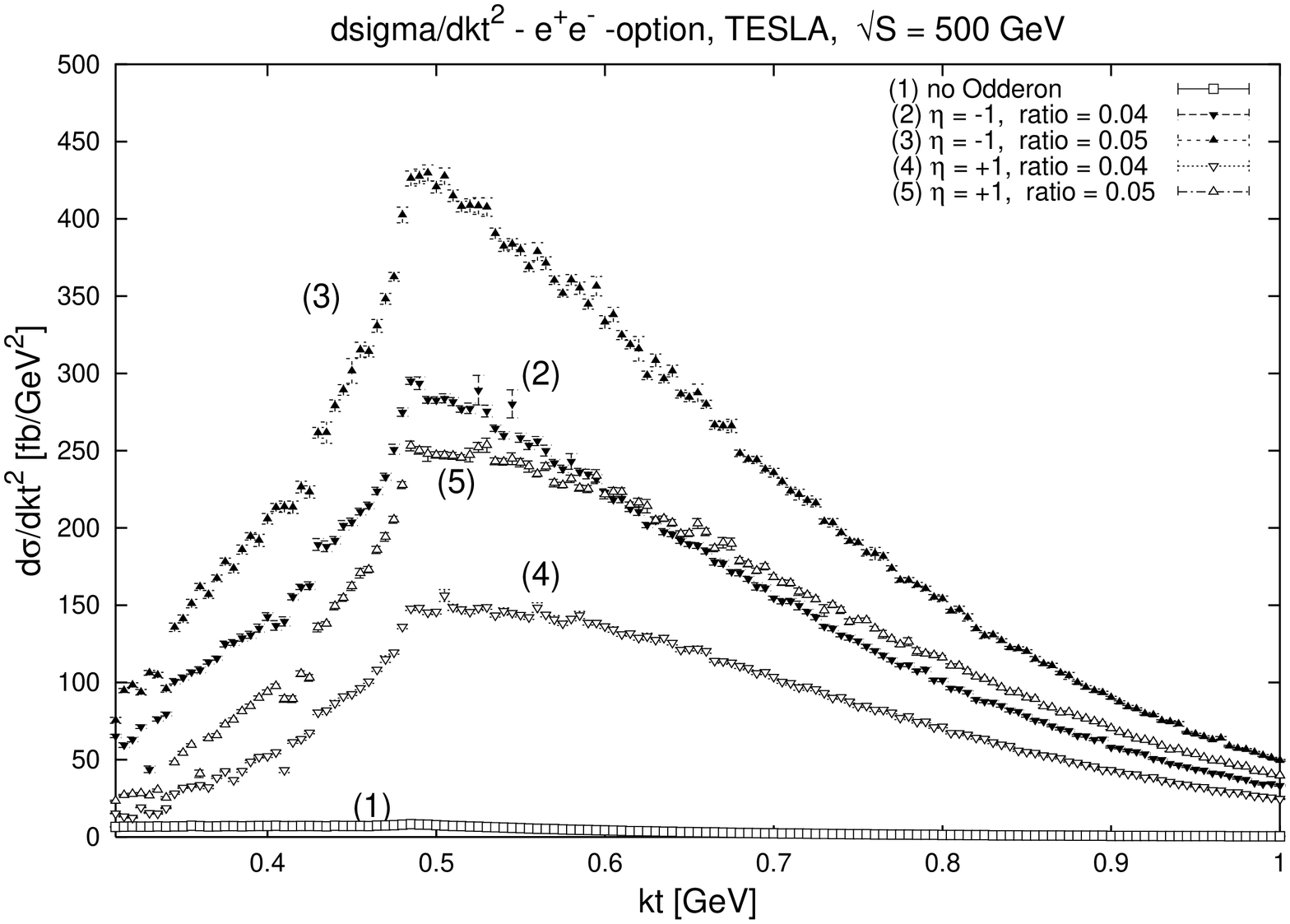}
\caption{comparison different values for $(\frac{\be_{\odd}}{\be_{\pom}})^{2}$ } 
\label{fig:39-1}                                      
\end{figure}

\begin{figure}
\centering
\includegraphics[angle=-90, width=0.95\textwidth]{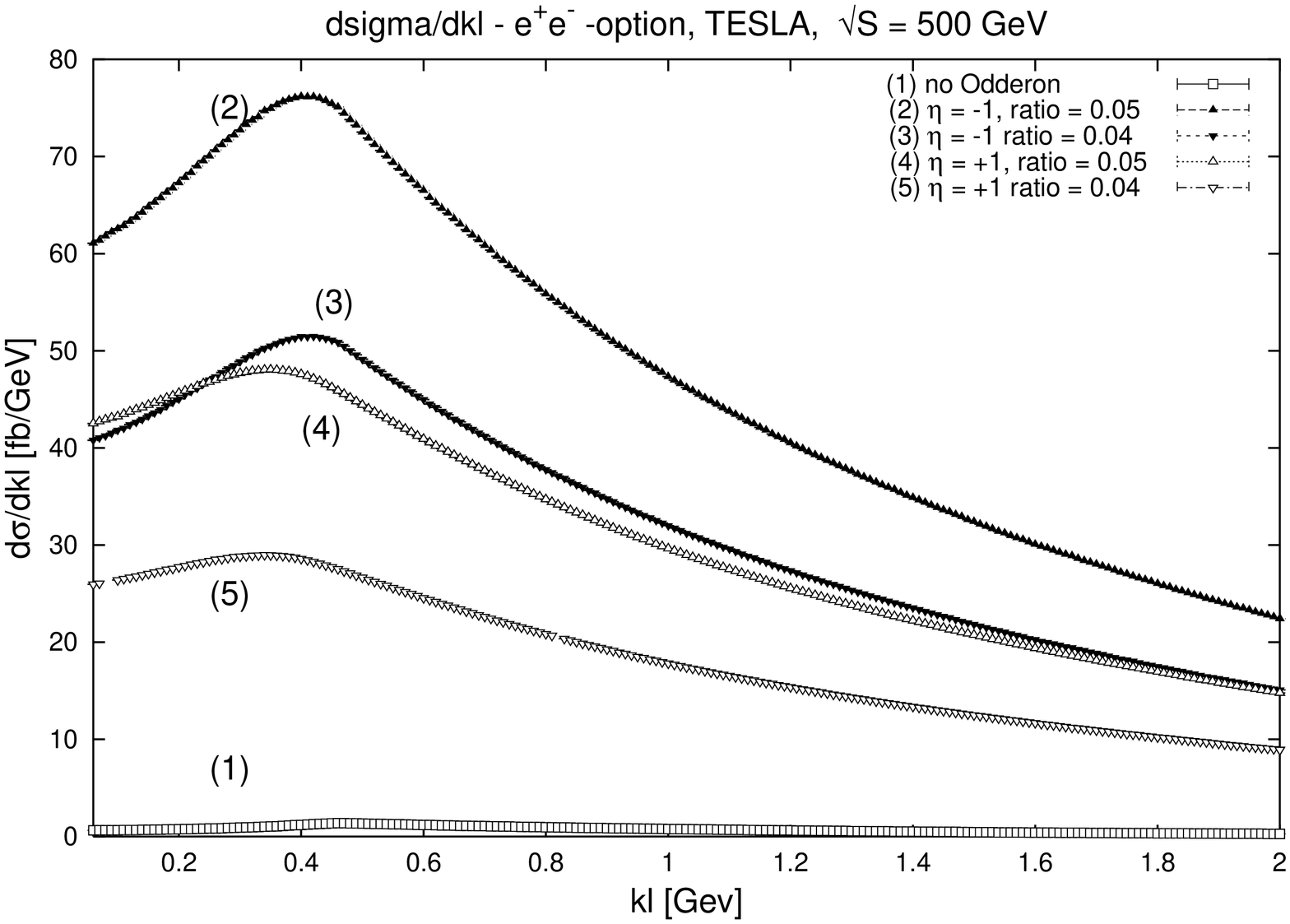}
\caption{comparison different values for $(\frac{\be_{\odd}}{\be_{\pom}})^{2}$}
\label{fig:39-2}
\end{figure}

For TESLA in the 500 GeV- version, we used \cite{Desch:2002}:

\begin{eqnarray*}
\sqrt{S_{lab}}\,=\,500\;\mbox{GeV};&|\cos\theta|_{max}\,=\,0.98;& E_{min}\,=\,0.5\;\mbox{GeV}\;.
\end{eqnarray*}
The results are displayed in figures \ref{fig:34} to \ref{fig:39-2}.

We notice that the results show again the same behavior as for LEP I and LEP II; however, due to  higher energy cuts, it becomes more difficult to distinguish the shapes of the differential cross sections for different $\eta_{\odd}\,$. Compare e.g. figures \ref{fig:28} and \ref{fig:34}; see also figure \ref{fig:1} for results without any detector cuts. So far, we can conclude that for collider experiments such as LEP and TESLA, while the presence of a nonperturbative Odderon described by (\ref{eq:Feynodd}) can be clearly distinguished from a scenario without an Odderon, statements about the phase and the intercept highly depend on the given angular and energy cuts of the detector. With the cuts for the single experiments given by section \ref{sec:detcut}, we therefore expect the best results for BaBar due to relatively small energy cuts.
The higher values for the cross section in the TESLA environment cannot be explained with a higher cm-energy as was the case when comparing results from LEP I and LEP II; the differences are mainly due to differences in the angular cuts. For TESLA, $|\cos \theta|_{max}\,=\,0.98$, while for OPAL $|\cos \theta|_{max}\,=\,0.81$, leading to a much larger reduction of the experimentally accessible kinematic region.

\subsection{TESLA - 500 GeV, photon collider}

For TESLA in the 500 GeV- version, we used \cite{Telnov:2002}:

\begin{eqnarray*}
\sqrt{S_{lab}}\,=\,500\;\mbox{GeV};&|\cos\theta|_{max}\,=\,0.98;& E_{min}\,=\,0.5\;\mbox{GeV}\,.
\end{eqnarray*}
The photon spectrum now differ from the spectrum in the $e^{+}e^{-}$ mode as described in chapter \ref{sec:lincol}; mainly, there is an extra peak at high energies in the photon spectra due to Compton scattering (see section \ref{sec:compspectra}). For the numerical calculation, we used the spectra from simulations by V. Telnov \cite{Telnov:2001}.

\begin{figure}
\centering
\includegraphics[angle=-90, width=0.95\textwidth]{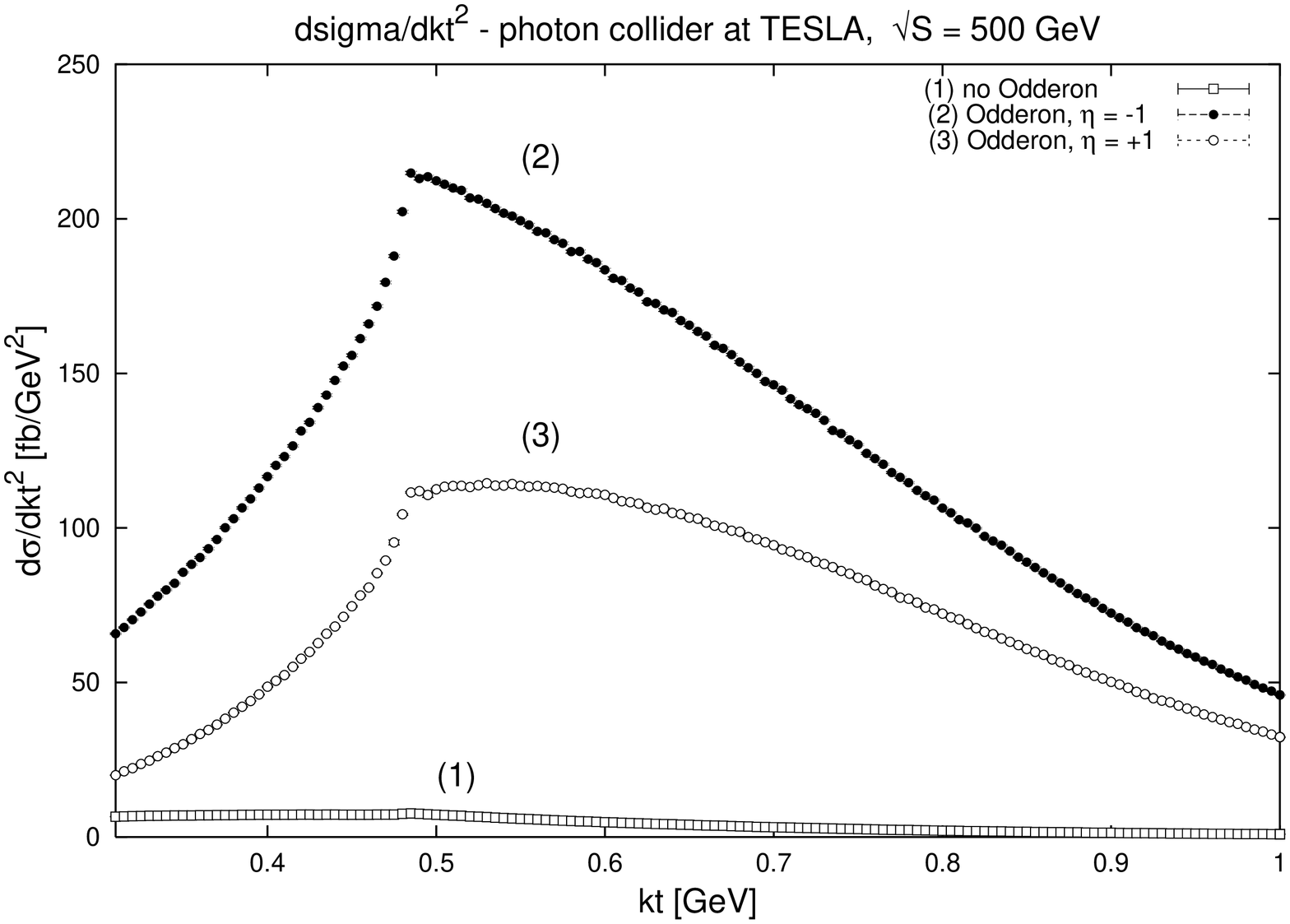}
\caption{comparison different values for $\eta_{\odd}$ } 
\label{fig:40}                                      
\end{figure}

\begin{figure}
\centering
\includegraphics[angle=-90, width=0.95\textwidth]{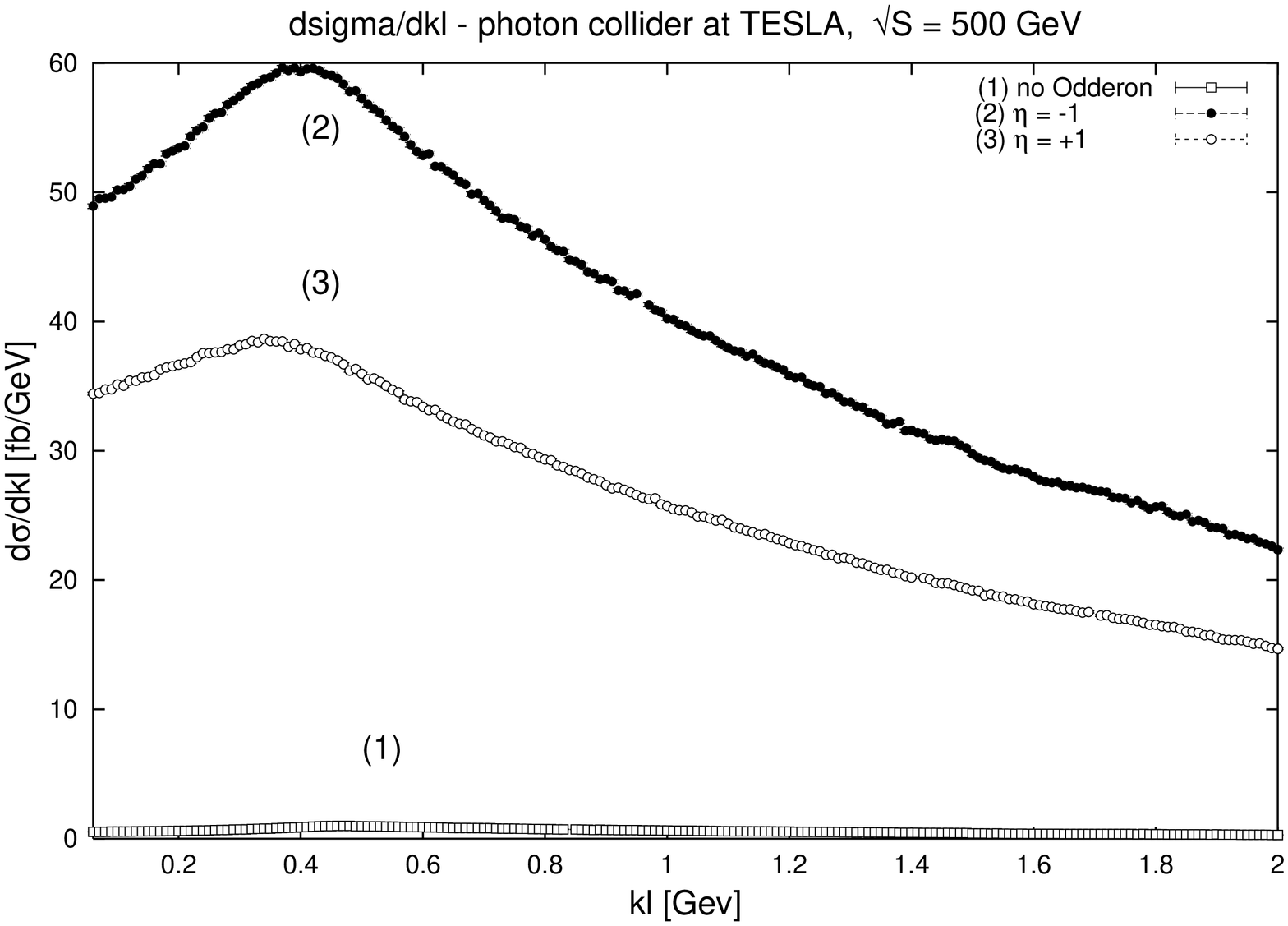}
\caption{comparison different values for $\eta_{\odd}$  }
\label{fig:41}
\end{figure}

\begin{figure}
\centering
\includegraphics[angle=-90, width=0.95\textwidth]{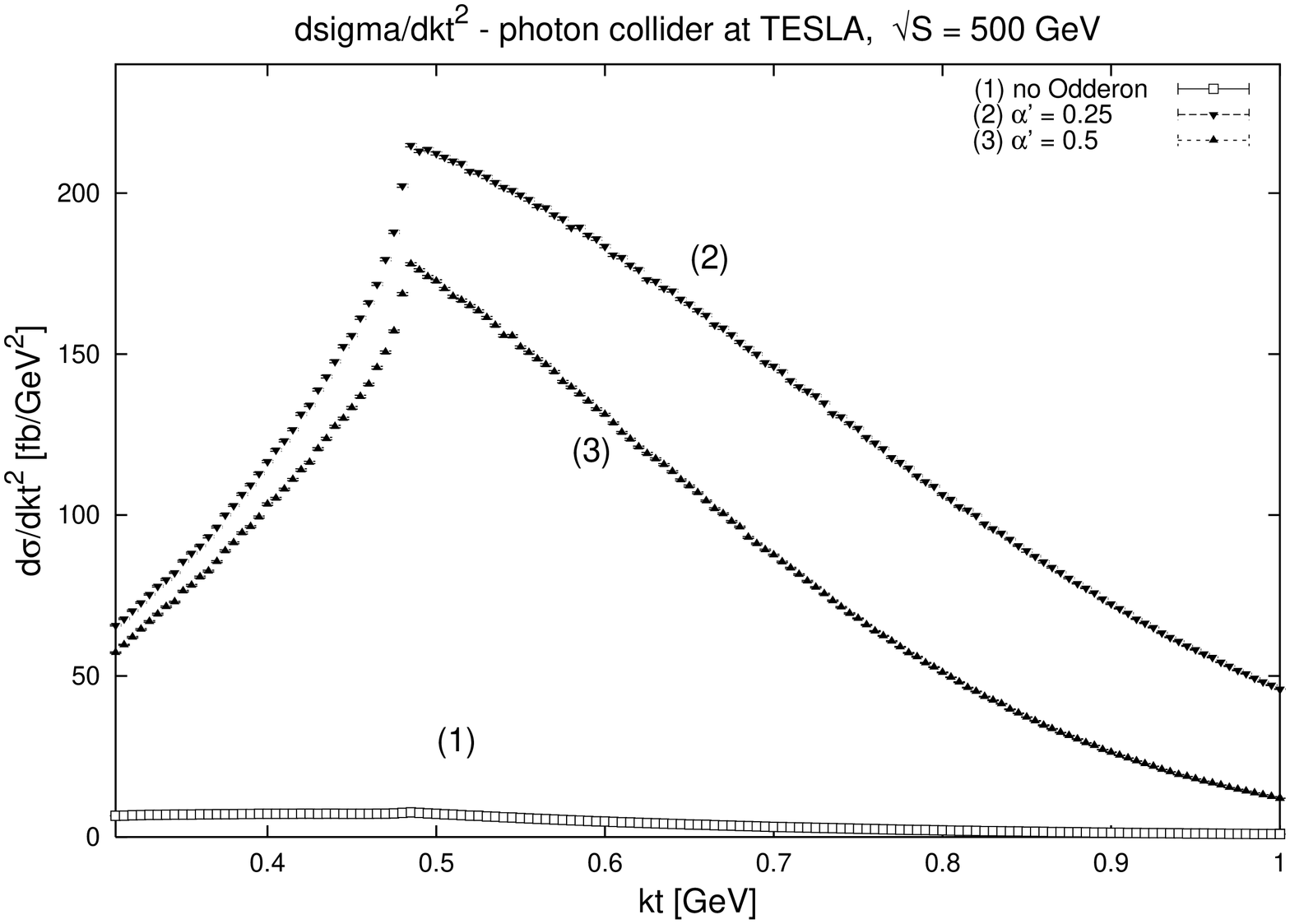}
\caption{comparison different values for $\al'$, $\eta_{\odd}=-1$ } 
\label{fig:44}                                      
\end{figure}

\begin{figure}
\centering
\includegraphics[angle=-90, width=0.95\textwidth]{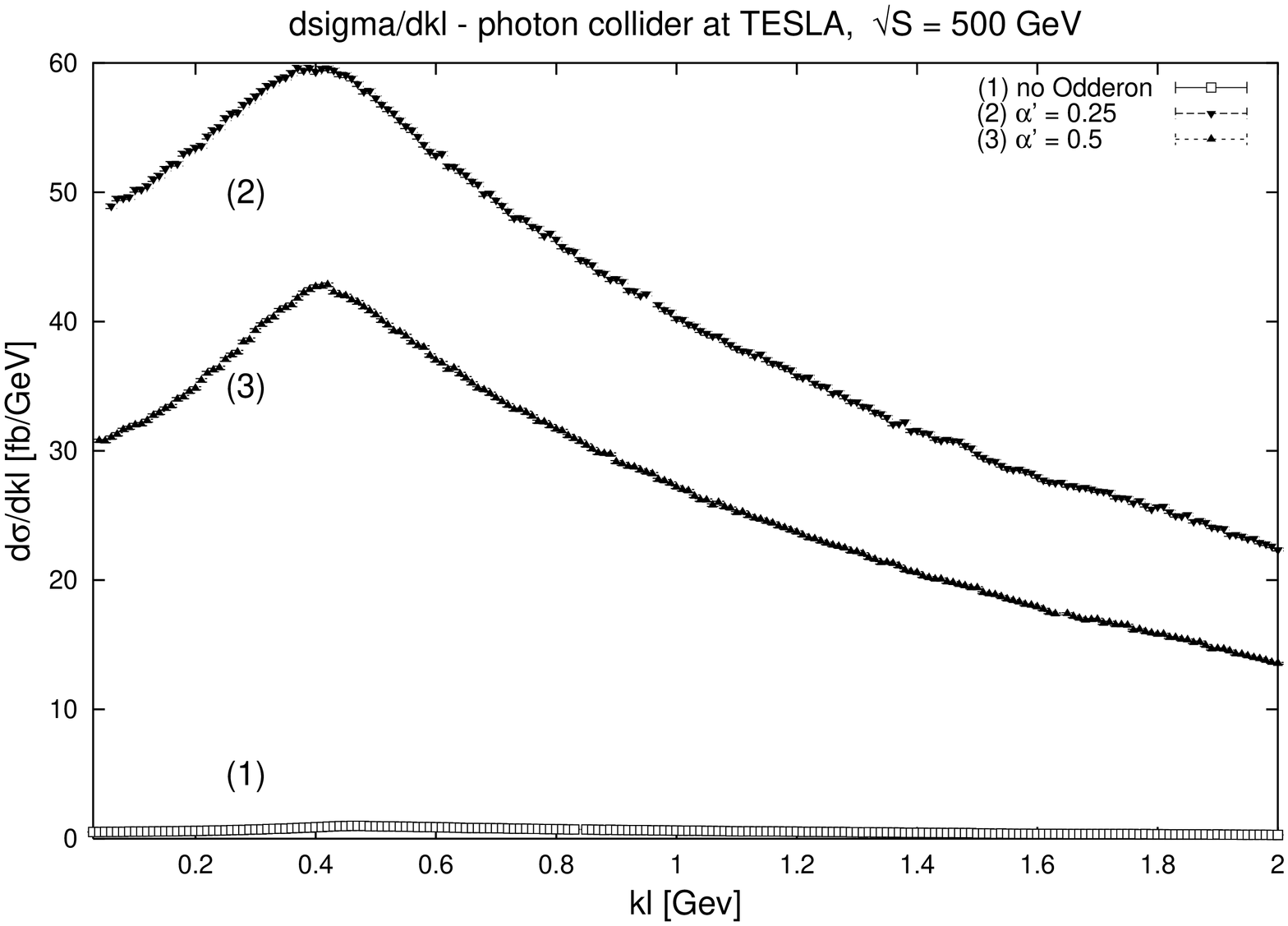}
\caption{comparison different values for $\al'$, $\eta_{\odd}=-1$}
\label{fig:45}
\end{figure}

\begin{figure}
\centering
\includegraphics[angle=-90, width=0.95\textwidth]{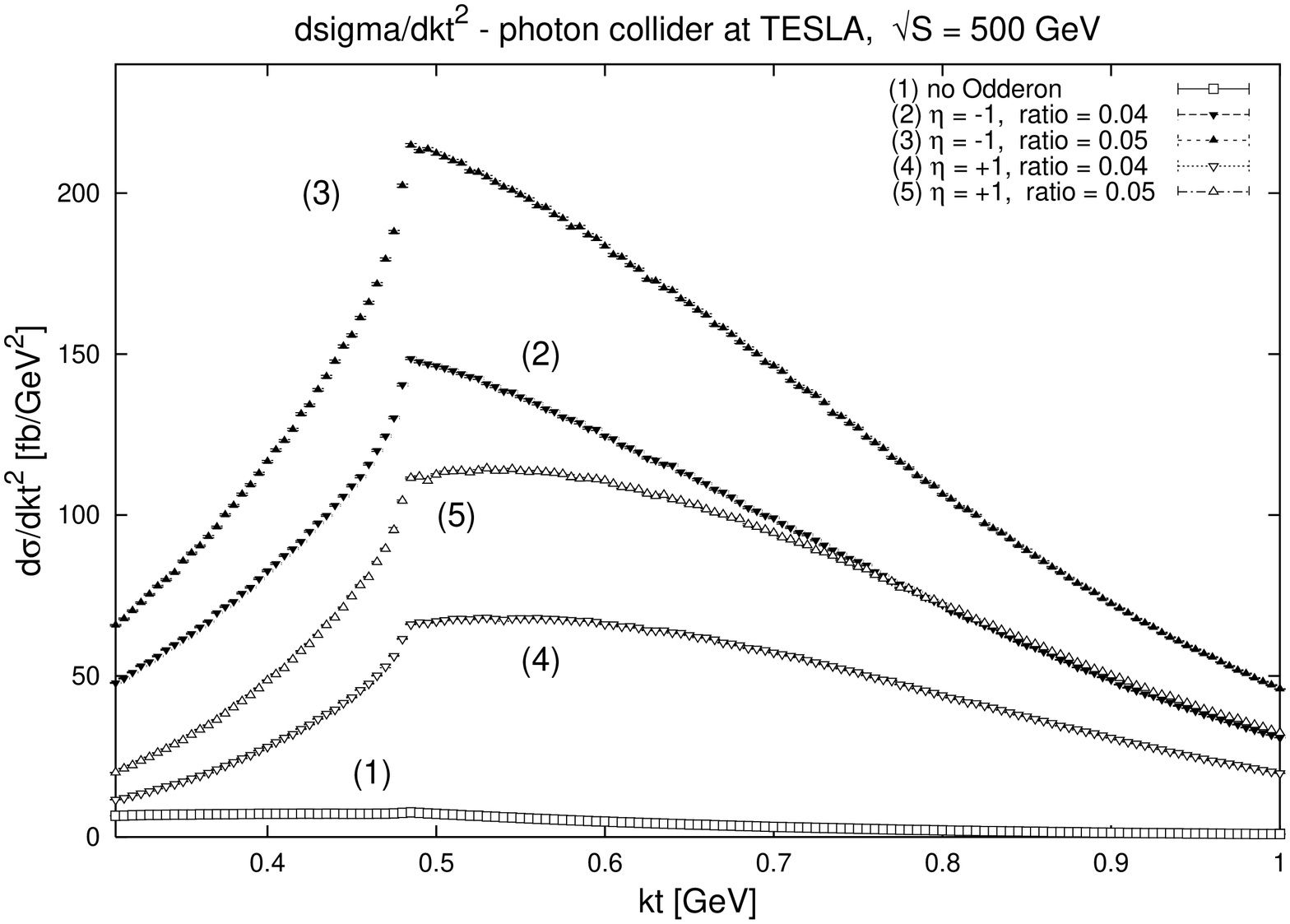}
\caption{comparison different values for $(\frac{\be_{\odd}}{\be_{\pom}})^{2}$ } 
\label{fig:45-1}                                      
\end{figure}

\begin{figure}
\centering
\includegraphics[angle=-90, width=0.95\textwidth]{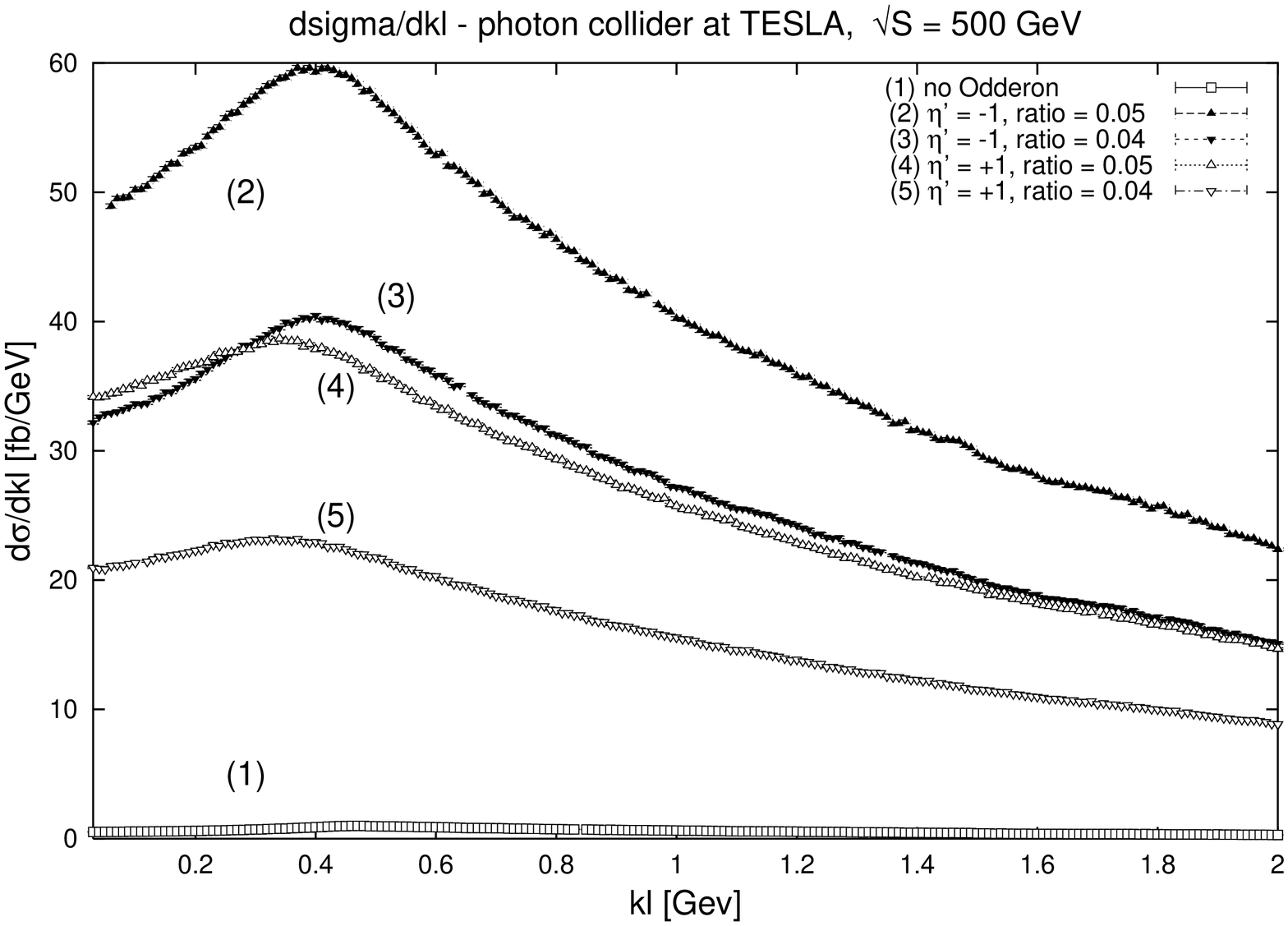}
\caption{comparison different values for $(\frac{\be_{\odd}}{\be_{\pom}})^{2}$}
\label{fig:45-2}
\end{figure}

The results are displayed in figures \ref{fig:40} to \ref{fig:45-2}; comparing them to the results obtained for the $e^{+}e^{-}$ option, the similarity of the differential cross sections suggests that the main contribution in the photon collider option comes from the low energy peak of the spectrum corresponding to beamstrahlung. Therefore, the photon collider option does not provide a significant advantage compared with the direct $e^{+}e^{-}$ option. The lower values for the cross sections are due to the inaccuracy of the spectra from the files used for the numerical integration; see section \ref{sec:compspectra}.

\subsection{BaBar}

For BaBar, we used the following parameters \cite{Aubert:2001tu}, \cite{Schieck:2002}:

\begin{eqnarray*}
\sqrt{S_{lab}}\,=\,10.58\;\mbox{GeV};&|\cos\theta|_{max}\,=\,0.96/0.77;& E_{min}\,=\,0.04\;\mbox{GeV}\,.
\end{eqnarray*}

\begin{figure}
\centering
\includegraphics[angle=-90, width=0.95\textwidth]{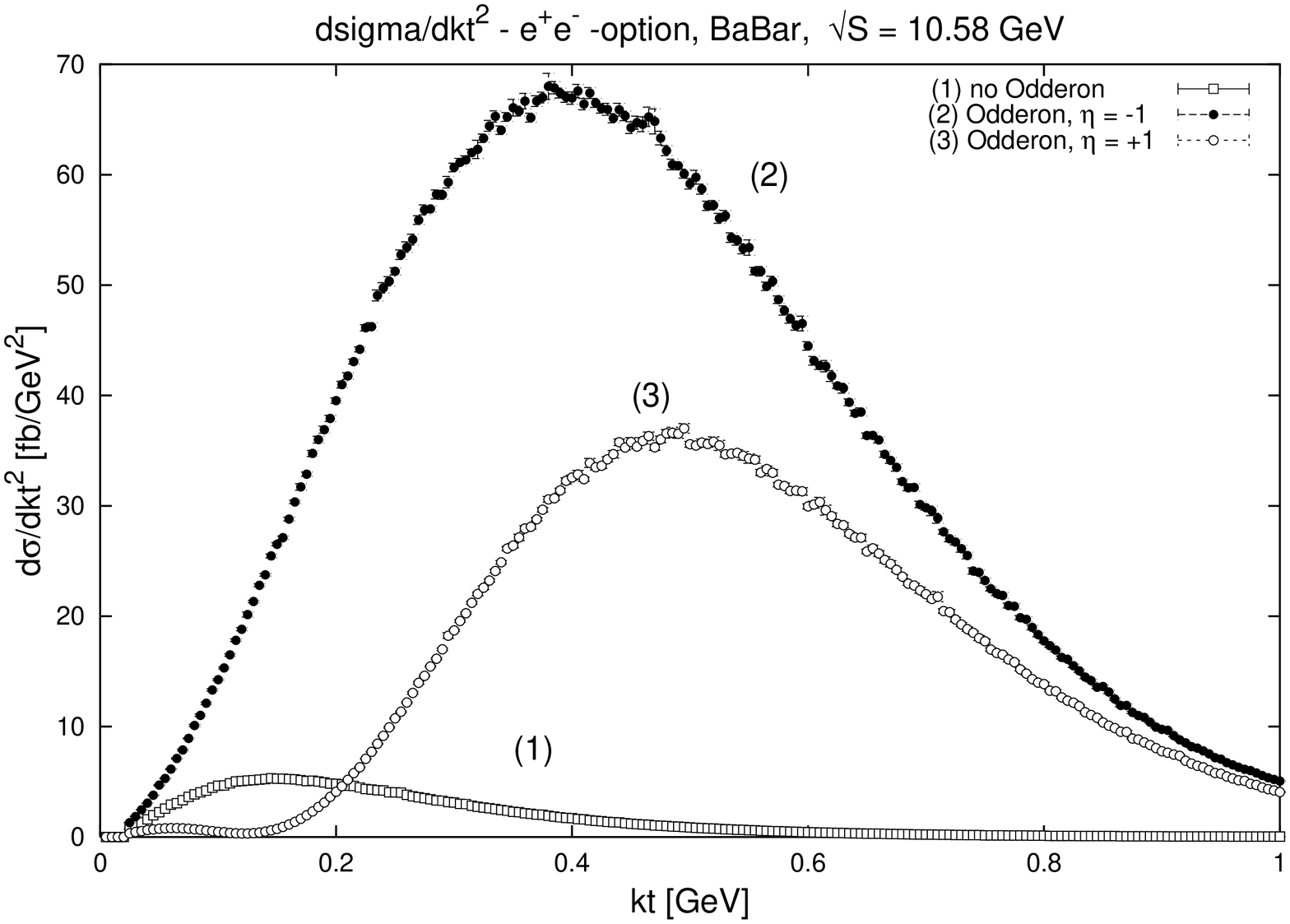}
\caption{comparison different values for $\eta_{\odd}$ } 
\label{fig:46}                                      
\end{figure}

\begin{figure}
\centering
\includegraphics[angle=-90, width=0.95\textwidth]{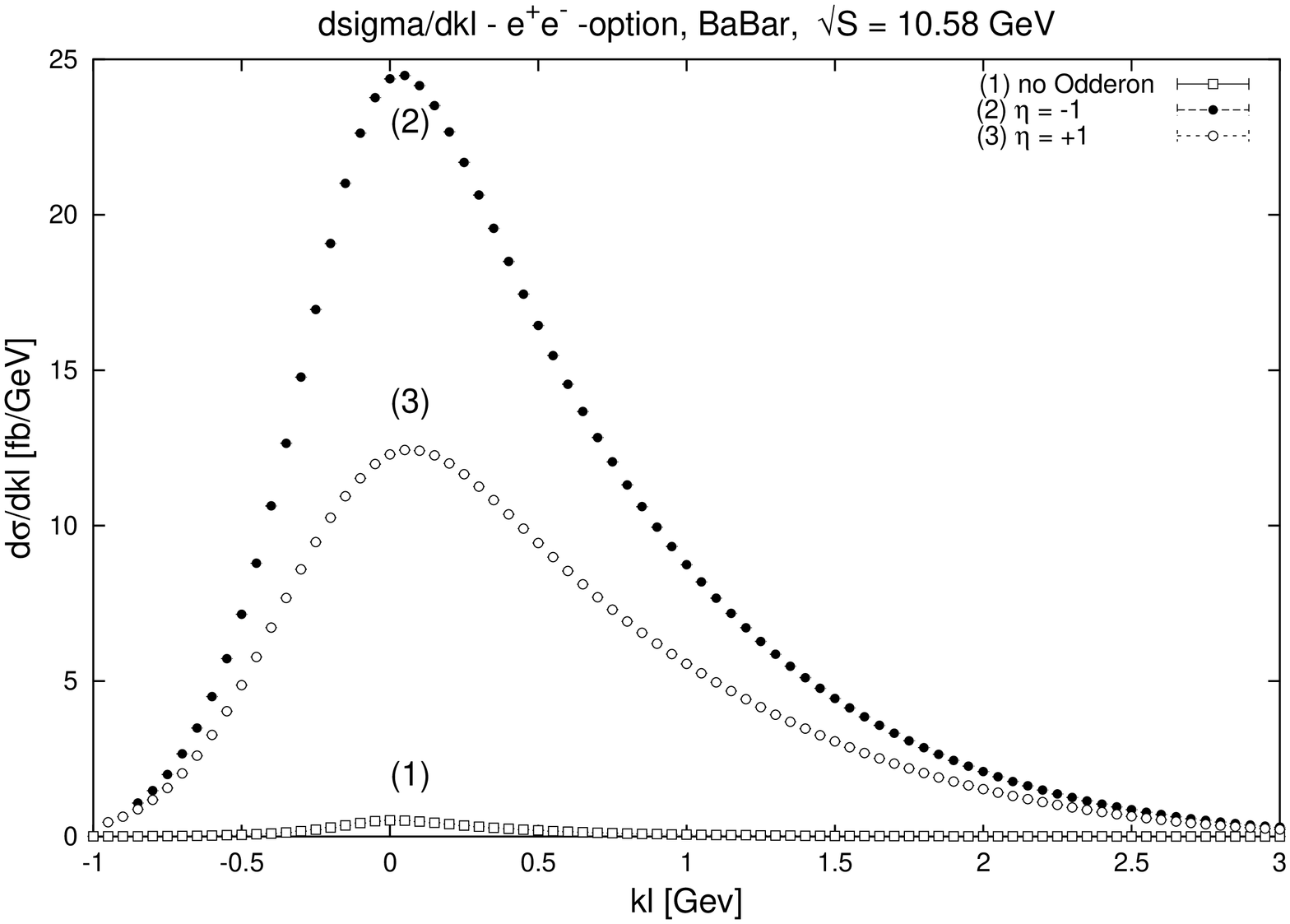}
\caption{comparison different values for $\eta_{\odd}$}
\label{fig:47}
\end{figure}

\begin{figure}
\centering
\includegraphics[angle=-90, width=0.95\textwidth]{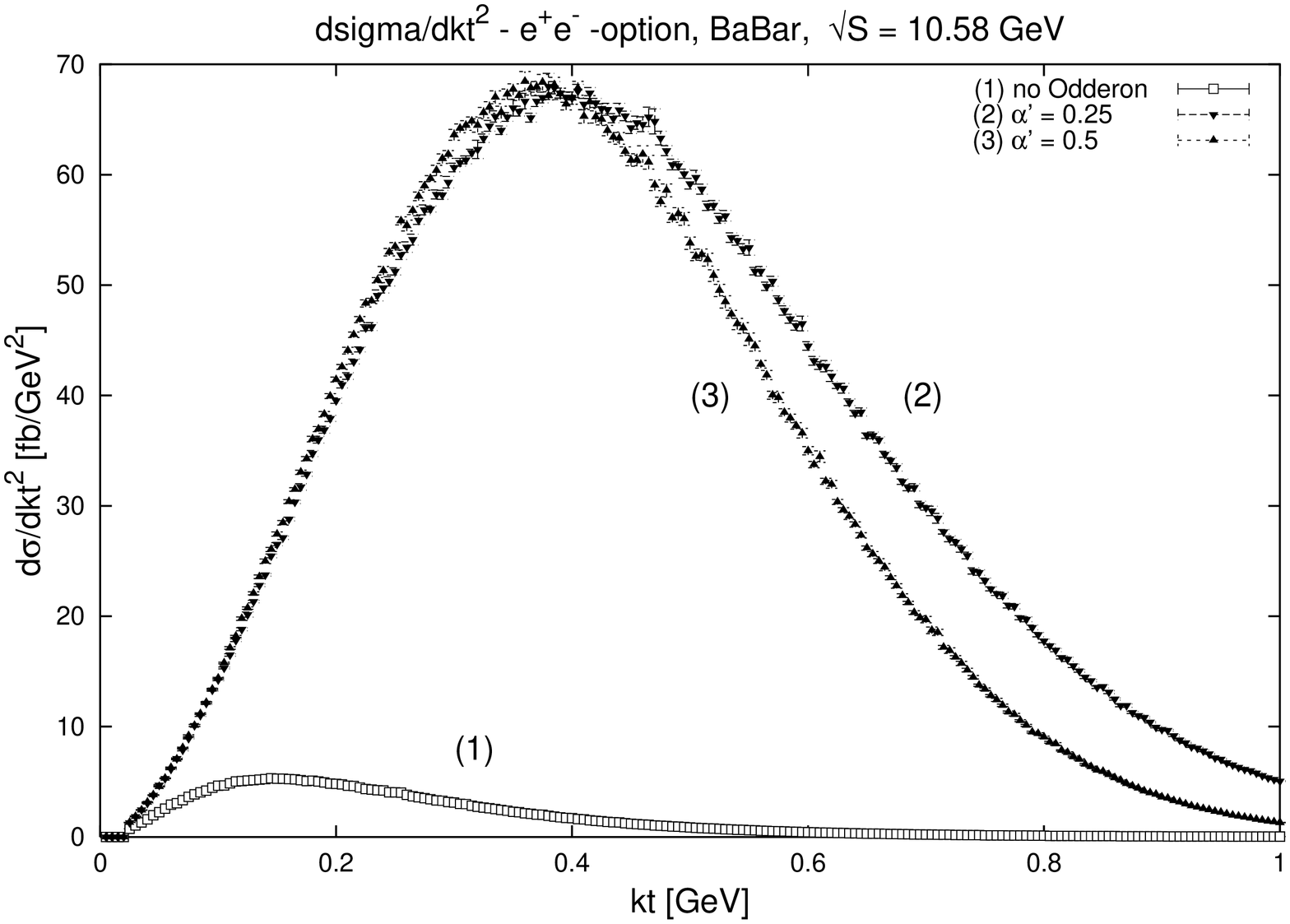}
\caption{comparison different values for $\al'$ } 
\label{fig:50}                                      
\end{figure}

\begin{figure}
\centering
\includegraphics[angle=-90, width=0.95\textwidth]{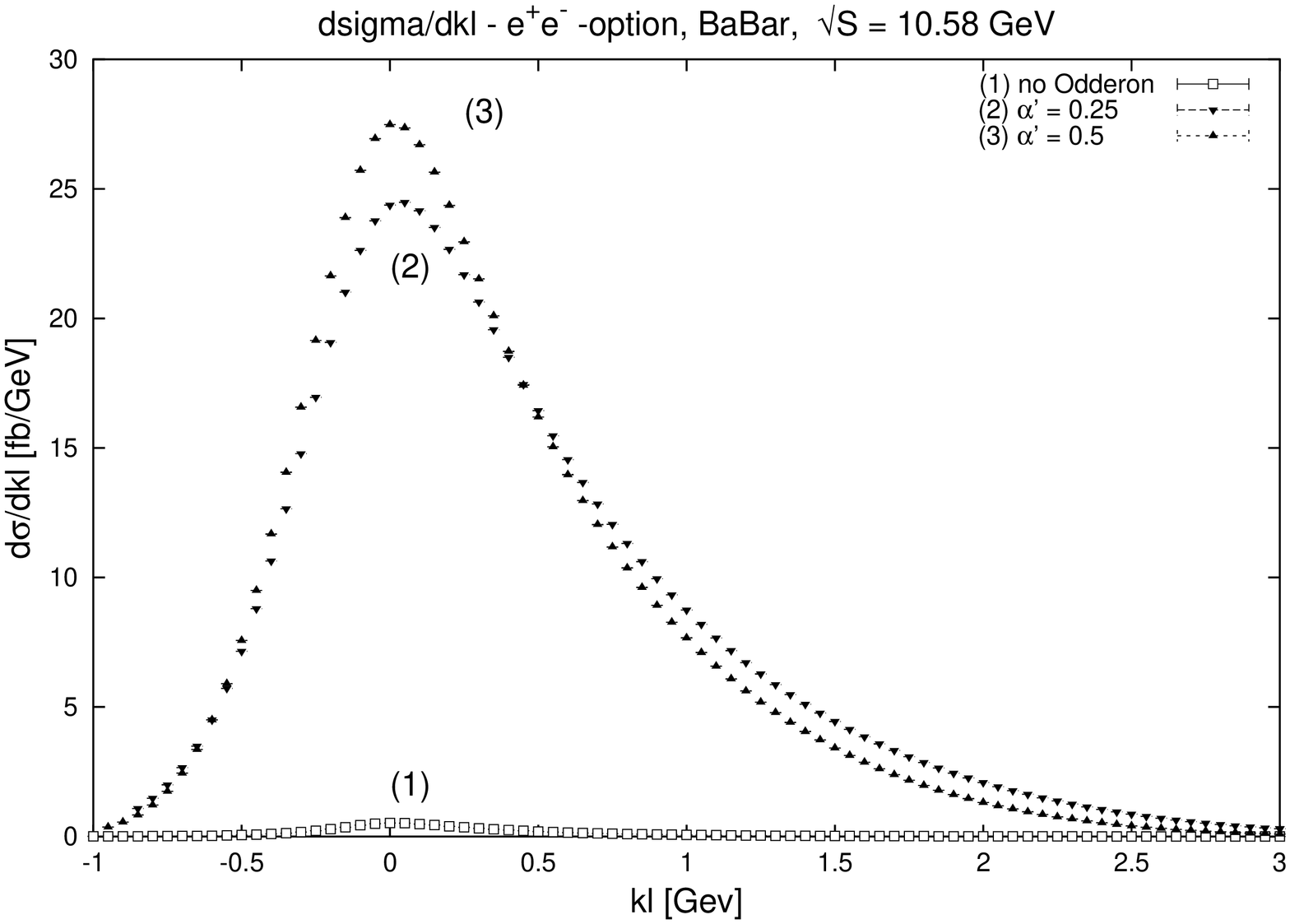}
\caption{comparison different values for $\al'$}
\label{fig:51}
\end{figure}

\begin{figure}
\centering
\includegraphics[angle=-90, width=0.95\textwidth]{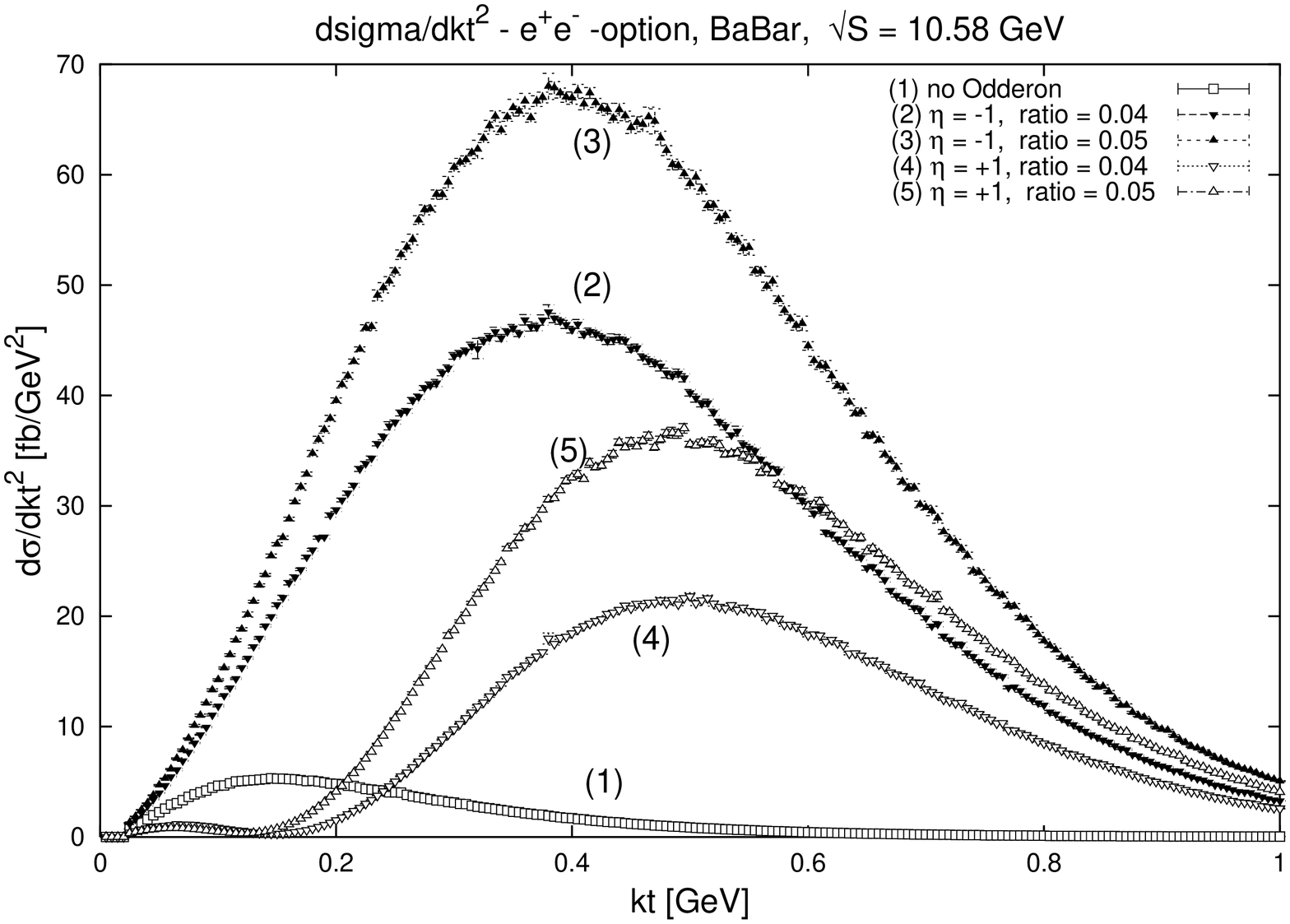}
\caption{comparison different values for $(\frac{\be_{\odd}}{\be_{\pom}})^{2}$ } 
\label{fig:52}                                      
\end{figure}

\begin{figure}
\centering
\includegraphics[angle=-90, width=0.95\textwidth]{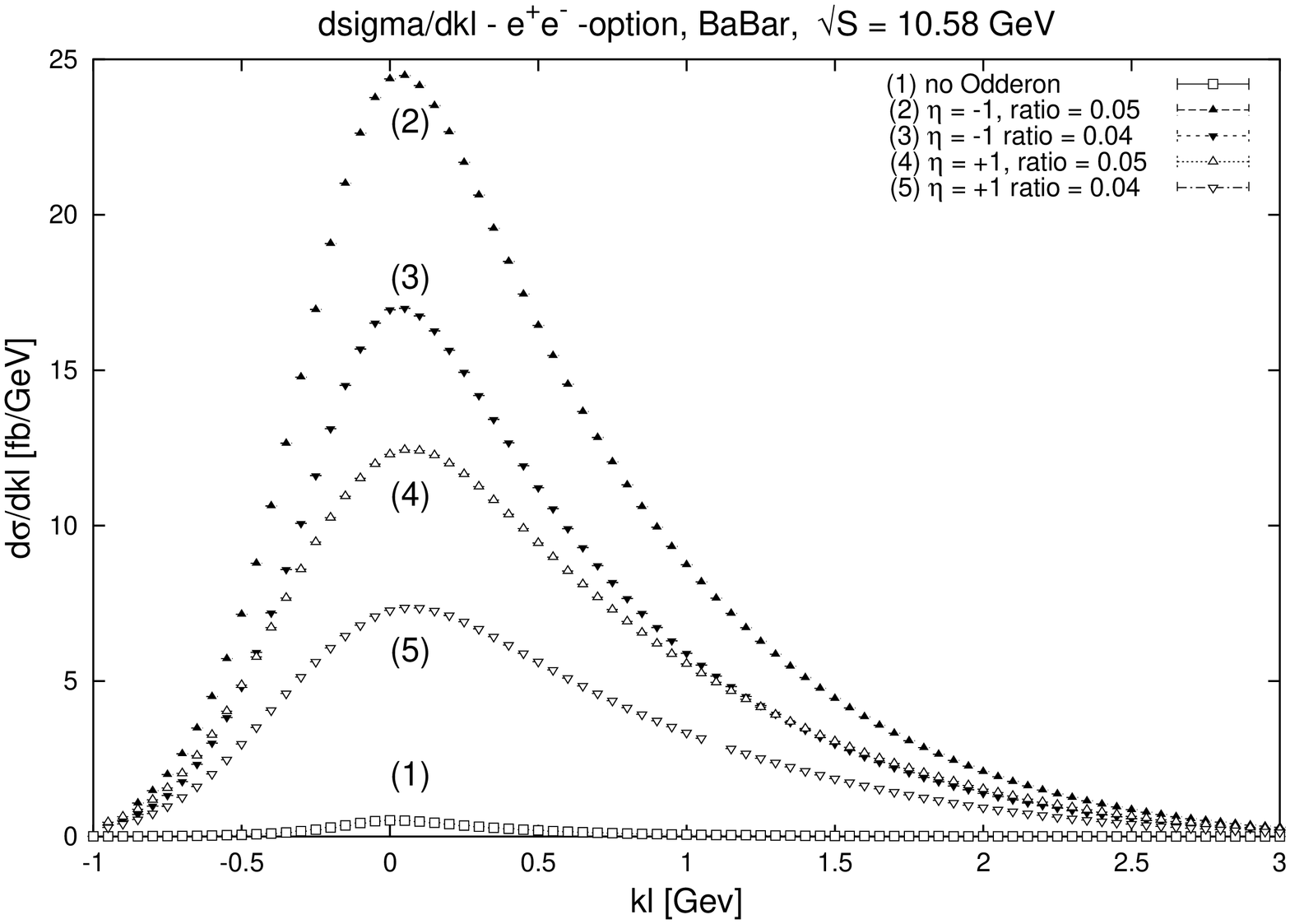}
\caption{comparison different values for $(\frac{\be_{\odd}}{\be_{\pom}})^{2}$}
\label{fig:53}
\end{figure}

The results are displayed in figures \ref{fig:46} to \ref{fig:53}. As expected, due to low energy cuts for the outgoing particles we obtain cross sections which behave similar to the cross sections without any energy detector cuts; for comparison, see figures \ref{fig:1} and \ref{fig:2}. Parameter variation leads to similar effects as for LEP and TESLA; see section \ref{sec:lepii} for a discussion.

In calculations for the cross section in the BaBar environment, we also had to take the ``boosted'' lab-system into account as described in sections \ref{sec:boostlab} and \ref{sec:detcut}\,.

\section{Total inclusive cross sections}

The result for the total inclusive cross section obtained from integration over $d|k_{t}|^{2}$ are given in table \ref{tab:totcross}; in the caculation, we respected the indistinguishability of the two pions according to (\ref{eq:siginc}).
\begin{table}[h!]
\begin{displaymath}
\centering
\begin{array}{|l|c|c|c|c|} \hline
\mbox{Experiment} &\rule[-2mm]{0mm}{7mm} \sqrt{S}\;\;\mbox{[GeV]} & \mbox{only photon} & \odd\,; \eta_{\odd}\,=\,-1 &\odd\,; \eta_{\odd}\,=\,+1 \\ \hline \hline   
OPAL {\scriptsize (LEP I)} & 92 & 0.26 & 25.9 & 17.62 \\ \hline
OPAL {\scriptsize(LEP II)} & 200 & 0.32 & 31.9 & 21.72 \\ \hline
TESLA (e^{+}e^{-}) & 500 & 1.44 & 97.40 & 62.34 \\ \hline
TESLA (\gamma\gamma) & 500 & 1.60 & 81.86 & 53.00 \\ \hline
BaBar & 10.58 & 0.41 & 16.06 & 9.43 \\ \hline
\end{array}
\end{displaymath}
\caption{Total cross sections; results in $fb$}
\label{tab:totcross}
\end{table}
For integration, we used that

\begin{equation}
\sigma_{t,\,inc}\;=\;\frac{1}{n_{i}!}\int\frac{d\sigma}{d|k_{t}|^{2}}(k_{t})\,d|k_{t}|^{2}
 \; = \;  \int \frac{d\sigma}{d|k_{t}|^{2}}(k_{t})\,k_{t}\,dk_{t}\,.
\end{equation}
with $n_{i}\,=\,2$.
Additionally, we only integrated in the limits displayed in the preceding figures, i.e. $k_{t}\,\leq\,1\;\mbox{GeV}$ and $k_{l}\,\leq\,2\;\mbox{GeV}$\footnote{For BaBar, the limits of integration over $k_{l}$ have to be modified.}. 
We now use the results given in table \ref{tab:totcross} to calculate expected event rates for the single experiments by using the relation

\begin{equation}
N\;=\Lc\,\sigma\,.\label{eq:NLsig}
\end{equation}
For luminosities, we used

\begin{eqnarray*}
\mbox{OPAL\footnotemark\,{\scriptsize (LEP I)}}&:& 64.5\; pb^{-1}y^{-1}\;\;\mbox{(1995)}\nonumber\\ 
\mbox{OPAL\addtocounter{footnote}{-1}\footnotemark\,{\scriptsize (LEP II)}}&:& 253\; pb^{-1}y^{-1}\;\;\mbox{(1999)}\nonumber\\ 
\mbox{TESLA}\footnotemark&:& 122.4\; pb^{-1}h^{-1}\nonumber\\
\mbox{BaBar}\footnotemark&:& 70\; fb^{-1}\;\;\mbox{(2/01 - 7/02)}\nonumber\
\end{eqnarray*}
\addtocounter{footnote}{-2}
\footnotetext{\cite{Lamont:2001rs}}\addtocounter{footnote}{+1}
\footnotetext{\cite{Brinkmann:2001qn}}\addtocounter{footnote}{+1}
\footnotetext{\cite{BaBar:2002}}
Notice the different units; for LEP and BaBar, the values correspond to actual integrated luminosities, while the luminosity for TESLA is given by an expected luminosity/second; therefore, the values given here correspond to values for expected beam-time at the maximum luminosity. As we used (\ref{eq:Lrel}) to calibrate the photon spectra from the photon collider, we have to use $\Lc_{e^{+}e^{-}}$ in (\ref{eq:NLsig}) for the photon collider event rates. The results are displayed in table \ref{tab:evrat}.

\begin{table}[h!]
\begin{displaymath}
\centering
\begin{array}{|l|c|c|c|c|} \hline
\mbox{Experiment} &\rule[-2mm]{0mm}{7mm} \sqrt{S}\;\;\mbox{[GeV]} & \mbox{only photon} & \odd\,; \eta_{\odd}\,=\,-1 &\odd\,; \eta_{\odd}\,=\,+1 \\ \hline \hline   
OPAL\,{\scriptsize (LEP I)}\; [y^{-1}] & 92 & 0.017 & 1.671 & 1.136 \\ \hline
OPAL\,{\scriptsize (LEP II)}\; [y^{-1}] & 200 & 0.081 & 8.071 & 5.495 \\ \hline
TESLA (e^{+}e^{-})\;[h^{-1}] & 500 & 0.176  & 11.922 & 7.630 \\ \hline
TESLA (\gamma\gamma)\;[h^{-1}] & 500 & 0.143 & 10.020  & 6.487 \\ \hline
BaBar\; [(1.5\,y)^{-1}]  & 10.58 & 28.7 & 1124.2 & 660.1 \\ \hline
\end{array}
\end{displaymath}
\caption{Event rates}
\label{tab:evrat}
\end{table}
We immediately see that, while investigations at LEP seem difficult due to low luminosities, TESLA as well as BaBar provide excellent environments to determine the existence of a non-perturbative Odderon described by (\ref{eq:Feynodd}) . Again, the differences between the $e^{+}e^{-}$ and the $\gamma\gamma$-collider option at TESLA can be explained by the inaccurate description of the photon spectrum; a more thorough investigation would probably lead to higher counting rates for the photon collider option similar to the $e^{+}e^{-}$ case.
Parameter variation roughly lead to the following results:

\begin{displaymath}
\begin{array}{l c c l}
\mbox{Variation of $\left(\frac{\be_{\odd}}{\be_{\pom}}\right)^{2}$}: & 0.05\rightarrow 0.04 & \sigma_{0.04}\,=\,0.64\,\sigma_{0.05}&\mbox{all experiments}\\
\mbox{Variation of $\al'$}: & 0.25\rightarrow 0.5 & \sigma_{0.5}\,=\,0.74\,\sigma_{0.25}&\mbox{LEPI, LEP II, TESLA} \\
 & & \sigma_{0.5}\,=\,0.95\,\sigma_{0.25} &\mbox{BaBar}\\
\end{array}
\end{displaymath}
The variation of $\left(\frac{\be_{\odd}}{\be_{\pom}}\right)^{2}$ as given above is expected to lead to a ration of roughly 0.64 (see section \ref{sec:lepi}); differences in the effects of the variation of $\al'$ for LEP/ TESLA and BaBar are due to the low energy cuts of BaBar: for $\frac{d\sigma}{d|k_{t}|^{2}}$, the variation mainly modifies the decrease of $\frac{d\sigma}{d|k_{t}|^{2}}$ after it has reached its maximum, i.e. it suppresses higher $k_{t}$ values. The high energy cuts cut off a larger part of the cross section in the LEP and TESLA environment. See figures \ref{fig:26}, \ref{fig:39}, and \ref{fig:50} for comparison.

\section{Parameter distinction}
We will quickly summarize the opportunities for parameter distinction for the effective Odderon described by (\ref{eq:Feynodd}). We obtain

\begin{itemize}
\item{for $\frac{d\sigma}{d|k_{t}|^{2}}$}\\
In general, the transverse differential cross section proves more valuable than the longitudinal one for parameter distinction. We can summarize:
\begin{enumerate}
\item{Odderon phase}\\ The Odderon phase can clearly be distinguished from the position of the maximum of the differential cross section: for $\eta_{\odd}=-1$, it is around $k_{t}=0.3\;\mbox{GeV}$, while for $\eta_{\odd}=+1$, the position is given by $k_{t}=0.5 \;\mbox{GeV}$. Variation of $\frac{\be_{\odd}}{\be_{\pom}}$ slightly vary the position of the maxima; however, this ratio is limited to approximately 0.05  according to (\ref{eq:limrho}) taking $\al_{\odd}(0)\,\approx\,1$; only higher ratios would lead to coincidence of the maxima for $\eta_{\odd}=-1$ and $\eta_{\odd}=+1$. The phase should therefore by easily determined by measuring the transverse differential cross section. However, this only holds if the energy cut of the detector stays below $0.5\;\mbox{GeV}$; for higher cuts, the position of the maxima can no longer be distinguished.
\item{intercept}\\
Intercept variations around $\al_{\odd}\,=\,1$ show little to no effect; the effects of intercept variation considered here will vanish in statistical errors in experiments. 
\item{slope}\\
The variation of $\al'_{\odd}$ from $0.25\;\mbox{GeV}^{-2}$ to $0.5\;\mbox{GeV}^{-2}$ lead to a faster decrease of the differential cross section; however, smaller variations which seem more probable taking the assumed similarity to the Pomeron into account will not lead to measurable modifications.
\item{relative coupling strength}\\
The relative Odderon coupling strength $\frac{\be_{\odd}}{\be_{\pom}}$ leads to obvious modifications for the transverse differential cross sections in magnitude as well as the position of the maximum. However, angular cuts equally significantly modify the magnitude of $\frac{d\sigma}{d|k_{t}|^{2}}$; if varying from the values used in this work, a new calculation with a ``standard'' value for $\frac{\be_{\odd}}{\be_{\pom}}$ is necessary.
\end{enumerate}
\item{from $\frac{d\sigma}{dk_{l}}$:}\\
In general, variation of various parameters lead to similar modifications of the longitudinal differential cross section; compare e.g figures \ref{fig:23}, \ref{fig:27}, and \ref{fig:27-2}. Therefore, the parameters describing the Odderon cannot be well distinguished by regarding $\frac{d\sigma}{dk_{l}}$ only.
\end{itemize}

\chapter[{\normalsize Summary and outlook}]{\LARGE Summary and outlook}

The aim of this work was to investigate the effects of Odderon exchange on the measurement of differential cross sections for the process $\gamma\gamma\longrightarrow\,\pi^{0}\pi^{0}$ at linear $e^{-}e^{+}$ colliders. The Odderon was described by an effective propagator closely following the ansatz for the non-perturbative Pomeron suggested by Donnachie and Landshoff \cite{Donnachie:1984hf} derived in connection with $pp$ and $p\bar{p}$ scattering. The parameters of the Odderon trajectory are taken in close resemblance to the Pomeron parameters, leading to values of $\al(0)\,=\,1$ and $\al'\,=\,0.25\, \mbox{GeV}^{-2}$. The $\gamma\gamma^{*}\pi^{0}$ coupling was taken in a standard form \cite{Brodsky:1981rp},\cite{Anisovich:1997hh},\cite{Radyushkin:1996pm}; the relative strength of the Pomeron-Odderon coupling was taken as a maximum value obeying the bound for $|\rho^{pp}(s)-\rho^{p\bar{p}}(s)|$ given by (\ref{eq:limrho}). We investigated the effects of phase, trajectory intercept, trajectory slope, and relative coupling strength variations as well as influences of detector cuts on the determination of the different parameters. Summarizing, we can draw the following conclusions:
\begin{itemize}
\item{}
The existence of a non-perturbative Odderon in the suggested form leads to obvious modifications of the differential cross sections for the process $\gamma\gamma\rightarrow\pi^{0}\pi^{0}$ at linear $e^{+}e^{-}$ colliders. Even with a low relative coupling strength $\frac{\be_{\odd}}{\be_{\pom}}$, $d\sigma/d|k_{t}|^{2}$ still differs from a scenario without Odderon contribution. The differences between $\eta_{\odd}=-1$ and $\eta_{\odd}=+1$ are equally obvious.
\item{}
In general, $d\sigma/d|k_{t}|^{2}$ proves more valuable than $d\sigma/dk_{l}$ for the determination of the different parameters describing the Odderon trajectory as well as the coupling.
\item{}
In a scenario without any detector cuts, the phase $\eta_{\odd}$ of the Odderon can be clearly determined from the shape of the transverse differential cross section. However, $\frac{d\sigma}{d|k_{t}|^{2}}$  becomes similar for both values of $\eta_{\odd}$ if $k_{t}\,\geq\,0.5\;\mbox{GeV}$; therefore, energy cuts of the detector play an important role. For TESLA, OPAL, and BaBar, $E_{min}\,\leq\,0.5\;\mbox{GeV}$; in these environments, the shapes for the different values of $\eta_{\odd}$ can still be distinguished.
\item{}
The variation of $\al_{\odd}(0)$ only slightly modifies the differential cross sections. In analogy to the Pomeron intercept, $\al_{\odd}(0)\,\approx\,1$; inspired by results from perturbative QCD \cite{Janik:1999ae}, we also looked at $\vare'\,=\,-0.04$. Effects for $|\vare'|\,\leq\,0.02$ as well as $\vare'=-0.04$ have been shown to be negligible.
\item{}
The variation of $\al'_{\odd}$ from $0.25\,\mbox{GeV}^{-2}$ to $0.5\,\mbox{GeV}^{-2}$ leads to a slight modification both $\frac{d\sigma}{d|k_{t}|^{2}}$ and $\frac{d\sigma}{dk_{l}}$; however, taking the assumed similarity of Pomeron and Odderon into account, small variations around $\al'_{\odd}\,=\,0.25\,\mbox{GeV}^{-2}$ seem more probable. The resulting effects should be hardly visible.
\item{} 
The strength of the relative Odderon-coupling can clearly be determined from differential as well as total cross sections; both are roughly proportional to $(\frac{\be_{\odd}}{\be_{\pom}})^{4}$. 
\item{}
All parameters can equally be determined in the LEPI, LEPII, TESLA $e^{+}e^{-}$, TESLA $\gamma\gamma$, and BaBar environment. Due to low energy cuts, BaBar offers the best opportunity for a detailed parameter investigation. For LEPI and LEPII, investigations seem difficult due to the large restriction of the kinematic regime by angular cuts as well as low luminosities. higher counting rates are expected at TESLA and BaBar.
\item{}
Compared with the $e^{+}e^{-}$ mode, the $\gamma\gamma$ mode at the TESLA collider does not really improve the possibilities for an analysis. The main contribution to both cross sections comes from the low-energy part of the photon spectrum roughly corresponding to the spectrum given by the DEPA; for the photon collider, we only expect a slightly higher cross section for processes induced by photons only actually leading to a worse background. However, the analysis should be redone with an improved description of the photon spectrum \cite{Ohl:1997fi}.
\end{itemize}

Similar analyses are possible for general pseudoscalar meson production as $\eta,\,\eta_{0},\,\eta_{c}$ \cite{Kilian:1998ew} as well as tensor mesons \cite{Berger:2000wt}, e.g. the $f_{2}(1270)$.

We conclude that analyses at linear colliders should provide clear information about the existence and phase of a non-perturbative Odderon described by (\ref{eq:Feynodd}) as well as the relative strength of the Odderon-coupling.   

\appendix

\chapter[\normalsize {Conversion of natural to SI units}]{\LARGE Conversion of natural to SI units}

All calculations have been done using

\begin{equation}
\hbar\,=\,c\,=1 \label{eq:natcon}
\end{equation}
in the system of so-called natural high-energy units. The SI-values of these constants are given by

\begin{equation}
\hbar  =  6.58\,\times\,10^{-22} \mbox{MeV s},\quad c  =  3\,\times\,10^{8}\, \frac{\mbox{m}}{\mbox{s}} \label{eq:sicon}
\end{equation}
Therefore, we have to convert our results for {\large $\frac{d\sigma}{dk_{t}}\,$} and {\large $\frac{d\sigma}{dk_{l}},\,$} which are obtained in terms of  GeV$^{-3}$, to  $\frac{\mbox{  cm}^{2}}{\mbox{ GeV}}$ or $\frac{\mbox{  pb}}{\mbox{ GeV}}$; this can be done by using the relation

\begin{displaymath}
1\,\mbox{m}\,\simeq\, \frac{1}{6.58\times3}\,\times 10^{17}\,\mbox{MeV}^{-1}\, =\,\frac{1}{1.97}\,\times\,10^{16}\,\mbox{GeV}^{-1}
\end{displaymath}
and therefore 

\begin{equation}
1\,\mbox{GeV}^{-3}\,=\,1\,\frac{\mbox{GeV}^{-2}}{\mbox{GeV}}\,\simeq 3.88\,\times\,10^{-32}\,\frac{\mbox{m}^{2}}{\mbox{GeV}}\,=\,  3.88\,\times\,10^{8}\,\frac{\mbox{pb}}{\mbox{GeV}} \label{eq:convun} 
\end{equation}
The same factor holds for the calculation of {\large$\frac{d\sigma}{d\left|k_{t}\right|^{2}}$}.

\newpage

\chapter[{\normalsize Standard notations in field theory}]{\LARGE Standard notations in field theory}\label{app:standnot}

We will provide a short list of standard notations in quantum field theory and refer to the literature (see e.g.\cite{Nachtmann:1990ta},  \cite{Peskin:1995ev},\cite{Itzykson:1980rh}) for a more detailed discussion.

\begin{itemize}
\item{quantized fields}\\
We distinguish three types of quantized fields:
\begin{itemize}
\item{scalar fields}\\
Scalar fields obey the Klein-Gordon equation; in literature, they are generally denoted by $\phi(x)$. The corresponding raising and lowering operators obey commutation rules\footnotemark : 
\begin{displaymath}
[a_{{\bf p}},a^{\dagger}_{{\bf p'}}]\;=\;(2\pi)^{3}\delta^{(3)}({\bf p}-{\bf p'})\,.
\end{displaymath}
${\bf p}$ and ${\bf p'}$ denote physical momenta.
\item{Dirac fields}\\
Dirac fields are usually denoted by $\psi(x)$ and $\bar{\psi}(x)$ with $\bar{\psi}(x)\,=\,\psi^{\dagger}\gamma^{0}$. The corresponding raising and lowering operators obey the anticommutation rules
\begin{displaymath}
\{a_{{\bf p}},a^{\dagger}_{{\bf p'}}\}\;=\;(2\pi)^{3}\delta^{(3)}({\bf p}-{\bf p'})\,.
\end{displaymath}
\item{gauge fields}\\
Local gauge symmetries in field theory require the introduction of gauge fields. We can distinguish two kinds of transformations: abelian gauge transformations as e.g. the $U(1)$ gauge transformation in electrodynamics, and non-abelian gauge-transformations as the $SU(2)$ transformation in Yang-Mills theories or the $SU(3)$ color transformation in QCD. We will restrict the discussion to the QED and QCD case. Here, the gauge fields are usually denoted by $A_{\mu}$ (QED) for the photon and $A^{a}_{\mu}$ (QCD) for the gluon, where $a$ in the additional index correlated to the color group.  Due to gauge invariance, these fields have to be quantized using the path integral formalism. Local gauge invariance implies the use of covariant derivatives 
\begin{displaymath}
D_{\mu}\;=\;\partial_{\mu}-i\,g\,A^{(a)}_{\mu}(x)t^{(a)}
\end{displaymath}
with $t^{(a)}$ being the representations of the generators of the corresponding Lie group obeying
\begin{displaymath}
[t^{a},t^{b}]\;=\;iC^{abc}t^{c}\,.
\end{displaymath}
The field-strength tensor $F^{a}_{\mu\nu}$ defined by

\begin{displaymath}
ig\,F_{\mu\nu}^{a}\;=\;\partial_{\mu}A^{a}_{\nu}-\partial_{\nu}A^{a}_{\mu}+g\,C^{abc}A^{b}_{\mu}A^{c}_{\nu}
\end{displaymath}
describes the gauge-field part of the Lagrangian in the form of $\Lc\,=\,-\frac{1}{4}\,F^{a,\mu\nu}F^{a}_{\mu\nu}$.
\end{itemize}
\item{Feynman rules for in- and outgoing particles and propagators}\\
In the following, $p$ will denote a particles 4-momentum vector, $m$ its mass, and $s$ its spin.
\begin{itemize}
\item{scalar fields}\\
For scalar fields, we use
\begin{eqnarray*}
\mbox{in-/ outgoing particle}&:&1\\
\mbox{propagator}&:&\frac{i}{p^{2}-m^{2}+i\epsilon}
\end{eqnarray*}
\item{Dirac fields}
\begin{eqnarray*}
\mbox{in-/ outgoing particle}&:&u_{s}(p)\;/\;\bar{u}_{s}(p)\\
\mbox{in-/ outgoing antiparticle}&:&\bar{v}_{s}(p)\;/\;v_{s}(p)\\
\mbox{propagator}&:&\frac{i(p\mkern-8.mu /+m)}{p^{2}-m^{2}+i\epsilon}\\
\end{eqnarray*}
\item{gauge fields}\\
\begin{eqnarray*}
\mbox{in-/ outgoing particle}&:&\epsilon_{\mu}(p)\;/\;\epsilon^{*}_{\mu}(p) \\
\mbox{propagator\footnotemark }&:&\frac{-i\,\delta^{ab}}{k^{2}+i\epsilon}\left(g^{\mu\nu}-(1-\xi)\frac{k^{\mu}k^{\nu}}{k^{2}}\right)\\
\end{eqnarray*}
\end{itemize}
\item{standard matrices}
\begin{itemize}
\item{Pauli matrices}
\begin{displaymath}
\mbox{algebra}\;:\; [\sigma^{i},\sigma^{j}]\,=\,2\,i\,\vare^{ijk}\,\sigma^{k}
\end{displaymath}
\begin{eqnarray*}
\sigma^{1}\;=\left(\begin{array}{cc}0&1\\1&0\end{array}\right) &, & \sigma^{2}\;=\left(\begin{array}{cc}0 & -i \\ i & 0\end{array}\right)\;, \\
\sigma^{3}\;=\left(\begin{array}{cc}1&0\\0&-1\end{array}\right)\;. & &
\end{eqnarray*}
For isospin considerations, the matrices are denoted $\tau^{i}$;  $\tau^{i}\equiv\sigma^{i}$.

\item{Dirac matrices}
\begin{eqnarray*}
\mbox{algebra} & : & \{\gamma^{\mu},\gamma^{\nu}\}\,=\,2\,g^{\mu\nu}\;,\\
\gamma^{5}&\equiv&i\gamma^{0}\gamma^{1}\gamma^{2}\gamma^{3}\;,\\
\sigma^{\mu\nu}&=&\frac{i}{2}\,[\gamma^{\mu},\gamma^{\nu}]\;.\\
\end{eqnarray*}
A standard representation is given by
\begin{eqnarray*} 
\gamma^{0}\;=\left(\begin{array}{cc} \mathbbm{1}&0\\0&-\mathbbm{1}\end{array}\right) &, & \gamma^{i}\;=\left(\begin{array}{cc}0 & \sigma^{i} \\ -\sigma^{i} & 0\end{array}\right)\;. 
\end{eqnarray*}
\item{Gell-Mann matrices}
\begin{eqnarray*}
\mbox{algebra} & : & [\lambda_{a},\lambda_{b}]\,=\,2 i\,f_{abc}\lambda_{c}\\
 & &\{\lambda_{a},\lambda_{b}\}\,=\frac{4}{3}\delta_{ab}+\,2 d_{abc}\lambda_{c}
\end{eqnarray*}
\begin{eqnarray*}
\lambda_{i}\;=\;\left(\begin{array}{cc} \sigma_{i} & \begin{array}{c} 0 \\0 \end{array} \\ \begin{array}{cc} 0 & 0 \end{array} & 0 \end{array}\right) &' & \lambda_{4}\;=\;\left(\begin{array}{ccc} 0 & 0 & 1 \\ 0 & 0 & 0 \\ 1 & 0 & 0 \end{array}\right)\;,\\
\lambda_{5}\;=\;\left(\begin{array}{ccc}0 & 0 & -i \\ 0 & 0 & 0 \\ i & 0 & 0 \end{array} \right) &, & \lambda_{6}\;=\;\left(\begin{array}{ccc} 0 & 0 & 0 \\ 0 & 0 & 1 \\ 0 & 1 & 0 \end{array}\right)\;,\\
\lambda_{7}\;=\;\left(\begin{array}{ccc} 0 & 0 & 0 \\ 0 & 0 & -i \\ 0 & i & 0 \end{array} \right) &, &  \lambda_{8}\;=\;\frac{1}{\sqrt{3}}\left(\begin{array}{ccc} 1 & 0 & 0 \\ 0 &0  & 0 \\ 0 & 0 & -2 \end{array}\right)\;.\\
\end{eqnarray*}
$f_{abd}$ and $d_{abc}$ are totally antisymmetric / symmetric structure constants of the SU(3) group; $\lambda_{i}$ goes over $i\,=1,2,3$.

\end{itemize}
\end{itemize}

\addtocounter{footnote}{-1}\footnotetext{Constants are normalization-dependent}
\addtocounter{footnote}{1}
\footnotetext{gauge dependent}

\chapter [{\normalsize Analytical spectrum for the photon collider }]{\LARGE Analytical spectrum for the photon collider}\label{chap:theophot} 

The theoretical spectrum for the photon collider respecting linear and non-linear Compton scattering is given by \cite{Galynskii:2000fk}:

\begin{equation}
N(x_{i})  =  \frac{k}{\sigma_{c}}\,\frac{d\sigma_{c}}{dx_{i}}\,  = \, \frac{k}{\sigma_{c}}\,\sum^{2}_{n=1}\left(F_{1\,n}(x_{i})+\lambda\,\lambda_{e}\,F_{2\,n}(x_{i})\right)\label{eq:ggspec}
\end{equation}
with

\begin{eqnarray*}
\sigma_{c} & = & \sum^{2}_{n=1}\:\int\limits_{0}^{x_{n_{max}}}\,dx\,\left(F_{1\,n}(x)+\lambda\,\lambda_{e}\,F_{2\,n}(x)\right)\\
x_{n_{max}} & = & n\,x_{c}/(n\,x_{c}+1+\xi^{2})\\
F_{1\,n}(x),\,F_{2\,n}(x) & : & \mbox{expansion terms of Bessel functions} \\
 & & \mbox{in the $n$-th mode}
\end{eqnarray*}
$k$ is the conversion-coefficient of the laser used for the Compton-scattering; $\lambda\,$ and $\lambda_{e}\,$ are the polarizations of the laser and initial electrons respectively. $x_{c}\,$ is given by (\ref{eq:sigcomp}), $\xi^{2}\,$ is the parameter associated with nonlinear QED effects in Compton-scattering (see section \ref{sec:gammacoll}). 
However, the analytical description of the spectrum only gives an adequate description of the high-energy region of the photon spectrum (valid for $x\,\geq\,0.4)$; therefore, it has not been used in the calculations of the differential cross sections in this work.
\newpage
\bibliographystyle{unsrt}
\bibliography{lit}

\begin{thebibliography}{10}

\bibitem{Regge:1959mz}
T.~Regge.
\newblock Introduction to complex orbital momenta.
\newblock {\em Nuovo Cim.}, 14:951, 1959.

\bibitem{Regge:1960zc}
T.~Regge.
\newblock Bound states, shadow states and mandelstam representation.
\newblock {\em Nuovo Cim.}, 18:947--956, 1960.

\bibitem{Hagiwara:2002pw}
K.~Hagiwara et~al.
\newblock Review of particle physics.
\newblock {\em Phys. Rev.}, D66:010001, 2002.

\bibitem{Donnachie:1984hf}
A.~Donnachie and P.~V. Landshoff.
\newblock p p and anti-p p elastic scattering.
\newblock {\em Nucl. Phys.}, B231:189, 1984.

\bibitem{Donnachie:2001xx}
A.~Donnachie and P.~V. Landshoff.
\newblock New data and the hard pomeron.
\newblock {\em Phys. Lett.}, B518:63--71, 2001.

\bibitem{Lukaszuk:1973nt}
L.~Lukaszuk and B.~Nicolescu.
\newblock A possible interpretation of p p rising total cross- sections.
\newblock {\em Nuovo Cim. Lett.}, 8:405--413, 1973.

\bibitem{Dosch:2002ai}
Hans~Gunter Dosch, Carlo Ewerz, and Volker Schatz.
\newblock The odderon in high energy elastic p p scattering.
\newblock {\em Eur. Phys. J.}, C24:561--571, 2002.

\bibitem{Rueter:1998gj}
Michael Rueter, H.~G. Dosch, and O.~Nachtmann.
\newblock Odd c-p contributions to diffractive processes.
\newblock {\em Phys. Rev.}, D59:014018, 1999.

\bibitem{Berger:2000wt}
E.~R. Berger, A.~Donnachie, H.~G. Dosch, and O.~Nachtmann.
\newblock Observing the odderon: Tensor meson photoproduction.
\newblock {\em Eur. Phys. J.}, C14:673--682, 2000.

\bibitem{Olsson:2001nm}
J.~Olsson.
\newblock Search for odderon induced contributions to exclusive meson
  photoproduction at hera.
\newblock 2001.

\bibitem{Kilian:1998ew}
W.~Kilian and O.~Nachtmann.
\newblock Single pseudoscalar meson production in diffractive e p scattering.
\newblock {\em Eur. Phys. J.}, C5:317--326, 1998.

\bibitem{Motyka:1998kb}
L.~Motyka and J.~Kwiecinski.
\newblock Possible probe of the {QCD} odderon singularity through the
  quasidiffractive eta/c production in gamma gamma collisions.
\newblock {\em Phys. Rev.}, D58:117501, 1998.

\bibitem{Nachtmann:1990ta}
O.~Nachtmann.
\newblock {\em ELEMENTARY PARTICLE PHYSICS: CONCEPTS AND PHENOMENA}.
\newblock Springer Verlag, 1990.

\bibitem{Peskin:1995ev}
Michael~E. Peskin and D.~V. Schroeder.
\newblock {\em An Introduction to quantum field theory}.
\newblock Addison-Wesley, 1995.

\bibitem{Cutkosky:1960sp}
R.~E. Cutkosky.
\newblock Singularities and discontinuities of feynman amplitudes.
\newblock {\em J. Math. Phys.}, 1:429--433, 1960.

\bibitem{Bak:1982}
J.~Bak and D.~J. Newman.
\newblock {\em Complex analysis}.
\newblock Springer, 1982.

\bibitem{Martin:1970}
A.~D. Martin and T.~D. Spearman.
\newblock {\em Elementary particle theory}.
\newblock North-Holland Publishing Company, 1970.

\bibitem{Barnett:1996hr}
R.~Michael Barnett et~al.
\newblock Review of particle physics. particle data group.
\newblock {\em Phys. Rev.}, D54:1--720, 1996.

\bibitem{Foldy:1963}
L.~F. Foldy and R.~F. Peierls.
\newblock Isotopic spin of exchanged systems.
\newblock {\em Phys. Rev.}, 130:1585--1589, 1963.

\bibitem{Pomeranchuk:1958}
I.~Ya. Pomeranchuk.
\newblock Equaility of the nucleon and antinucleon total interaction cross
  section at high energies.
\newblock {\em Sov. Phys.JETP}, 7:499--501, 1958.

\bibitem{Forshaw:1997dc}
J.~R. Forshaw and D.~A. Ross.
\newblock {\em Quantum chromodynamics and the pomeron}.
\newblock Cambridge Univ. Pr., 1997.

\bibitem{Joynson:1975az}
David Joynson, Elliot Leader, Basarab Nicolescu, and Cayetano Lopez.
\newblock Nonregge and hyperregge effects in pion - nucleon charge exchange
  scattering at high-energies.
\newblock {\em Nuovo Cim.}, A30:345, 1975.

\bibitem{Bouquet:1976xz}
A.~Bouquet, B.~Diu, E.~Leader, and B.~Nicolescu.
\newblock Problems in the phenomenological analysis of cross-section
  differences: sigma p p - sigma p n and sigma anti-p p - sigma anti-p n.
\newblock {\em Nuovo Cim.}, A31:411, 1976.

\bibitem{Itzykson:1980rh}
C.~Itzykson and J.~B. Zuber.
\newblock {\em QUANTUM FIELD THEORY}.
\newblock Mcgraw-hill, 1980.

\bibitem{Gilman:1970vi}
Frederick~J. Gilman, Jon Pumplin, A.~Schwimmer, and Leo Stodolsky.
\newblock Helicity conservation in diffraction scattering.
\newblock {\em Phys. Lett.}, B31:387, 1970.

\bibitem{Landshoff:1971pw}
P.~V. Landshoff and J.~C. Polkinghorne.
\newblock The dual quark-parton model and high energy hadronic processes.
\newblock {\em Nucl. Phys.}, B32:541--556, 1971.

\bibitem{Jones:1970}
H.~F. Gilman.
\newblock s-channel helicity conservation,fixed poles and dip mechanism.
\newblock {\em Lett. Nuov. Cim..}, 4:545, 1970.

\bibitem{Bosted:1992rq}
P.~Bosted et~al.
\newblock Measurements of the electric and magnetic form-factors of the proton
  from q**2 = 1.75-gev/c**2 to 8.83-gev/c**2.
\newblock {\em Phys. Rev. Lett.}, 68:3841--3844, 1992.

\bibitem{Arnold:1988us}
R.~G. Arnold et~al.
\newblock Measurements of transverse quasielastic electron scattering from the
  deuteron at high momentum transfers.
\newblock {\em Phys. Rev. Lett.}, 61:806, 1988.

\bibitem{Collins:1977}
P.D.B Collins.
\newblock {\em An introduction to Regge theory and high energy physics}.
\newblock Cambridge Univ. Pr., 1977.

\bibitem{Gell-Mann:1962}
M.~Gell-Mann and M.~L. Goldberger.
\newblock Elementary particles of conventional field theory as regge poles.
\newblock {\em Phys. rev. Lett.}, 9:275--277, 1962.

\bibitem{McCoy:1976ff}
B.~M. McCoy and Tai~Tsun Wu.
\newblock Theory of fermion exchange in massive quantum electrodynamics at
  high-energy. i.
\newblock {\em Phys. Rev.}, D13:369--378, 1976.

\bibitem{Mason:1976sq}
A.~L. Mason.
\newblock Factorization and hence reggeization in massive q.e.d.
\newblock {\em Nucl. Phys.}, B104:141, 1976.

\bibitem{Sen:1983xv}
Ashoke Sen.
\newblock Asymptotic behavior of the fermion and gluon exchange amplitudes in
  massive quantum electrodynamics in the regge limit.
\newblock {\em Phys. Rev.}, D27:2997, 1983.

\bibitem{Tyburski:1976mr}
Lawrence Tyburski.
\newblock Reggeization of the fermion - fermion scattering amplitude in
  nonabelian gauge theories.
\newblock {\em Phys. Rev.}, D13:1107, 1976.

\bibitem{Frankfurt:1975}
L.~L. Frankfurt and V.~E. Sherman.
\newblock Reggeization of the vector particles and the vacuum singularity in
  the renormalizable yang-mills type theories.
\newblock {\em Sov. J. Nucl. Phys.}, 23:581--587, 1975.

\bibitem{Lipatov:1976zz}
L.~N. Lipatov.
\newblock Reggeization of the vector meson and the vacuum singularity in
  nonabelian gauge theories.
\newblock {\em Sov. J. Nucl. Phys.}, 23:338--345, 1976.

\bibitem{Kuraev:1976ge}
E.~A. Kuraev, L.~N. Lipatov, and Victor~S. Fadin.
\newblock Multi - reggeon processes in the yang-mills theory.
\newblock {\em Sov. Phys. JETP}, 44:443--450, 1976.

\bibitem{Balitsky:1978ic}
I.~I. Balitsky and L.~N. Lipatov.
\newblock The pomeranchuk singularity in quantum chromodynamics.
\newblock {\em Sov. J. Nucl. Phys.}, 28:822--829, 1978.

\bibitem{Bartels:1980pe}
Jochen Bartels.
\newblock High-energy behavior in a nonabelian gauge theory. 2. first
  corrections to t(n--->m) beyond the leading lns approximation.
\newblock {\em Nucl. Phys.}, B175:365, 1980.

\bibitem{Kwiecinski:1980wb}
J.~Kwiecinski and M.~Praszalowicz.
\newblock Three gluon integral equation and odd c singlet regge singularities
  in qcd.
\newblock {\em Phys. Lett.}, B94:413, 1980.

\bibitem{Braun:1998mg}
M.~A. Braun, P.~Gauron, and B.~Nicolescu.
\newblock Direct calculations of the odderon intercept in the perturbative
  {QCD}.
\newblock {\em Nucl. Phys.}, B542:329--345, 1999.

\bibitem{Janik:1999ae}
R.~A. Janik and J.~Wosiek.
\newblock The perturbative odderon intercept.
\newblock 1999.

\bibitem{Bartels:1999yt}
Jochen Bartels, L.~N. Lipatov, and G.~P. Vacca.
\newblock A new odderon solution in perturbative qcd.
\newblock {\em Phys. Lett.}, B477:178--186, 2000.

\bibitem{deAlfaro:1973}
V.~DeAlfaro et~al.
\newblock {\em Currents in hadron physics}.
\newblock North-Holland Publishing Company, 1973.

\bibitem{Halzen:1984mc}
F.~Halzen and Alan~D. Martin.
\newblock {\em QUARKS AND LEPTONS: AN INTRODUCTORY COURSE IN MODERN PARTICLE
  PHYSICS}.
\newblock Wiley, 1984.

\bibitem{Gell-Mann:1960np}
Murray Gell-Mann and M~Levy.
\newblock The axial vector current in beta decay.
\newblock {\em Nuovo Cim.}, 16:705, 1960.

\bibitem{Nambu:1960xd}
Yoichiro Nambu.
\newblock Axial vector current conservation in weak interactions.
\newblock {\em Phys. Rev. Lett.}, 4:380--382, 1960.

\bibitem{Chou:1961}
K.-C. Chou.
\newblock On the pseudovector current and lepton decays of baryons and mesons.
\newblock {\em Sov. Phys. JETP}, 12:492--497, 1961.

\bibitem{Bertlmann:1996xk}
R.~A. Bertlmann.
\newblock {\em Anomalies in quantum field theory}.
\newblock Clarendon, 1996.

\bibitem{Adler:1969gk}
Stephen~L. Adler.
\newblock Axial vector vertex in spinor electrodynamics.
\newblock {\em Phys. Rev.}, 177:2426--2438, 1969.

\bibitem{Lepage:1979zb}
G.~Peter Lepage and Stanley~J. Brodsky.
\newblock Exclusive processes in quantum chromodynamics: Evolution equations
  for hadronic wave functions and the form-factors of mesons.
\newblock {\em Phys. Lett.}, B87:359--365, 1979.

\bibitem{Lepage:1980fj}
G.~Peter Lepage and Stanley~J. Brodsky.
\newblock Exclusive processes in perturbative quantum chromodynamics.
\newblock {\em Phys. Rev.}, D22:2157, 1980.

\bibitem{Brodsky:1981rp}
Stanley~J. Brodsky and G.~Peter Lepage.
\newblock Large angle two photon exclusive channels in quantum chromodynamics.
\newblock {\em Phys. Rev.}, D24:1808, 1981.

\bibitem{Anisovich:1997hh}
V.~V. Anisovich, D.~I. Melikhov, and V.~A. Nikonov.
\newblock Photon meson transition form factors gamma pi0, gamma eta and gamma
  eta' at low and moderately high q**2.
\newblock {\em Phys. Rev.}, D55:2918--2930, 1997.

\bibitem{Radyushkin:1996pm}
A.~V. Radyushkin and R.~Ruskov.
\newblock Qcd sum rule calculation of $\gamma\gamma~*\to\pi~0$ transition form
  factor.
\newblock {\em Phys. Lett.}, B374:173--180, 1996.

\bibitem{Budnev:1975}
V.~M. Budnev et~al.
\newblock The two-photon particle production mechanism.
\newblock {\em Phys. Rep.}, 15:182--282, 1975.

\bibitem{Ginzburg:1983vm}
I.~F. Ginzburg, G.~L. Kotkin, V.~G. Serbo, and V.~I. Telnov.
\newblock Colliding gamma e and gamma gamma beams based on the single pass
  accelerators (of vlepp type).
\newblock {\em Nucl. Instr. Meth.}, 205:47, 1983.

\bibitem{Ginzburg:1984yr}
I.~F. Ginzburg, G.~L. Kotkin, S.~L. Panfil, V.~G. Serbo, and V.~I. Telnov.
\newblock Colliding gamma e and gamma gamma beams based on the single pass e+
  e- accelerators. 2. polarization effects. monochromatization improvement.
\newblock {\em Nucl. Instr. Meth.}, A219:5--24, 1984.

\bibitem{Badelek:2001xb}
B.~Badelek et~al.
\newblock Tesla technical design report, part vi, chapter 1: Photon collider at
  tesla.
\newblock 2001.

\bibitem{Aitchison:1972}
I.~J. Aitchison.
\newblock {\em Relativistic quantum mechanics}.
\newblock Macmillan, 1972.

\bibitem{Telnov:2001}
http://www.desy.de/~telnov/ggtesla/spectra/.

\bibitem{Berestetzki:1980}
Lifschitz E.~M. Berestetzki, W.~B. and L.~P. Pitajewski.
\newblock {\em Quantum electrodynamics}.
\newblock Nauka, 1980.

\bibitem{Telnov:1995hc}
V.~Telnov.
\newblock Principles of photon colliders.
\newblock {\em Nucl. Instrum. Meth.}, A355:3--18, 1995.

\bibitem{Noble:1987yz}
Robert~J. Noble.
\newblock Bremsstrahlung from colliding electron - positron beams with
  negligible disruption.
\newblock {\em Nucl. Instr. Meth.}, A256:427, 1987.

\bibitem{Ohl:1997fi}
Thorsten Ohl.
\newblock Circe version 1.0: Beam spectra for simulating linear collider
  physics.
\newblock {\em Comput. Phys. Commun.}, 101:269--288, 1997.

\bibitem{Galynskii:2000fk}
Mikhail~V. Galynskii, Eduard Kuraev, Michael Levchuk, and Valery Telnov.
\newblock Nonlinear effects in compton scattering at photon colliders.
\newblock {\em Nucl. Instrum. Meth.}, A472:267--279, 2001.

\bibitem{Desch:2002}
private communication with K.Desch, Klaus.Desch@desy.de.

\bibitem{Telnov:2002}
private communication with V.Telnov, V.I.Telnov@inp.nsk.su.

\bibitem{Lillich:2002}
private communication with J.Lillich, lillich@sct.physik.uni-freiburg.de.

\bibitem{Aubert:2001tu}
B.~Aubert et~al.
\newblock The babar detector.
\newblock {\em Nucl. Instrum. Meth.}, A479:1--116, 2002.

\bibitem{Schieck:2002}
private communication with J.Schieck, schieck@mppmu.mpg.de.

\bibitem{Tapprogge:1996}
S.~Tapprogge.
\newblock {\em Diffraktive Phaenomene in der Elektron-Proton Streuung bei
  HERA}.
\newblock PhD thesis, Universitaet Heidelberg, Germany, 1996.

\bibitem{Nachtmann:1991ua}
O.~Nachtmann.
\newblock Considerations concerning diffraction scattering in quantum
  chromodynamics.
\newblock {\em Annals Phys.}, 209:436--478, 1991.

\bibitem{Armesto:1997gg}
N.~Armesto and M.~A. Braun.
\newblock On the odderon intercept in qcd.
\newblock 1997.

\bibitem{Lamont:2001rs}
M.~Lamont.
\newblock Twelve years of lep.
\newblock Presented at IEEE Particle Accelerator Conference (PAC2001), Chicago,
  Illinois, 18-22 Jun 2001.

\bibitem{Brinkmann:2001qn}
(ed.~) Brinkmann, R. et~al.
\newblock Tesla: The superconducting electron positron linear collider with an
  integrated x-ray laser laboratory. technical design report. pt. 2: The
  accelerator.
\newblock DESY-01-011.

\bibitem{BaBar:2002}
http://www.slac/stanford.edu/BFROOT/www/Detector/Operations/ Operations.html.

\end{thebibliography}

\end{document}